%% file: orpheus_thesis.tex
\documentclass [11pt, proquest] {uwthesis}[2020/02/24]
 
\usepackage[backend=biber, sorting=none]{biblatex}
\usepackage{alltt}  %
\usepackage{siunitx}
\usepackage{graphicx}
\usepackage{caption}
\usepackage{subfig}
\usepackage{tikz}
\usepackage{placeins}
\usepackage{amsmath, amsthm, amssymb}
\usepackage{booktabs}
\usepackage{setspace}
\usepackage[utf8]{inputenc}
\usepackage{lscape}
\usepackage{textcomp}
\usepackage{MnSymbol,wasysym}
\usepackage{gensymb}
\usepackage{todonotes}
\usepackage{hyperref}
\usepackage{xcolor}
\usepackage{physics}
\hypersetup{
    colorlinks,
    linkcolor={red!50!black},
    citecolor={blue!50!black},
    urlcolor={blue!80!black}
}

\graphicspath{{figures/}{figures/analysis_plots/}{figures/simulations/}{figures/electronics/}{figures/intro_chapter/}}

\newcommand{\snr}{\text{SNR}}
\newcommand{\tem}{TEM$_{00-18}$}
\newcommand{\fm}{f_{00-18}}

\newcommand{\tbar}{\bar{\Theta}}
\newcommand{\upq}{U_{PQ} (1)}
\newcommand{\ap}{A^{\prime}}
\newcommand{\betaterm}{\frac{\beta}{\beta+1}}
\newcommand{\gagg}{g_{a\gamma \gamma}}
\newcommand{\veff}{V_{eff}}
\newcommand{\cost}{\langle \cos^2\theta \rangle_T}
\newcommand{\ldb}{\lambda_{\text{dB}}}
\newcommand{\dfdt}{\frac{df}{dt}}

\begin{document}

%
%

\prelimpages
 
%
%
\Title{A Search for Wavelike Dark Matter with Dielectrically-loaded Multimode Cavities}
\Author{Raphael Cervantes}
\Year{2021}
\Program{Physics}

\Chair{Gray Rybka}{Associate Professor}{Physics}
\Signature{Leslie Rosenberg}
\Signature{Masha Baryakhtar}

\copyrightpage

\titlepage

%
%

%
%

\setcounter{page}{-1}
\abstract{%
  Dark matter makes up 85\% of the matter in the universe and 27\% of its energy density, but we don't know what comprises dark matter. There are several compelling candidates for dark matter that have wavelike properties, including axions and dark photons. Wavelike dark matter can be detected using ultra-sensitive microwave cavities. The ADMX experiment uses a cylindrical cavity operating at the fundamental mode to search for axions in the few \SI{}{\mu eV} mass range. However, the ADMX search technique becomes increasingly challenging with increasing axion mass. This is because higher masses require smaller-diameter cavities, and a smaller cavity volume reduces the signal strength. Thus, there is interest in developing more sophisticated resonators to overcome this problem. 

  The ADMX-Orpheus experiment uses a dielectric-loaded Fabry-Perot cavity to search for axions and dark photons with masses approaching \SI{100}{\mu eV}. Orpheus maintains a large volume by operating at a higher-order mode, and the dielectrics shape the electric field so that the mode couples more strongly to the axion and dark photon.

  This thesis describes the development and commissioning of ADMX-Orpheus to search for dark photons with masses between \SI{65.5}{\mu eV} and \SI{69.3}{\mu eV}. This thesis includes

  \begin{itemize}
    \item Motivation for why dielectric cavities are suitable for detecting axions and dark photons around \SI{100}{\mu eV}.
    \item The design and characterization of such a cavity, including the resonant frequency, quality factor, and detection volume of the mode of interest.
    \item The mechanical design, electronics, and data acquisition system for the Orpheus experiment.
    \item The inaugural search for dark photons \SI{65.5}{\mu eV} and \SI{69.3}{\mu eV} and the resulting excluded parameter space.
    \item Plans for upgrading Orpheus to search for axions in a similar mass range.
  \end{itemize}
}
 
%
%
\tableofcontents
\listoffigures
\acknowledgments{
  I thank my advisor, Gray Rybka, for six years of continuous and accommodating advising. He was a well of knowledge, experience, intuition, creativity, resourcefulness, and equanimity that I drew from as I built a dark matter detector from scratch. He also put me on various exciting projects in both ADMX and Project~8 before I settled on ADMX-Orpheus. He tolerated my ambitious timelines that I could never meet.

  I thank everyone who has worked on Orpheus. I thank Parashar Mohapatra and James Sinnis for letting me advise and supervise them. There was always a lot to do and a lot of problems to solve. Most of the time, I didn't know what I was doing and guessed my way to a functional haloscope. They took all of this in stride. I thank Jihee Yang, Charles Hanretty, and Grant Leum for helping me put together the vacuum and cryogenic systems; Nick Du for helping me with temperature sensors, Yujin Park for designing the first version of the superheterodyne; Ben LaRoque and Noah Oblath for helping me set up and develop the DAQ; Eric Smith and Gary Holman for facility support; Rich Ottens and Seth Kimes for designing the dipole magnet and much of the mechanics; the CENPA and Physics machine shop for fabricating many components for the Orpheus resonator and insert; and Ryan Roehnelt for his excellent mechanical advice as I commissioned the insert.

  I thank the ADMX group for creating a welcoming and fun work environment. I have many fond memories, in and out of the lab, that I will always cherish. In particular, I will always remember the night at the Unicorn in Capitol Hill and 1UP Arcade in Denver. At the risk of favoritism, I thank Nick Du and Chelsea Bartram for years of friendship and emotional support. And I thank them for the wonderful hikes out in the Pacific Northwest. I valued everyone else's friendship, but our time together was more brief. Finally, I thank Leslie Rosenberg for his advice and for understanding realistic timelines. There was a time when I was falling apart because an Indium seal surface wasn't machined to spec. Leslie sat me down and gave me a homegrown orange. I will always remember that orange.

  I thank Masha Baryakhtar for agreeing to be on the reading committee, for her helpful feedback, and for her patience while explaining dark photon cosmology.

  I thank CENPA for being a great place to work. I enjoyed learning about all the cool tabletop experiments probing fundamental physics.

  I thank the author, Anne Lamott, for giving sound life advice. I repeated the mantra ``bird by bird" as problems mounted while commissioning Orpheus. Her essay about first drafts helped me power through this thesis.

  I thank my roommates for being cool (especially during quarantine), for taking care of my dog while I was on travel, and for letting me have a dog. I thank my dog, Seraphine, for her enthusiastic companionship these past five years. She made me a better person and taught me that love is more of a verb than a noun.

  I thank my hometown friends. In particular, I thank Rose Soto, Jesse Mu\~noz, and Karen Quirarte for helping me through a family crisis.

  I thank my family for their continual love and support. They keep asking me how my classes are going or when I'll get married, but that's ok. 

  I thank my dad for a good 28 years of love and support. I miss you, and I wish you could have seen me get my Ph.D.

}

%
%
\dedication{\begin{center}To my father Ruben Cervantes.\end{center}}

%

%
%

\textpages

 
\input{introduction.tex}

\input{haloscopes.tex}

\input{orpheus_rf_design_characterization.tex}

\input{cryogenic_mechanical_design.tex}

\input{experimental_design_operation.tex}

\input{analysis.tex}

\input{future_directions.tex}

\input{conclusion.tex}
%
%
%
\printbibliography
\appendix
\raggedbottom\sloppy
\input{appendix.tex}
\end{document}

%% file: introduction.tex
\chapter{Introduction}
\section{Dark Matter Problem}
Dark matter is the non-luminous, non-absorbing matter that makes up about 84.4\% of the matter of the universe and 26.4\% of its mass\cite{Zyla:2020zbs}. The Lambda cold dark matter ($\Lambda$CDM) model describes dark matter as feebly interacting, non-relativistic, non-interacting, and stable on cosmological timescales. The evidence for dark matter is abundant and includes the galactic rotation curves\cite{Rubin:1982kyu,10.1093/mnras/249.3.523}, gravitational lensing\cite{1998gravitational_lensing,10.1093/mnras/stw3385}, the bullet cluster\cite{2004bulletcluster}, and the cosmic microwave background\cite{2020Planck}. However, despite the overwhelming evidence for the abundance of non-luminous matter in our universe, not much is known about the nature of dark matter. We don't know what makes up dark matter. Many hypothetical particles could make up dark matter. These candidates include WIMPS, sterile neutrinos, chameleons, axions, axion-like particles, and dark photons. Dark matter may be made up of a combination of many of these particles (a dark sector).

This thesis focuses on a set of dark matter candidates that have wavelike properties. For a particle to be wavelike, the de Broglie wavelength $\ldb$ is much greater than the inter-particle spacing. In other words, if the number of particles $N_{dB}$ inside a de Broglie volume $\ldb^3$ is very large, the set of particles is best described as a classical wave\cite{hui2021wave}. The dark matter density in our halo is fixed to $\rho\sim \SI{0.45}{GeV/cm^3}$. So the smaller the dark matter mass, the larger $N_{dB}$. Wavelike dark matter must be bosonic because fermions cannot occupy the same phase space.

A subset of dark matter candidates can be wavelike: axions, axion-like particles (ALPs), and dark photons. Axions are a particularly compelling candidate because they solve an outstanding problem in particle physics known as the Strong CP problem. ALPs and dark photos aren't as motivated as the axion but are still well-motivated for other reasons and can be searched for with the same axion detector and even the same data.

I will first describe axions because Orpheus was developed in the context of an axion search. But much of the discussion applies readily to ALPs and dark photons. After my exposition on axions, I will focus on the other two candidates, emphasizing how they are different from axions.

\section{Axions as Dark Matter}
The axion is a hypothetical particle that, if shown to exist, would solve both the strong CP problem and the dark matter problem. This section explains the motivation for axions, axion physics, and axion cosmology. To summarize, the axion is a result of extending the Standard Model with a field that makes the strong force conserve CP symmetry. The new field results in a new particle called the axion, which would have been produced copiously in the early universe. If the axion is light enough, it would survive on cosmological time scales and interact feebly with normal matter. This makes the axion a compelling dark matter candidate. Astrophysical observations has suggested that the axion mass can range from $\SI{e-22}{eV}$ to $\SI{e-2}{eV}$.

\subsection{Strong CP Problem}
The Standard Model of particle physics describes the interactions of elementary particles through the strong force, weak force, and electromagnetic force. Underlying Standard Model physics is the conservation of symmetries. For example, all standard model processes must obey charge, parity, and time reversal (CPT) symmetry, meaning that the physics of a system remains the same if one were to simultaneously invert the charges, invert the spatial coordinates, and reverse the direction of time. Even though CPT symmetry is conserved in the Standard Model, CP symmetry (charge conjugation and parity symmetry) is violated. This CP violation is caused by the complex quark masses that contain arbitrary phases. It's also expected that the strong force also violates CP symmetry. The CP violation is described by a term in the QCD Lagrangian, ${ L_{\tbar} = -\tbar \left ( \frac{\alpha_S}{8\pi} \right )G^{\mu \nu a} \tilde{G}^a_{\mu \nu} }$, where $\alpha_s$ is the strong coupling constant, $G^{\mu \nu a}$ is the gluonic field strength, $\tilde{G}^a_{\mu \nu}$ is its dual, and $\bar{\Theta}$ is an angular parameter that ranges from $-\pi$ to $\pi$. More precisely, $\bar{\Theta} = \theta - \arg \det \mathcal{M}$, where $\theta$ is a phase from the QCD vacuum and $\mathcal{M}$ is the quark mass matrix. It's expected that $\tbar$ be of $\mathcal{O}(1)$. 

\begin{figure}
	\centering
	\includegraphics[width=0.2\textwidth]{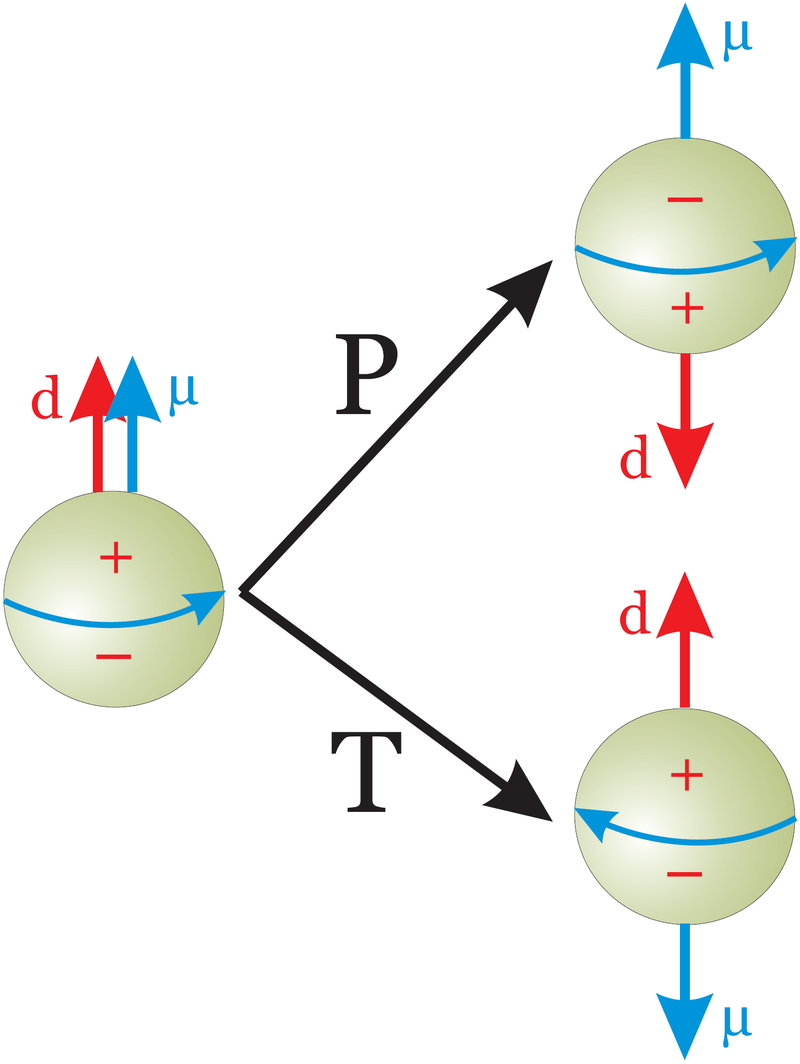}
	\caption{A neutron with an electric dipole moment. This system violates both time-reversal symmetry and parity symmetry, and therefore violates charge-parity reversal symmetry.\cite{knecht_nedm}\label{fig:nedm}}
\end{figure}

Strong CP violation would manifest itself as a neutron electric dipole moment (nEDM). Figure~\ref{fig:nedm} demonstrates how neutron with an EDM would violate CP symmetry. If time is reversed, the magnetic dipole moment would be flipped, but the electric dipole moment would stay the same. If parity is inverted, the electric dipole moment would flip, but the magnetic dipole moment would stay the same. Thus a neutron with an EDM would violate P symmetry and T symmetry. Because CPT symmetry holds, a neutron EDM would violate CP symmetry. However, the neutron dipole moment has not been observed. The upper bound on a possible nEDM is measured to be $|d_n| < 1.8\cdot 10^{-26}e \cdot cm$\cite{Abel_2020}. This limits any possible strong CP violation to $\tbar < 10^{-10}$. The Standard Model doesn't explain why $\tbar$ is unnaturally small. This is the strong CP problem.

One might struggle to understand why the strong CP problem is actually a problem. Why can't $\tbar=0$? Isn't the idea of naturalness just a human construct? The absence of CP violation is troubling because ${ \tbar << 1 }$ would require that the random phases in the quark mass matrix conspire to make $ {\theta - \arg \det \mathcal{M} << 1 }$. One should keep in mind that the random phases in the quark mass matrix (and the neutrino mixing matrix) are responsible for CP violation in the weak force, and that the weak force is maximally CP violating. So ${ \theta - \arg \det \mathcal{M} = 0 }$ would require that the CP violation from the strong force (the $\theta$ contribution) and weak force (the $\arg \det \mathcal{M}$ contribution) cancel each other exactly, even though the strong force and weak force have no relation to each other. The absence of a neutron dipole moment is also strange since this would require that the neutron's valence quarks and ever-fluctuating sea quarks are always arranged to not have an electric dipole moment. One ought to demand a mechanism for this coincidental arrangement, but the Standard Model does not provide one. 

\subsection{Peccei-Quinn Mechanism to Solve Strong CP problem}
A few solutions have been proposed to solve the Strong CP problem. One elegant solution is to make the up quark massless. A massless up quark would mean $\det \mathcal{M} = 0$, implying that $\arg \det \mathcal{M}$ is unphysical. Quantum fields can then be redefined, and $\theta$ can be rotated to remove the CP violating term in the QCD Lagrangian. However, measurements and lattice QCD simulations have determined the mass of the up quark to be $m_u \sim \SI{2}{MeV}$, making the massless up quark solution untenable. Another proposed solution is the Nelson-Barr mechanism, but it requires a fine-tuning more perverse than $\tbar = 0$\cite{DineTalk}.

\begin{figure}
	\centering
	\begin{minipage}[t]{0.4\textwidth}
		\centering
		\includegraphics[height=0.25\textheight]{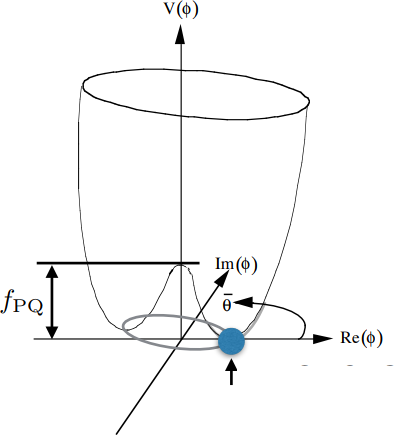}
		\caption{The effective potential for the axion field. It resembles a tilted Mexican hat.\cite{daw2018search}\label{fig:pq_mechanism}}
	\end{minipage}\hspace{2cm}
	\begin{minipage}[t]{0.4\textwidth}
		\centering
		\includegraphics[width=\textwidth]{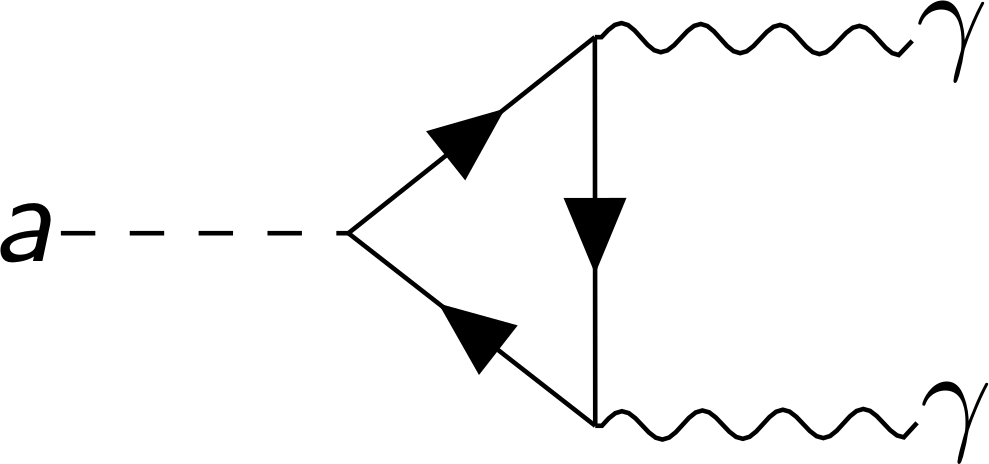}
		\caption{The Feynmann interaction for axions coupling to photons.\label{fig:axion_feynman}}
	\end{minipage}
\end{figure}

The most popular solution to the strong CP problem is the Peccei-Quinn mechanism. In this solution, one extends the standard model by adding a complex scalar field $\psi$ with a $\upq$ symmetry (meaning that the physics of this field is invariant under a rotational transformation $\psi \rightarrow \psi \exp(i \alpha)$). When the system cools below a critical temperature, the field undergoes a phase transition and the $\upq$ symmetry is spontaneously broken. The potential associated with this field becomes a Mexican hat potential and the angular mode is a Goldstone boson called the axion\footnote{It might help to know that particles in QFT are excitations of a quantum field's eingenmodes.}. Once the temperature approaches the QCD scale, the field starts interacting with gluons and the $\upq$ symmetry is explicitly broken. The Mexican hat potential becomes tilted and now has a minimum potential (as shown in Figure~\ref{fig:pq_mechanism}). The axion now has mass. The CP violating term in the Lagrangian is modified to $L_{\Theta} = \left ( \frac{a}{f_A}-\bar{\Theta}\right ) \left ( \frac{\alpha_S}{8\pi} \right )G^{\mu \nu a} \tilde{G}^a_{\mu \nu} $, where $f_a$ is the Peccei-Quinn energy scale, and $a$ is the axion field. The axion particle naturally relaxes to the potential minimum where the CP violation goes away. The effective potential for the axion field is $V(a) = f^2_a m^2_a(T) \left( 1 - \cos \frac{a}{f_a}\right)$.  The mass of the axion is formally defined as the Taylor exansion of the effective potential, $m_a^2 = \eval{\pdv[2]{V(a)}{a}}_{a=f\tbar}$. QCD calculations show that $m_a = 5.691(51)\left(\frac{\SI{e9}{GeV}}{f_a}\right)\si{meV}$.

Even though axions were invented to solve the strong CP problem, they happen to also be an excellent candidate for dark matter. A proper dark matter candidate would have existed before the epoch of recombination, would be stable on cosmological timescales, and would be feebly interacting with normal matter. As will be explained in Section~\ref{sec:cosmology}, cold (i.e., non-relativistic) axions would have been produced copiously in the early universe. Also, if the axion mass is small enough, then axions from the early universe would still be around today. That is because the axion oscillation period is inversely proportional to the axion mass\footnote{Near the potential minimum, $V\propto m_a^2 a^2$. A spring's potential is $V\propto k x^2$, and the oscillation period is $T\propto\frac{1}{\sqrt{k}}$. $m_a$ is analogous to $k$, so the axion oscillation period is $T\propto\frac{1}{m_a}$.}, so a light axion would still be oscillating today, even in the presence of a damping mechanism. Axions also have a feeble interaction with normal matter, especially when they are light. For axions interacting with photons (Figure~\ref{fig:axion_feynman}), the coupling constant is $g_{a\gamma\gamma} = \left(0.203(3)\frac{E}{N} - 0.39(1)\right) \frac{m_a}{\si{GeV}}$, where $\frac{E}{N}\sim \mathcal{O}(1)$. That means if $m_a \sim \SI{1}{\mu eV}$, then $g_{a \gamma \gamma} \sim 10^{-15}$.

To recap, the axion field is hypothesized to solve the strong CP problem by enabling the CP violating term to naturally relax to zero. It turns out that the axion is an excellent dark matter candidate because it would have been produced copiously in the early universe, would be stable on cosmological timescales, and would have feeble interactions with normal matter.

\subsection{Axion Cosmology}\label{sec:cosmology}
\subsubsection{Sources of Cosmological Axions}
There are three methods of generating cosmological axions: thermal production, the misalignment mechanism, and radiation from topological defects.

Thermal (or relativistic) axions would have been produced through the scattering of the cosmic plasma. One dominant process would be $\pi + \pi \leftrightarrow \pi + a$. The relic axion number density and temperature would be comparable to that of one species of neutrinos and would contribute to hot dark matter in the universe. However, the dark matter primarily responsible for driving galactic structure is known to be cold. Thus axions must be produced athermally through the misalignment mechanism and radiation of topological defects.

The misalignment mechanism describes how axions are produced from the oscillation of the axion field about its potential minimum. The following is a history of how these oscillations started. The early universe was hot ($\sim \SI{e32}{K}$). When the universe cools below the PQ energy scale, the $\upq$ symmetry is spontaneously broken. The effective potential takes the shape of a Mexican hat and the axion field takes on random initial values. As the universe cools down further and approaches the QCD energy scale, the $\upq$ symmetry is explicitly broken. The axion field evolves according to the equation 
\begin{equation}
	\ddot \theta + 3 H(T) \dot \theta + m_a^2(T) \theta = 0
\end{equation}
where $\theta \equiv a / f_a$ is the defined misalignment angle, $H(T)$ is the Hubble constant as a function of temperature\footnote{equivalently as a function of time since universe's temperature evolves with time.}, and $m_a(T)$ is the axion mass as a function of temperature. 

The temperature dependence of the axion mass is described by 
\begin{equation}
	m_a(T) = 
	\begin{cases}
		C m_a(T=0)(\Lambda_QCD/T)^4 & \text{for  $T \gtrsim \Lambda_{QCD}$}\\
		m_a(T=0) & \text{for $T \lesssim \Lambda_{QCD}$}
	\end{cases}
\end{equation}
The mass dependence on temperature makes sense because the effective potential is also temperature dependent. The mass of a field can be thought of as describing the steepness of the paraboloid near the potential minimum. The parabaloid becomes steeper as the universe approaches the QCD scale and the effective potential begins to tilt.  

The axion field evolves like a harmonic oscillator with damping due to the Hubble expansion\footnote{More technically and possibly more correct, the field evolved according to the Klein-Gordon equation on a Friedmann-Robertson-Walker metric.}. Thus the axion oscillates coherently around the potential minimum. This coherent oscillation forms a zero momentum condensate that forms cold dark matter. The mass-energy density of axions today related to the misalignment production is $\Omega_{a,mis} \sim 0.15 \left( \frac{f_a}{\SI{e12}{GeV}}\right)^{7/6}\theta^2_i$, where $\theta_i$ is the initial misalignment angle.

Axions can also be produced through topological defects in the axion field, such as axion strings and domain walls. A cosmic string is a topological defect formed during a symmetry breaking phase transition when the topology of a vacuum manifold is not simply connected. In particular, an axion string is a linear structure where the misalignment angle $\theta$ changes by $2\pi$ when physically traversing the string. A domain wall is a surface where $\theta = \pi$, and each axion string has one domain wall ending on it. This is all terribly abstract. To give a more intuitive explanation of these defects, axion strings are where the axion field has wrapped around itself in a way that has caused a discontinuity in the axion field. Quantum fields do not like discontinuities, so the axion field will want to time-evolve in a way that removes these discontinuities. But removing these discontinuities around the strings would require changing the axion field values throughout the entire universe. That would require an infinite amount of energy and therefore would be untenable. So axion strings are metastable and will disappear once they radiate away axions. 

\subsubsection{PQ Symmetry Breaking and Inflation}
\begin{figure}
	\centering
	\includegraphics[width=0.7\textwidth]{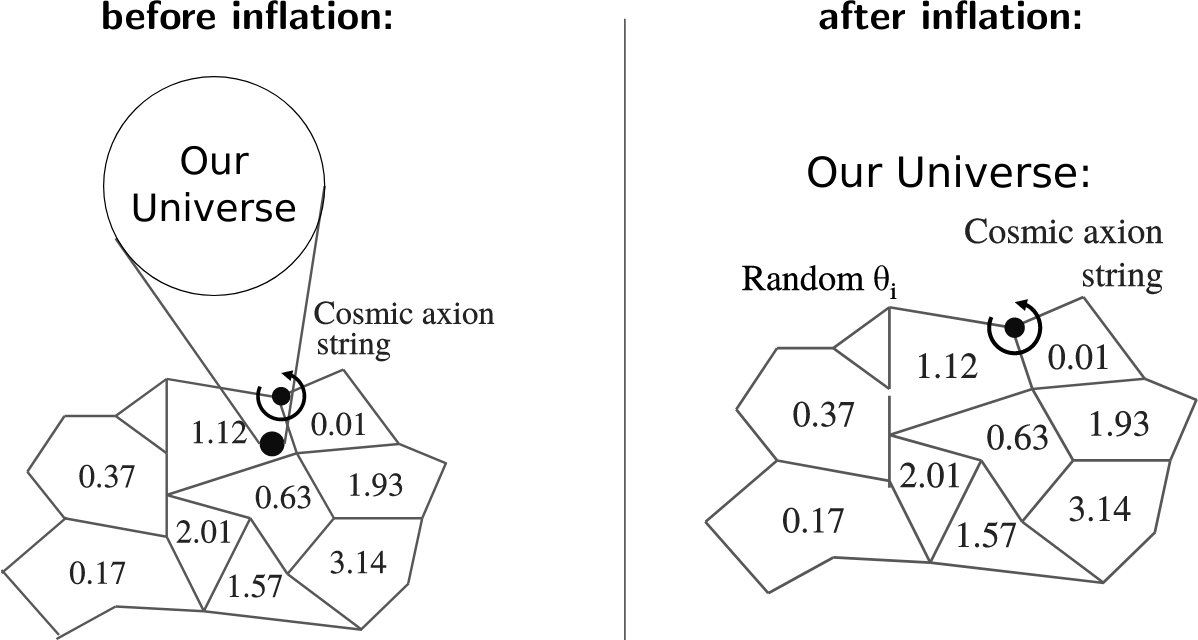}
	\caption{Left: PQ symmetry is broken before inflation and never restored. At inflation, each ``patch'' of initial axion value is blown up. Our observable universe sits in one of those patches. Thus our observable universe has only experienced one value for the axion (with some quantum perterbations). Right: PQ symmetry was broken after inflation. The axion field starts with random values throughout the observable universe. Our universe will contain topological defects like axion strings\cite{KnirckMadmax}.\label{fig:pq_symmetry_breaking_scenarios}}
\end{figure}

There are two scenarios to consider for cosmic axion production: PQ symmetry is broken before inflation and is never restored, and PQ symmetry is broken after inflation (Figure~\ref{fig:pq_symmetry_breaking_scenarios}).

If PQ symmetry was broken before symmetry breaking, then each random patch of theta is inflated. Our observable universe would be inside a single value of theta. The axion strings would inflate away and so the only sources of axions are either through thermal production or through the misalignment mechanism. The energy density would be $\Omega_{a,mis} \sim 0.15 \left( \frac{f_a}{\SI{e12}{GeV}}\right)^{7/6}\theta^2_i$. If one posits that the axions make up all of dark matter, then $\Omega_{dm} \sim 0.15 \left( \frac{f_a}{\SI{e12}{GeV}}\right)^{7/6}\theta^2_i$. $\Omega_{dm}$ is known, so the axion mass would be fixed by the initial misalignment angle. There will however be a variance in the initial angle due to quantum fluctuations, $\left< \sigma_{\theta}^2 \right> = \frac{H_I}{2\pi f_a}$, where $H_I$ is the Hubble constant at the time of inflation. Quantum fluctuations of the axion field during inflation lead to isocurvature density fluctuations which get imprinted as temperature fluctuations in the cosmic microwave background\footnote{The CMB is remarkably uniform but has tiny fluctuations in its power spectrum. Adiabatic fluctuations are density variations in all forms of matter and energy, which have equal over/under number densities. Isocurvature fluctuations are density variations in which the number density variations for one component do not correspond to the number density variations in other components.}. The resulting isocurvature power spectrum is $\Delta_a(k) = \frac{H_I^2}{\pi^2}\theta_i^2 f_a^2$.

If PQ symmetry was broken after inflation, then at the time of symmetry breaking, the observable universe would be filled with different random values of theta. The axion energy density will be averaged over the Hubble volume, such that $\Omega_{a,mis} \sim 0.15 \left( \frac{f_a}{\SI{e12}{GeV}}\right)^{7/6}\left<\theta^2_i\right> \sim 2.07 \left( \frac{f_a}{\SI{e12}{GeV}}\right)^{7/6}$. If one posits that axions make up all of dark matter, then this suggests that $m_a > \si{\mu eV}$.  The initial state of the axion field doesn't affect present-day cosmology. However, axion strings and domain walls will contribute to the axion matter in the universe.

\subsection{Axion Mass Predictions}

Direct detection requires looking for axions with a specific range of mass and coupling strengths. A wide range of axion masses would solve the strong CP problem. Furthermore, axions are difficult to detect because of their feeble interaction with normal matter. The most sensitive axion searches use resonators that enhance the signal power at the cost of only being able to search a very narrow range of parameter space. Thus it behooves the physics community to use whatever clues nature has provided to reduce the search parameter space.

\begin{figure}
	\includegraphics[width=\textwidth]{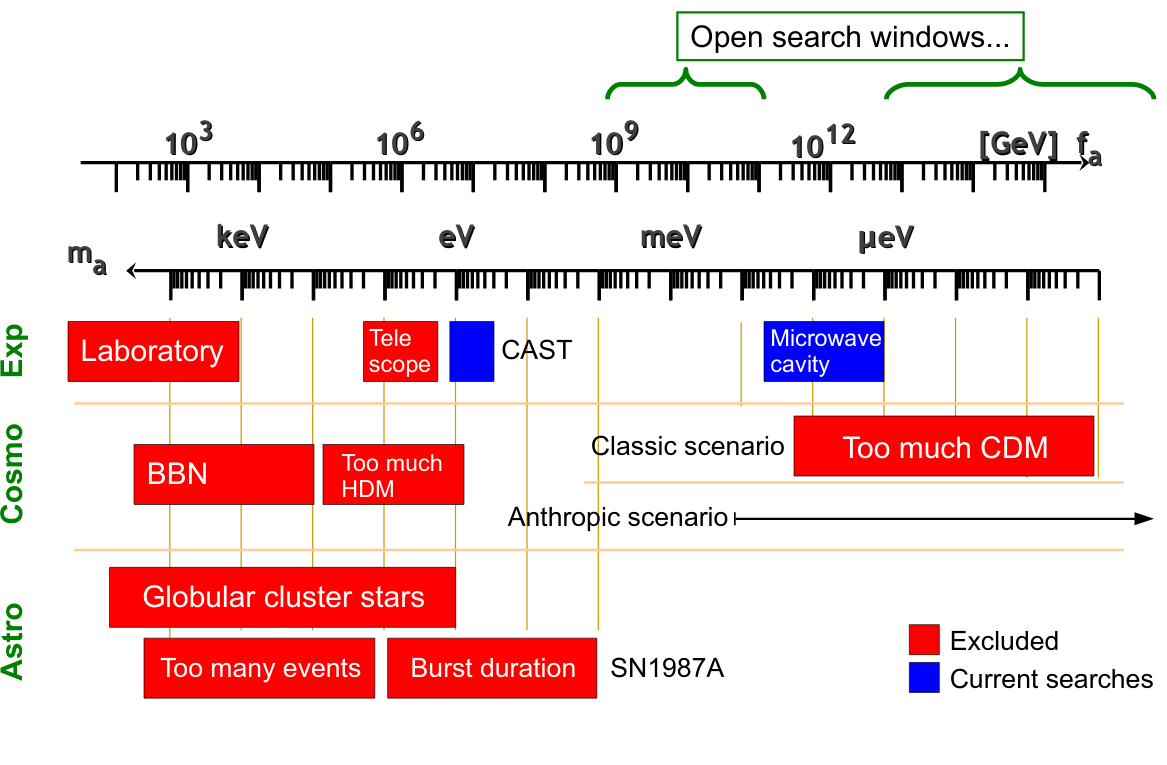}
	\caption{A figure from 2010 that shows how astrophysical sources place limits on axion mass. \cite{Wong}\label{fig:astro_limits}}
\end{figure}

\begin{figure}
	\includegraphics[width=\textwidth]{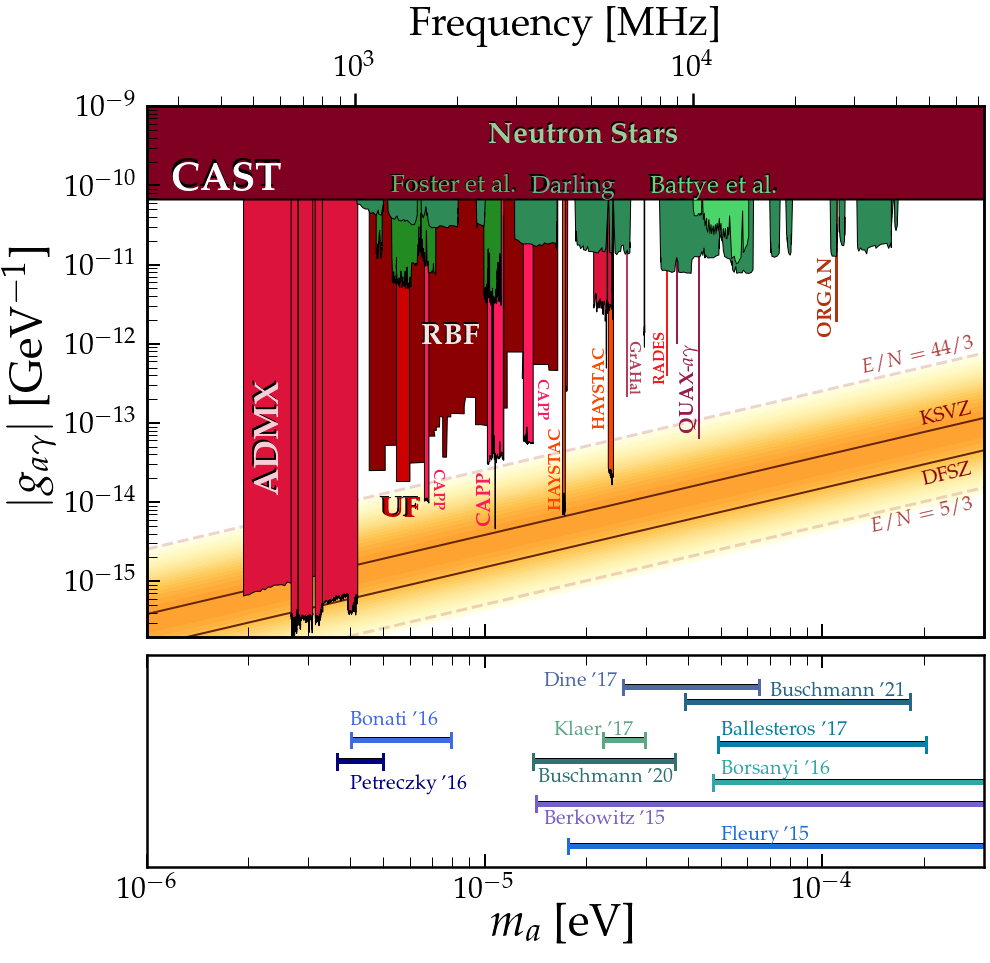}
  \caption{The excluded parameter space for axions, zoomed into microwave cavity haloscopes. The bottom plot shows different theoretical predictions for post-inflationary axions. A lot of the tensions have to do with how much do topological defects contribute to the axion population. Figure courtesy of Ciaran O'Hare\cite{ciaran_o_hare_2020_3932430}.}
  \label{fig:postinflation_axion_theory}
\end{figure}

Astrophysical sources, such as white dwarves, neutron stars, red giants, and supernovas, can be used to limit the axion mass parameter space.  The lower limit on the axion mass is $m_a > \SI{e-22}{eV}$\cite{ferreira2020ultralight}. Otherwise, the de Broglie wavelength would be bigger than galaxies, which would have a noticeable effect on galactic structure. Astrophysical objects would also produce axions thermally, and this production is a source of energy loss. For axions to efficiently transport energy, their mean free path must be comparable to that of the astrophysical object. If the axion coupling strength is too large, then the axions have a hard time escaping these bodies. If the axion coupling strength is too small, then astrophysical bodies cannot produce them efficiently. If astrophysical objects cool at the rate predicted by standard model processes, one can place an upper limit on the axion coupling strength. The nuclear physics is tedious for the purposes of this thesis, so I will just state limits. Stars in the horizontal branch limit the axion-photon coupling strength to $|g_{a\gamma\gamma}| < \num{6.6e-11}$. Red giants on the horizontal branch limit the axion-electron coupling to $|g_{aee}| < \num{4.3e-13}$. The best limit is from the detection of neutrinos from supernova SN1987a. Sufficiently light axions would the copiously produced by this supernova and would steal energy away from neutrinos, reducing the duration of the neutrino burst. The burst duration limits the strength of axions coupling to normal matter to $g_{ann}^2+0.29g_{app}^2 + 0.27g_{ann}g_{app}< \num{3.25e-18}$. This translates to a conservative mass limit of $m_a < \SI{16}{meV}$. Figure~\ref{fig:astro_limits} shows how astrophysics has limited the possible axion mass ranges. 
If PQ symmetry breaking happened before inflation, then $\SI{e-22}{eV}<m_a<\SI{e-2}{eV}$. However, if we posit that PQ symmetry breaking happened after inflation, the lower limit becomes far more stringent. However, there are topological complications such as axion strings and domain walls. Lattice QCD is used to predict the contributions of axion string radiation. Figure~\ref{fig:postinflation_axion_theory} shows the range of published mass predictions. 

\section{ALPs and Dark Photons}
Axion-like particles (ALPs) have the same properties as axions, except their photon coupling $\gagg$ is now independent of the ALP mass. ALPs do not solve the strong CP problem, but they can make up all of the dark matter. ALPs are predicted to arise generically from low-energy effective field theories emerging from string theory\cite{Zyla:2020zbs}.

The dark photon\footnote{Also known as a hidden photon or paraphoton.} (DP) story is similar to the axion story. The dark photon is a vector boson associated with an added Abelian U(1) symmetry to the Standard Model\cite{essig2013dark,Ghosh:2021ard,caputo2021dark}. The dark photon is analogous to the Standard Model (SM) photon in that the SM photon is also a vector boson associated with an Abelian U(1) gauge symmetry. The dark photon interacts with the SM photon through kinetic mixing~\cite{HOLDOM198665,HOLDOM1986196}
\begin{align}
  \mathcal{L} = -\frac{1}{4}F_1^{\mu \nu}F_{1\mu \nu} + -\frac{1}{4}F_2^{\mu \nu}F_{2\mu \nu} + \frac{1}{2}\chi F_1^{\mu \nu}F_{2\mu \nu} + \frac{1}{2} m_{\ap}^{2} A^{\prime 2}
  \label{eqn:dp_mixing_lagrangian}
\end{align}
where $F_1^{\mu \nu}$ is the electromagnetic field tensor, $F_2^{\mu \nu}$ is the dark photon field tensor, $\chi$ is the kinetic mixing angle, $m_{\ap}$ is the DP mass, and $\ap$ is the DP gauge field. The consequence is that the dark photon and SM photon can oscillate into each other (reminiscent of neutrino oscillations). Unlike axions, dark photons and SM photons do not need the presence of a magnetic field to mix. Also, unlike axions, the dark photon kinetic mixing angle is independent of the dark photon mass.

If the dark photon has a small kinetic mixing, i.e., small coupling to the SM particles, then it is stable on cosmological timescales and makes for a compelling dark matter candidate. The rate that dark photons decay into three photons is
\begin{align}
  \Gamma_{\ap \rightarrow 3\gamma} = (\SI{4.70e-8}{})\times \alpha^3 \alpha^{\prime} \frac{m_{\ap}^9}{m_e^9}
\end{align}
where $\alpha$ is the fine structure constant, $m_e$ is the mass of the electron, and $\alpha^{\prime}\equiv\frac{(e\chi)^2}{4\pi}$ is the dark photon counterpart to the fine structure constant\cite{PhysRevD.78.115012}. The dark photon lifetime is about the same as the age of the universe if $m_{A^{\prime}} (\alpha^{\prime})^{1/9} < \SI{1}{keV}$. For dark photons with $m_{\ap} \approx \SI{e-4}{eV}$ and $\chi < \SI{e-12}{}$, this condition is easily met.

Several mechanisms would produce cosmic dark photons. One of these mechanisms is a misalignment mechanism that is similar to that for axions. However, the dark photon, due to its vector nature, will have its energy density redshifted as the universe expands\footnote{In the same way SM photons redshift during cosmic expansion.} and wouldn't contribute to the energy density of matter $\Omega_m$. For the misalignment mechanism to generate the correct relic abundance, a nonminimal coupling to gravity needs to be invoked. This leads to instabilities in the longitudinal DP mode, which would lead to DP with a fixed polarization within a cosmological horizon\cite{caputo2021dark}. Dark photons may also be produced by quantum fluctuations during inflation\cite{PhysRevD.93.103520}. These quantum fluctuations seed excitations in the dark photon field that result in the cold dark matter observed today in the form of coherent oscillation of this field. The predicted mass from this mechanism is $m_{A^{\prime}}\approx\SI{e-5}{eV}\times\left (\frac{\SI{e14}{GeV}}{H_I}\right )^4$, where $H_I$ is the Hubble constant during inflation. Measurements of the CMB tensor to scalar ratio constrain $H_I < \SI{e14}{GeV}$\cite{2016Planck}. This mechanism is the most minimal model for producing dark photon dark matter and makes the search for $m_{A^{\prime}}>\SI{e-5}{eV}$ well-motivated. Dark photons can also be produced from topological defects like cosmic strings. Dark photons may also be produced thermally through processes like $e^- + \gamma \rightarrow e^- + A^{\prime}$\cite{PhysRevD.101.063030}. However, due to the small coupling to SM particles, this process would be inefficient and not produce the observed abundance of dark matter.

%% file: haloscopes.tex
\chapter{Dielectric Haloscopes for Detecting Axions and Dark Photons}

This chapter describes how ADMX uses microwave cavities to search for $\mathcal{O}(\SI{1}{\mu eV})$ axions, how the same technique gets harder to apply at higher axion masses, and how dielectric haloscopes can be used for higher mass searches. 

This section focuses heavily on axions because that's the context for which Orpheus was developed. Most of the time, the following discussion applies to dark photons. The main difference between detecting either is in the polarization of the detected photons. I will talk about the detection of axions and remark about dark photons when appropriate.

This chapter assumes the reader understands resonant cavities.

\section{Haloscopes to search for wavelike dark matter}\label{sec:haloscope_search}
An axion can be detected by detecting its feeble coupling to the SM photon. Axions mix with photons in a magnetic field (similar to how neutrinos of different flavors mix in free space). If this mixing happens inside a resonant cavity, and if that resonant frequency matches the photon frequency, the axion signal power is resonantly enhanced\cite{PhysRevLett.51.1415}.

The Lagrangian that describes an axion coupling to two photons is

\begin{align}
  \mathcal{L}_{a\gamma\gamma} = \frac{1}{4} \gagg a F_{\mu\nu}\tilde{F}_{\mu\nu} = -\gagg a \vb{E}\vdot\vb{B}
\end{align}

From the Lagrangian, the derived axion modified Maxwell's equations are~\cite{2017MADMAXtheory, PhysRevD.99.055010}

\begin{align}
  &\div{\vb{E}} = \rho - \gagg \vb{B}\vdot \grad{a},\\
  &\div{\vb{B}} = 0,\\
  &\curl{\vb{E}} = -\dot{\vb{B}},\\
  &\curl{\vb{B}} = \dot{\vb{E}} + \vb{J} - \gagg(\vb{E}\cross \grad{a} - \dot{a}\vb{B}),\\
  &\ddot{a} - \laplacian{a} + m_a^2 a = \gagg \vb{E} \vdot \vb{B}.
\end{align}

This can be simplified by noting that $\grad{a}\approx 0$. This approximation works because de~Broglie wavelength of the axion is much larger than the size of the detector. For an axion with $m_a = \SI{80}{\mu eV}$ and $\langle v^2 \rangle^{1/2} = \SI{270}{km/s}$, the de~Broglie wavelength is $\ldb \approx \SI{17}{m}$. Orpheus is about \SI{15}{cm}, so the axion field is approximately constant throughout the detector volume.

Also note that $\gagg \dot{a}\vb{B}$ is on equal footing with $\vb{J}$. So the axion field can act like a current source that excites electromagnetic fields. Treatments of currents exciting cavity modes can be found in Jackson Chapter 8\cite{Jackson} and Pozar 4th ed. Chapter 6\cite{pozar}.

\begin{figure}
  \centering
  \includegraphics[width=\textwidth]{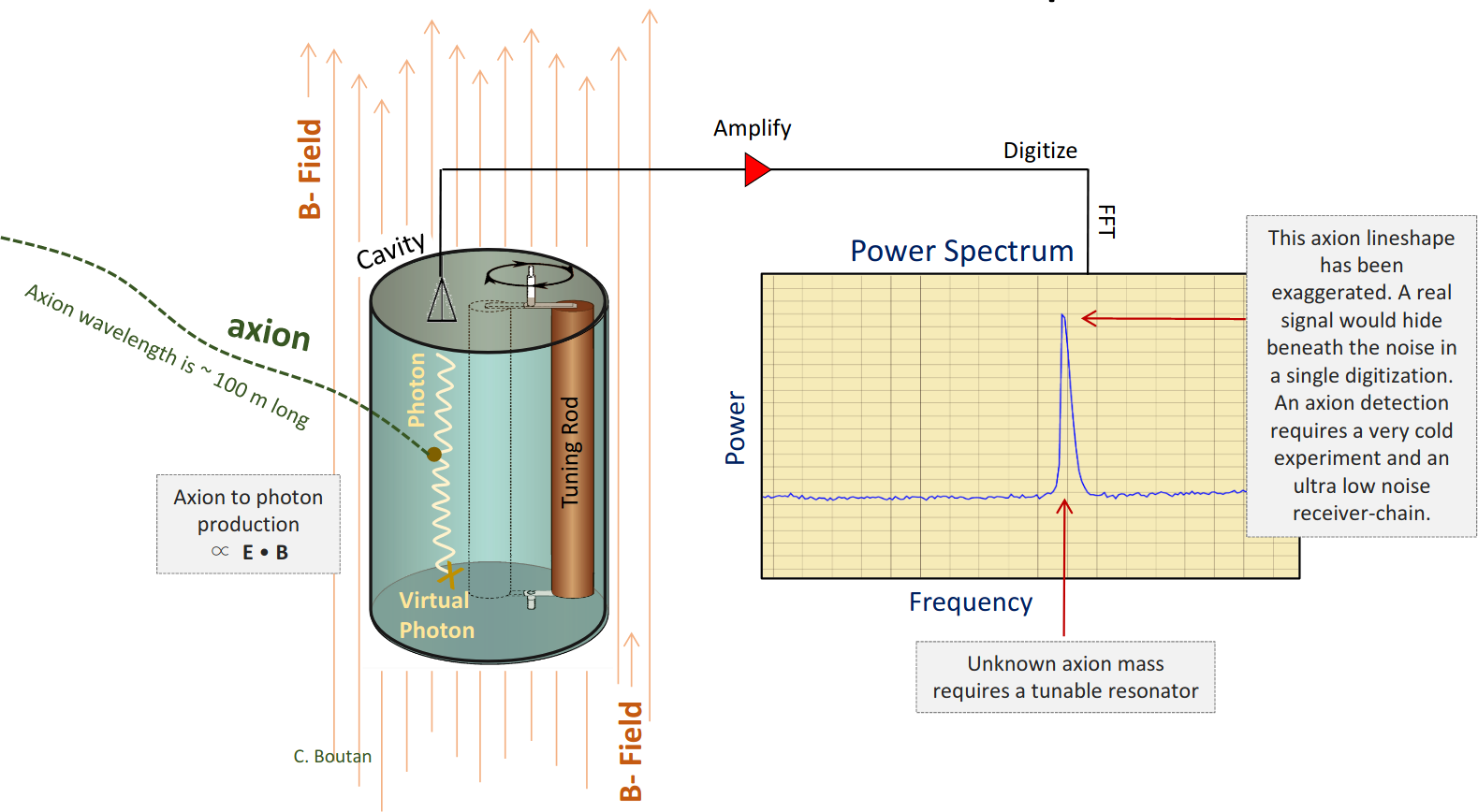}
  \caption{Haloscope search for axions. An axion, when it sees a strong magnetic field, can convert to a photon. If the photon frequency matches the resonant frequency of the cavity, the axion signal power is resonantly enhanced. A haloscope search for dark photons works the same way, except that a dark photon can convert to an SM photon without the presence of a magnetic field. Figure courtesy of Christian Boutan.}
  \label{fig:haloscope}
\end{figure}

This coupling is feeble, but the resulting signal can be made detectable if the interacting magnetic field is very strong and if the signal is resonantly enhanced by a high-Q cavity\footnote{The cavity allows for the axion signal to build up (think of a photon of having about Q bounces before it dissipates).}. This signal can be detected with low noise electronics as a spectrally narrow power excess over the noise background, as shown in Figure~\ref{fig:haloscope}. 

I will outline the steps to deriving the equation for the axion signal power inside a cavity. First, derive the wave equation from Maxwell's equations. Assume harmonic time dependence and write $\vb{E}$ as the sum of cavity eigenmodes, i.e., rewrite the wave equation in k-space such that $\vb{E} = \sum_i \vb{E}_i \exp(-i(\omega t - \vb{k}_i\vdot\vb{x}))$. The k-space wave equation should have a damping term that's related to the cavity Q. Solve for the electric field amplitude of the relevant cavity mode and find the time-averaged energy stored $U = \int dV \frac{1}{2} \epsilon |E_0|^2$. 
For a signal to be detected, the cavity has to be coupled to an external receiver. From the energy stored in the cavity, the unloaded Q of the cavity, and the coupling of the receiver to the cavity, one can derive the power dissipated in the receiver, i.e., the signal power. Let

\begin{align}
  \frac{1}{Q_L} = \frac{1}{Q_0} +\frac{1}{Q_e}
\end{align}
where $Q_L$ is the loaded Q of the cavity, $Q_0$ is the unloaded Q, and $Q_e$ characterizes how much energy gets dissipated in the external load. The cavity coupling is $\beta \equiv \frac{Q_0}{Q_e}$. The power drawn into the external load is $P_{a}=\omega \frac{U}{Q_e}$. Putting all this together, the resulting signal power is (in natural units)

\begin{align}
  & P_a = \frac{\gagg^2}{m_a} m_a \rho_a B_0^2 V_{eff} Q_L \betaterm  L(f, f_0, Q_L) \\
  & L(f, f_0, Q_L) = \frac{1}{1+4\Delta^2} \hspace{1cm} \Delta \equiv Q_L \frac{f-f_0}{f_0}\\
  & V_{eff} = \frac{\left |\int dV \vb{B}_0\vdot \vb{E}_a \right |^2}{B_0^2 \int dV \epsilon_r |E_a|^2}
\end{align}

The derivation was merely verbally outlined, and the math is left as an exercise to the reader. The reader is encouraged to read other sources like~\cite{brubaker2018results} for a more thorough treatment.

Axion searches look for a power excess over a thermal noise floor $P_n$. $P_n$ is the average of thousands of power spectrum measurements and consequently follows the Central Limit Theorem, so $\sigma_{P_n} = \frac{P_n}{\sqrt{N}}$, where N is the number of averaged spectra. The number of spectra is $N = b \Delta t$, where $b$ is the frequency bin width and $\Delta t$ is the total digitization time. The SNR of a haloscope signal is
\begin{align}
  \snr &= \frac{Ps}{\sigma_{P_n}} = \frac{Ps}{P_n}\sqrt{N} = \frac{Ps}{P_n}\sqrt{b\Delta t }
\end{align}

Keeping in mind that the cavity bandwidth is $\Delta f = \frac{f_0}{Q_L}$, the instantaneous scan rate is 

\begin{align}
  \dv{f}{t} = \frac{\Delta f}{\Delta t} = \frac{f_0 Q_L}{b}\left (\frac{\gagg^2 \rho_a B_0^2 \veff \beta}{\snr m_a T_n(\beta+1)}\right )^2 
\end{align}
ADMX has successfully applied this technique to search for DFSZ axions with masses between \SI{2.7}{\mu eV} (\SI{650}{MHz}) and \SI{3.3}{\mu eV} (\SI{800}{MHz}) and continues to push the search to higher masses~\cite{PhysRevLett.120.151301, PhysRevLett.124.101303, admxcollaboration2021search}. For Run 1B, ADMX employed a \SI{136}{L} cylindrical cavity. The cavity was immersed in a \SI{7.6}{T} magnetic field and operated at the TM$_{010}$ mode such that the axion's electric field maximally aligned with the external magnetic field (the axion's electric field looks like that of a parallel plate capacitor). For Run 1B, $Q_L = 30000$ and $\dfdt = \SI{543}{MHz/yr}$.

\section{Haloscopes for Dark Photons}
The dark photon story is similar but with some key differences.

The dark photon interacts with the SM photon through kinetic mixing. The Lagrangian for this interaction is shown in Equation~\ref{eqn:dp_mixing_lagrangian}. 
 The consequence is that the dark photon and SM photon will oscillate into each other. One can derive the electromagnetic field produced by the dark photon to be~\cite{caputo2021dark}

\begin{align}
  |\vb{E_0}| = \left | \frac{\chi m_{X}}{\epsilon} \vb{X}_0 \right |
\end{align}
where $\vb{X}$ is the dark photon field. One should note that the SM photon polarization is determined by the dark photon polarization rather than some external magnetic field. This polarization can be misaligned with the probing mode of the cavity.

The resulting dark photon signal power in a cavity is~\cite{Ghosh:2021ard} (in natural units)

\begin{align}
  & P_{S} = \eta \chi^2 m_{\ap} \rho_{\ap} V_{eff} Q_L \betaterm L(f, f_0, Q_L)  \\
  & V_{eff} = \frac{\left (\int dV \vec{E}(\vec{x}) \vdot \vec{X}(\vec{x})\right )^2}{\int dV \epsilon_r |\vec{E}(\vec{x})|^2|\vec{X}(\vec{x})|^2}
\end{align}

Let $\theta$ be the angle between the cavity field and the dark photon field, such that $\vec{X}\vdot\vec{E} = |\vec{X}||\vec{E}|\cos\theta$. The $\veff$ can be rewritten as

\begin{align}
V_{eff} = \frac{\left (\int dV |\vec{E}(\vec{x})| | \vec{X}(\vec{x})|\right )^2  }{\int dV \epsilon_r |\vec{E}(\vec{x})|^2|\vec{X}(\vec{x})|^2}\langle \cos^2\theta \rangle_T
\end{align}
where $\cost$ is the time-averaged $\cos^2\theta$ value. The actual value of $\cost$ depends on the dark photon cosmology and the detector design and orientation. Often I will write $\veff = V_{eff,max} \cost$ because it allows me to rescale my dark photon limits to any cosmology.

\FloatBarrier
\section{Scaling the Haloscope Concept to Higher Frequencies}
ADMX is currently using this haloscope method to look for axions around a few \SI{}{\mu eV} with great success. Unfortunately, this haloscope design becomes increasingly difficult to implement at higher frequencies. Increasing mass corresponds to higher frequency photons. Operating at the TM$_{010}$ mode would require smaller-diameter cavities, and a smaller cavity volume reduces the signal strength. The volume would scale by $V_{eff} \propto f^{-2}$, or $V_{eff} \propto f^{-3}$ if one wanted to keep the same aspect ratio. Furthermore, the decreased volume-to-surface ratio decreases Q, further decreasing the signal. Q also reduces as a function of frequency because of the anomalous skin effect, so $Q_u \propto f^{-2/3}$. The smaller $Q_u$ also makes it more difficult for the receiver to couple critically. The quantum noise limit also increases linearly with $f$. So putting all these effects together, for a single closed cavity operating at the lowest order mode with a quantum noise limited amplifier, the axion signal power scales optimistically as $P_s \propto f^{-2.66}$ and the scan rate scales optimistically as $\dv{f}{t} \propto f^{-7.66}$. This unfavorable frequency scaling motivates the design of more sophisticated resonators\footnote{Someone at a workshop called axions above \SI{10}{GHz} axions from hell. Orpheus came out of hell but lost someone he loved in the process.}.

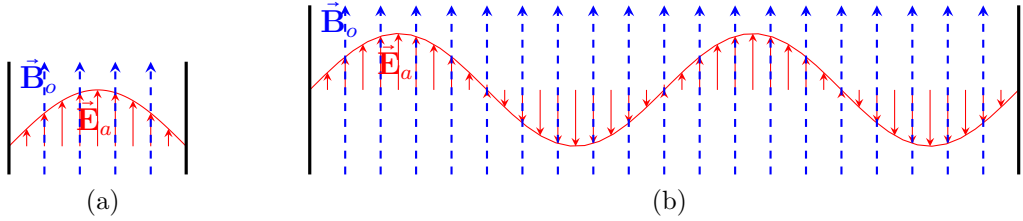
\begin{figure}[h]
  \centering
  \subfloat[]{\input{small_resonator.tikz}}\hfil
  \subfloat[]{\input{multimode_resonator.tikz}}
  \caption{Traditional haloscope signal power scale poorly with frequency. (a) Small cavity operating at the fundamental mode, e.g., TM$_{001}$ mode. The volume of such a cavity scales as $\sim f^{-3}$ for the same aspect ratio. (b) Operating a large cavity at higher frequencies requires operating at higher order modes. The effective volume $\veff \propto \int dV \vb{E}_a \vdot \vb{B}_o \approx 0$. }
\end{figure}

One can think about keeping a large volume and operating at a mode higher than the TM$_{010}$. But then portions of $\va{E}_a$ are anti-aligned with $\va{B}_o$, and $\qty|\int dV {\va{B}_o\vdot \va{E}_a }| \approx 0$. The effective volume approaches zero even though the physical volume is large. Thus, there is little benefit to operating an empty cylindrical cavity at a higher-order mode.

One can also combine many cylindrical cavities. This is the plan for future ADMX runs. However, once the frequency approaches \SI{15}{GHz}, the wavelength is about \SI{1}{cm^3}. To have $\veff = \SI{1}{L}$, one would need to coherently power combine about 2000 cylindrical cavities. It would be difficult to instrument a \SI{1}{cm^3} cavity, and it would be unfeasibly complex to instrument 2000 cavities in a coordinated way. 

\section{Multimode Dielectric Haloscopes and Orpheus Conceptual Design}
\begin{figure}[h]
  \centering
  \input{dielectric_resonator.tikz}
  \caption{A multimode dielectrically-loaded cavity. Dielectrics are placed where the electric field is anti-aligned with the external magnetic field. Dielectrics suppress electric fields, so $V_{eff} \propto \int dV \vec{E} \vdot \vec{B} > 0$.}
  \label{fig:dielectric_haloscope}
\end{figure}
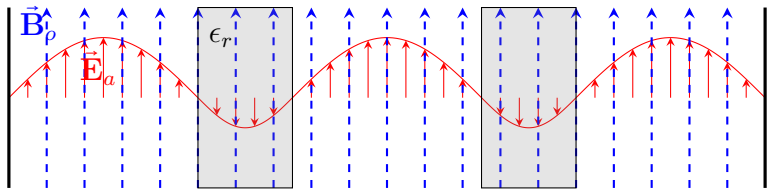

High-order modes can couple to the axion when dielectrics are placed inside of the resonator. Dielectrics suppress electric fields. If dielectrics can be placed where the electric field is anti-aligned with the magnetic field, then the overlap between the axion's electric field and the external magnetic field is greater than zero (see Figure~\ref{fig:dielectric_haloscope}). Thus the effective volume can become arbitrarily large, and the axion signal power is greater than what it would have been for a cylindrical cavity operating at the TM$_{010}$ mode. Overall, dielectric resonators can be operated at higher-order modes while maintaining coupling to the axion and can be made arbitrarily large, making them suited for higher-frequency searches.

Orpheus will implement this dielectric haloscope concept to search for axions around \SI{70}{\mu eV}. Orpheus\footnote{Orpheus was originally designed to have a spatially alternating magnetic field rather than a periodic dielectric structure\cite{PhysRevD.91.011701}. However, this alternating magnetic field design is difficult to scale to many Tesla.} is a dielectrically loaded Fabry-Perot open cavity placed inside of a dipole magnet. Dielectrics are placed every fourth\footnote{I could have also designed Orpheus so that the dielectrics were at every other half-wavelength. Intuitively, this would have resulted in a much greater $\veff$. But that didn't happen for various practical reasons that will be addressed in the next few sections. To summarize, it's because it would have been harder to design a cryo-compatible mechanical structure with the dielectric plates that close together, and I was having a hard time simulating the mode of interest\footnote{The axion-coupling mode or the dark photon coupling mode, depending on the context.} for the entire tuning range. I managed to make Orpheus work when the dielectrics were spaced every fourth of a half-wavelength and kept forging ahead with what worked. Orpheus could have worked if I placed dielectrics at every other half-wavelength, but it would have been harder to implement, and the tuning range would have likely been more limited.} of a half-wavelength to suppress the electric field where it is anti-aligned with the dipole field. Orpheus is designed to search for axions around \SI{16}{GHz} with over \SI{1}{GHz} of tuning range. The cavity tunes by changing its length, and the dielectrics are automatically adjusted to maintain even spacing throughout the cavity.

There are several benefits to the open resonator design. Less metallic walls lead to less ohmic losses and a higher Q. Less metallic walls also mean fewer resonating modes. This leads to a sparse spectrum and fewer mode crossings, making it easier to maintain the mode of interest. 	

However, this experiment has many challenges. First, the optics must be designed to maintain good axion coupling for over \SI{1}{GHz} of tuning range. This includes choosing the right radius of curvature for the Fabry-Perot mirrors and appropriate dielectric thicknesses. Another challenge is that a dipole magnet is required to have $B_{0}$ aligned with the TEM electric fields. Dipoles are about ten times more expensive than solenoids. In addition, the mechanical design for such a cavity is complicated because there are many moving parts that have to work in a cryogenic environment. Finally, both diffraction and dielectric losses will decrease the resonator Q.

\section{Other related experiments}
Dielectric haloscopes are a growing field. Without going into detail, there is MADMAX that uses a very similar periodic dielectric structure~\cite{Brun2019}. Baryakhtar, et.\ al., has developed a similar concept for the infrared and optical range~\cite{PhysRevD.98.035006}. Quiskamp, et.\ al., has also developed a cylindrical cavity with a periodic dielectric structure~\cite{PhysRevApplied.14.044051}. This design has the advantage of working in a solenoid rather than a dipole magnet. RADES is not a dielectric haloscope but is a multimode cavity designed to work in a dipole magnet~\cite{melcon2021results}.

%% file: small_resonator.tikz
\begin{tikzpicture}[>=stealth, scale=0.75]
  \draw [domain=0:pi, samples=50, color=red] plot(\x, {sin(\x r});
  \foreach \x in {1, 2, ..., 9}
  \draw[->, color=red] (\x/10*pi,0) -- (\x/10*pi,{sin(\x/10*pi r)});
  \foreach \x in {1, 2, ..., 4}
  \draw[->, color=blue, dashed, thick] (\x/5*pi,-0.5) -- (\x/5*pi,1.5);
  \draw[very thick] (0, -0.5) -- (0,1.5) (pi, -0.5) -- (pi,1.5);
      	\draw(1,0.5) node[red, right] {$\va{E}_a$};
  \draw (0, 1.25) node[blue, right] {$\va{B}_{o}$};
\end{tikzpicture}

%% file: multimode_resonator.tikz
\begin{tikzpicture}[>=stealth, scale =0.75]
  \draw [domain=0:4*pi, samples=50, color=red] plot(\x, {sin(\x r});
  \foreach \x in {1, 2, ..., 39}
  \draw[->, color=red] (\x/10*pi,0) -- (\x/10*pi,{sin(\x/10*pi r)});
  \draw[very thick] (0, -1.5) -- (0,1.5) (4*pi, -1.5) -- (4*pi,1.5);
  \foreach \x in {1, 2, ..., 19}
  \draw[->, color=blue, dashed, thick] (\x/5*pi,-1.5) -- (\x/5*pi,1.5);
    	  \draw(1,0.5) node[red, right] {$\va{E}_a$};
  \draw (0, 1.25) node[blue, right] {$\va{B}_{o}$};
\end{tikzpicture}

%% file: dielectric_resonator.tikz
\begin{tikzpicture}[>=stealth, scale =0.8]
  \draw [domain=0:pi, samples=50, color=red] plot(\x, {sin(\x r});
  \foreach \x in {1, 2, ..., 9}
  \draw[->, color=red] (\x/10*pi,0) -- (\x/10*pi,{sin(\x/10*pi r)});
  \draw [domain=0:pi/2, samples=50, color=red, shift={(pi,0)}] plot(\x, {-0.5*sin(2*\x r});
  \foreach \x in {1, 2, ..., 4}
  \draw[->, color=red, shift={(pi,0)}] (\x/10*pi,0) -- (\x/10*pi,{-0.5*sin(2*\x/10*pi r)});
  \draw [domain=0:pi, samples=50, color=red, shift={(3*pi/2,0)}] plot(\x, {sin(\x r});
  \foreach \x in {1, 2, ..., 9}
  \draw[->, color=red, , shift={(3*pi/2,0)}] (\x/10*pi,0) -- (\x/10*pi,{sin(\x/10*pi r)});
  \draw [domain=0:pi/2, samples=50, color=red, shift={(5*pi/2,0)}] plot(\x, {-0.5*sin(2*\x r});
  \foreach \x in {1, 2, ..., 4}
  \draw[->, color=red, shift={(5*pi/2,0)}] (\x/10*pi,0) -- (\x/10*pi,{-0.5*sin(2*\x/10*pi r)});
  \draw [domain=0:pi, samples=50, color=red, shift={(3*pi,0)}] plot(\x, {sin(\x r});
  \foreach \x in {1, 2, ..., 9}
  \draw[->, color=red, , shift={(3*pi,0)}] (\x/10*pi,0) -- (\x/10*pi,{sin(\x/10*pi r)});

  \draw[very thick] (0, -1.5) -- (0,1.5) (4*pi, -1.5) -- (4*pi,1.5);
  \foreach \x in {1, 2, ..., 19}
  \draw[->, color=blue, dashed, thick] (\x/5*pi,-1.5) -- (\x/5*pi,1.5);
  \draw[fill=gray, fill opacity=0.2] (pi, -1.5) rectangle (3*pi/2, 1.5);
  \draw[fill=gray, fill opacity=0.2] (5*pi/2, -1.5) rectangle (3*pi, 1.5);
  \draw(1,0.5) node[red, right] {$\va{E}_a$};
  \draw (0, 1.25) node[blue, right] {$\va{B}_{o}$};
  \draw (pi, 1) node[right] {$\epsilon_r$};
\end{tikzpicture}

%% file: orpheus_rf_design_characterization.tex
\chapter{Orpheus RF Design and Characterization}\label{ch:rf_design}

This chapter describes the RF characterization of Fabry-Perot cavity with dielectrics evenly-spaced at every fourth half-wavelength\footnote{Section~\ref{sec:cavity_mechanics} describes how the cavity was built.}. Measurements are taken to extract the resonant frequency, quality factor, and cavity coupling coefficient of the mode of interest. These measurements are compared to Finite Element Analysis simulations. Simulations are also used to determine $\veff$ and tolerance for misplaced dielectrics. The chapter ends with describing different cavity coupling schemes and why the current coupling configuration was chosen.

\section{The Empty Fabry-Perot Cavity}
Intuition for the dielectrically-loaded Fabry-Perot cavity builds on the well-established intuition for the empty Fabry-Perot cavity. The empty cavity modes have analytical solutions. But an analytical solution to the dielectrically-loaded case is infeasible\footnote{Intuition from geometric optics suggest that the solutions from the empty cavity if the optical length is modified from the dielectrics. That's a good way to guess the mode frequencies. But it turns out that the resonant frequencies and electric field deviate quite strongly from what one would suppose from geometric optics. Therefore, simulations are needed to get the electric field.}, and simulations are required to understand the electric field. One way to corroborate the simulations is to study the empty cavity in detail. If the mode frequencies from the measurement, simulation, and analytical calculation match within experimental/simulation uncertainty, then the simulation and measurement techniques are validated and can be trusted for the dielectrically-loaded case.

\subsection{Theoretical Basis}
\begin{figure}[ht]
  \centering
  \subfloat[]{\includegraphics[height=0.3\textheight]{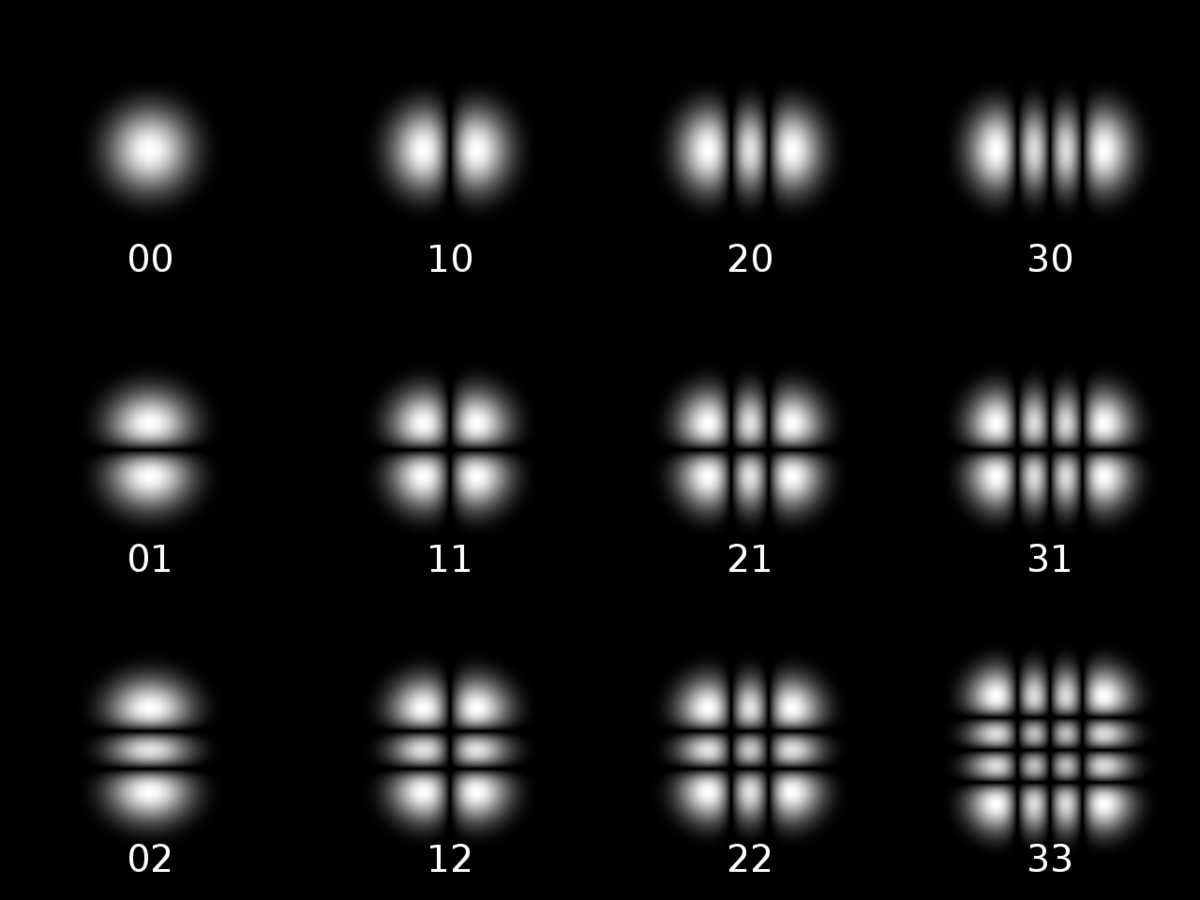}}
  \subfloat[]{\includegraphics[height=0.3\textheight]{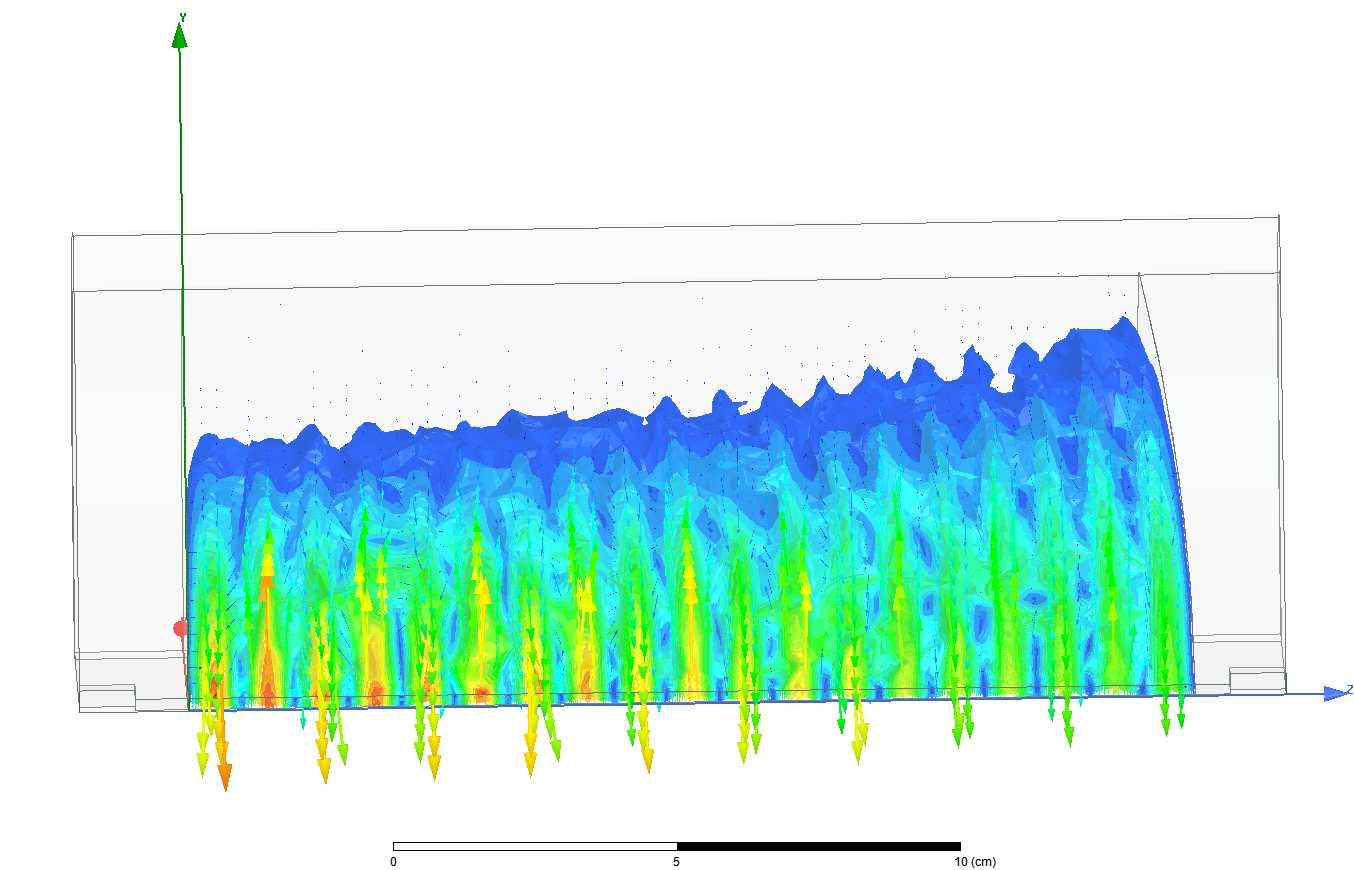}}
  \caption{(a) The Hermite-Gaussian modes of a Fabry-Perot cavity. Figure from Kogelnik\cite{1447049}. (b) FEA simulation of the \tem mode in an empty Fabry-Perot cavity. }
  \label{fig:empty_gaussian}
\end{figure}

The Fabry-Perot cavity modes are described by TEM Gaussian modes, as shown in Figure~\ref{fig:empty_gaussian}. The details and derivations are in references~\cite{1447049, Clarke_1982}. I will outline the solutions and main ideas relevant to Orpheus.

The traveling wave Gaussian beam electromagnetic wave is described by 
\begin{align}
  u_{mn} = \frac{w_0}{w} \left [H_m\left ( \sqrt{2} \frac{x}{w} \right ) H_y\left ( \sqrt{2} \frac{y}{w} \right ) \exp \left ( \frac{-\rho^2}{w^2} \right ) \right ] \exp\left ( -i (kz-\varphi ) - \frac{i k \rho^2}{2r} \right )
\end{align}
where $k$ is the wavenumber, $z$ is the position coordinate along the beam axis, $\rho$ is the transverse distance from the beam axis, $w(z)$ characterizes the width of the beam, $w_0$ is the beam waist (the smallest width of the beam), $r$ is the radius of curvature of the wavefront, $H_m$ and $H_n$ are the Hermite polynomials that characterize modes in the transverse direction. Finally, $\varphi$ is the Gouy phase. It's a phase change additional to the free space propogation that results from the wavefront radius of curvature making the phase velocity $v_p > c$. The Gouy phase is derived to be 

\begin{align}
  \varphi = (1+m+n) \tan^{-1}(z/z_0)
\end{align}

Note that for the fundamental Gaussian beam, $H_0=1$. The beam width and wavefront radius of curvature are hyperbolic functions that depend on $z$.

\begin{align}
  w^2 &= w_0^2 [1+(z^2/z_0^2)]\\
  r &= z[1+(z_0^2/z^2)]\\
  z_0 &= kw_0^2/2
\end{align}

Inside the Fabry-Perot cavity, the resonant condition for the Gaussian beam is 
\begin{gather}
    \frac{f}{f_0} = (q+1) + \frac{1}{\pi} (m+n+1) \arccos\sqrt{(1-L/R_1)(1-L/R_2)}\\
    f_0 = \frac{c}{2D}
  \label{eqn:fp_frequencies}
\end{gather}
where $L$ is the length of the cavity, $R_1$ and $R_2$ are the radii of curvature of the mirrors, and q is the mode number along the axis of the cavity. If $q=0$, then there is one half-wavelength along the axis and no nodes. If $q=18$, then 18 half-wavelengths along the axis and 18 nodes. For Orpheus, one of the mirrors is flat, so $R_2 = \infty$ and ${\frac{f}{f_0} = (q+1) + \frac{1}{\pi} (m+n+1) \arccos\sqrt{1-L/R_1}}$.

The stability condition for the Fabry-Perot cavity is
\begin{align}
  0 \leq (1-L/R_1) (1-L/R_2) \leq 1.
\end{align}
The condition is derived in~\cite{1447049} and can be intuited from ray tracing multiple bounces between the mirrors.

A stable beam with a good Q can be achieved if the cavity operates in the near-confocal configuration. For a symmetric cavity where $R_1 = R_2 = R_0$, the near confocal geometry describes a configuration where the mirrors are placed at each others focus, i.e. $L = R_0/2$. 

For good Q in the near-confocal regime, the Fresnel number should be\cite{Clarke_1982}
\begin{align}
  N = \frac{a_1 a_2}{\lambda L} > 1
\end{align}
where $a_1, a_2$ are the diameters of the mirrors.

The quality factor of the cavity is\cite{Dunseith_2015}
\begin{align}
  Q = 2\pi f_n \frac{2L}{c}\frac{1}{\beta}
\end{align}
where $f_n$ is the resonant frequency and $\beta$ is the fraction of power lost from the cavity in each round trip\footnote{Not to be confused with the cavity coupling coefficient. The term only appears in this section, so I will keep it as is to stay consistent with the literature.}. $\beta$ is comprised of the ohmic losses in the mirror, diffraction, and dielectric losses if dielectrics are present. The diffraction loss caused by the spillover of the mode around the edges of the mirrors is $\exp(-\frac{a^2}{2w_m^2})$, where $w_m$ is the beam size at the mirror. 

The cavity modes are excited through a rectangular waveguide and aperture. The magnetic field $\vb{H}$ in the waveguide will fringe through the aperture and appear like a magnetic dipole moment\cite{pozar, Dunseith_2015, 241661aperturecoupling}\footnote{Pozar 4th ed. Section 4.8 has a good illustration.}. The effective magnetic moment of the hole is~\cite{Dunseith_2015}
\begin{align}
  &\vb{m}_h = \alpha_M \vb{H}_h\\
  &\alpha_M = 4/3 f_t r_h^3\nonumber\\
  & f_t = \exp \left (\frac{-2\pi A_m t}{\lambda_c} \sqrt{1-(\lambda_c/\lambda)^2} \right )\nonumber\\
  & \lambda_c = 3.412 r_h\nonumber\\
  & A_m t = 1.0064t + 0.0818r_h \quad \text{for} \quad t>0.2r_h\nonumber
\end{align}
where $\vb{H}_h$ is the magnetic field at the hole, t is the thickness of the hole, and $f_t$ is a factor that accounts for the thickness of the hole.

For achieve critical coupling, the radius of the hole $r_h$ should be~\cite{Dunseith_2015}
\begin{align}
  r_h = \left (\frac{9\pi w_m^2 L a^2}{256 Q_0 k_g f_t^2}\right )^{1/6}\\
  \label{eqn:critical_coupling_hole}
\end{align}
where $k_g$ is the wavenumber in the waveguide.

The cavity can also be understood using network theory. The Orpheus cavity will have a strongly-coupled port and a weakly-coupled port\footnote{Sometimes colloquially called the strong port and the weak port.}. So it can be thought of as a two-port network. Networks can be fully described by the scattering matrix\cite{pozar}. The scattering matrix element is defined as 
\begin{align}
  S_{ij} = \eval{\frac{V_i^{-}}{V_j^{+}}}_{V_k^+=0\text{ for } k\neq j}
\end{align}
where $V_j^+$ is the amplitude of the voltage wave incident on port $j$ and $V_i^-$ is the amplitude of the voltage wave reflected from port $i$. $S_{11}$ is the reflection coefficient $\Gamma$ of port 1, and $S_{21}$ is the transmission coefficient $T$ from port 1 to port 2. 

Near resonance of the cavity, i.e., near the mode frequency, the transmitted and reflected power can be derived to be Lorentzian (Section~\ref{sec:reflection_derivation}):

\begin{align}
  &\abs{T}^2 = \frac{\delta y}{1+4\Delta^2} + C_1\label{eqn:lorentzian_fits}\\
  &\abs{\Gamma}^2 = -\frac{\delta y}{1+4\Delta^2} + C_2 \nonumber\\
  &\Delta \equiv Q\frac{f-f_0}{f_0}\nonumber
\end{align}
where $T$ is the transmission coefficient, $R$ is the reflection coefficient, $\delta_y$ is the depth of the Lorentzian, and $C_1$ and $C_2$ are constant offsets.

\subsection{Simulation of the empty cavity}\label{sec:empty_simulation}
The mode structure of the cavity can be simulated using Finite Element Analysis. The Orpheus mode structure was simulated using ANSYS HFSS. These simulations are resource-intensive because the cavity is electrically large, and there aren't any simplifying assumptions other than two symmetry boundaries. It took me over a year to learn to simulate Orpheus for the entire tuning range. I will outline my simulation setup so that it may help others. 

The simulation for the empty cavity is shown in Figure~\ref{fig:empty_gaussian}. I model only a quarter of the cavity (in the +X,+Y quadrant). I place a Perfect H symmetry plane where the electric field is parallel to the boundary and a Perfect E symmetry plane where the electric field is perpendicular to the boundary. The cavity is surrounded by vacuum, and there exists a cylindrical radiation boundary \SI{1}{cm} away from the mirrors. 

I apply curvilinear meshing to all curved surfaces. For the curvilinear meshing, I use a resolution of 7 (maximum is 9). An artificial cylindrical volume surrounds the beam axis that helps the simulation converge faster. The cylindrical volume connects the weakly-coupled port aperture and the strongly-coupled port aperture. Because of the high-resolution curvilinear meshing, the initial mesh starts with more mesh elements near the beam axis where they are needed.

To simulate the electric field, I use a driven modal solution. The eigenmode solver is untenable because of the large density of modes, particularly due to the subresonances of the dielectrics. The majority of these modes are irrelevant. Using the driven modal solution filters out modes that don't couple well to the aperture. I used the iterative solver instead of the direct solver to save RAM. I ask for a Mixed Order Basis and a Maximum Delta S of 0.001 for the convergence condition. 

For simulating a mode, I use the Fast Sweep, which seems to be optimized for narrowband simulations. The frequency sweep is centered around the mode frequencies predicted by Equation~\ref{eqn:fp_frequencies} and will span $\sim \SI{30}{MHz}$. After the frequency sweep is completed, $|S_{21}|^2$ is plotted to find the resonant frequency. The electric field is plotted at the simulated resonant frequency to check that it is the expected Gaussian TEM mode.

To find all the resonances between \SI{15}{GHz} and \SI{18}{GHz}, I will use an interpolating sweep. The interpolating sweep is better suited for simulating multiple resonances over a broader range. However, the interpolating sweep doesn't simulate any field data. Therefore, for plotting the electric field, I recommend using the fast sweep and focusing the frequency range around the mode of interest.

The simulated \tem mode is shown in Figure~\ref{fig:empty_gaussian}.

\begin{figure}[ht]
  \centering
  \includegraphics[width=0.5\textwidth]{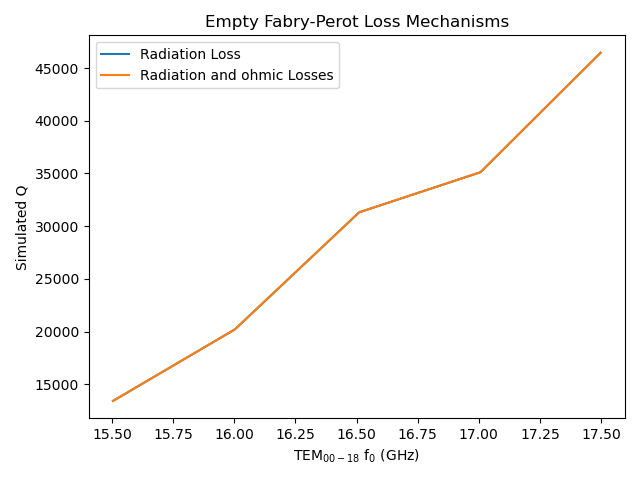}
  \caption{A comparison of simulated quality factor of the \tem mode with different sets of loss mechanisms. The blue curve corresponds to a cavity with perfectly-conducting mirrors. The orange curve corresponds to a cavity with resistive aluminum mirrors. They are nearly identical, suggesting that the dominant loss is diffraction.} 
  \label{fig:empty_fp_losses}
\end{figure}

Simulations can help determine the dominant loss mechanism in the cavity. For this study, two simulations were performed, one where the mirrors are perfect conductors, and another simulation where mirrors are made of aluminum with a conductivity of \SI{38e6}{siemens}. Figure~\ref{fig:empty_fp_losses} shows that the simulated $Q$ are nearly identical, suggesting that the ohmic losses are negligible compared to diffraction losses. One can also calculate the beam size at the mirrors for the \tem mode, calculate the diffraction of the mode spilling around the cavity, and find that it overwhelmingly dominates the ohmic losses.

\FloatBarrier
\subsection{Measurement of the empty cavity}\label{sec:empty_measurement}
\begin{figure}[ht]
  \centering
  \includegraphics[height=0.3\textheight]{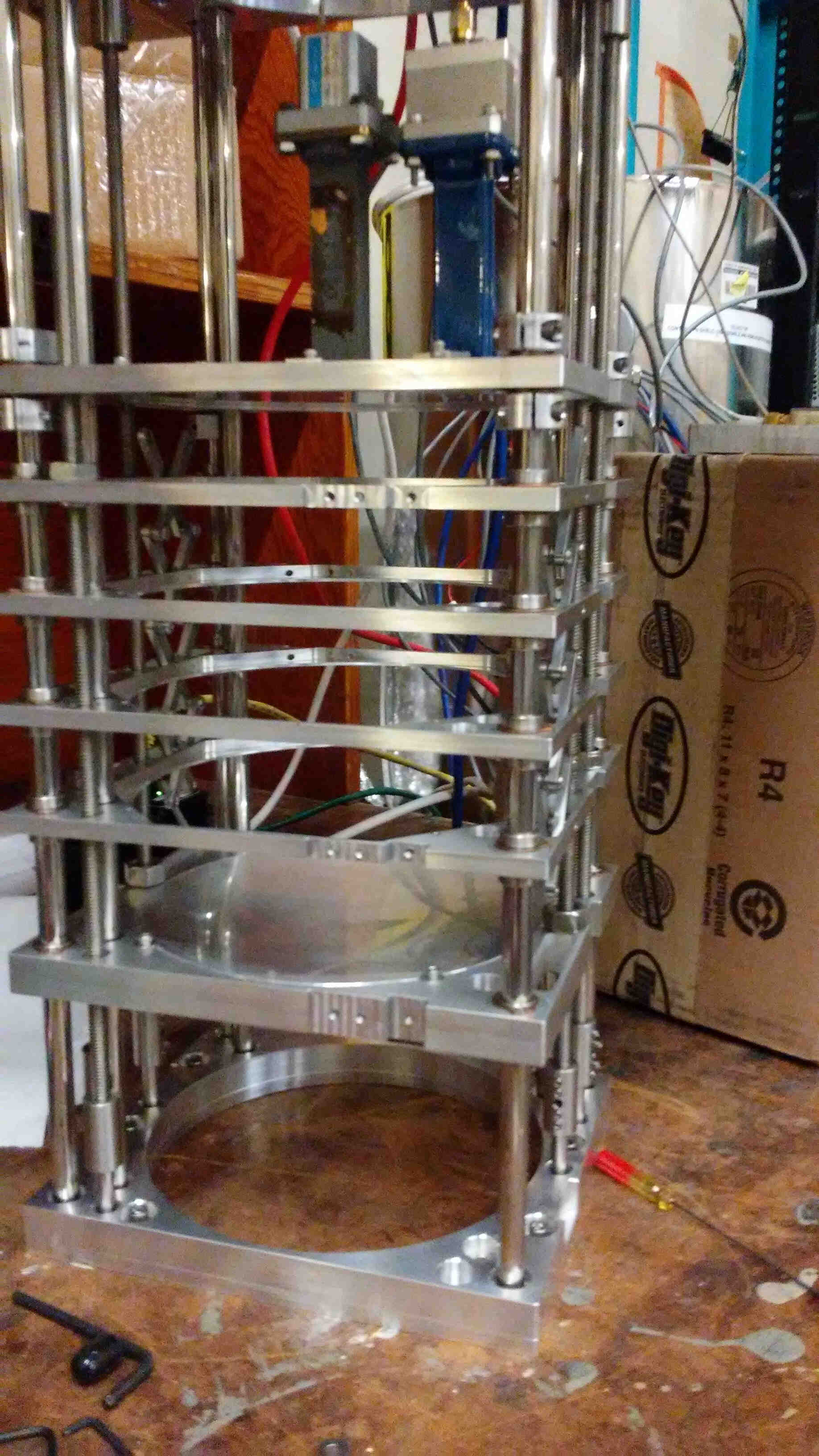}
  \caption{The empty Fabry-Perot cavity consists of a flat mirror and a curved mirror. The cavity coupling scheme shown in this picture is out of date. This picture is just meant to show the cavity without dielectric plates.}
  \label{fig:empty_fp}
\end{figure}

Now I measure the mode structure of a tabletop empty Fabry-Perot cavity, shown in Figure~\ref{fig:empty_fp}\footnote{This picture is out of date. But I can't seem to find a better photo of the empty cavity.}. The mechanical implementation of the cavity is described in Section~\ref{sec:cavity_mechanics}. Assuming the cavity is built, the modes can be characterized using a Vector Network Analyzer (VNA), which can measure $S_{21}$ and $S_{11}$. The Orpheus test stand uses a Keysight E5063A. 

\begin{figure}[ht]
  \centering
  \includegraphics[height=0.25\textheight]{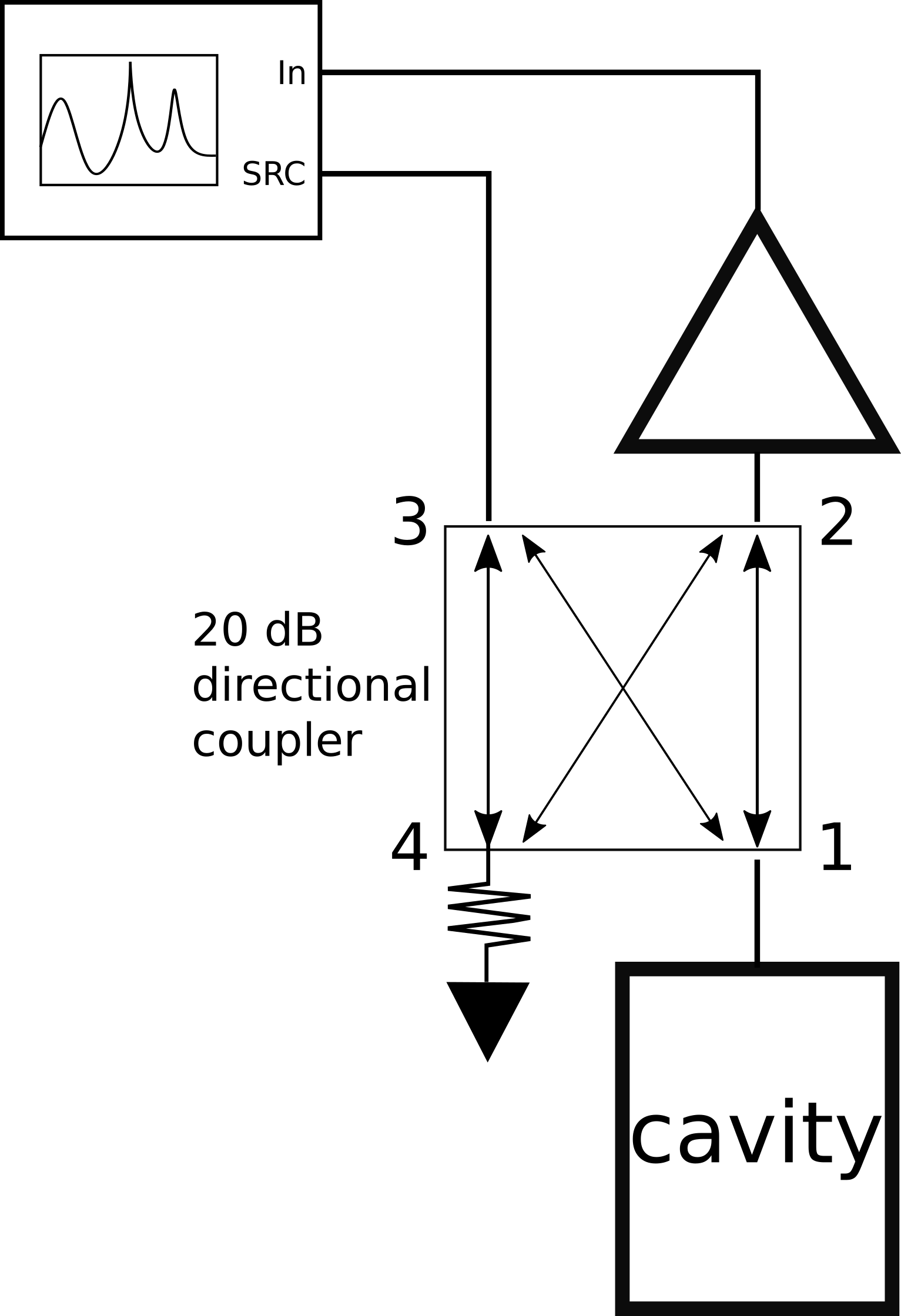}
  \caption{The reflection coefficient of the cavity's strongly-coupled port is measured using a directional coupler. $S_{11}$ can't be measured directly because the amplifier is a non-reciprocal device. The directional coupler allows an injected signal to bypass the amplifier to reach the cavity's strongly-coupled port. The signal from the VNA travels from port 3 of the directional coupler to port 1 (with a \SI{20}{dB} attenuation), bounces off the cavity's strongly-coupled port, then reaches port 2 of the directional coupler. The signal is amplified and then reaches the VNA's port 2.  }
  \label{fig:reflection_measurement}
\end{figure}

The reflection coefficient is needed to extract the cavity coupling coefficient of the strongly-coupled port. However, the strongly-coupled port is connected to an amplifier, which is a one-directional device. Thus the reflection measurement needs to bypass the amplifier to reach the strongly-coupled port. The amplifier is bypassed using the directional coupler, as shown in Figure~\ref{fig:reflection_measurement}. To measure the transmission and reflection coefficient of the cavity, the VNA makes an $S_{21}$ measurement, where the port numbers correspond to the VNA. For a transmission measurement, the VNA injects a signal into the weakly-coupled port. For a reflection measurement, the VNA injects a signal into the directional coupler. I may still refer to the reflection measurement as an $S_{11}$ measurement. If so, port 1 corresponds to the cavity's strongly-coupled port instead of the port on the VNA.

For these measurements, the aperture diameter is \SI{0.2130}{in.}. Equation~\ref{eqn:critical_coupling_hole} suggests a hole larger than the waveguide dimension. This is untenable. So I started with a hole size of \SI{0.18}{in} and increased the size until I got close to critical coupling for some frequency range.

\begin{figure}[ht]
  \centering
  \subfloat[]{\includegraphics[width=0.47\textwidth]{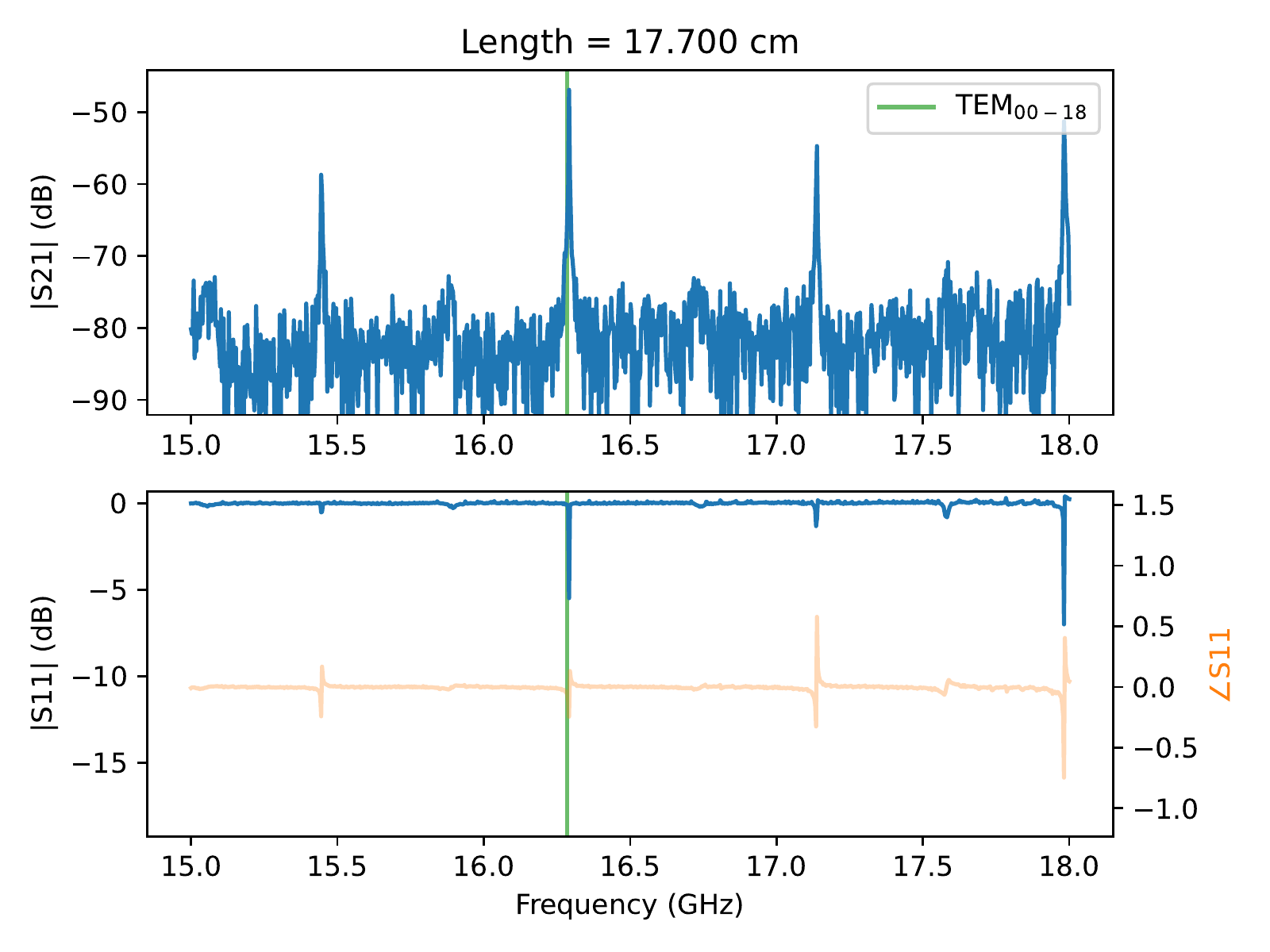}}\hfil
  \subfloat[]{\includegraphics[width=0.47\textwidth]{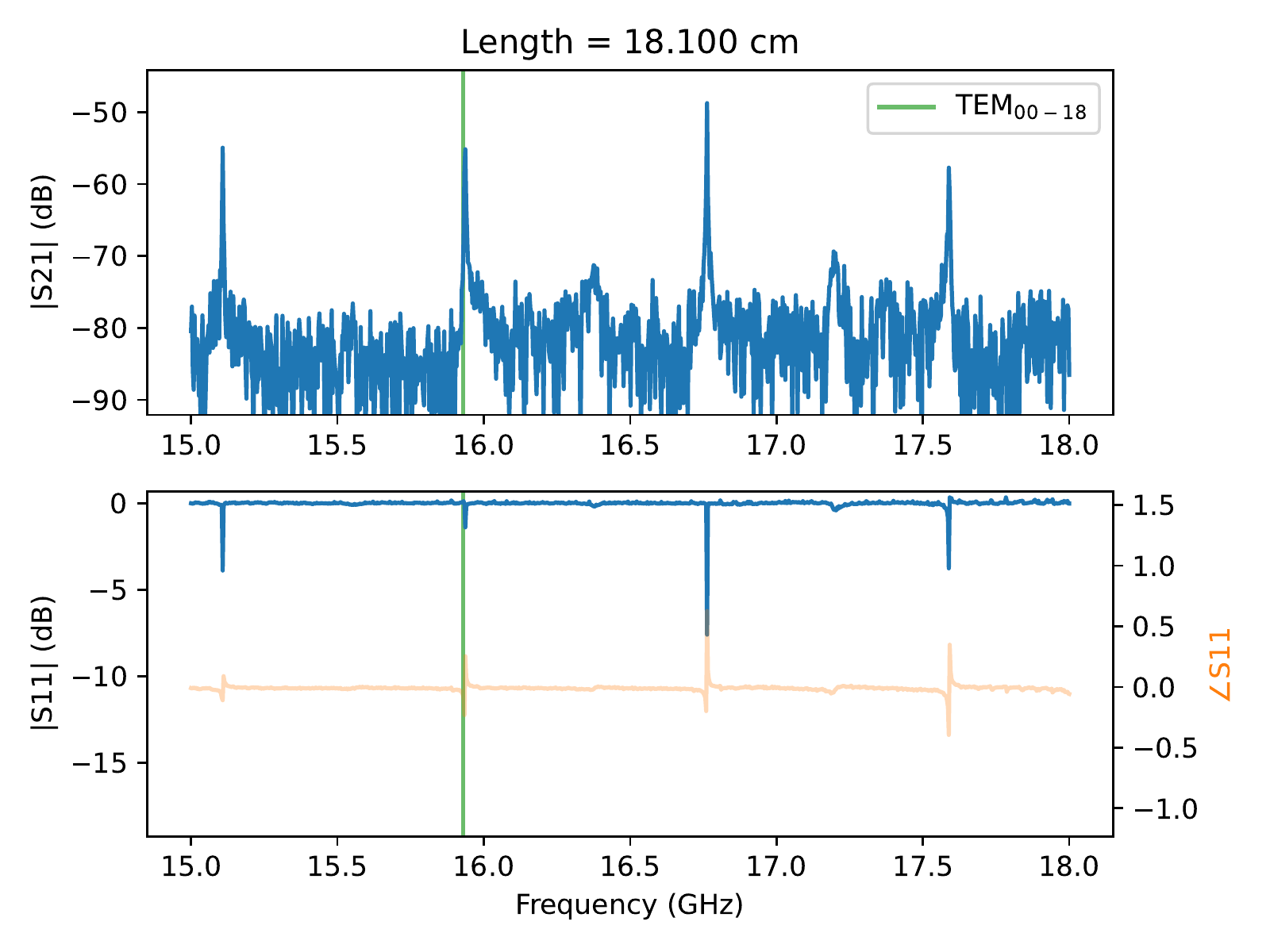}}
  \caption{The measured scattering parameters S21 and S11 of an empty Fabry-Perot cavity as a function of frequency for two different cavity lengths. The prominent peaks are Lorentzian and correspond to different Gaussian modes.}
  \label{fig:tabletop_empty_fp_widescan}
\end{figure}

The transmission and reflection measurements are used to characterize the mode structure of the cavity. First, ``wide scan'' measurements are taken, where the transmission and reflection coefficients are measured from \SI{15}{GHz} to \SI{18}{GHz}. The wide scan measurements are repeated for cavity lengths between \SI{16}{cm} and \SI{19.2}{cm} so that the \tem mode is tuned from \SI{18}{GHz} to \SI{15}{GHz}. Figure~\ref{fig:tabletop_empty_fp_widescan} shows the wide scan for different cavity lengths. 

\begin{figure}[ht]
  \centering
  \subfloat[]{\includegraphics[width=0.5\textwidth]{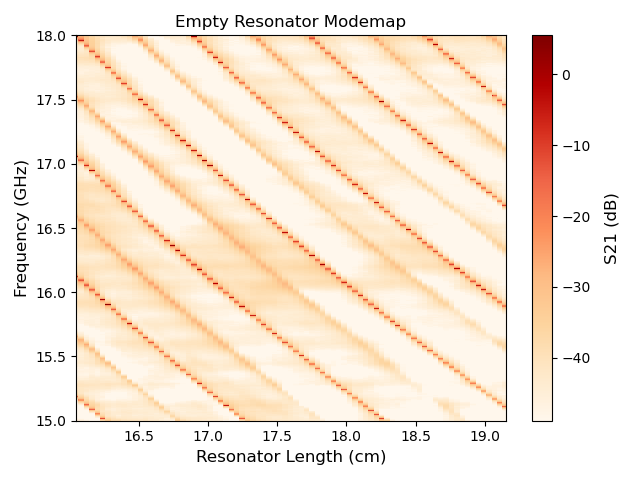}}
  \subfloat[]{\includegraphics[width=0.5\textwidth]{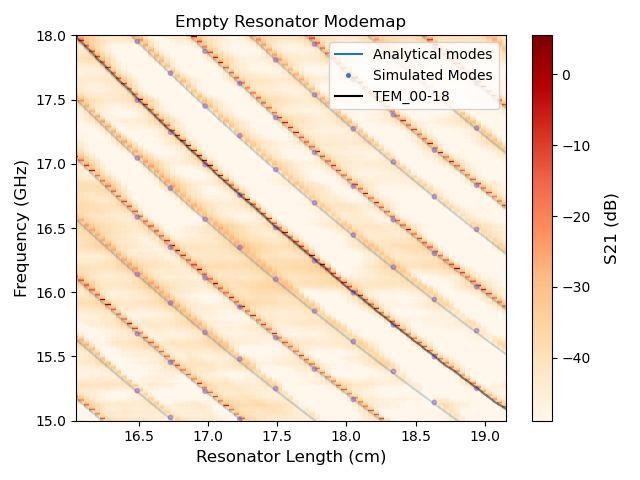}}
  \caption{(a) Measured mode map for an empty Fabry-Perot Cavity. This is the measured scattering parameter $S_{21}$ as a function of frequency and cavity length. The dark lines correspond to the modes of the cavity where transmission is highest. (b) The analytical TEM modes and the simulated modes are overlaid on the measurement. The agreement between the analytical formula, FEA simulation resonances, and measured resonances are within a few MHz.}
  \label{fig:empty_modemap}
\end{figure}

The wide scan measurements can be combined into a 2D heat map to form a mode map, as shown in Figure~\ref{fig:empty_modemap}. The mode map illustrates the mode structure of the cavity for the entire tuning range and demonstrates the mode frequencies decrease with increasing cavity length, as is expected from Equation~\ref{eqn:fp_frequencies}. The simulated modes and analytical modes are then overlaid on the mode map. All three solutions agree within a few \si{}{MHz}, and the disagreement can be attributed to a systematic uncertainty in the absolute cavity length. The fact that the measured and simulated modes agree with the analytical modes lends great confidence to both the simulation and measurement techniques. This confidence is important because the dielectrically-loaded case doesn't have analytical solutions for comparison. So the simulations need to be trusted to correctly identify the \tem mode. $\veff$ calculated from simulations also need to be trusted because it's infeasible to measure the electric field in the cavity.

\begin{figure}[ht]
  \centering
  \subfloat[]{\includegraphics[width=0.47\textwidth]{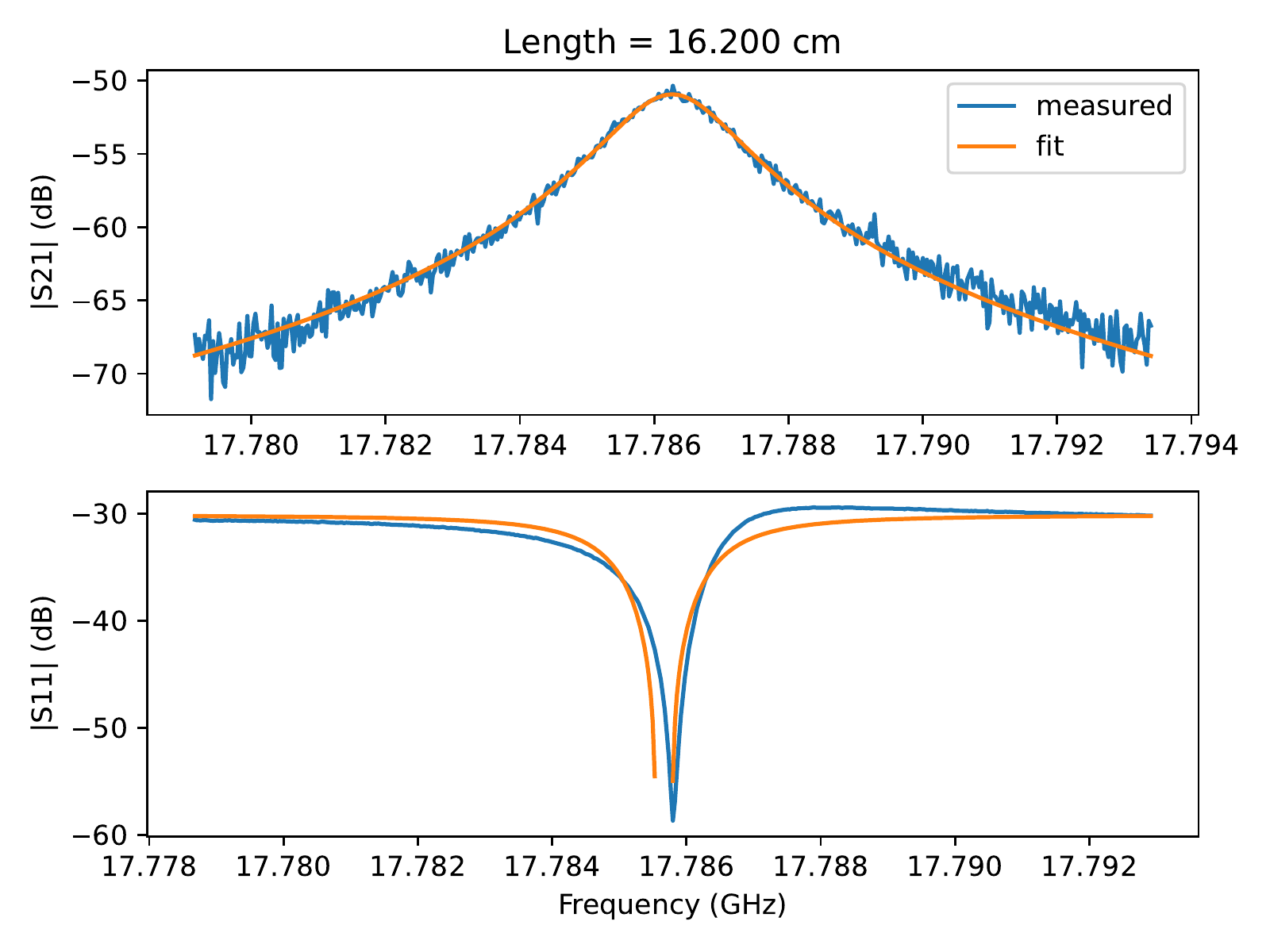}}\hfil
  \subfloat[]{\includegraphics[width=0.47\textwidth]{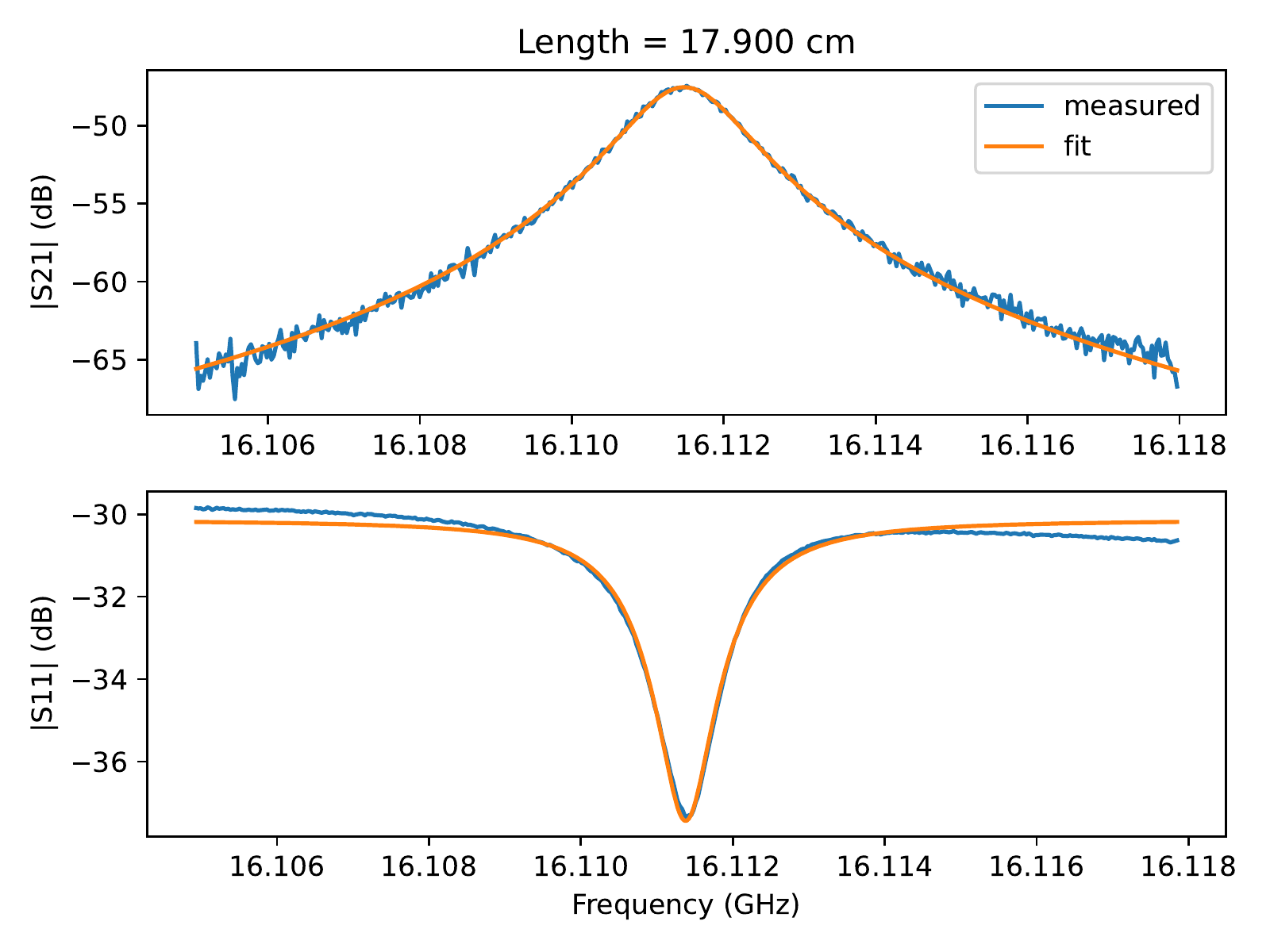}}
  \caption{$S_{21}$ and $S_{11}$ of an empty Fabry-Perot cavity near the \tem mode for two different cavity lengths. The transmitted power is a Lorentzian. The reflected power is approximately Lorentzian, but effects discussed in Section~\ref{sec:reflection_coupler} distort the Lorentzian. Because of the skew in the data, the fitted Lorentzian in (a) predicts a small negative reflected power at resonance, which is unphysical}
  \label{fig:tabletop_empty_fp_narrowscan}
\end{figure}

Other important parameters for the cavity modes are the loaded quality factor and the cavity coupling coefficient. To extract these parameters, ``narrow scans'' are taken, where the VNA measures the $S_{21}$ and $S_{11}$ parameters within a few Q-widths of the \tem mode, as shown in Figure~\ref{fig:tabletop_empty_fp_narrowscan}. These narrow scans are fitted to the Lorentzian functions in Equation~\ref{eqn:lorentzian_fits}. From the Lorentzian fits, one can extract $f_0$ and $Q_L$. The cavity coupling coefficient $\beta$ is extracted from the value of the reflection coefficient on resonance and the phase change on resonance (See~\ref{fig:gamma_angle} for determining whether the cavity is undercoupled or overcoupled based on the phase change.).

\begin{align}
 \beta = 
  \begin{cases} 
   \frac{1-{\abs{\Gamma_c(f_0)}}}{1+\abs{\Gamma_c(f_0)}} & \text{if } \beta \leq 1 \\
   \frac{1+\abs{\Gamma_c(f_0)}}{1-\abs{\Gamma_c(f_0)}} & \text{if } \beta \geq 1
  \end{cases}
\end{align}

For a chi-square minimization fit, the uncertainty in the reflected and transmitted power is derived to be (Section~\ref{sec:reflection_fit_procedure}).
\begin{align}
  \sigma_{P}^2 = 4\sigma_V^2 P 
\end{align}

\begin{figure}[ht]
  \centering
  \subfloat[]{\includegraphics[width=0.5\textwidth]{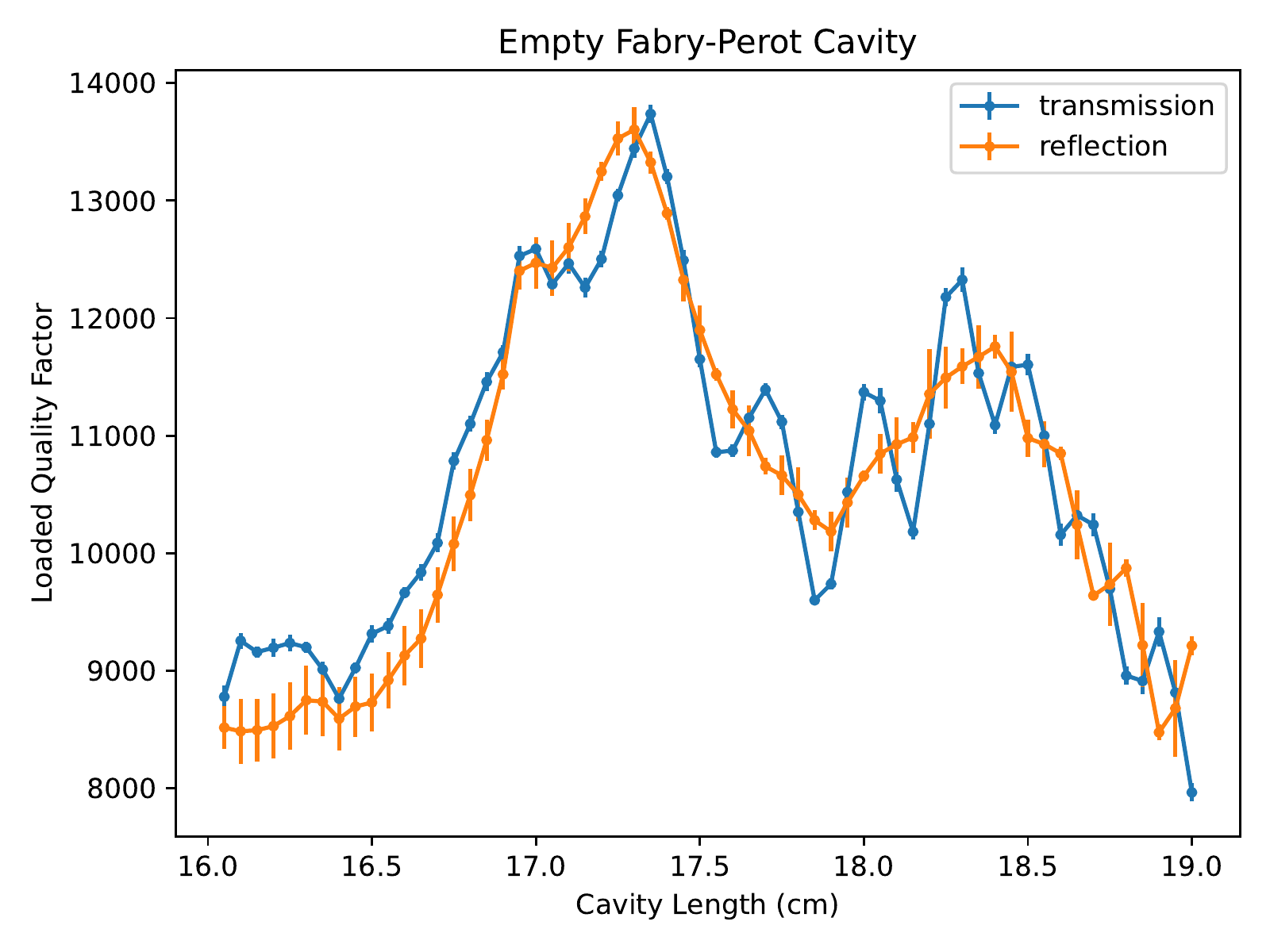}}
  \subfloat[]{\includegraphics[width=0.5\textwidth]{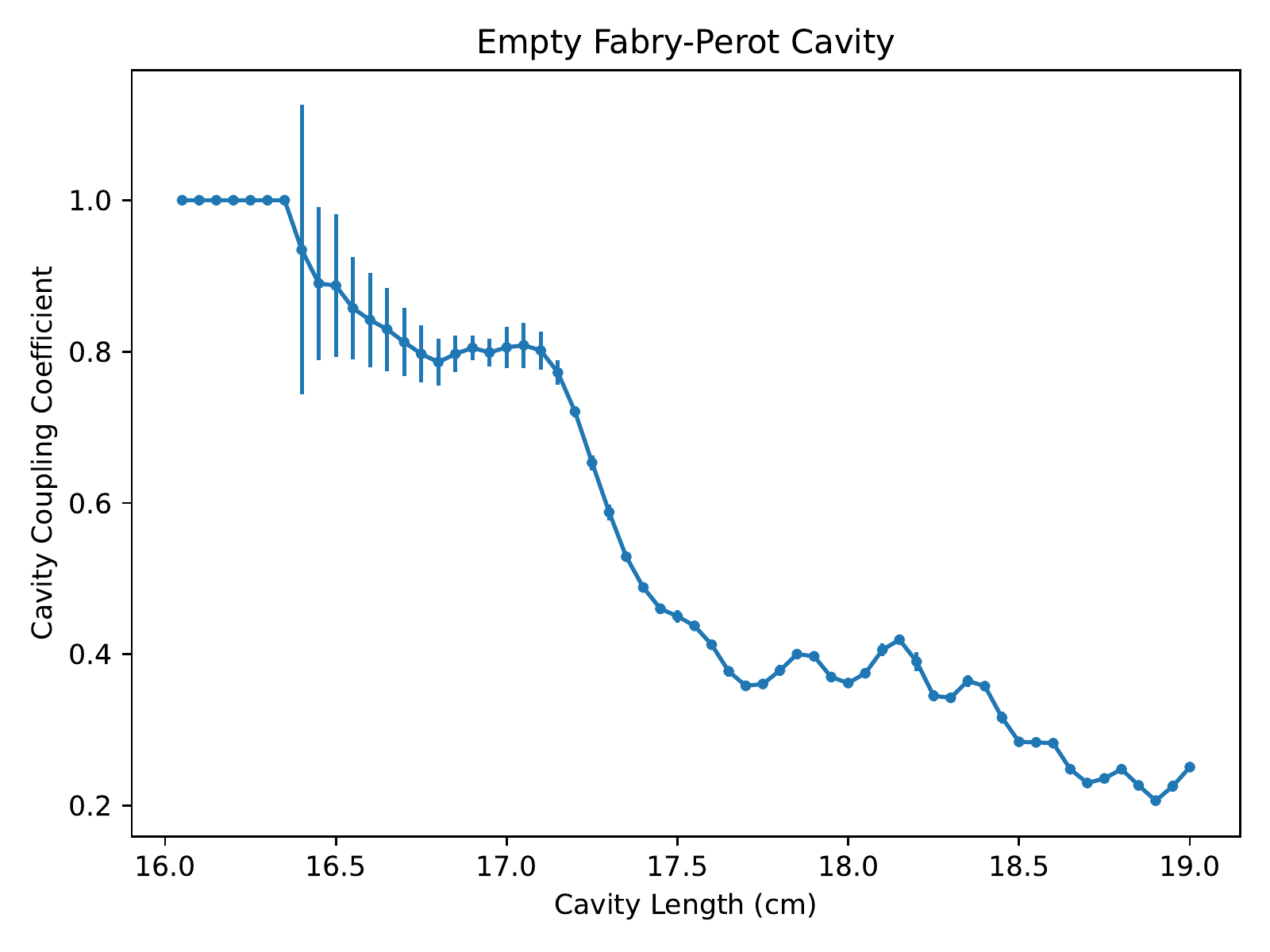}}
  \caption{The loaded quality factor and cavity coupling coefficient of the \tem mode as a function of cavity length.}
  \label{fig:empty_q_beta}
\end{figure}

The extracted $Q_L$ and $\beta$ are shown in Figure~\ref{fig:empty_q_beta}. $Q_L$ should be the same whether it's extracted from the transmission measurement or reflection measurement. The measured discrepancy between the quality factors is caused because of interference effects inside the directional coupler. Section~\ref{sec:reflection_coupler} derives a skewed Lorentzian function that more fully captures the physics of the directional coupler. Figures~\ref{fig:complicated_fit} and~\ref{fig:fitted_qs} show that the more sophisticated reflection model fits the data better, which results in the reflection $Q_L$ matching the transmission $Q_L$. The cost of implementing this model is that there is significant degeneracy in the rest of the parameters and the cavity coupling coefficient is difficult to interpret. So for the rest of the analysis, the straightforward Lorentzian model from Equation~\ref{eqn:lorentzian_fits} is used to extract $\beta$.

\FloatBarrier
\section{Adding dielectrics to a Fabry-Perot Cavity}
Dielectrics are added at every fourth half-wavelength to increase $\veff$. The \tem mode is engineered to be the mode of interest. Much of the formalism and techniques built up for the empty cavity apply to the dielectrically-loaded case.

\subsection{Choice of Dielectrics}
99.5\% alumina sheets were purchased from Superior Technical Ceramic because they offered high-purity alumina at custom dimensions. The alumina plates are about $\SI{5.97}{in}\times\SI{5.97}{in}\times\SI{3}{mm}$. The plates are octagonal to approximate a circular shape, but the straight lines are easier to machine. The dielectrics are \SI{3}{mm} thick because that is approximately half a wavelength inside a $\epsilon_r = 9.8$ dielectric at \SI{16.5}{GHz}.

\subsection{Simulated TEM$_{00-18}$ Mode}\label{sec:orpheus_simulations}
The same simulation setup used in Section~\ref{sec:empty_simulation} applies to simulating the dielectrically-loaded case. A major difference is that there aren't analytical solutions for the cavity mode frequencies. An excellent first guess is to use the free space solutions in Equation~\ref{eqn:fp_frequencies} and replace $L$ with the optical length. For some cavity lengths, this initial guess was correct within \SI{10}{MHz}. But for other frequencies, this guess was off by over \SI{100}{MHz}, and I had to spend a lot of time searching for the mode.

\begin{figure}[ht]
  \begin{tabular}{r r r}
    \includegraphics[height=70mm]{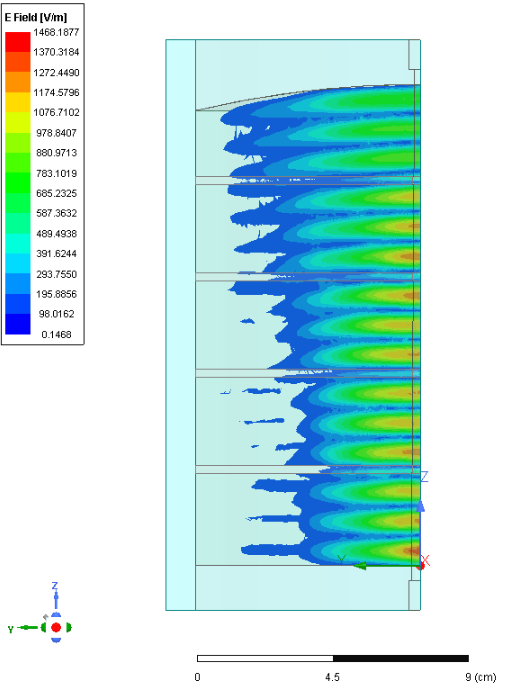}& \includegraphics[height=70mm]{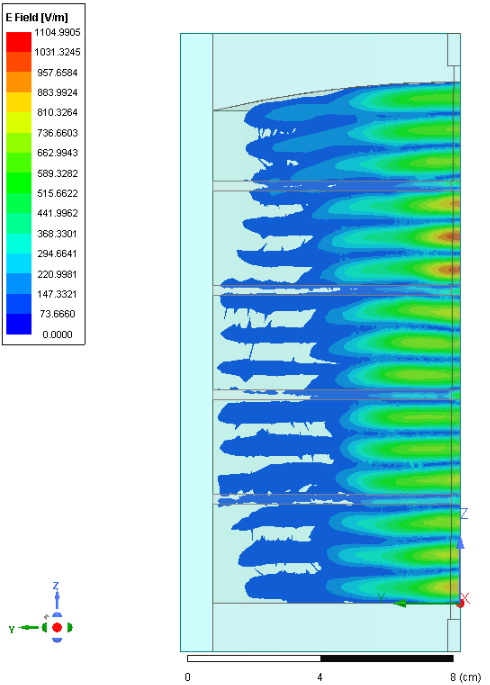} & \includegraphics[height=70mm]{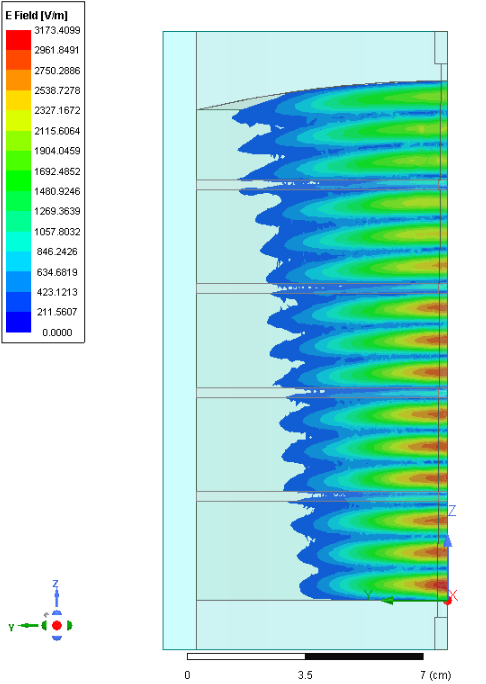} \\
    $\fm = \SI{15.23}{GHz}$ & $\fm = \SI{15.43}{GHz}$ & $\fm = \SI{15.73}{GHz}$\\ 
    $L=\SI{16.3}{cm}$ & $L=\SI{16.0794}{cm}$ & $L=\SI{15.75}{cm}$\\ 
    \includegraphics[height=70mm]{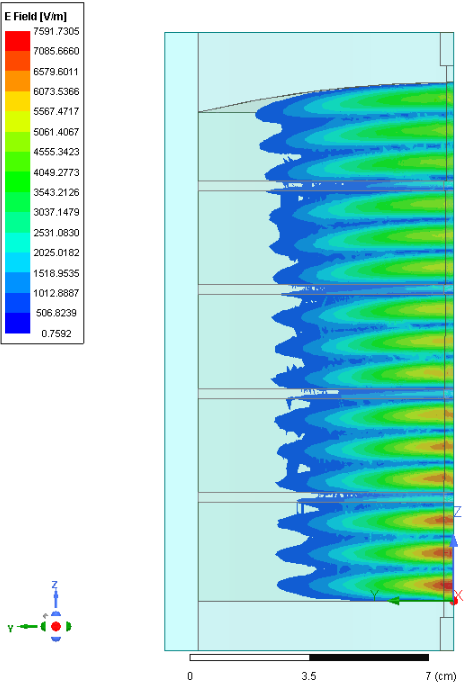} & \includegraphics[height=70mm]{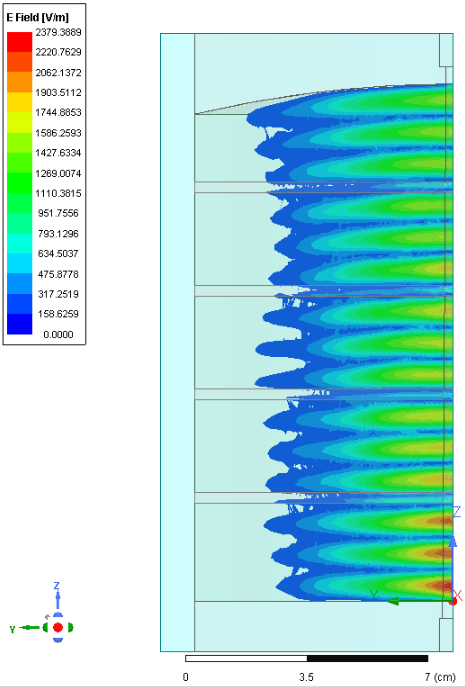} & \includegraphics[height=70mm]{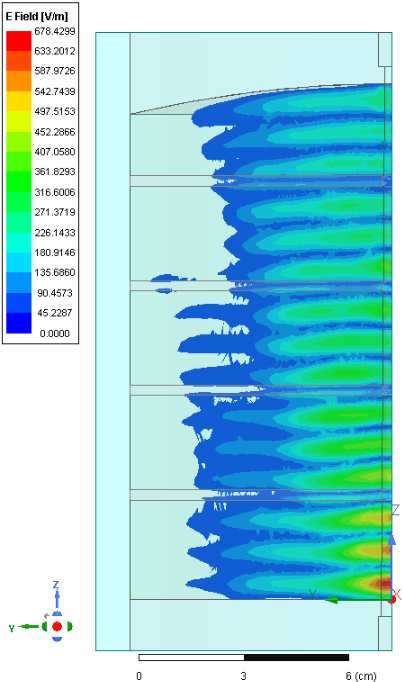}\\
    $\fm = \SI{15.97}{GHz}$ & $\fm = \SI{16.19}{GHz}$ & $\fm = \SI{16.35}{GHz}$\\ 
    $L=\SI{15.4919}{cm}$ & $L=\SI{15.27}{cm}$ & $L=\SI{15.06}{cm}$\\ 
  \end{tabular}
  \caption{The simulated electric field magnitude of the \tem mode with different cavity lengths. There are mode crossings around \SI{15.4}{GHz} and \SI{16.3}{GHz}.} 
  \label{fig:orpheus_simulations1}
\end{figure}

\begin{figure}[ht]
  \begin{tabular}{r r r}
    \includegraphics[height=70mm]{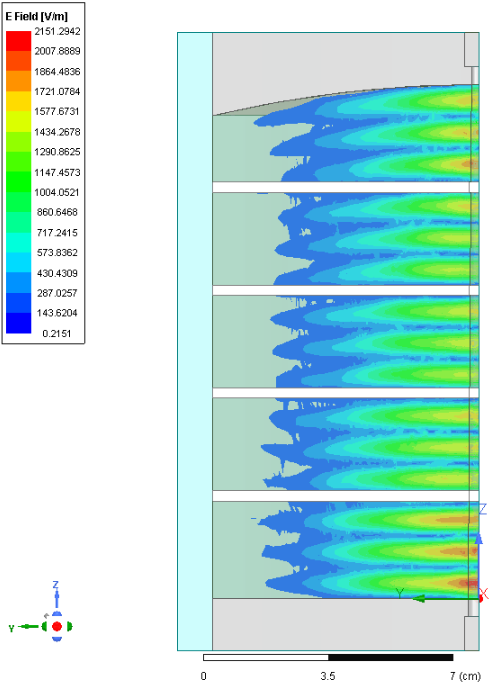}& \includegraphics[height=70mm]{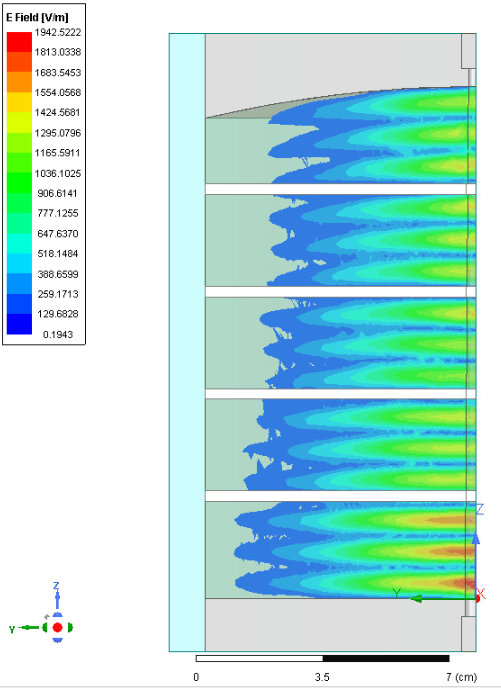} & \includegraphics[height=70mm]{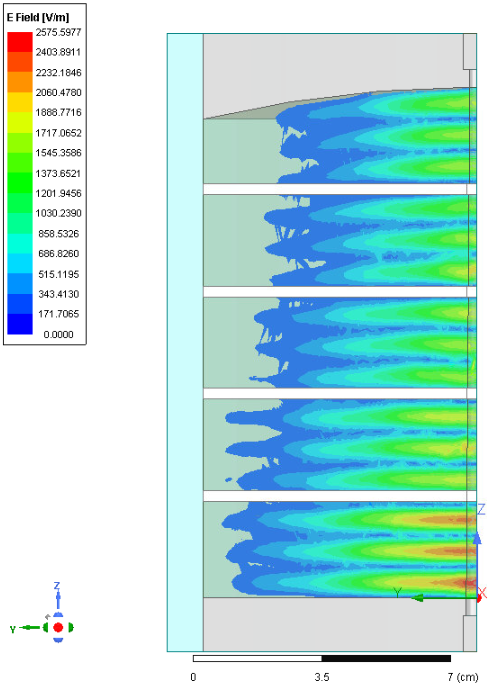} \\
    $\fm = \SI{16.72}{GHz}$ & $\fm = \SI{17.07}{GHz}$ & $\fm = \SI{17.32}{GHz}$\\ 
    $L=\SI{14.744}{cm}$ & $L=\SI{14.4219}{cm}$ & $L=\SI{14.2}{cm}$\\ 
    \includegraphics[height=70mm]{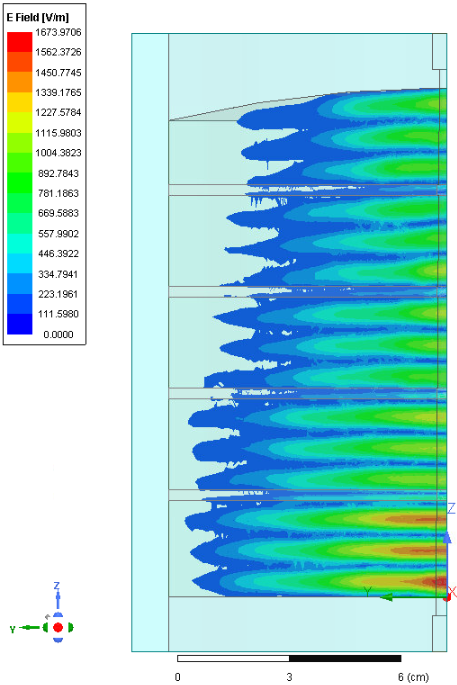} &\includegraphics[height=70mm]{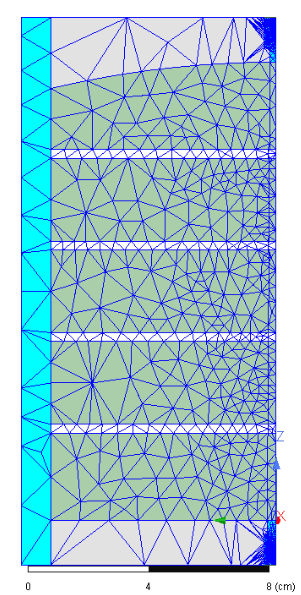}&\\
    $\fm = \SI{17.63}{GHz}$ & FEA mesh & \\ 
    $L=\SI{13.9318}{cm}$ &  & \\ 
  \end{tabular}
  \caption{The simulated electric field magnitude of the \tem mode with different cavity lengths. The field profile starts to deviate from what is expected from physical optics. One intuits from physical optics that the waist of the beam would be on the flat mirror instead of the curved mirror. But that is not the case.} 
  \label{fig:orpheus_simulations2}
\end{figure}

The \tem mode fields are simulated and shown in Figures~\ref{fig:orpheus_simulations1} and~\ref{fig:orpheus_simulations2}. The fields resemble their free space Gaussian counterparts in the empty cavity case with a few notable exceptions. At higher frequencies, the beam waist is at the curved mirror instead of the flat mirror. This is likely a result of the properties of the dielectric that's not considered by physical optics. The other notable difference is that the fields deviate from the Gaussian profile for frequencies near \SI{16.35}{GHz} and \SI{15.43}{GHz}. This is the result of a mode crossing seen in Figure~\ref{fig:tabletop_orpheus_modemap}. In a mode crossing, two modes become degenerate, and the field shape is distorted. The new field shape is a linear combination of the two modes, only one of which couples to the axion or dark photon. 

\begin{figure}[ht]
  \begin{tabular}{c c}
    \includegraphics[width=65mm]{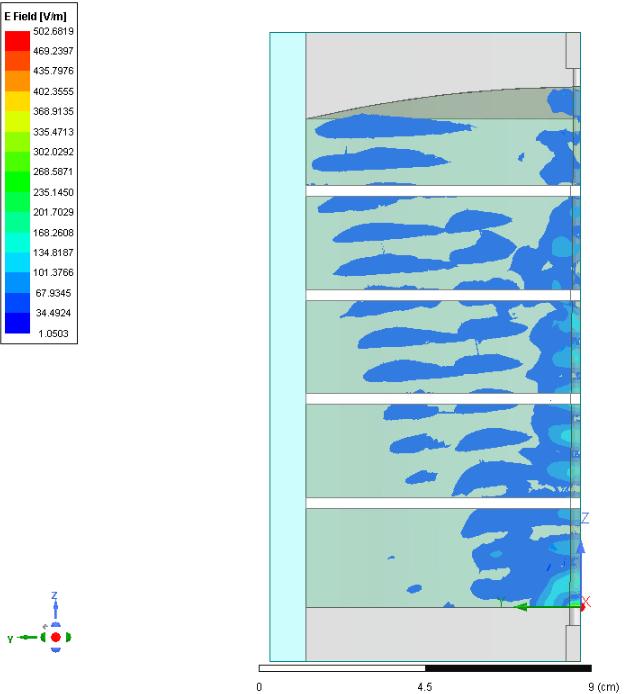} & \includegraphics[width=65mm]{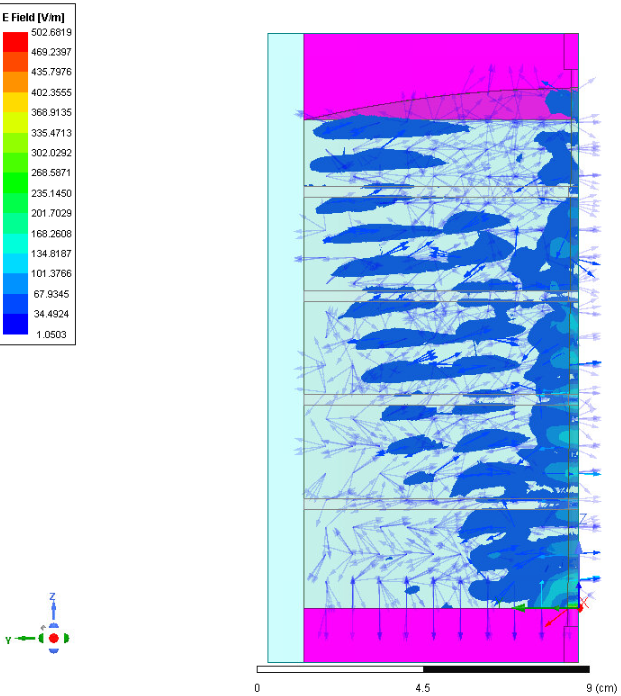}
  \end{tabular}
  \caption{Intruder mode while it's separated from the \tem mode. Here, the cavity length is L = 14.4219 cm. The \tem resonance is at $\fm=\SI{17.07}{GHz}$ while the intruder mode frequency is at $\SI{16.86}{GHz}$. They will cross around $L\approx\SI{14.95}{cm}$, $\fm\approx\SI{16.4}{GHz}$. Left: Magnitude of the intruder mode electric field. Right: Complicated vector field superimposed on the magnitude of the electric field. The field is complicated, but it looks like some mixture of a TM mode (based on the field pointing into the mirror) and a TEM mode (based on the vector field being transverse to the axis for much of the interior of the cavity). TM modes don't tune with the changing cavity length, which may explain why the intruder mode tunes slower than the \tem mode. } 
  \label{fig:field_modecrossing}
\end{figure}

The intruding mode (or set of modes) is shown in Figure~\ref{fig:field_modecrossing}. This is the mode at \SI{16.86}{GHz} when the cavity is $L=\SI{14.4219}{cm}$. The measured mode map (Figure~\ref{fig:tabletop_orpheus_modemap}) shows that this mode tunes more slowly than the \tem mode. The vector field suggests that this mode is some combination of a TM mode (from the arrows that point into the mirror) and a TEM mode (from the arrows that point perpendicular to the beam axis). TEM modes tune with the cavity length, but TM modes do not. Hence, the intruding mode tunes more slowly than the more pure TEM modes.

\begin{figure}[ht]
  \centering
  \subfloat[]{\includegraphics[height=0.4\textheight]{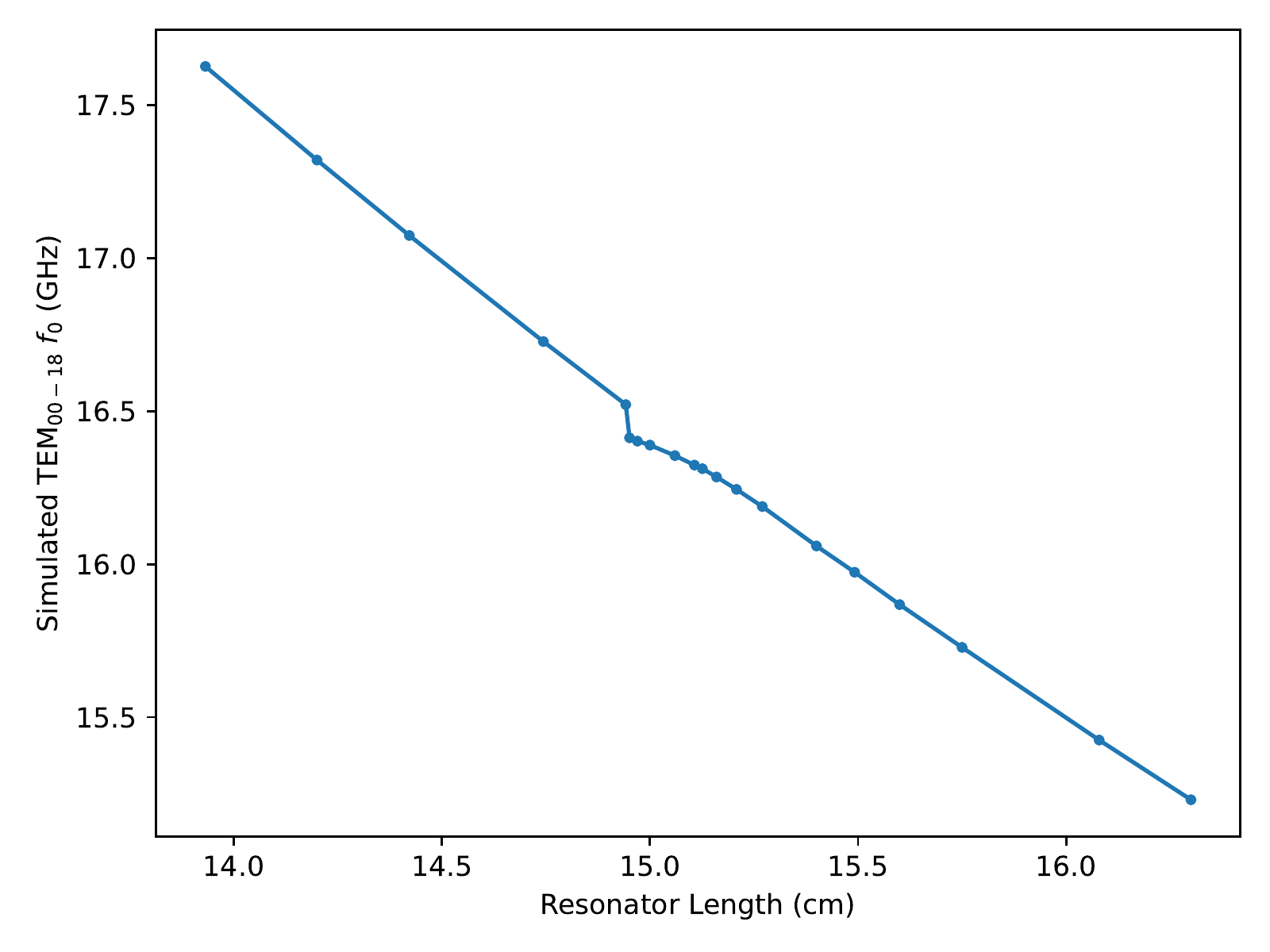}}\\
  \subfloat[]{\includegraphics[height=0.4\textheight]{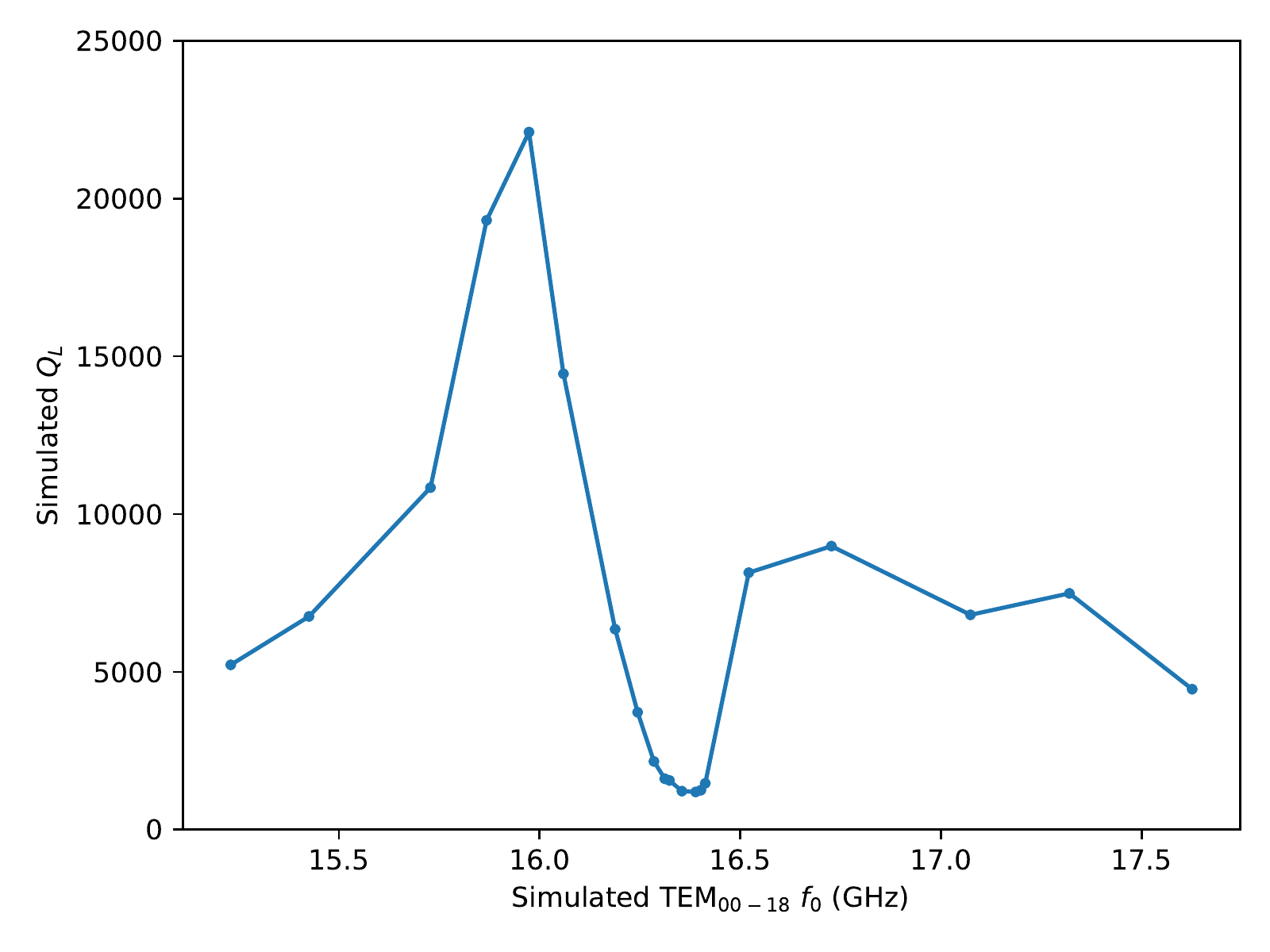}}
  \caption{The simulated frequency and loaded quality factor for the \tem mode. (a) The simulations struggles to simulate the mode crossing, hence the discontinuity in resonant frequency at around $\SI{16.4}{GHz}$. Perhaps we need higher accuracy requirements. (b) The simulated loaded quality factor. In our simulations, the cavity is severely undercoupled, so this is close to the unloaded quality factor.}
  \label{fig:simulated_f_q}
\end{figure}

\begin{figure}[ht]
  \centering
  \subfloat[]{\includegraphics[height=0.4\textheight]{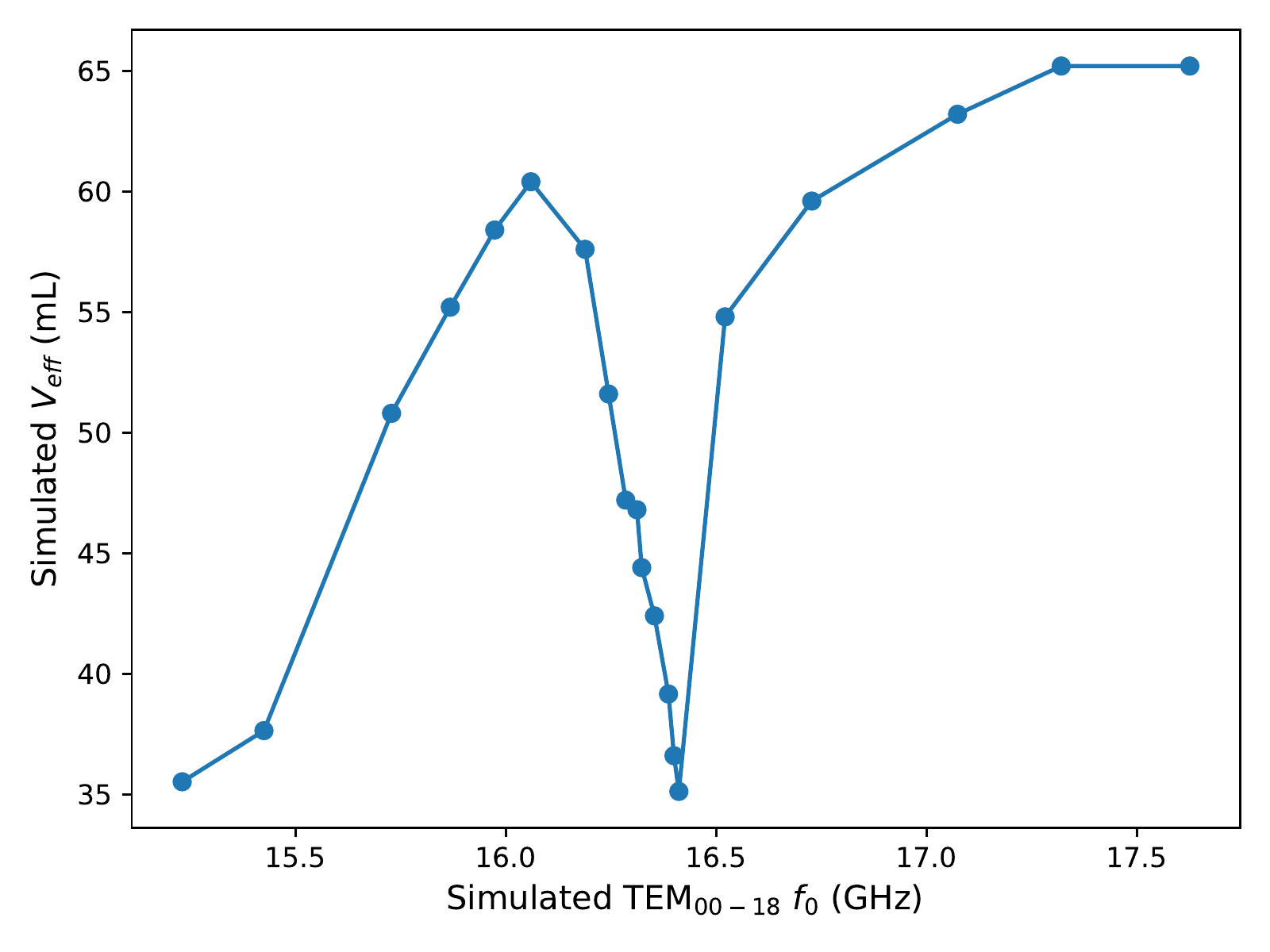}}\\
  \subfloat[]{\includegraphics[height=0.4\textheight]{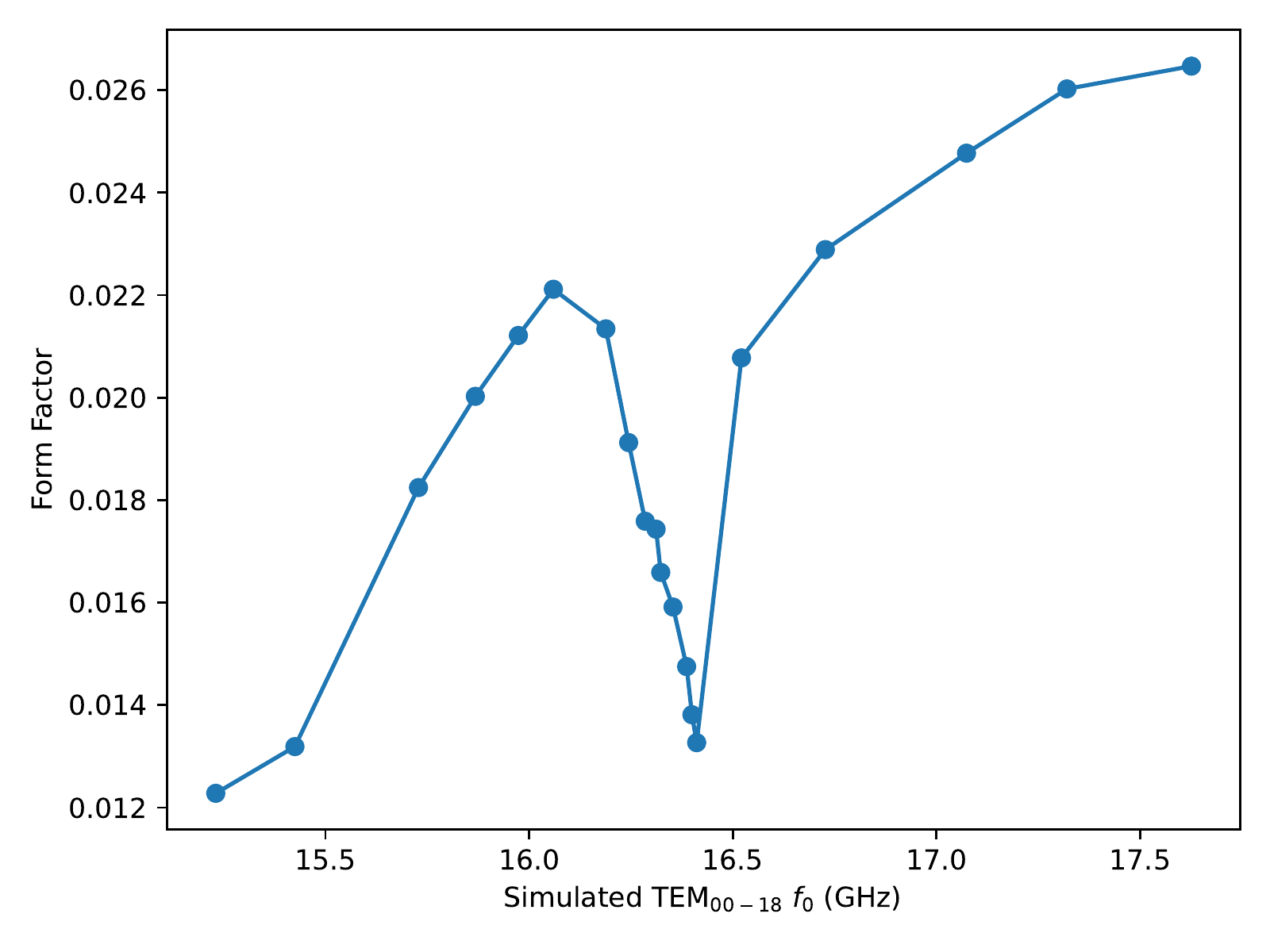}}
  \caption{The effective volume and form factor for the \tem mode. The sharp dip at $\SI{16.4}{GHz}$ corresponds to the mode crossing.}
  \label{fig:simulated_veff}
\end{figure}

The simulated resonant frequency and quality factor are plotted in Figure~\ref{fig:simulated_f_q}. The simulated cavity is very undercoupled, so $Q_L \approx Q_u$. The simulated $\veff$ and form factor are plotted in Figure~\ref{fig:simulated_veff}. All plots have a notch at \SI{16.4}{GHz} at the location of the mode crossing. The discontinuity in the resonant frequency is actually caused by the simulation not accurately modeling the mode crossing. The mode map in Figure~\ref{fig:tabletop_orpheus_modemap} shows that the \tem mode tunes very smoothly, even through the mode crossing. 

The form factor in Figure~\ref{fig:simulated_veff} shows that only about 2\% of the cavity volume is used to detect axions. That's not good, but the cavity properties haven't been optimized for maximal $V_{eff}$ and $Q_L$. Changing mirror curvatures, dielectric thicknesses, dielectric positioning, and adding more dielectrics will increase the form factor.

The width of the notch in the simulated Q matches the width of the notch of the simulated $\veff$. This suggests that the measured Q can be used as a proxy for the width of the mode crossing.

\FloatBarrier
\subsection{$V_{eff}$ Tolerance to Misplaced Dielectrics.}\label{sec:position_error}
Previous simulations assumed the dielectrics are evenly spaced inside the cavity. But $V_{eff}$ is an important parameter that can only be obtained through simulation. Thus, it's essential to understand the uncertainty in $V_{eff}$ and how much it tolerates positioning errors of the dielectrics. Understanding this tolerance will be crucial for the data analysis (Section~\ref{sec:parameter_extraction}).

Let the position error $\delta$ be defined as the deviation from the evenly spaced configuration (actual position - intended position). Orpheus is oriented vertically, such that the flat mirror is on the top and the curved mirror is on the bottom, as shown in Figure~\ref{fig:orpheus_cavity_cad}. A positive position error means the dielectric plate is closer to the curved mirror than what would be intended by the evenly spaced configuration. This sign convention is consistent with the direction of increasing or decreasing cavity length (the flat mirror is fixed, and the curved mirror moves to adjust the cavity length). The middle two dielectric plates are constrained by the scissors jacks and stay evenly spaced between the top and bottom dielectric plates. 

The dielectric plate closest to the curved mirror is the bottom dielectric plate, and the plate closest to the flat mirror is the top dielectric plate. This is reverse from the simulation shown in Figure~\ref{fig:orpheus_simulations1}, where the flat mirror is on the bottom, and the curved mirror is on top. However, the physical cavity (Figures~\ref{fig:tabletop_orpheus} and~\ref{fig:orpheus_real}) has the curved mirror on the bottom, and I will stick with this convention.

\begin{figure}[ht]
  \centering
  \subfloat[]{\includegraphics[width=0.3\textwidth]{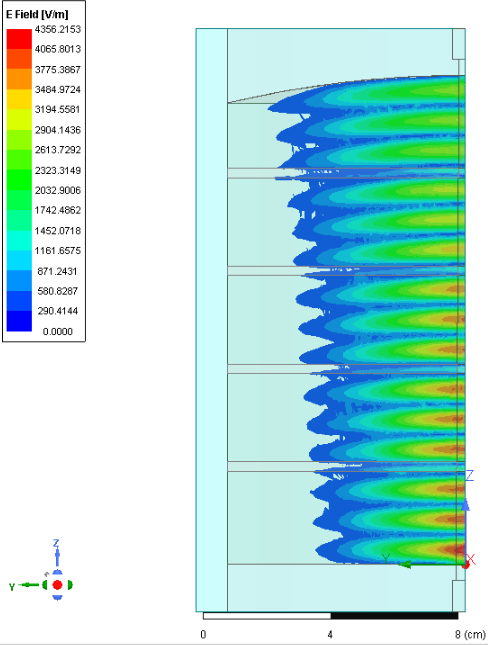}}\hfil
  \subfloat[]{\includegraphics[width=0.3\textwidth]{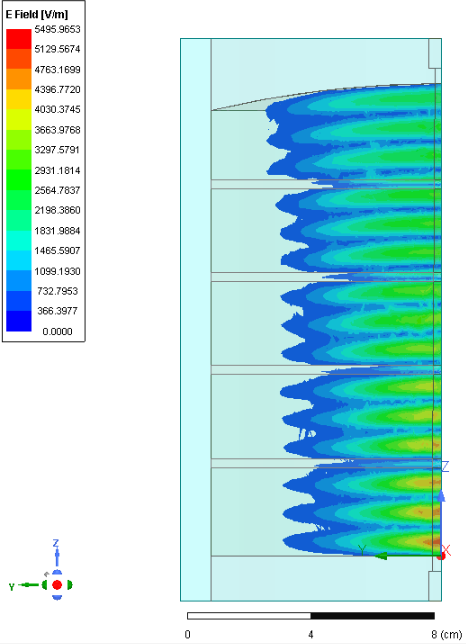}}
  \caption{The effect of dielectric plate position error on the electric field, simulated at about \SI{15.82}{GHz}. (a) No position error. (b) $\delta_{top} = \SI{-0.52}{mm}$, $\delta_{bottom} = \SI{-2}{mm}$. Note the $\delta_{bottom}$ corresponds to the bottom dielectric plate closest to the curved mirror, so it appears as the uppermost plate in this figure.}
  \label{fig:field_error_position}
\end{figure}

Figure~\ref{fig:field_error_position} compares the electric field with and without any error position. The mode is robust against error positions as much as \SI{2}{mm}. 

\begin{figure}[ht]
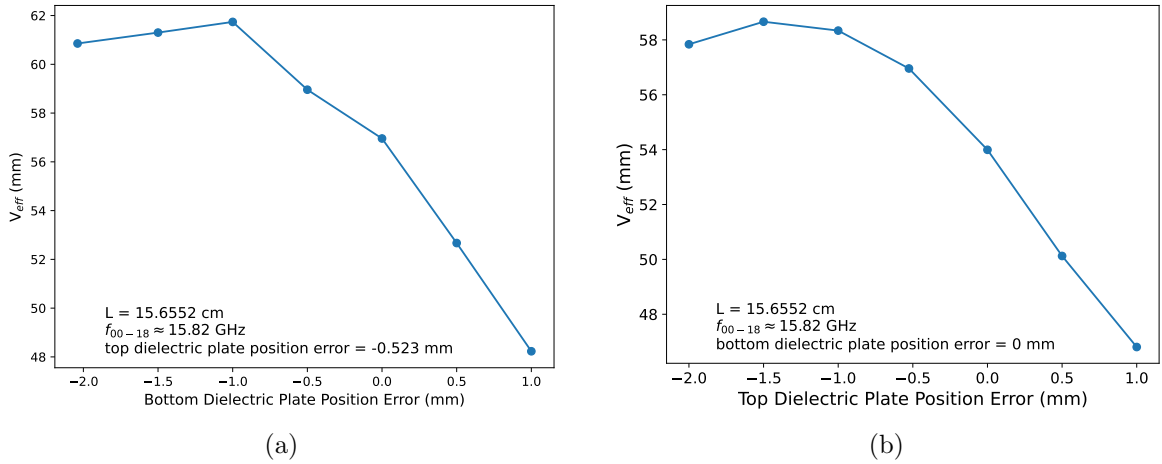

  \centering
  \subfloat[]{\includegraphics[width=0.46\textwidth]{perturbation_studies_veff_bottom_dielectric_plate.pdf}}\hfil
  \subfloat[]{\includegraphics[width=0.46\textwidth]{perturbation_studies_veff_top_dielectric_plate.pdf}}
  \caption{The effect of position error on the effective volume. This study suggests that it would actually be more optimal to deviate from the evenly-spaced configuration. At \SI{15.82}{GHz}, it appears that the effective volume can increase by more than 8\% by moving the dielectrics further away from the curved mirror.}
  \label{fig:veff_position_err}
\end{figure}

Figure~\ref{fig:veff_position_err} shows how the error position affects $V_{eff}$ at a fixed cavity length of $L = \SI{15.66}{cm}$. Surprisingly, $\veff$ improves with negative position error. Comparing the unperturbed and perturbed cases in Figure~\ref{fig:field_error_position} provides intuition for why $V_{eff}$ increases with negative position error. For the unperturbed case, the portion of the wavefront radially farther away from the beam axis gets pushed towards the dielectrics, reducing $\veff$. Pulling the dielectrics away from the curved mirror also pulls it away from the curved wavefront. As I think about it, the negative position errors give the mode more room to breathe.
For positive position errors, the curved wavefront gets pushed more into the dielectrics, and $V_{eff}$ decreases.

It appears that an position error of $\approx \SI{-1}{mm}$ optimizes $V_{eff}$. So not only are perturbations tolerated, they are actually beneficial.

\begin{figure}[ht]
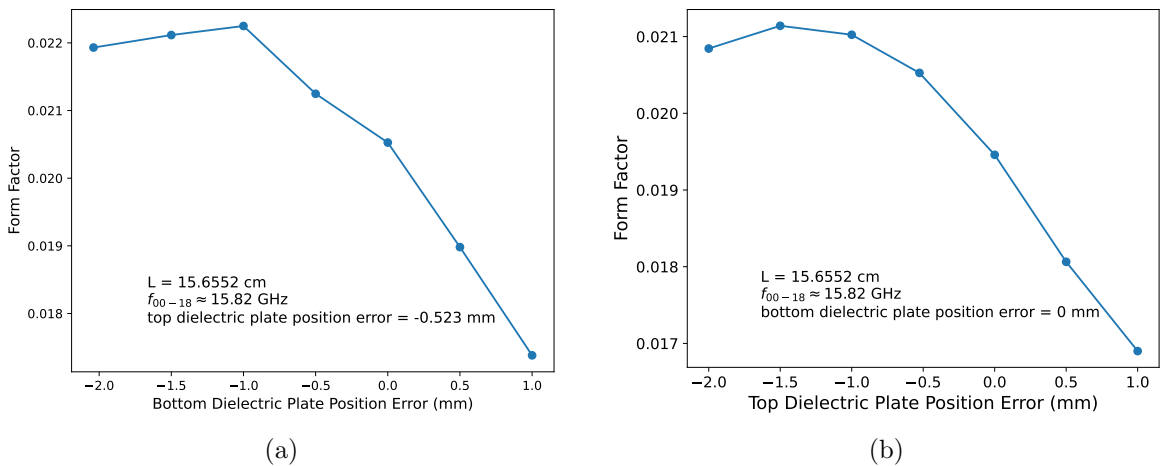

  \centering
  \subfloat[]{\includegraphics[width=0.46\textwidth]{perturbation_studies_ff_bottom_dielectric_plate.pdf}}\hfil
  \subfloat[]{\includegraphics[width=0.46\textwidth]{perturbation_studies_ff_top_dielectric_plate.pdf}}
  \caption{The effect of position error on the form factor.}
\end{figure}

\FloatBarrier

\subsection{Characterization of TEM$_{00-18}$ Mode}\label{sec:tem_characterization}
\begin{figure}[ht]
  \centering
  \includegraphics[height=0.35\textheight]{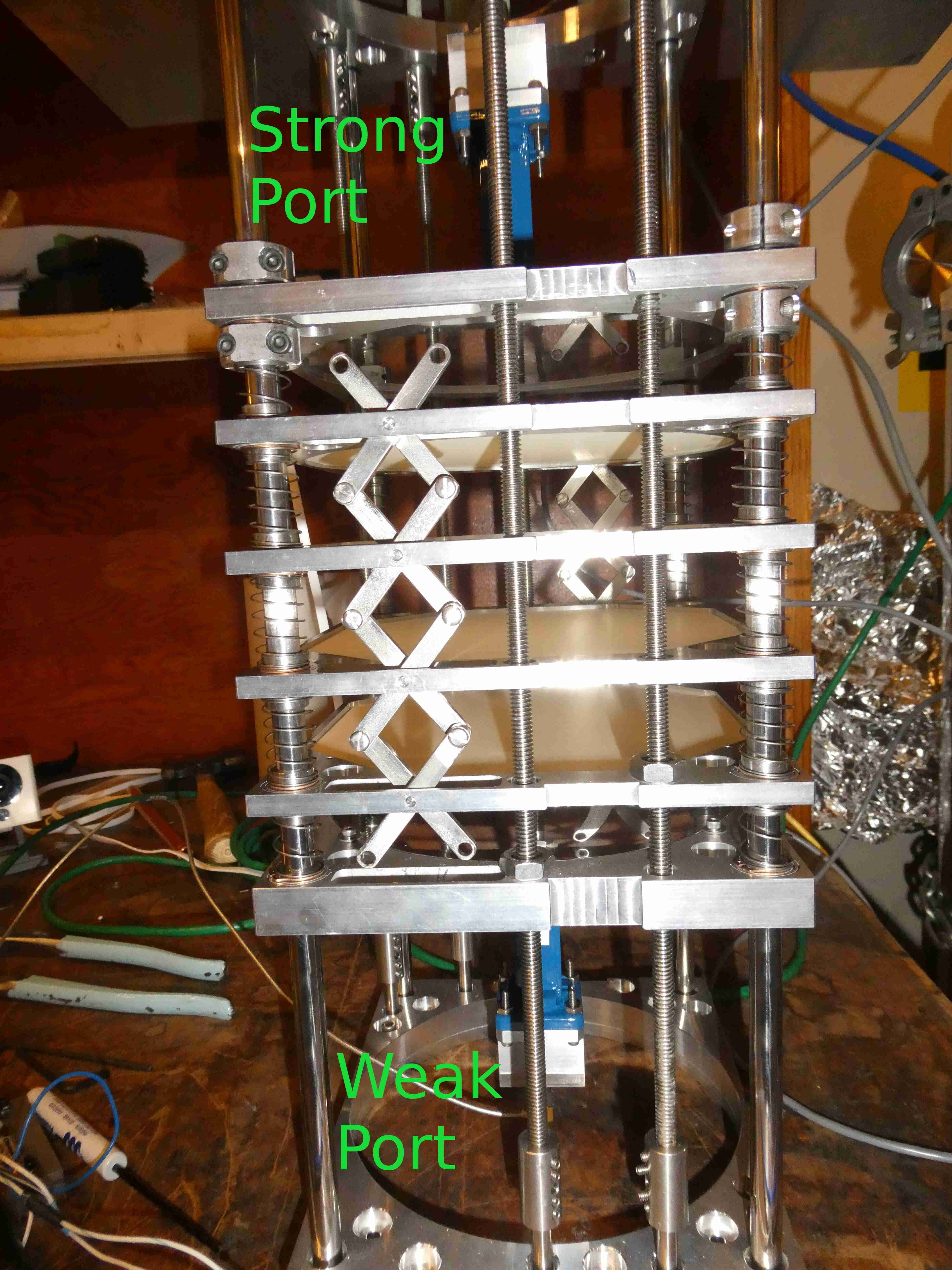}
  \caption{The tabletop Orpheus cavity.}
  \label{fig:tabletop_orpheus}
\end{figure}

Figure~\ref{fig:tabletop_orpheus} shows a tabletop dielectric-loaded Fabry-Perot cavity. The \tem mode is characterized in the same way as the empty cavity (Section:\ref{sec:empty_measurement}). The transmission and reflection coefficients are measured, and $Q_L$ and $\beta$ are extracted from those measurements.

\begin{figure}[ht]
  \centering
  \subfloat[]{\includegraphics[width=0.47\textwidth]{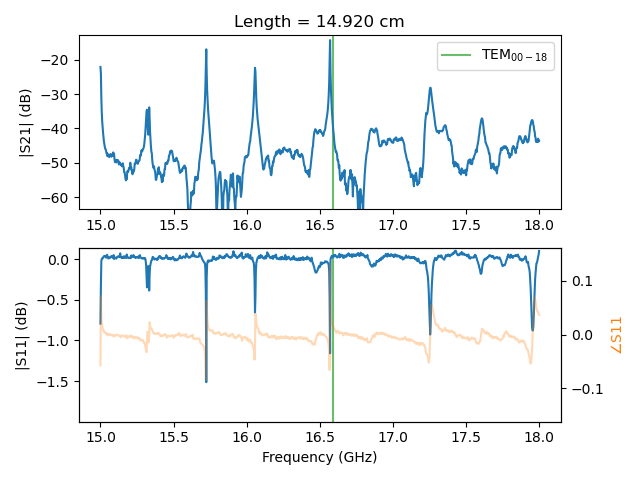}}\hfil
  \subfloat[]{\includegraphics[width=0.47\textwidth]{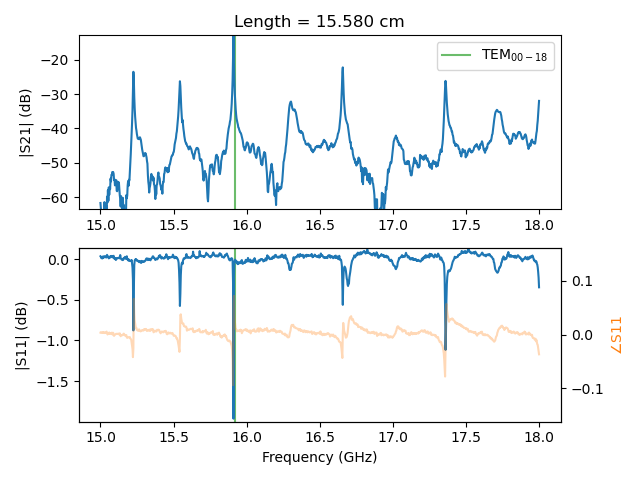}}
  \caption{The measured scattering parameters $S_{21}$ and $S_{11}$ of an dielectrically-loaded Fabry-Perot cavity as a function of frequency for a fixed cavity length. The prominent peaks are Lorentzian and correspond to different Gaussian modes.}
  \label{fig:tabletop_orpheus_widescan}
\end{figure}

\begin{figure}[ht]
  \centering
  \subfloat[]{\includegraphics[width=0.47\textwidth]{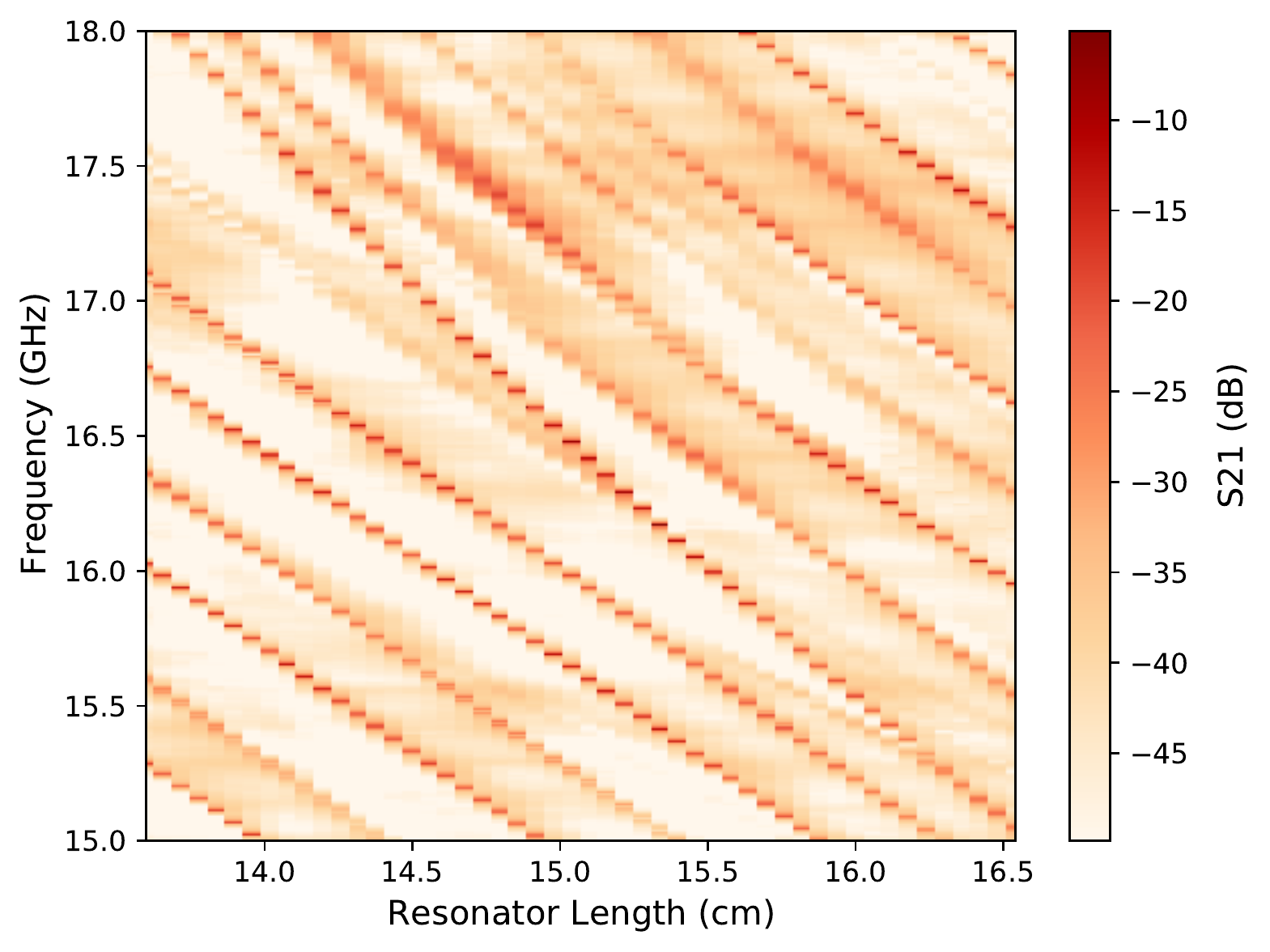}}\hfil
  \subfloat[]{\includegraphics[width=0.47\textwidth]{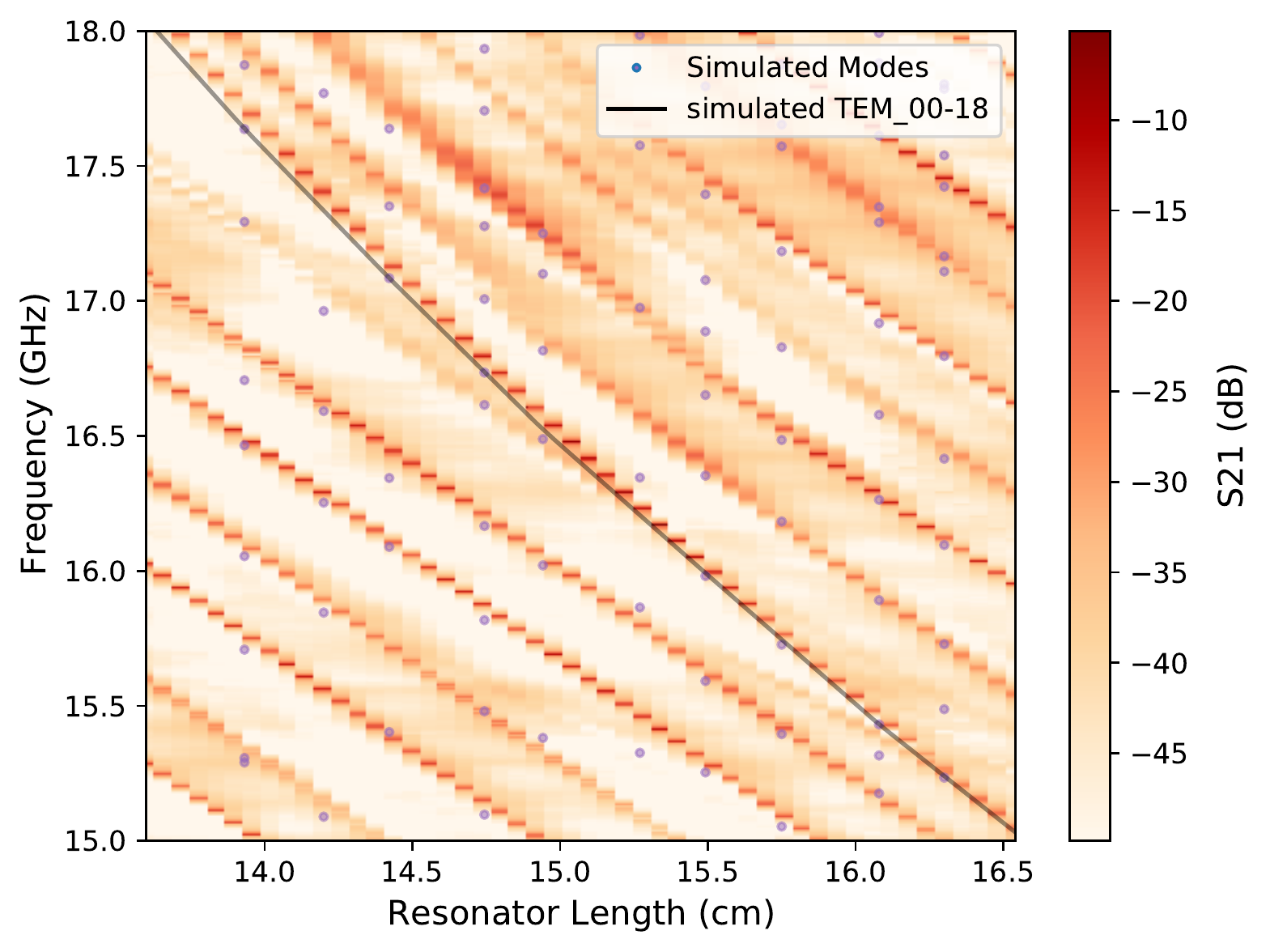}}
  \caption{Measured mode map for dielectrically-loaded Fabry-Perot cavity. This is the measured scattering parameter $S_{21}$ as a function of frequency and cavity length. The dark lines correspond to the modes of the cavity where transmission is highest. (b) The simulated modes are overlaid on the measurement.}
  \label{fig:tabletop_orpheus_modemap}
\end{figure}

The wide scan measurements are plotted in Figure~\ref{fig:tabletop_orpheus_widescan}. They are more feature-rich than the empty case because there are more resonating substructures in the cavity\footnote{The dielectrically-loaded cavity can be thought of as a series of resonators.}. The wide scan measurements can be combined to form the mode map shown in Figure~\ref{fig:tabletop_orpheus_modemap}. 

\begin{figure}[ht]
  \centering
  \subfloat[]{\includegraphics[width=0.47\textwidth]{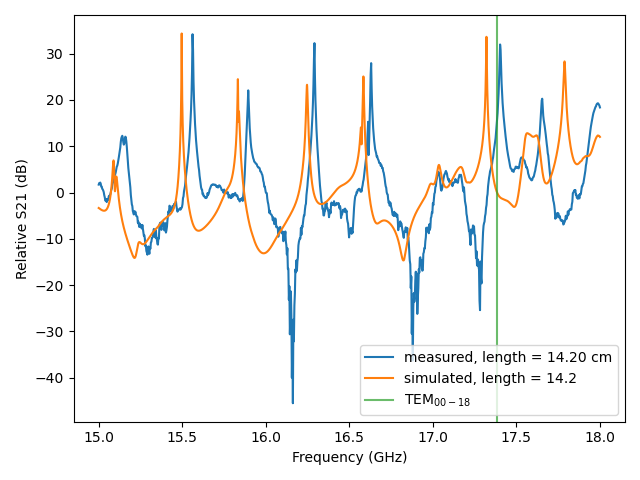}}\hfil
  \subfloat[]{\includegraphics[width=0.47\textwidth]{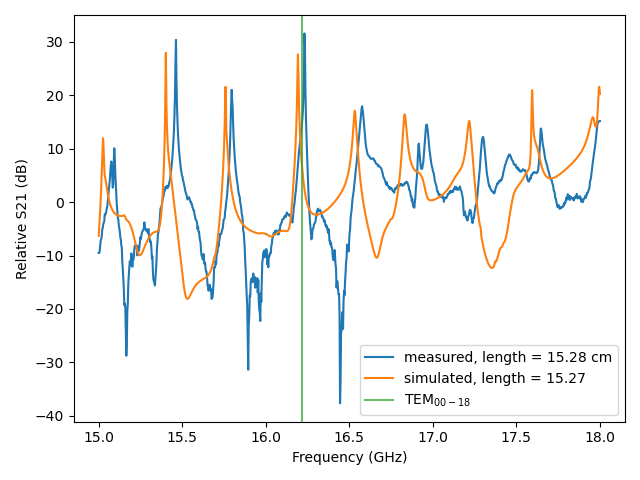}}
  \caption{The measured transmission wide scans are compared to the simulated wide scans. The general shape between the two matches. There is some discrepancy in the location of the resonances. This may be improved by demanding more accuracy from the wideband simulation, but this improving the agreement doesn't have much bearing on the science results.}
  \label{fig:sim_meas_orpheus}
\end{figure}

The measured wide scans can be compared to the simulated wide scans, as shown in Figure~\ref{fig:sim_meas_orpheus}. The simulated modes are then overlaid on the mode map. With some discrepancies, the simulated modes match the measured mode structure well. There is some deviation at higher frequencies and smaller cavity lengths. This discrepancy has several possible explanations:
\begin{enumerate}
  \item There is a fixed systematic uncertainty in the cavity length, which results in a larger relative uncertainty for smaller cavity lengths.
  \item The simulation assumes a fixed dielectric constant. But the dielectric constant may decrease notably at higher frequencies.
  \item I am using a single interpolating sweep to simulate the cavity modes from \SI{15}{GHz} to \SI{18}{GHz}. The simulation mesh may not have enough resolution for the interpolating sweep to accurately locate the cavity modes at higher frequencies.
  \item The simulated geometry doesn't capture the entire cavity geometry. For example, I don't simulate the metallic dielectric plate holders, scissor jacks, or guide rails. All these components may shift the resonant frequencies. But I believe this effect to be subdominant because the empty cavity simulations agreed so well with measurement and analytical predictions. 
\end{enumerate}
Regardless, it is clear that the simulated mode structure follows the same patterns as the measured mode structure. 

\begin{figure}[ht]
  \centering
  \subfloat[]{\includegraphics[width=0.47\textwidth]{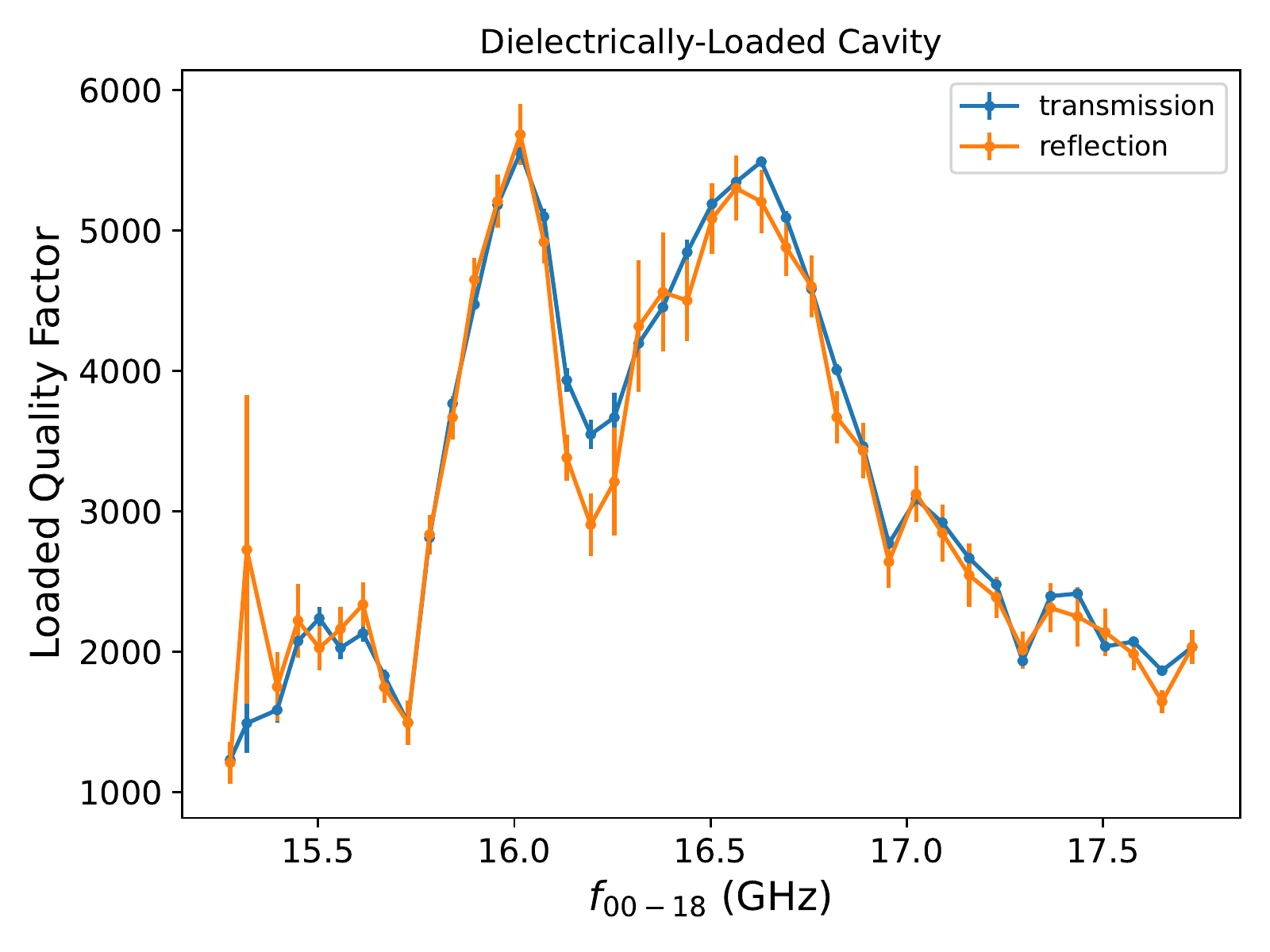}}\hfil
  \subfloat[]{\includegraphics[width=0.47\textwidth]{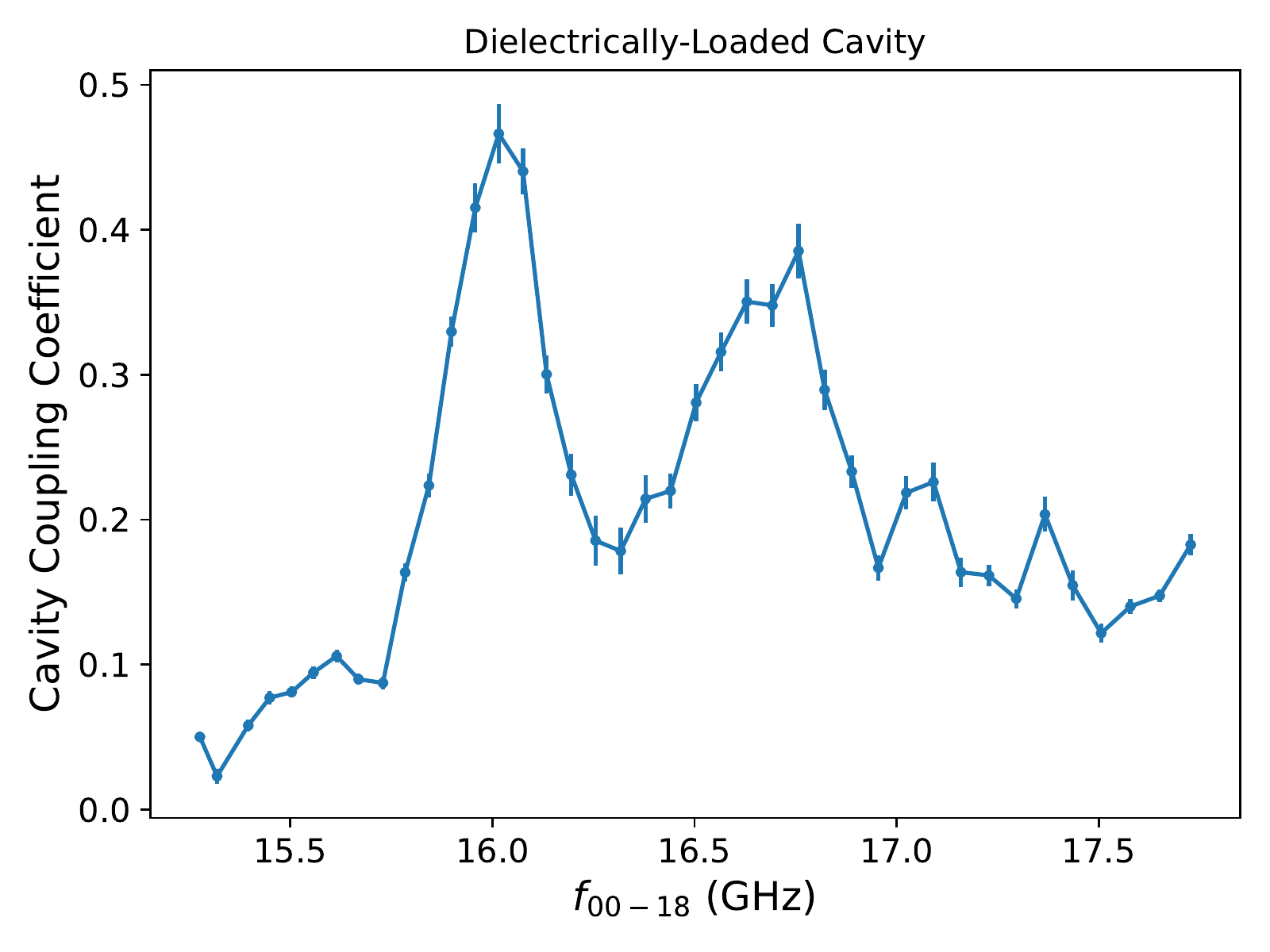}}
  \caption{The measured loaded quality factor and cavity coupling coefficient of the \tem mode as a function of cavity length. Measurements are done in air at room temperature. }
  \label{fig:tabletop_q_coupling}
\end{figure}

The quality factor and cavity coupling coefficient for the \tem mode are shown in Figure~\ref{fig:tabletop_q_coupling}. At room temperature, $Q_L \sim 5000$ and the $\beta \sim 0.4$. There is a notch in the middle of the tuning range, at about \SI{16.4}{GHz}. This notch corresponds to the mode crossing seen on the mode map in Figure~\ref{fig:tabletop_orpheus_modemap} and is corroborated by simulations discussed in Section~\ref{sec:orpheus_simulations}. 

This measurement also suggests Orpheus has a natural bandwidth. The loaded Q drops off below \SI{16}{GHz} and above \SI{17}{GHz}. This makes sense because the dielectric thickness is chosen to be about $\lambda/2$ thick at \SI{16.5}{GHz}. The more the dielectric thickness deviates from $\lambda/2$, the more destructively interfering the dielectrics become. If the dielectric thickness is $\lambda/4$, then the wave destructively interferes in the dielectric and would not transmit through the cavity. My intuition tells me that the more dielectrics there are in the cavity and the higher the dielectric constant, the smaller the cavity bandwidth. This system is reminiscent of a cavity filter.

At cryogenic temperatures, the dielectric losses are expected to drop considerably. I expect the cryogenic quality factor to be similar to that of the empty cavity in Figure~\ref{fig:empty_q_beta}. I predict that the quality will double. Since $\beta \propto Q_0$, I expect the cavity coupling to also double. With many caveats, that is roughly what is observed in the cryogenic measurement (Figures~\ref{fig:dpsearch_q_v_freq} and~\ref{fig:dpsearch_beta_v_freq}).

\begin{figure}[ht]
  \centering
  \includegraphics[height=0.4\textheight]{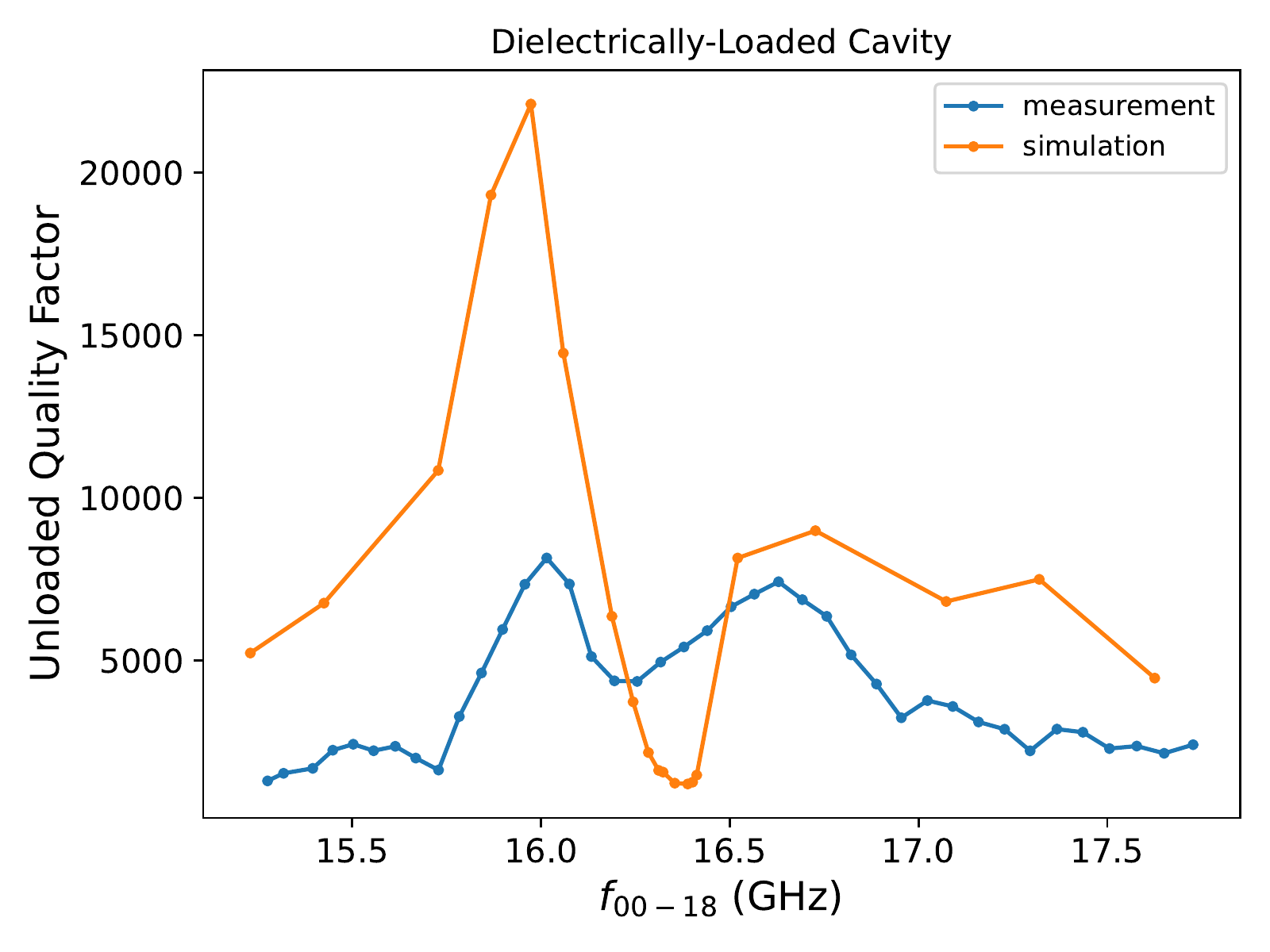}
  \caption{The simulated $Q_u$ is compared to the measured $Q_u$. The quality factor is notoriously difficult to simulate, so the values don't match. But the general shape between the two curves are similar.}
  \label{fig:orpheus_simulated_v_measured_Qu}
\end{figure}

The simulated $Q_u$ can be compared to the measured $Q_u$, as shown in Figure~\ref{fig:orpheus_simulated_v_measured_Qu}. While the values between the two don't match, the general shape matches\footnote{The mismatch is to be expected because simulating quality factors is notoriously difficult. One has to make a lot of assumptions about material properties and alignment. To have a more accurately simulated $Q_u$, one may have to include small features like screws and gaps into the simulation. This makes the mesh size untenably large for a desktop computer.}. The notch width of the measured $Q$ matches that of the simulated $Q$. And the notch in simulated Q matches the notch in $\veff$. This suggests that notches in the measured quality factor can be used as a proxy for notches in the $\veff$. This is important because $\veff$ can't be measured directly.

\section{Coupling to Orpheus}\label{sec:orpheus_coupling}
Ultimately, power is extracted from the Orpheus cavity with aperture coupling and with a WR62 waveguide. The aperture is about \SI{0.213}{in} in diameter and \SI{0.15}{in} thick. There are several alternatives and modifications to this scheme that were explored but not pursued.

\subsection{Coaxial antenna vs. waveguide aperture coupling}
The downside to aperture coupling is that the hole size is fixed for the entire tuning range, and the coupling coefficient can't be adjusted in-situ.

\begin{figure}[ht]
  \centering
  \subfloat{\includegraphics[width=0.46\textwidth]{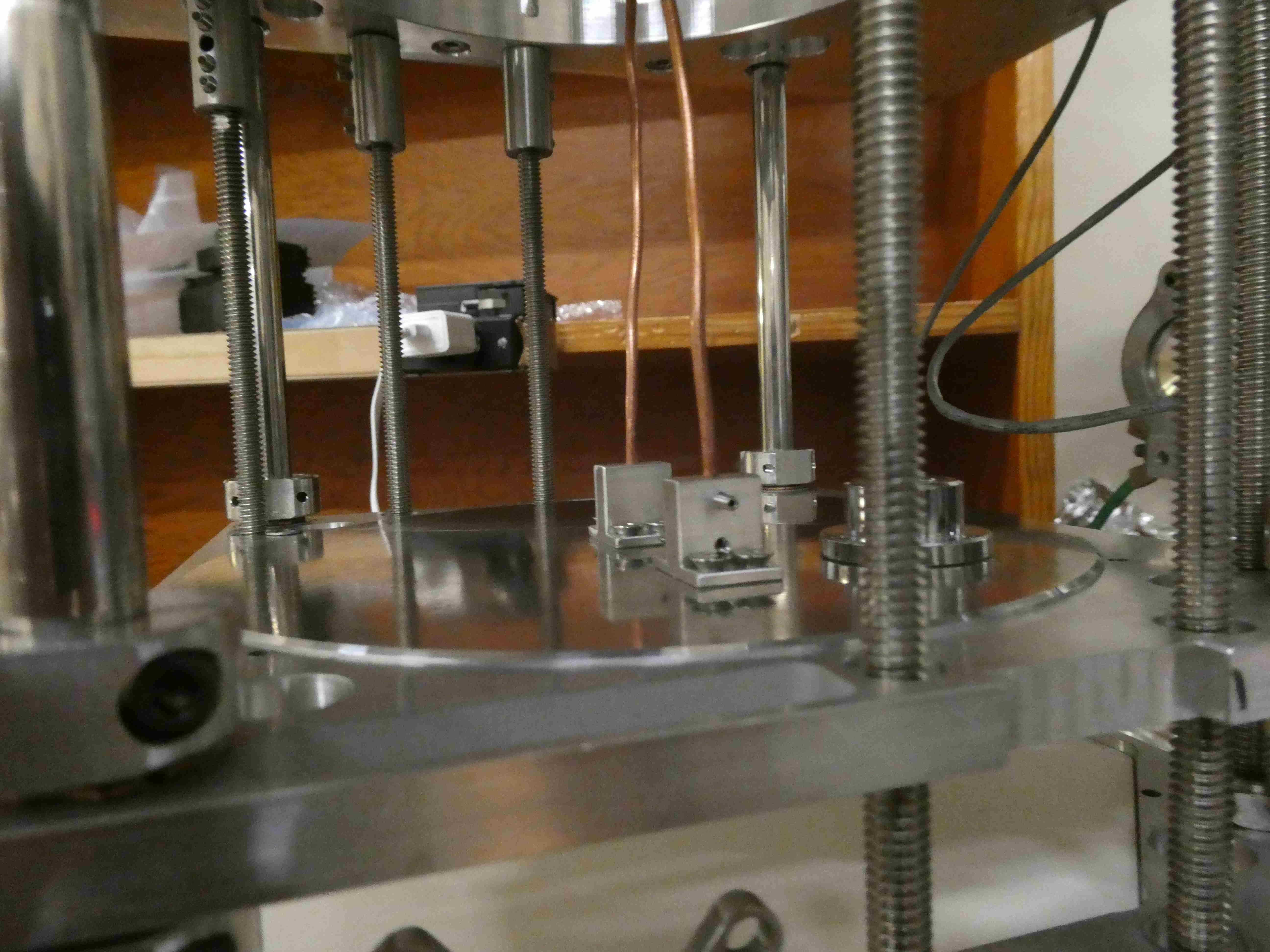}}\hfil
  \subfloat{\includegraphics[width=0.46\textwidth]{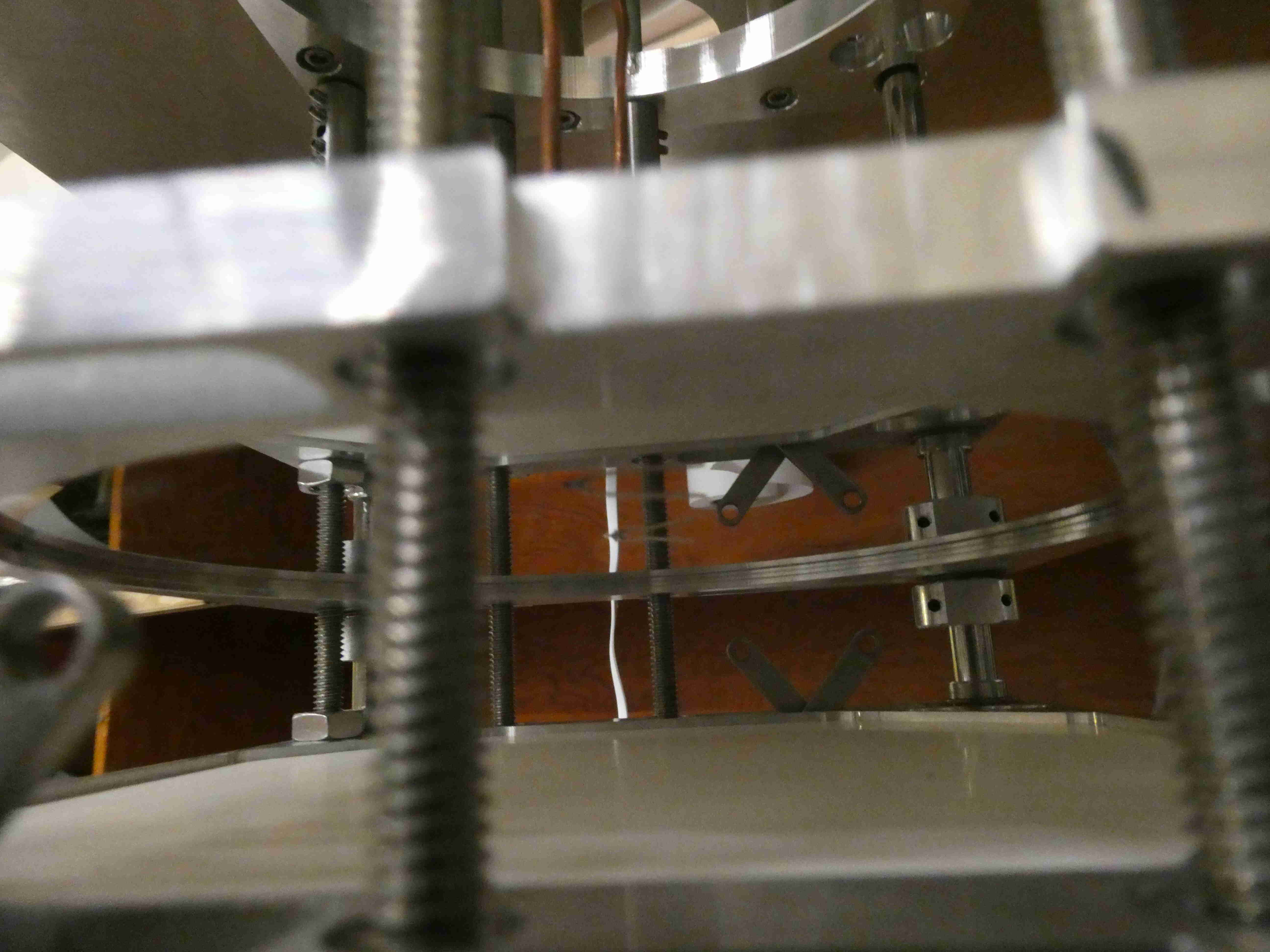}}
  \caption{Two dipole antennas were inserted into the cavity through the flat mirror and bent so they are transverse to the axis of the cavity. The antenna in the center of the mirror corresponds to the strongly-coupled port; the antenna further from the center corresponds to the weakly-coupled port.}
  \label{fig:antenna_setup}
\end{figure}

Using a coaxial antenna to couple to the cavity is attractive because there are well-established ways to adjust the coupling; a motor could adjust the insertion depth of the antenna. An antenna setup is shown in Figure~\ref{fig:antenna_setup}. Two holes are drilled on the flat mirror for the antenna ports. The weakly-coupled antenna is placed \SI{3}{in} away from the strongly-coupled antenna at the center.

Unfortunately, the antennas performed worse than aperture coupling. The quality factor was worse by a factor of two or three. Even more unfortunate, the degenerate modes split and were shifted by about \SI{100}{MHz} from where they are predicted to be. In contrast, the modes measured by aperture coupling are consistent with what is expected. It is evident that this antenna coupling mechanism strongly perturbs the cavity mode structure.

Antenna coupling is also harder to simulate than aperture coupling. This is because the aperture coupling affords me two symmetry boundaries to reduce the needed computing power. However, an antenna would only allow me at most one symmetry boundary.

It's possible that antenna coupling may have worked better if the antennas were inserted from the side of the cavity instead of through the mirror. But it would have been challenging to design a cavity coupling adjustment mechanism.

For all these complications, coaxial antenna coupling was not pursued any further.

\FloatBarrier
\begin{figure}[ht]
  \centering
  \subfloat[Empty Fabry-Perot Cavity]{\includegraphics[width=0.49\textwidth]{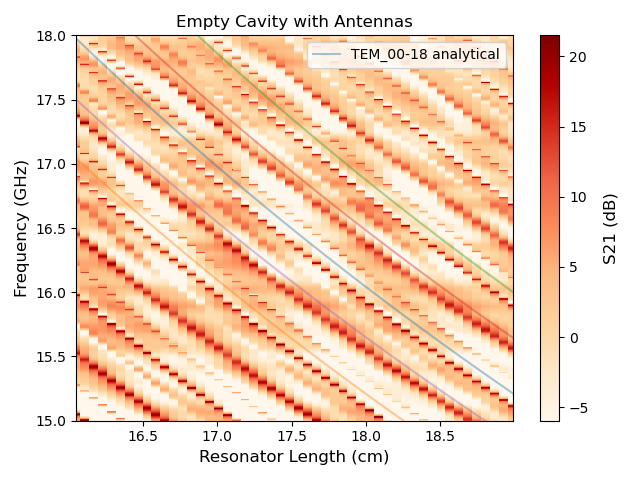}}\hfil
  \subfloat[Dielectrically-loaded Fabry-Perot Cavity]{\includegraphics[width=0.49\textwidth]{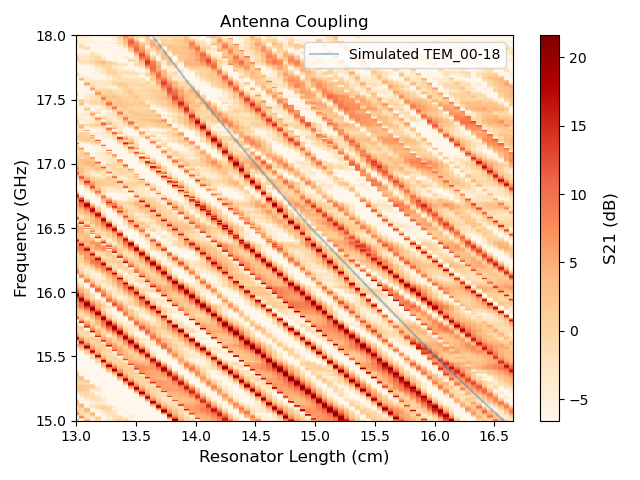}}
  \caption{The mode map of an empty Fabry-Perot cavity and dielectrically-loaded Fabry-Perot cavity that's coupled using coaxial antennas. (a) The doubly-degenerate TEM modes are split into two. The modes with the polarization that most closely lines up with the orientation of the dipole antennas are pushed lower in frequency by about \SI{100}{MHz} compared to where they are expected to be from the analytical formula. The other polarization is pushed to higher frequencies than the analytical expectation and couples more weakly to the antennas. (b) The \tem mode is lower than the simulated prediction by about \SI{100}{MHz}.}
\end{figure}

\subsection{How thin should the hole be?}
\begin{figure}[ht]
  \centering
  \subfloat[]{\includegraphics[height=0.25\textheight]{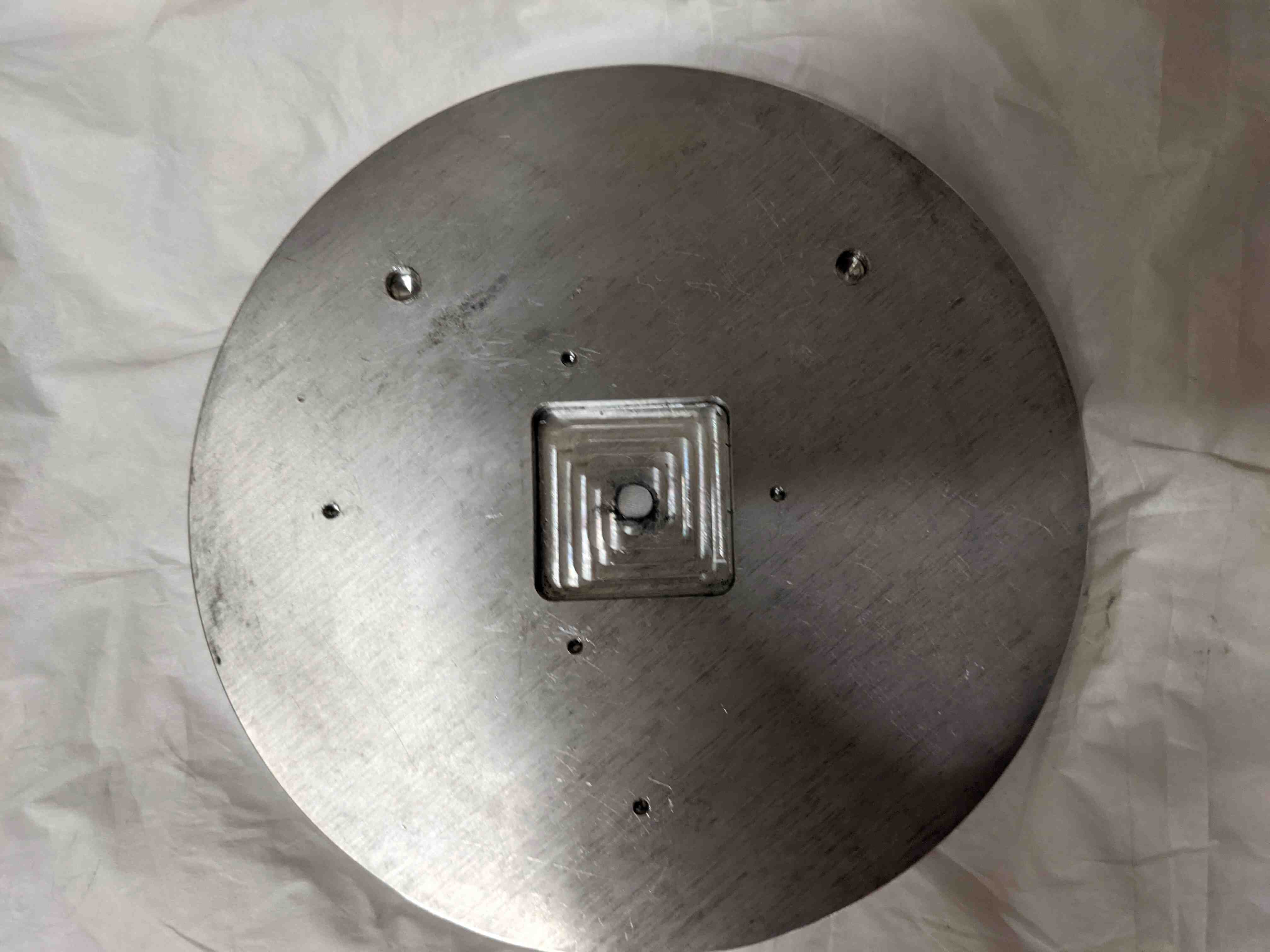}}\hfil
  \subfloat[]{\includegraphics[height=0.25\textheight]{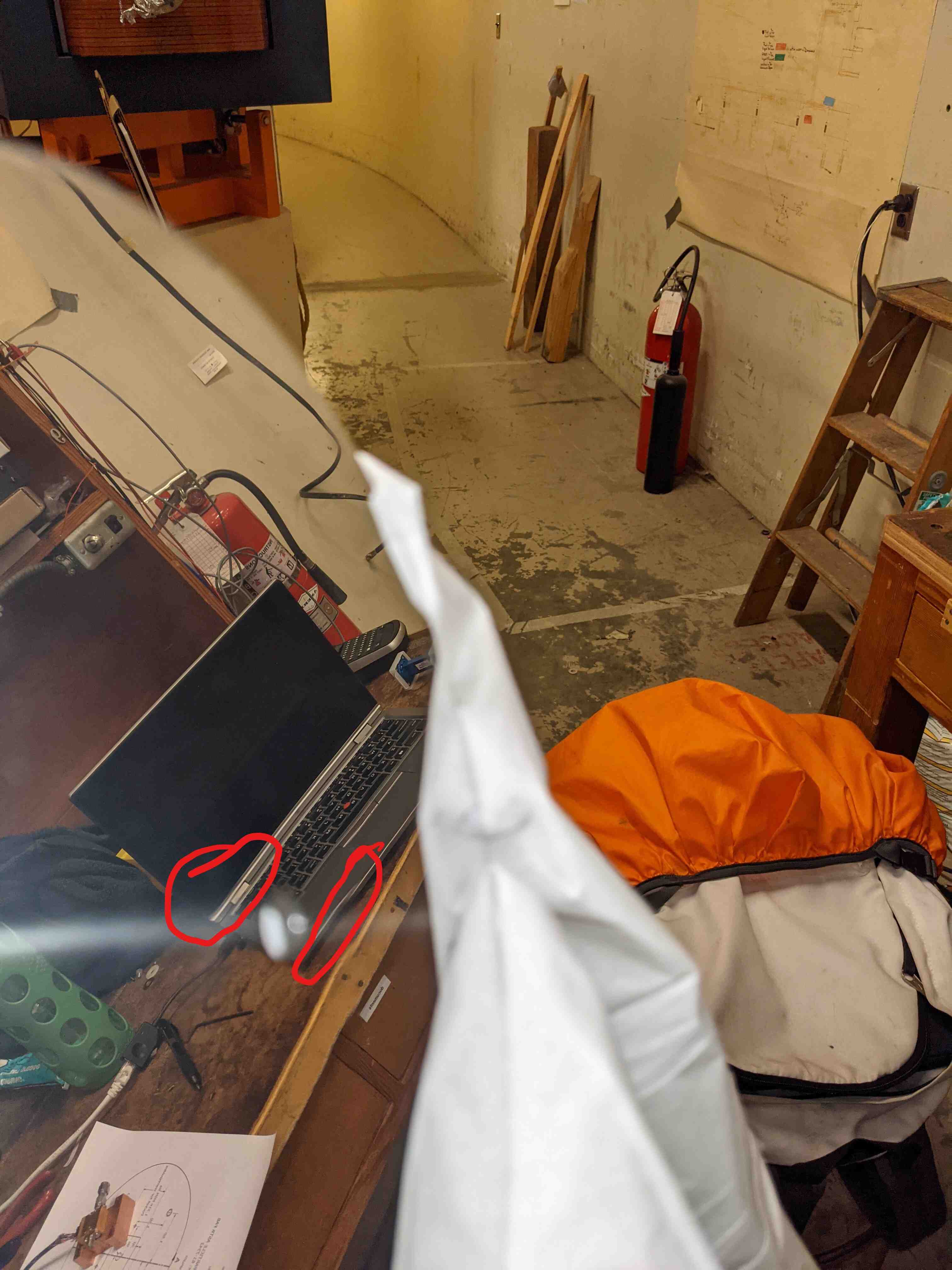}}
  \caption{The waveguide fits into a mirror pocket shown in (a). The mirror thickness at the pocket is \SI{0.15}{in.}. (b) The thin mirror surface improves the cavity coupling, but resulted in warping of the mirror surface around the beam axis. This likely reduced the Q.}
  \label{fig:waveguide_pocket}
\end{figure}

The cavity coupling can be increased by making the hole as thin as possible while maintaining a sturdy mirror surface. To understand why thinner apertures increase coupling, one can examine Equation~\ref{eqn:critical_coupling_hole}. One can also think of the exponential decay of evanescent modes. The aperture thickness is reduced by machining a pocket for a rectangular waveguide into the back of the mirror, as shown in Figure~\ref{fig:waveguide_pocket}. The mirror thickness at the pocket is about \SI{0.15}{in.}. The thin surface improves the coupling substantially. Unfortunately, the thin mirror surface is warped in a subtle way around the beam axis, likely leading to a reduced quality factor.

\subsection{Diagnosing and avoiding crosstalk between ports}
Crosstalk occurs when photons from one port reach the other port without interacting with the cavity. Crosstalk is problematic because it can interfere with the cavity signals and distort the Lorentzian peaks. 

Crosstalk makes it challenging to measure $\fm$ and $Q_L$ because the transmission and reflections no longer look like Lorentzians around the resonances. In fact, sometimes, a resonance in a transmission looks like a dip instead of a peak. 

\begin{figure}[ht]
  \centering
  \subfloat[]{\includegraphics[height=0.20\textheight]{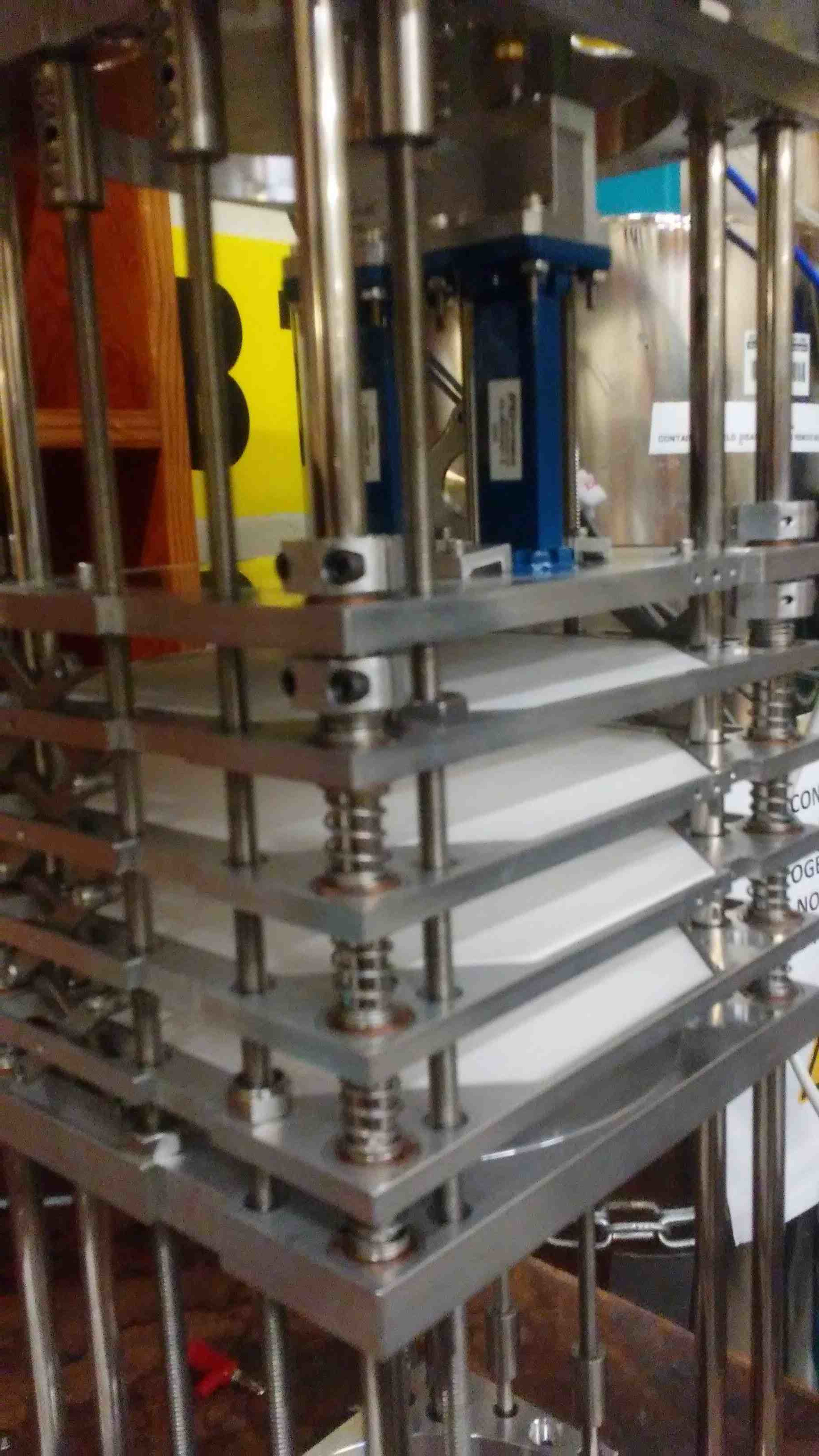}}\hfil
  \subfloat[]{\includegraphics[height=0.25\textheight]{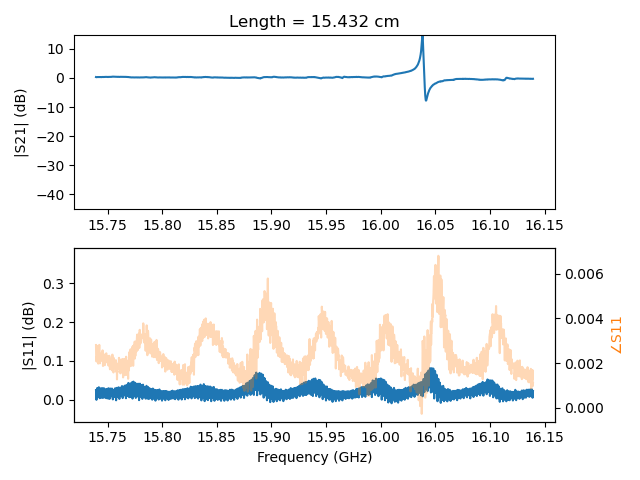}}\hfil
  \subfloat[]{\includegraphics[height=0.25\textheight]{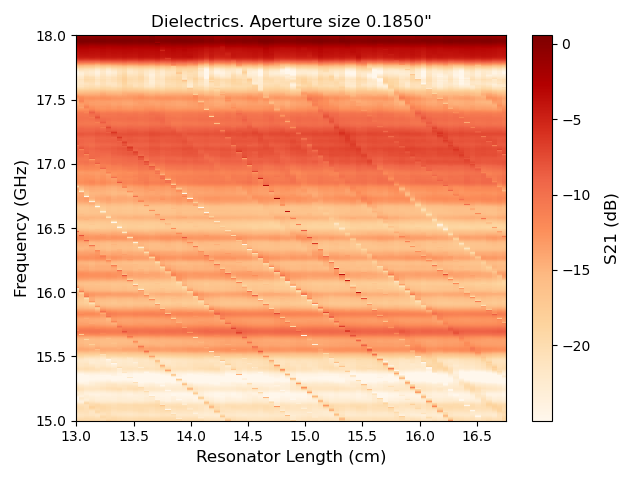}}
  \caption{Having the strongly-coupled port and weakly-coupled port right next to each other resulted in significant crosstalk. The crosstalk interferes with the cavity signal, which causes the Lorentzian peaks to distort. In the most perverse case, the crosstalk and cavity signal interfere perfectly destructively, causing a dip in the $S_{21}$ measurement where a Lorentzian peak is expected.}
  \label{fig:modemap_crosstalk}
\end{figure}

The first Orpheus design had the weakly-coupled port and strongly-coupled port on the flat mirror \SI{3}{in.} apart from each other. The corresponding transmission measurements and mode map are shown in Figure~\ref{fig:modemap_crosstalk}. The effects of crosstalk are evident when narrow scan measurement and mode map of Figure~\ref{fig:modemap_crosstalk} are compared to Figures~\ref{fig:tabletop_empty_fp_narrowscan} and~\ref{fig:tabletop_orpheus_modemap}, where the aperture ports are on opposite ends of the cavity and crosstalk is negligible. Crosstalk between the strongly-coupled port and weakly-coupled must be mitigated for proper characterization of the cavity.

\begin{figure}[ht]
  \centering
  \subfloat[]{\includegraphics[width=0.49\textwidth]{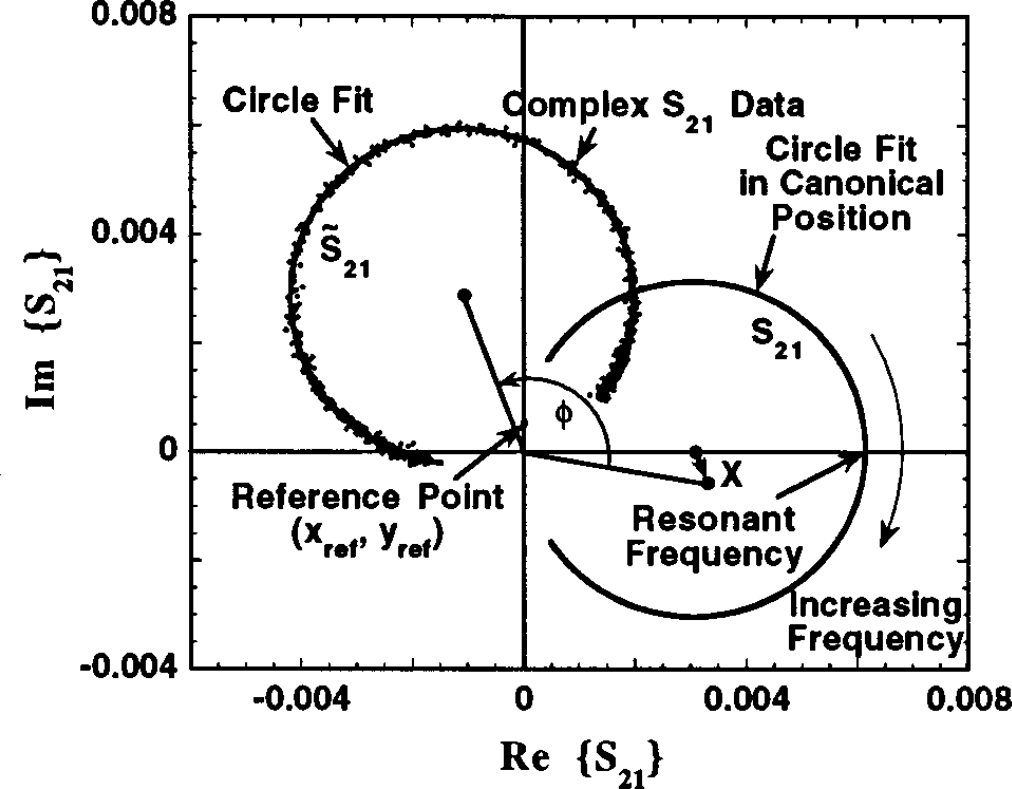}}\hfil
  \subfloat[]{\includegraphics[width=0.49\textwidth]{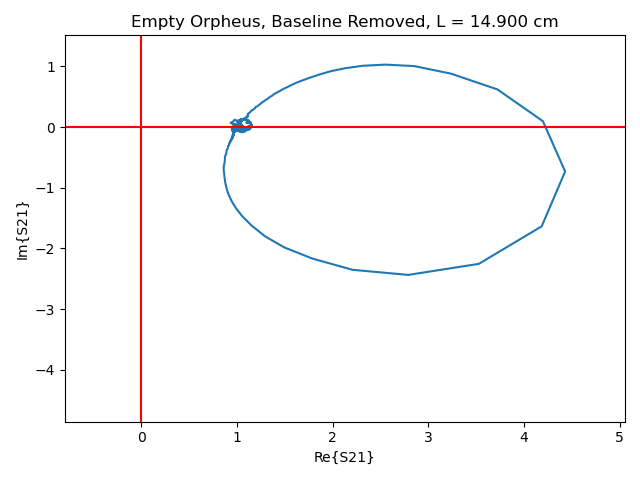}}
  \caption{(a) $S_{21}$ can be plotted in the complex plane. A cavity without any crosstalk or phase shift will have the $S_{21}$ in the canonical position. Crosstalk will cause the q-circle to be shifted from this canonical position\cite{doi:10.1063/1.368498_qcircle}. (b) This shift is observed for the measured $S_{21}$ for the configuration when the strongly-coupled and weakly-coupled ports were next to each other.}
  \label{fig:qcircle}
\end{figure}

Reference \cite{doi:10.1063/1.368498_qcircle} details how to understand and diagnose crosstalk. One can plot $S_{21}$ near resonance on the complex plane, as shown in Figure~\ref{fig:qcircle}. This is known as a Q circle. Crosstalk will shift the Q circle away from its canonical position. This crosstalk shift was observed for all transmission measurements when the ports were next to each other. 

\begin{figure}[ht]
  \centering
  \subfloat[]{\includegraphics[height=0.3\textheight]{coupling_ports_on_both_sides_compressed.jpg}}\hfil
  \subfloat[]{\includegraphics[height=0.3\textheight]{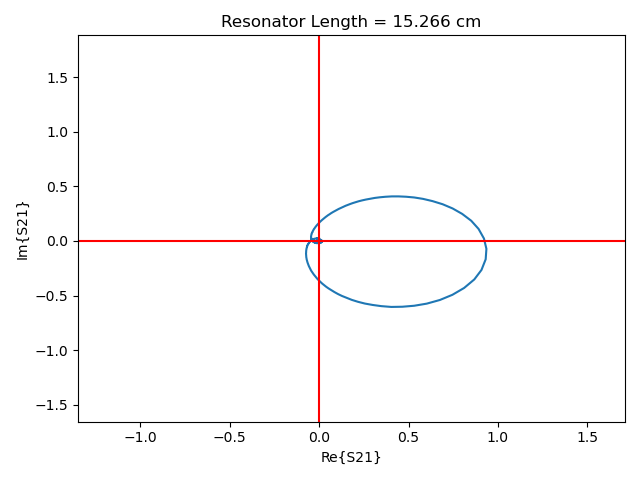}}
  \caption{Moving the weakly-coupled port to the other side of the cavity mitigated crosstalk.}
  \label{fig:no_crosstalk_qcircle}
\end{figure}

Moving the weakly-coupled port to the curved mirror mitigated crosstalk, as shown by the Q circle on Figure~\ref{fig:no_crosstalk_qcircle}.

\FloatBarrier
\subsection{Adjustable cavity coupling}
\begin{figure}[ht]
  \centering
  \includegraphics[height=0.25\textheight]{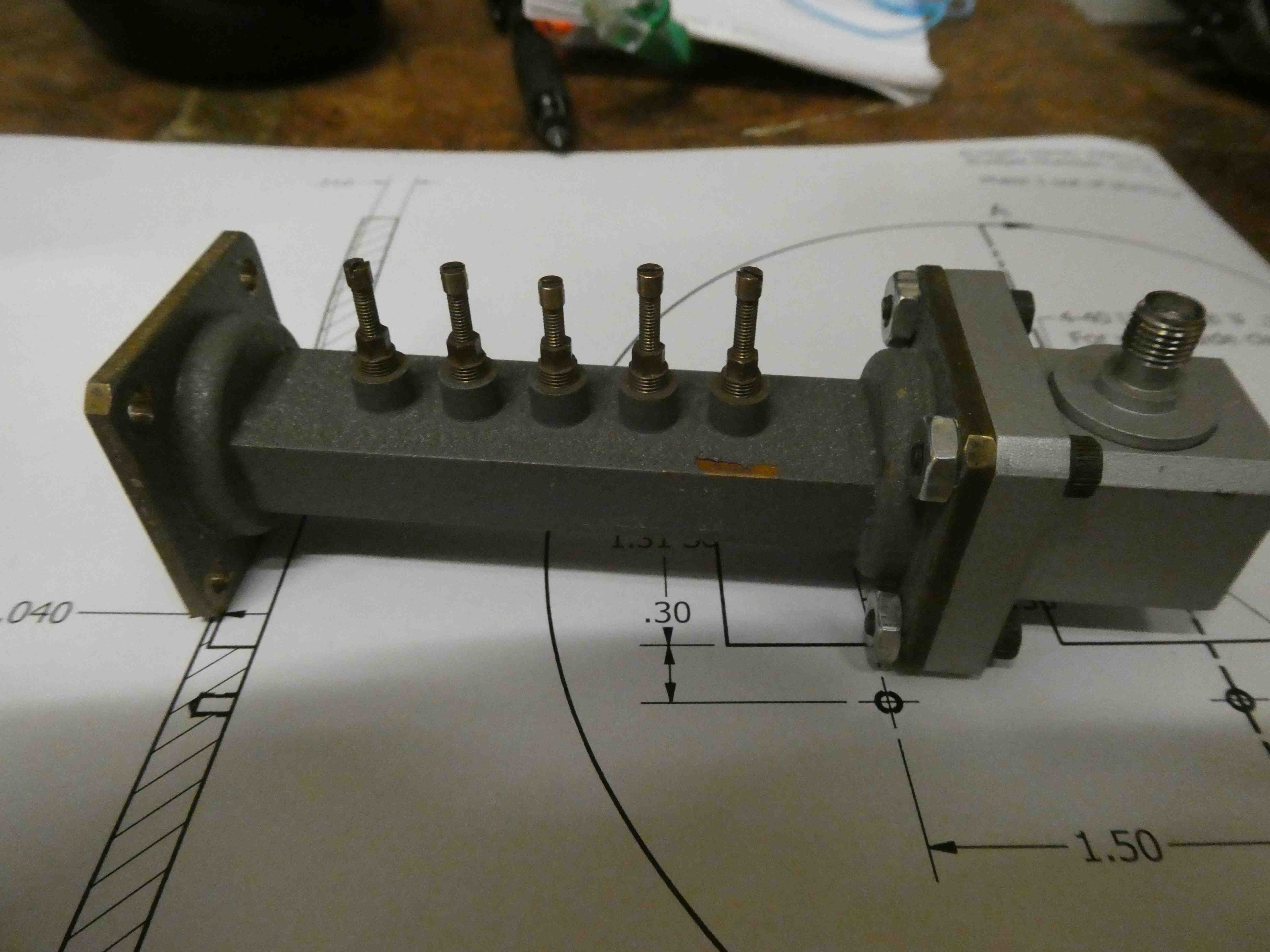}
  \caption{Waveguide tuner used to attempt to create an impedance matching network. I was unsuccessful in using it in that manner.}
  \label{fig:waveguide_tuner}
\end{figure}

Theoretically, impedance matching networks can be designed to improve coupling to the Orpheus cavity without relying solely on the size of the aperture. Pozar 4th ed. Ch.5 describes how to use single-stub tuning and double-stub tuning to match impedances. There exists waveguide tuners (shown in Figure~\ref{fig:waveguide_tuner}) that are analogous to double-stub tuners\cite{waveguide_tuner}. The waveguide tuner has a series of stubs spaced \SI{0.35}{in} apart, and these stubs can be inserted into the waveguide to change its impedance. The stub makes the waveguide either more inductive or more capacitive, depending on how far it's inserted into the waveguide. One could use a Smith chart and the double-stub tuning recipe outlined in Pozar to achieve an impedance matched network. Unfortunately, I just never got this to work. Adjusting the stubs seemed to just change the susceptibility displayed on the Smith chart without ever moving the impedance closer to the center of the chart. I tried to make this work for more than two weeks and moved on. 

%% file: cryogenic_mechanical_design.tex
\chapter{Orpheus Cryogenic and Mechanical Design}
Previous chapters discussed the conceptual design and electrodynamics of an open cavity with periodically-placed dielectrics. It is now time to talk about the practical implementation of such a cavity. There must be a mechanical structure that moves the mirrors and dielectric plates, and the mechanical structure must work in a vacuum and cryogenic environment. The cavity should also work as a room-temperature, tabletop setup for easy RF characterization and rapid prototyping. The machining tolerances need to be tight enough to maintain good alignment and achieve good $Q$, but there needs to be enough slop\footnote{Also known as backlash, lash, or play. Mechanical systems need it to prevent jamming.} in the system so that the tuning mechanism works even with small amounts of misalignment.

\section{Orpheus Cavity Mechanical Design}\label{sec:cavity_mechanics}
\begin{figure}
  \centering
  \includegraphics[height=0.7\textheight]{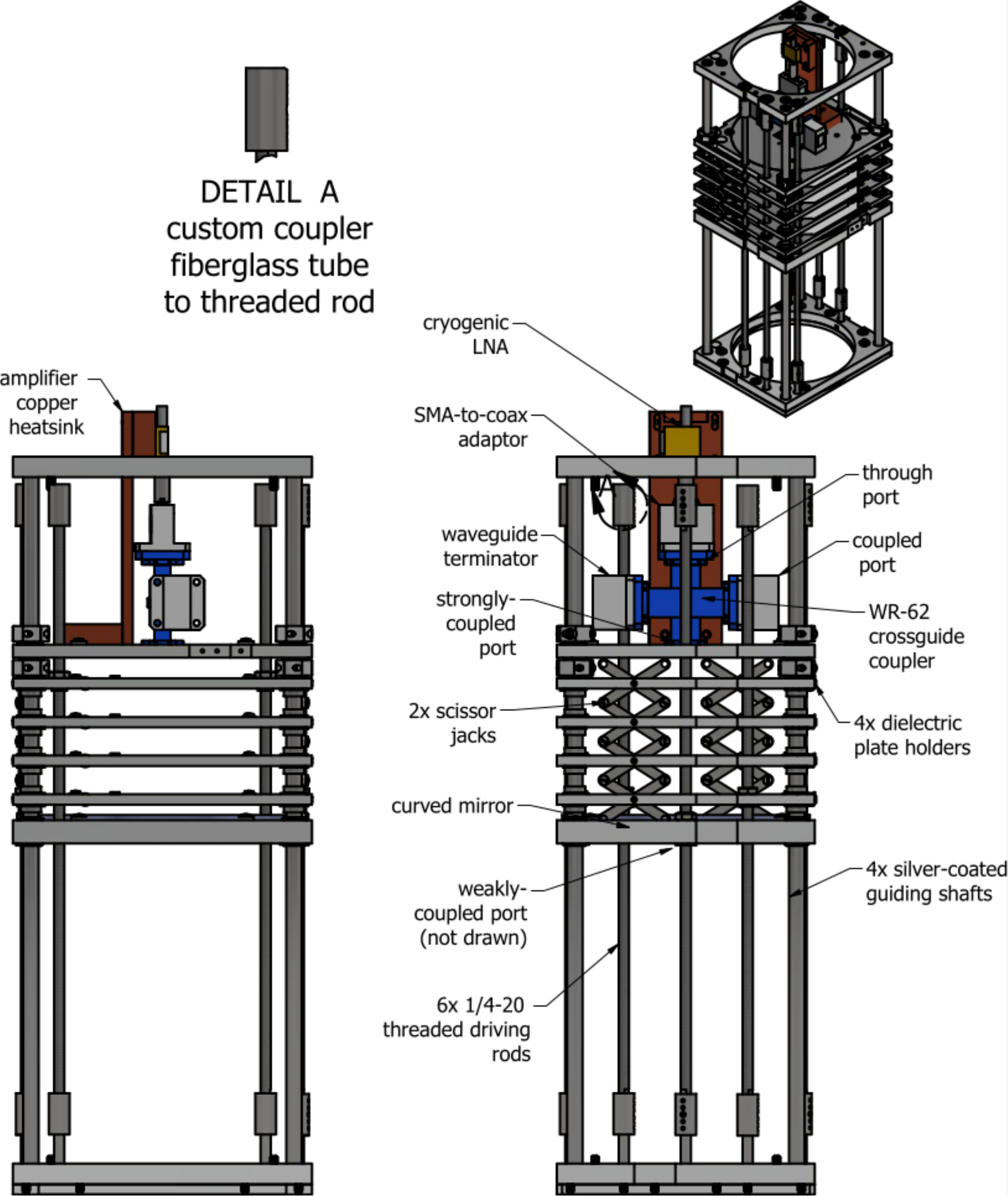}
  \caption{CAD model for the Orpheus cavity.}
  \label{fig:orpheus_cavity_cad}
\end{figure}

Most of the Orpheus mechanics were designed by Richard Ottens. I prototyped some of the mechanical concepts and made adjustments, and afterward, I commissioned the UW Physics machine shop to fabricate the cavity components in November 2019. I received the completed parts around February 2020\footnote{For those reading this in the far future, COVID-19 lockdowns started around March 2020.}. The Orpheus cavity CAD model is shown in Figure~\ref{fig:orpheus_cavity_cad}. I will describe the main design choices.

\begin{figure}
  \centering
  \includegraphics[height=0.4\textheight]{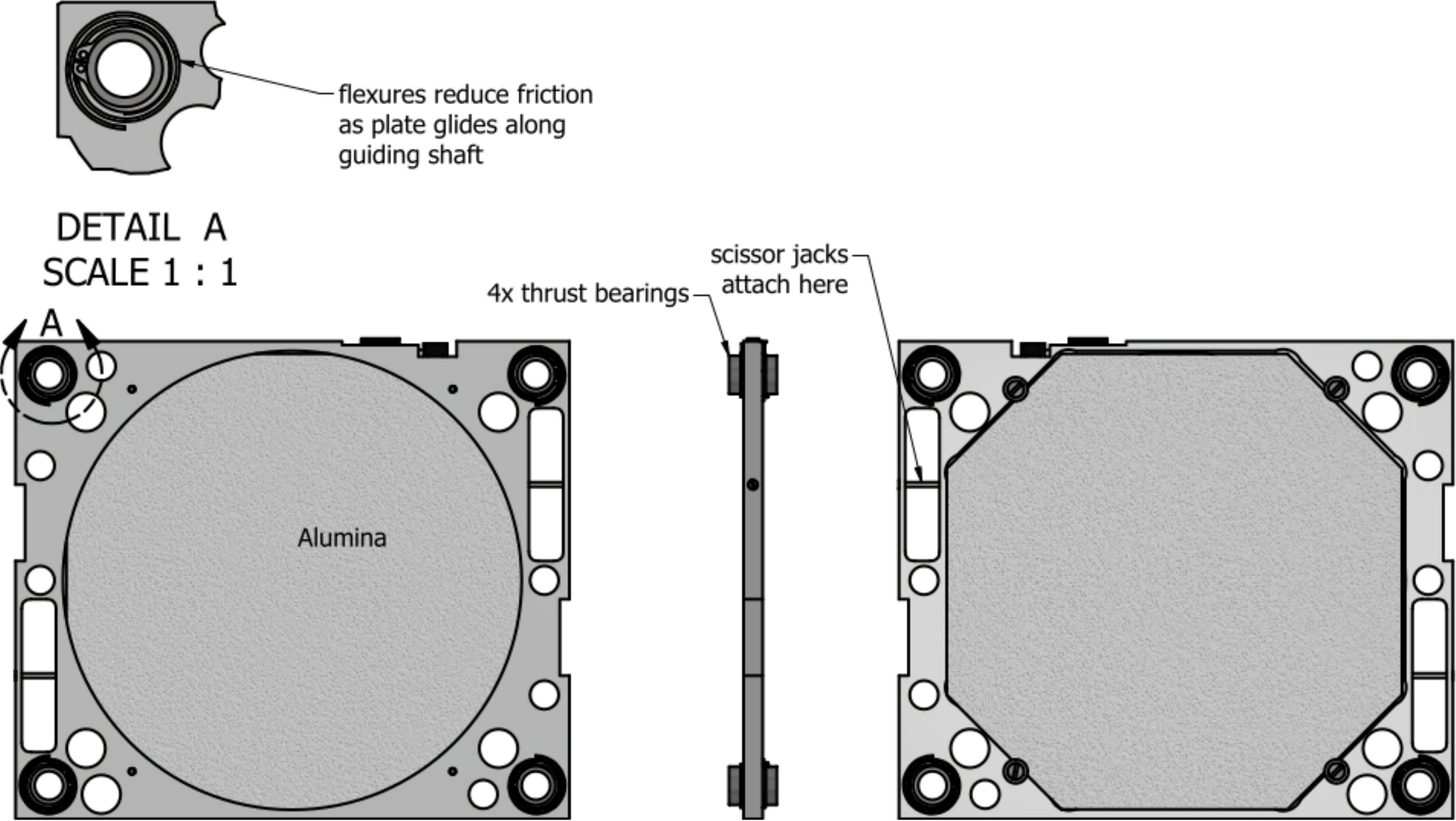}
  \caption{CAD model for one of the dielectric plates.}
  \label{fig:orpheus_plate_cad}
\end{figure}

\begin{figure}
  \centering
  \includegraphics[height=0.3\textheight]{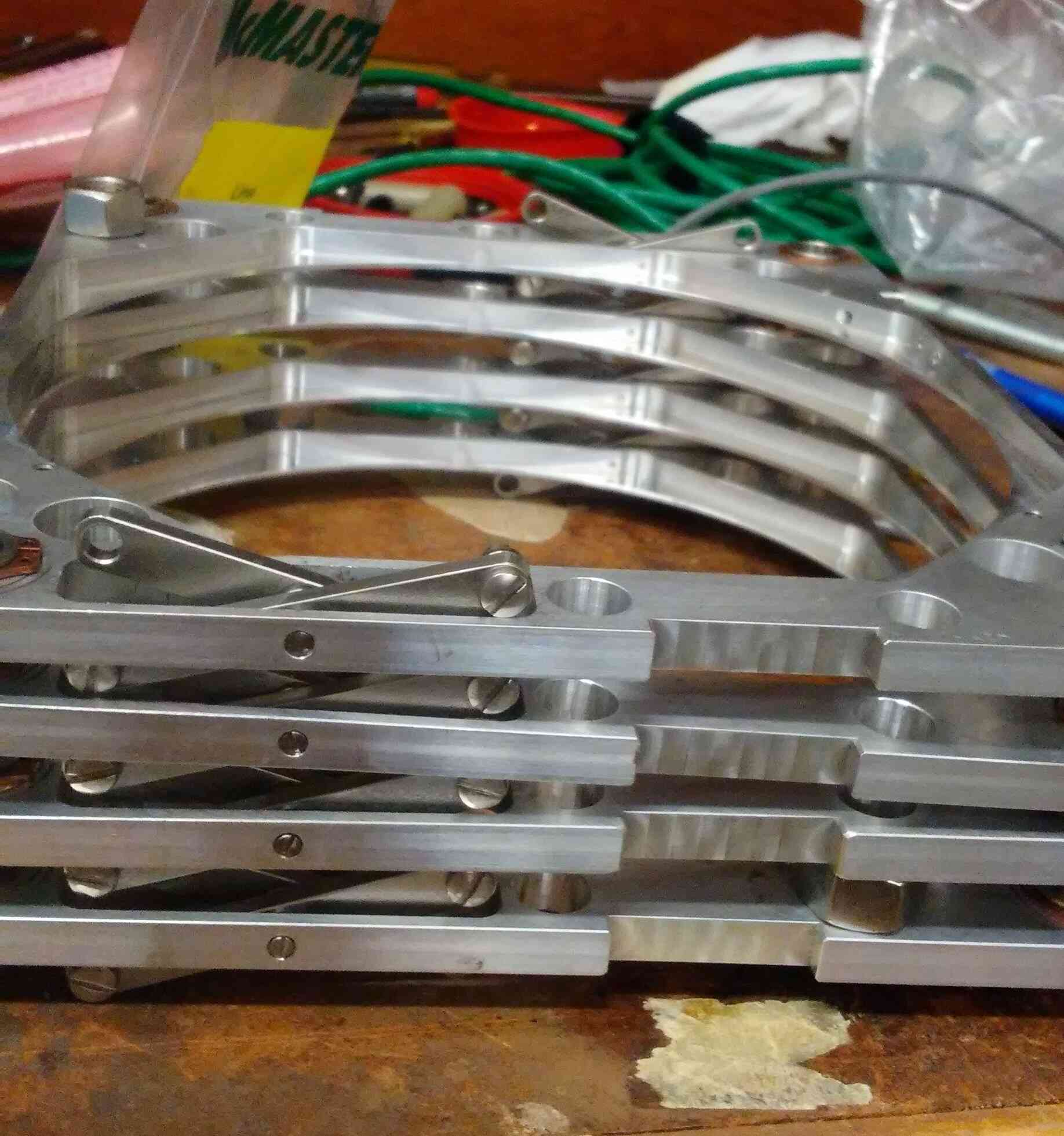}
  \caption{The dielectric holder assembly without the dielectrics. There is a lip inside the octagon where the dielectrics freely rest. Note that the thrust bearings in this photo are \SI{0.25}{in.} long are are shorter than the final bearing length of \SI{0.615}{in}.}
  \label{fig:dielectric_assembly_lip}
\end{figure}

The mirror and alumina plates are held in place by aluminum holders (Figure~\ref{fig:orpheus_plate_cad}). They rest on a lip and rely on gravity for mechanical stability\footnote{This is not a viable choice if the cavity needs to rest horizontally.}. The lip can be seen in the dielectric plate assembly in Figure~\ref{fig:dielectric_assembly_lip}. The dielectric plates sit inside the aluminum holder with lots of slack to accommodate the differential thermal expansion coefficients during cooldown. If the tolerances were tight at room temperature, the aluminum would contract faster and crush the alumina\footnote{Having this much slop isn't a viable option if the dielectrics were curved and needed to be centered on the axis of the cavity.}. Nothing is clamping down the alumina because the clamping mechanism would contract more quickly than the alumina and crush it. The mirrors are made out of aluminum, so there is no differential expansion between the mirror holders. Tighter tolerances are viable and even desired to keep the coupling holes in the axis of the cavity.

The aluminum holders containing the curved mirror, top dielectric plate, and bottom dielectric plate are moved vertically by a pair of 1/4-20'' stainless steel threaded rods. This results in a tuning mechanism with three degrees of freedom. The middle two dielectric plates are constrained to be spaced evenly between the top and bottom dielectric plate because they are attached to a pair of scissor jacks. The scissor jacks are connected to all dielectric plates, but not the mirrors. All the plates slide along four guide rails. To compensate for any possible misalignment that would cause the plates to jam, each aluminum holder has spiral cut flexures around the bearing surface.

Cavity materials were chosen to accommodate thermal contractions. Alignment is maintained after cooldown because all vertical structures are made from stainless steel, and all horizontal structures are made from aluminum. Bearings are made out of stainless steel so that the fitting tolerance between the shaft and bearing remains the same after cooldown\footnote{It would be bad if you had a stainless steel shaft and you made your sleeve bearing out of aluminum. Even if your sleeve bearing slides freely along your shaft at room temperature, the bearing will shrink more than the stainless steel after cooldown. Now your aluminum bearing is gripping the stainless steel shaft, and it's completely jammed.}. However, having stainless steel thrust bearings rub against the stainless steel guide rails would lead to galling\footnote{In general, having similar materials slide against each other causes both sliding surfaces to scratch each other, leading to galling and possibly cold welding}. All bearing surfaces are coated with silver to prevent galling and reduce friction.

Without further measures, the friction between the bearing and the guide rail would cause the plates to tilt as the cavity tunes. Two measures are taken to mitigate this tilting. First, the thrust bearings (\SI{0.615}{in.}) are much longer than the dielectric plate thickness (\SI{0.25}{in}). The bearing length gives the plates less leeway to tilt. Second, a weak compression spring is placed around the guide rails in between the dielectric plates and mirror plates. For this experiment,  LP 024L 03 S316 from Lee Spring was used. It has a spring constant of \SI{0.37}{lb/in}. If the springs are too stiff, then they deform the spiral flexures and cause large position errors.

\begin{figure}
  \centering
  \subfloat{\includegraphics[height=0.25\textheight]{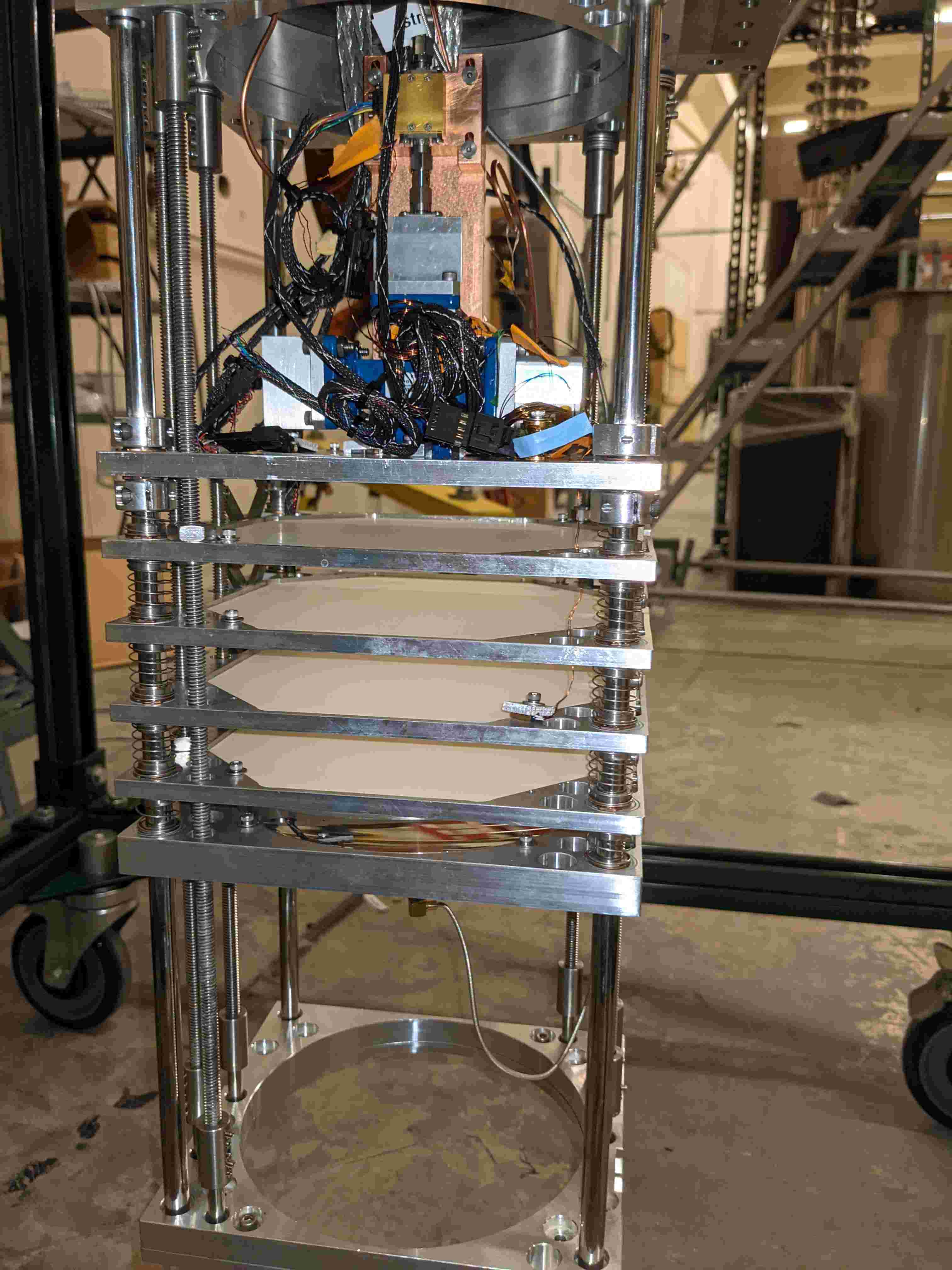}}\hfil
  \subfloat{\includegraphics[height=0.25\textheight]{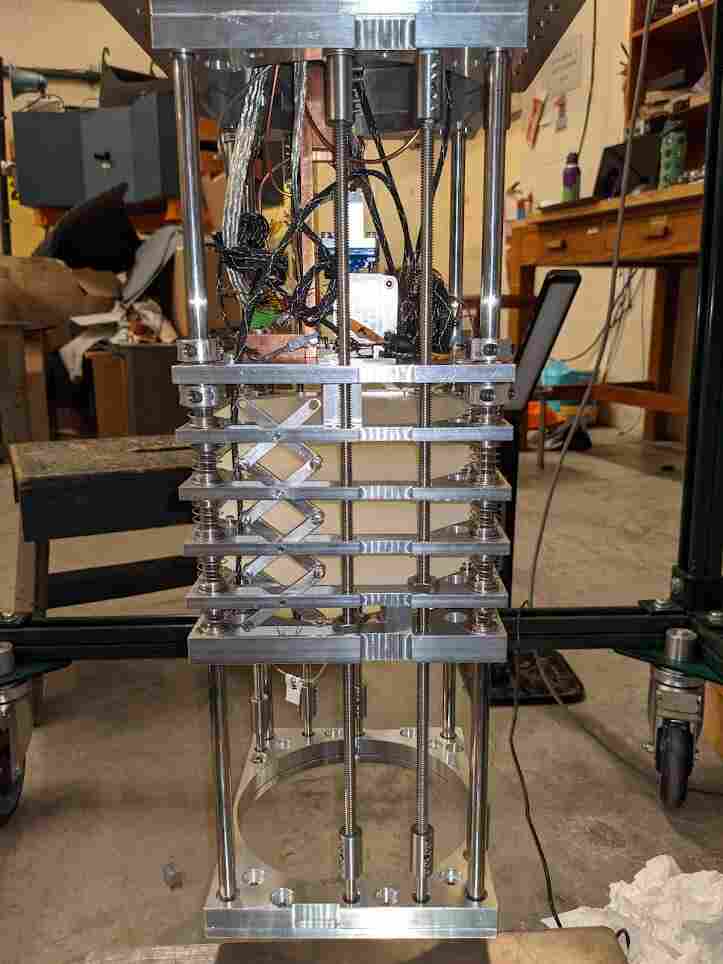}}\\
  \caption{Orpheus cavity attached to the cryogenic insert.}
  \label{fig:orpheus_real}
\end{figure}

The real cavity is shown in Figure~\ref{fig:orpheus_real}. The tuning mechanism performed to satisfaction in the tabletop setup. Measurements with a circular bubble level showed that plates tuned with negligible tilting. The position error was formally studied with a cavity that didn't have the springs. The measurements showed that the position error was often around \SI{0.2}{mm} and rarely exceeded \SI{0.4}{mm}. Simulations in Section~\ref{sec:position_error} show that $\veff$ is tolerant to this magnitude of position error. 

Overall, the cavity tunes accurately and maintains good alignment while tuning.

\FloatBarrier
\section{Insert}
\begin{figure}
  \centering
  \includegraphics[height=0.7\textheight]{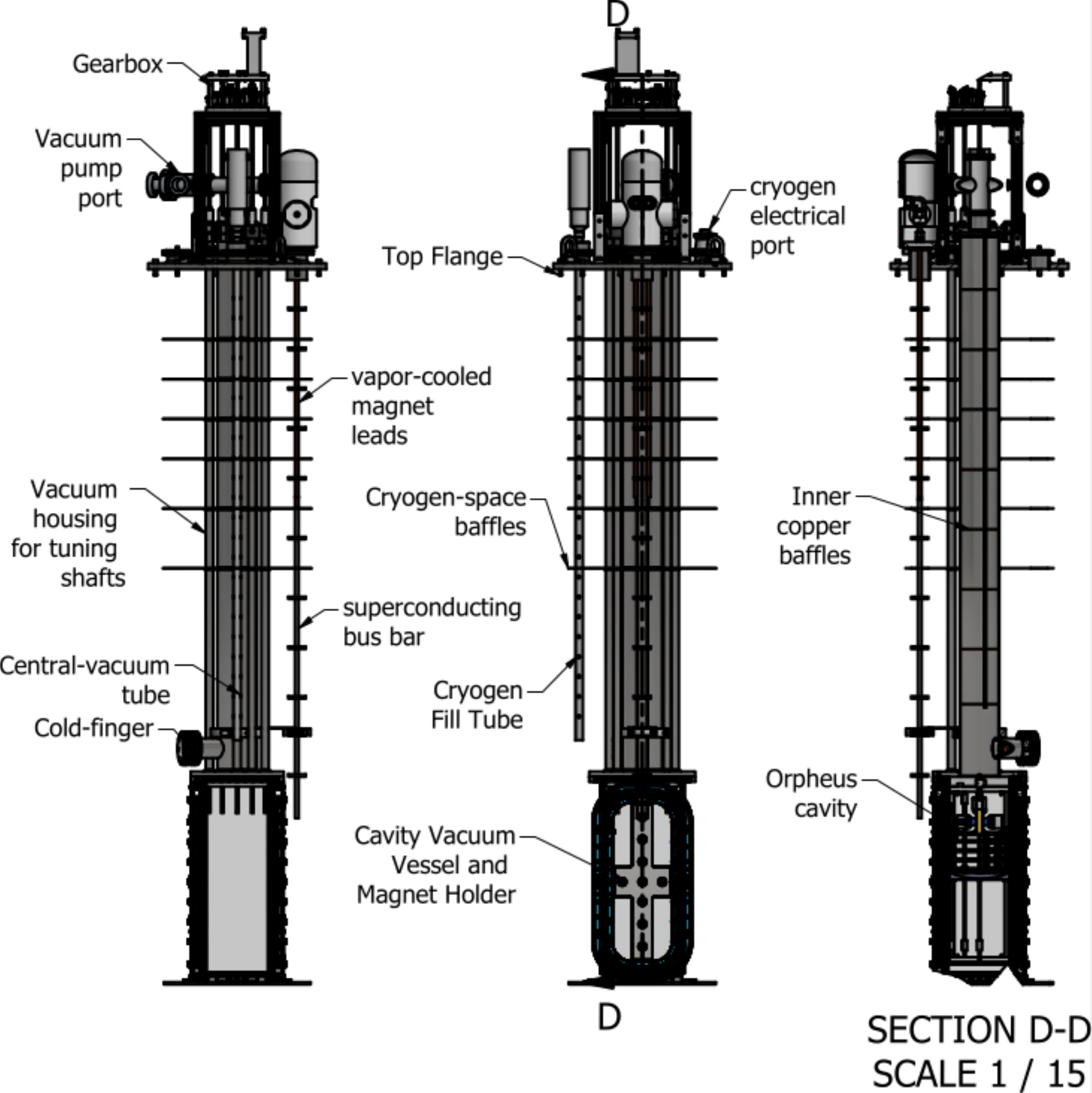}
  \caption{Orpheus cryogenic insert.}
  \label{fig:insert_cad}
\end{figure}

\begin{figure}
  \centering
  \includegraphics[height=0.4\textheight]{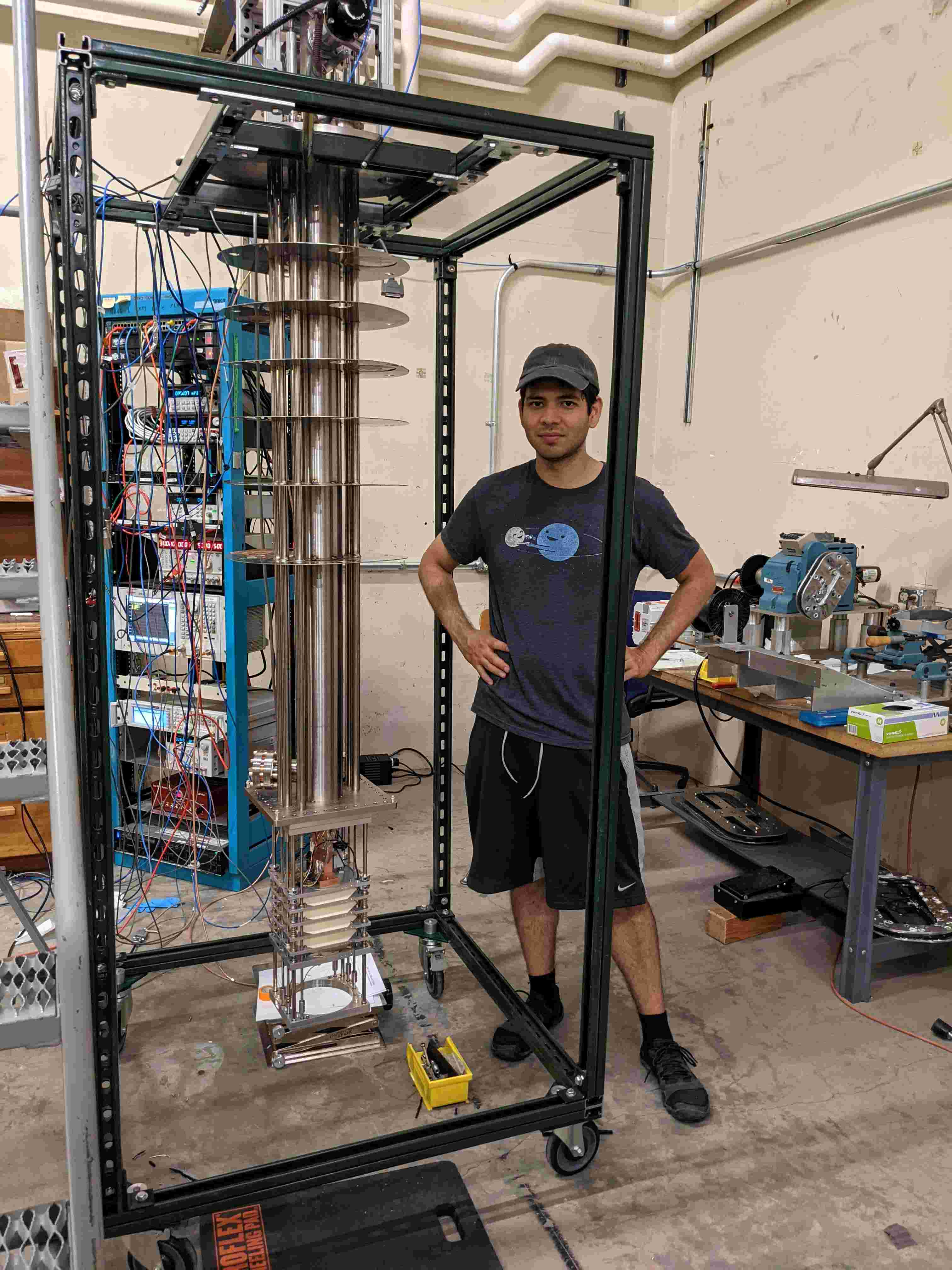}
  \caption{Cryogenic insert in real life.}
  \label{fig:insert_real}
\end{figure}

The cavity is the core of the experiment, but the cavity needs to be cooled down to \SI{4}{K} under vacuum. The cavity is kept in a vacuum instead of directly under cryogen because the cavity isn't designed to remain mechanically stable under a boiling fluid\footnote{The cavity modes were not stable when I tried to dunk it in liquid nitrogen. I've since been told that liquid helium can be very quiet. So that might be a viable option, but was not pursued in this experiment.}. The cavity is to remain cold for about a week to allow enough time to scan the tuning range. Room temperature stepper motors tune the cavity, and the stepper motors need to remain far away from the magnet. The insert in Figure~\ref{fig:insert_cad} and Figure~\ref{fig:insert_real} were designed to meet these requirements. I will describe the main components from the top down.

\begin{enumerate}
  \item Motor stage shown in Figure~\ref{fig:gearbox}. Each stepper motor (Applied Motion Products STM23S-2EE) drives a pair of miter gears that transmit power to the vertical shafts that are connected to the vacuum rotary feedthroughs on the top flange.
    \begin{figure}[h]
      \centering
      \subfloat{\includegraphics[height=0.25\textheight]{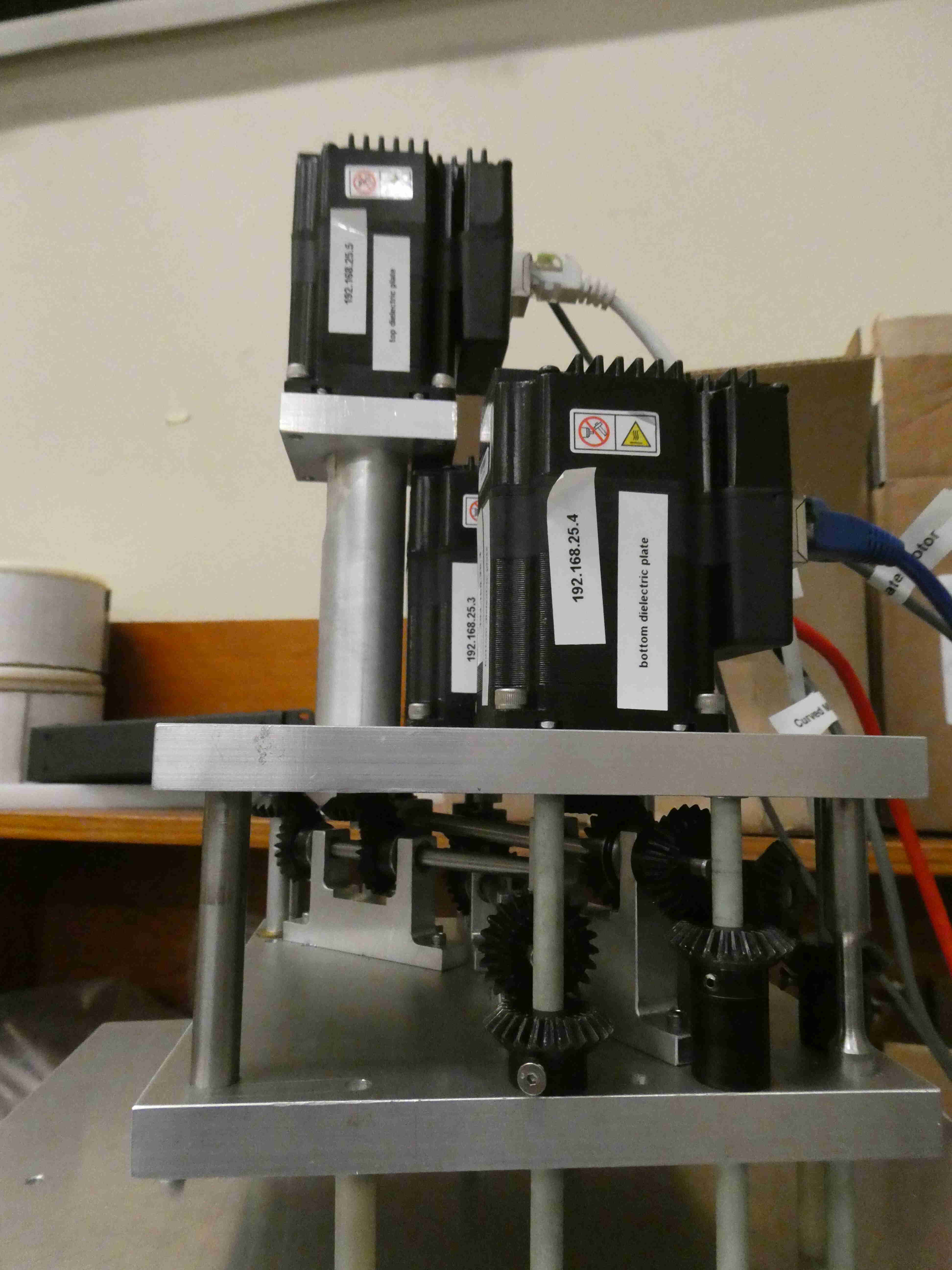}}\hfil
      \subfloat{\includegraphics[height=0.25\textheight]{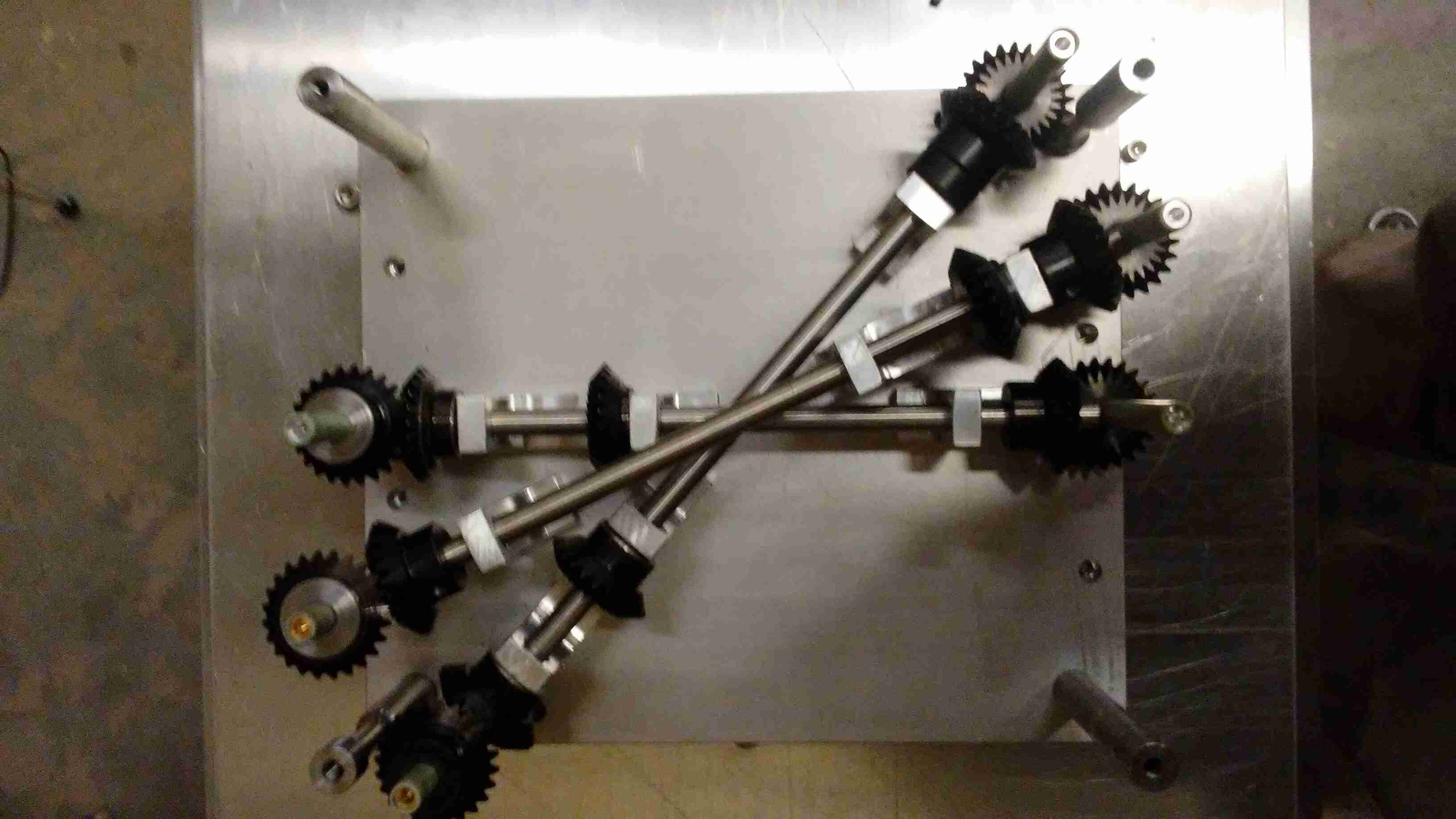}}
      \caption{Orpheus gearbox}
      \label{fig:gearbox}
    \end{figure}

  \item Top flange that sits on top of the dewar (Figure~\ref{fig:top_flange}). Has many of the vacuum and cryogenic ports listed below.
    \begin{figure}[h]
      \centering
      \includegraphics[height=0.7\textheight]{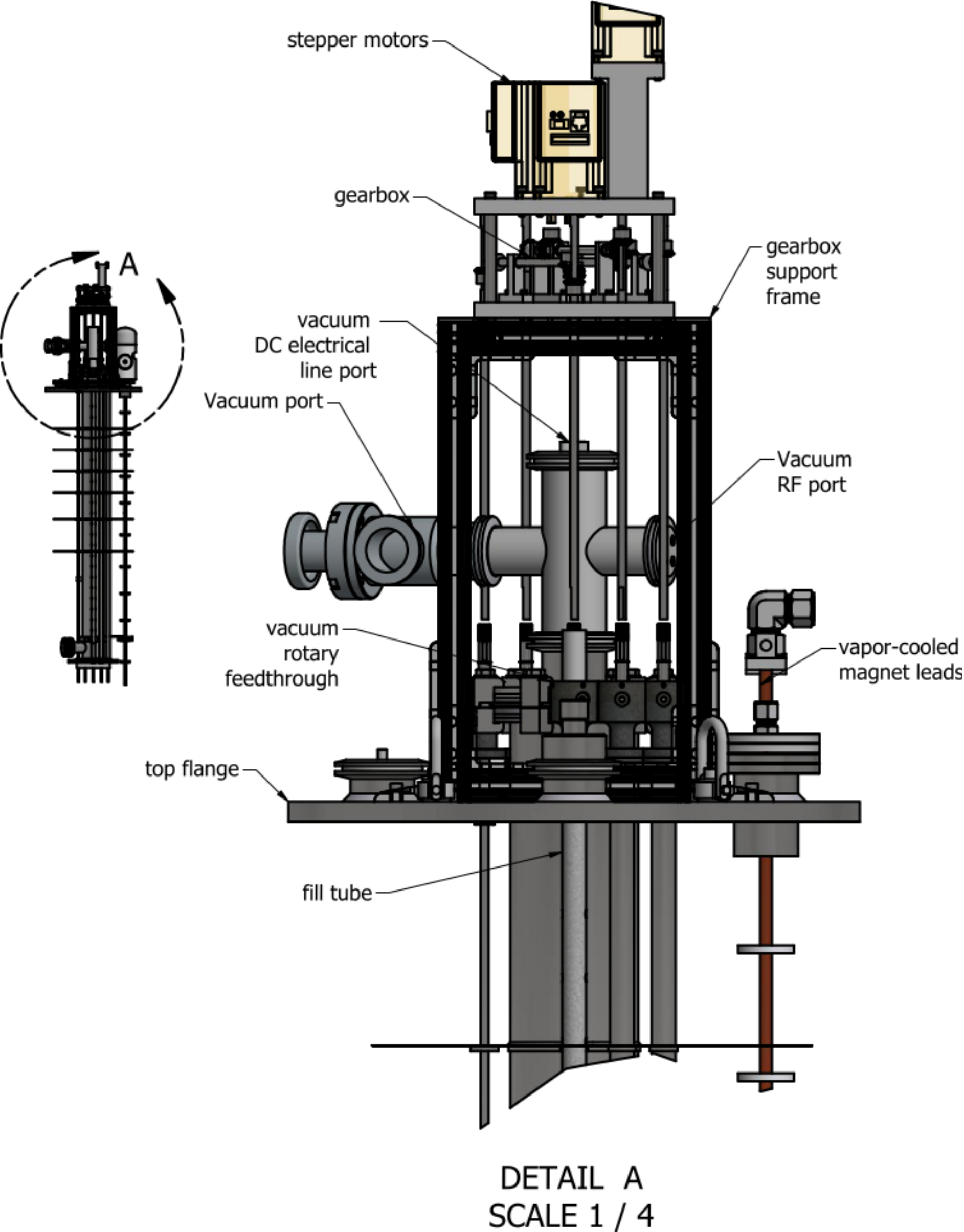}
      \caption{Insert top flange with all the vacuum ports, cryogen ports, and motor stage.}
      \label{fig:top_flange}
    \end{figure}

    \begin{itemize}
      \item Six MDC Precision 670000 rotary motion feedthroughs.
      \item AMI vapor-cooled magnet leads with superconducting busbars. They have a \SI{8}{L/day} boil off.
      \item Kurt-Lesker IFDRG197018B 19 pin electrical feedthrough. 
      \item Isocross for vacuum port, 51 pin micro type-D port (MDC 9163004), and RF coaxial port.
      \item cryogen fill port
      \item cryogen vent port
    \end{itemize}
  \item \SI{300}{L} Dewar with a boil \SI{18}{L/day} boil off.
  \item 50'' long, 4'' OD stainless steel tube. 
  \item Six 50'' long, 0.75'' ID stainless steel tubes to house motor shafts in vacuum space.
  \item Six motor shafts. Each motor shaft consists of a stainless steel tube coupled to a G10 fiberglass tube. The fiberglass tube reduces thermal contact to the top flange. The stainless steel tube is not as thermally isolating as the fiberglass but has less elasticity and would lead to less mechanical backlash.
  \item Copper plug to act as cold finger. Used to thermally sink cryogen space to cavity top.
  \item Rectangular experimental support plate welded to the bottom of the tube. The cavity attaches to the experimental support plate.
  \item Cavity vacuum vessel. The dipole magnet attaches to this vessel. 
\end{enumerate}

\FloatBarrier
\section{Issues with the Tuning Mechanism}\label{sec:bad_shafts}
\begin{figure}
  \centering
  \subfloat[]{\includegraphics[height=0.3\textheight]{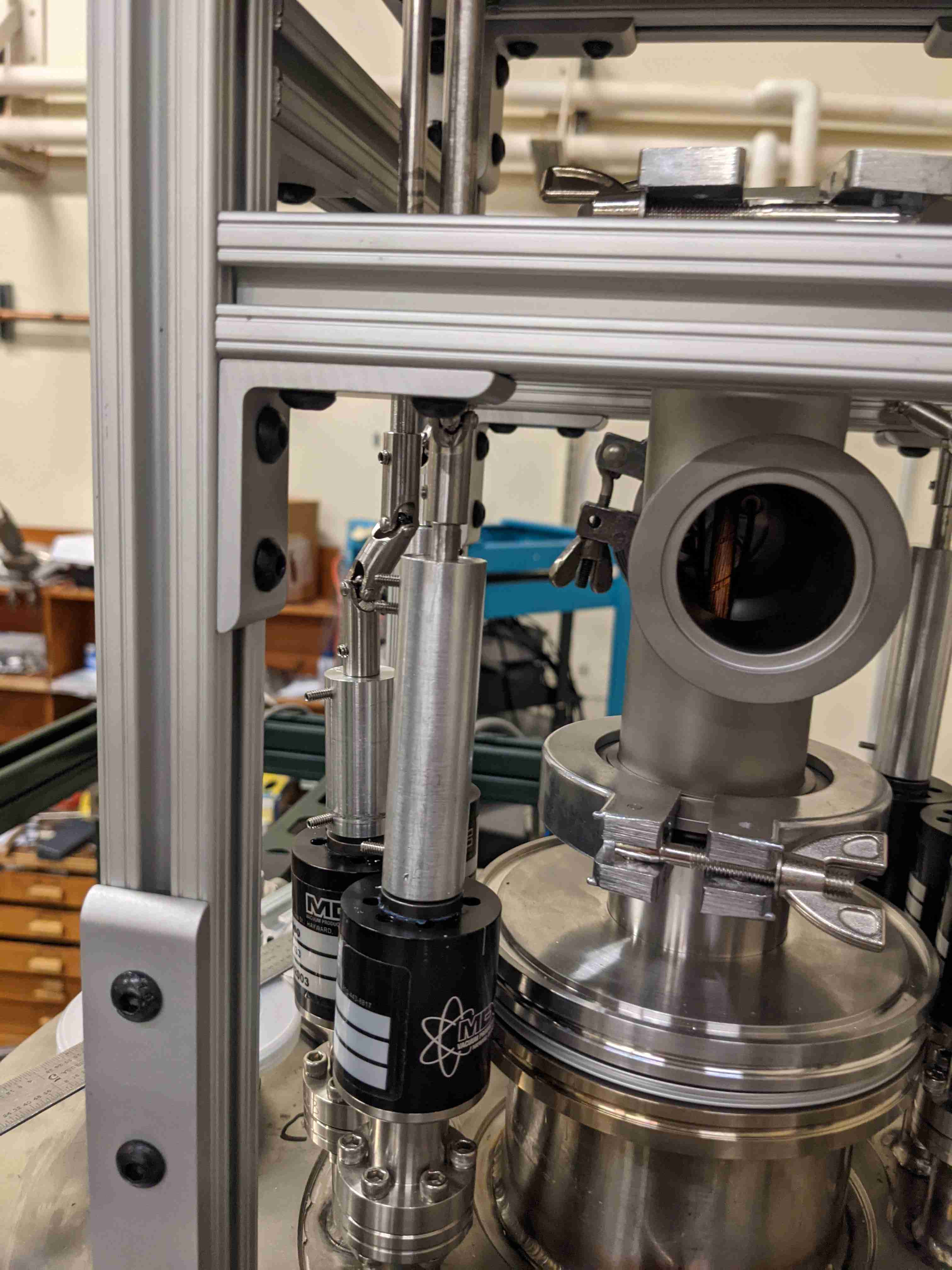}}\hfil
  \subfloat[]{\includegraphics[height=0.3\textheight]{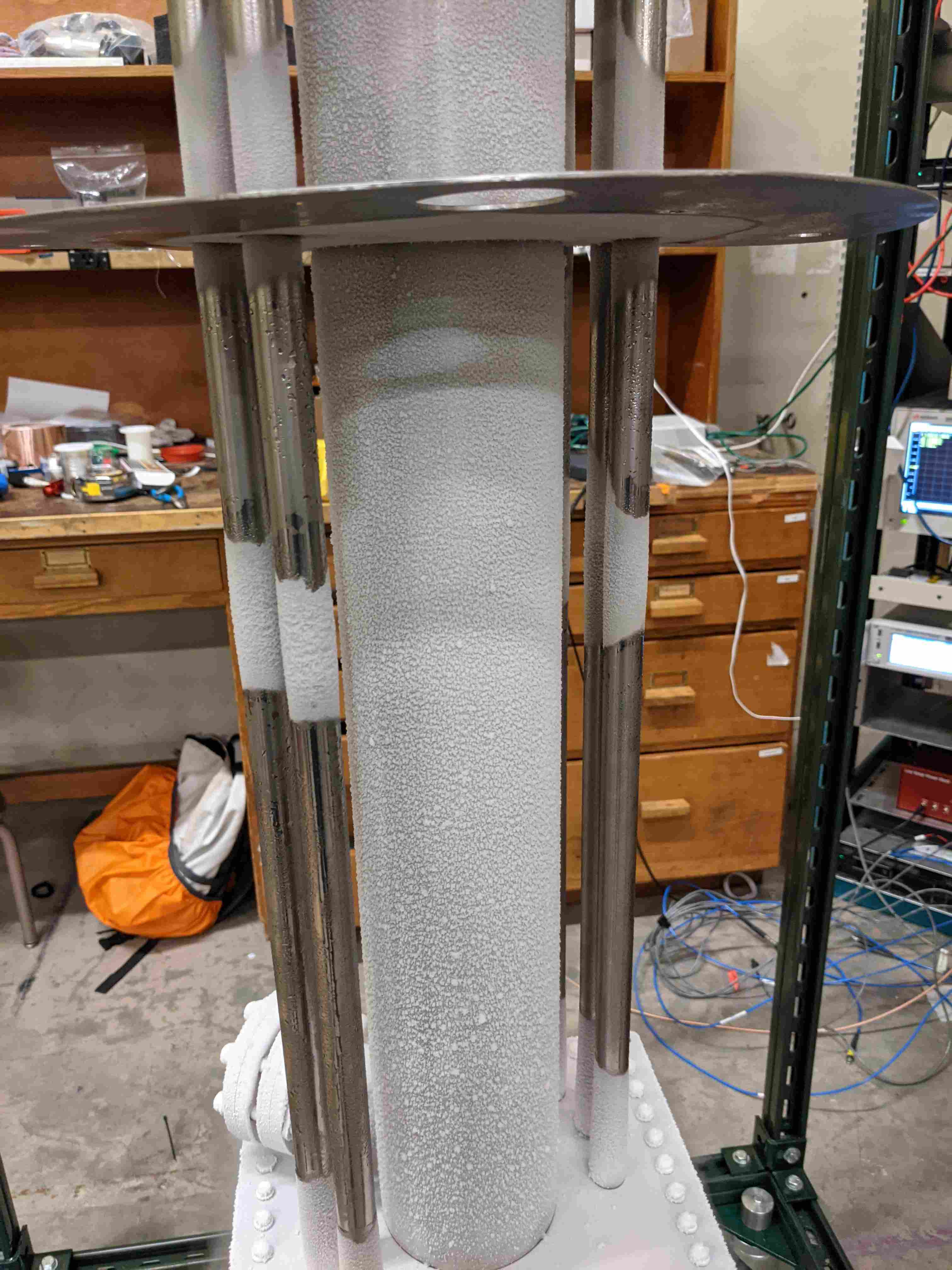}}
  \caption{(a)The vacuum ports for the rotary feedthroughs are misaligned. This contributed to a cascade of effects that caused the motors to stall frequently after the cooldown. (b) The vacuum shaft consists of a stainless steel tube connected to a G10 fiberglass tube connected by a spiral shaft coupling with a 5/8'' outer diameter. The vacuum housing tube has a 3/4'' inner diameter, and the misalignment from the top flange caused the shaft coupler to touch the vacuum walls. The point of contact is made obvious by the localized frost marks.}
  \label{fig:orpheus_misalignment}
\end{figure}

There were several shortcomings of this tuning mechanism, as implemented on Sept. 2021. The most severe is in the friction caused by shaft misalignments. One of the most prominent sources of misalignment is in the vacuum port where the vacuum rotary feedthroughs attach (Figure~\ref{fig:orpheus_misalignment}). The vacuum ports were unfortunately welded with a lot of tilt. This meant that the gearbox shafts were very misaligned from the rotary feedthroughs. Double universal joints were used to compensate for the misalignment. However, these double u-joints would apply a radial force on the gearbox shaft during rotation, which caused the miter gear to move closer or farther away from the mating gear. The changing mounting distance caused the mating gears to mesh poorly, leading to friction, inefficient power transfer, and motor stalls. A flexible coupler with a spring joint may have mitigated this issue since it wouldn't have provided as much radial force, but vendors were out of stock at the time.

The vacuum port misalignment had propagating effects into the vacuum space. The tilted vacuum rotary feedthrough caused the vacuum shafts to also be tilted. The vacuum shafts consist of a 1/4'' OD stainless steel tube connected to a 1/4'' OD fiberglass tube connected by a 5/8'' OD spiral flexible shaft coupler. The stainless steel vacuum housing tube has a 3/4'' inner diameter, so the misalignment caused the flexible coupler to rub against the inner walls of the stainless steel tube. This touch might have been especially problematic during cooldown if residual moisture was frozen onto the coupler. The touch was evident when the insert was pulled out of the dewar, as seen by the frost marks were localized around the location of the flexible coupler (Figure~\ref{fig:orpheus_misalignment}). 

The effects of the misalignment can propagate even further. If the vacuum shafts are misaligned, some shafts may need to be longer than others to reach the resonator. Because the vacuum shafts may have different lengths, they would have contracted by different amounts during cooldown. That could lead to the curved mirror and dielectric plate becoming tilted after cooldown, and the tilted plates would experience more friction.

Besides the myriad misalignment issues, there is no mechanism that mitigates backlash. The gear ratios in the gearbox are all one-to-one. The only increase in gear ratio is in the transfer of power to the threaded rods, and the resulting gear ratio is about 16. The backlash is approximately a quarter of a turn. This amounts to sub-millimeter systematic uncertainty in the absolute length of the cavity and relative positions of the dielectric plates. This systematic uncertainty caused by the backlash, while undesirable, can be mitigated by always tuning continuously in one direction.

\section{Thermal Gradient Issues}\label{sec:thermal_issues}
There are also multiple issues with the thermal design. The resonator is cooled because the flat mirror and copper amplifier heatsink is connected to a cold finger. The mirror is then supposed to cool down the rest of the cavity. However, the flat mirror can lose thermal contact with the rest of the cavity, and this thermal connection is lost. The contact is lost if the flat mirror is lifted by the contracting copper braids or copper RF coaxial cables. 

The flat mirror also has a higher heat leak than the rest of the cavity because it is connected to two coaxial cables that connect straight to the room-temperature top flange. In contrast, the curved mirror at the bottom only has one coaxial cable attached. The flat mirror also has a higher heat load since it is cooling down the cryogenic amplifier.

The dielectric plates and mirrors are also somewhat thermally isolated from each other as only the stainless steel provides thermal connections. 

The combination of these effects causes the temperature of the flat mirror to diverge significantly from the curved mirror during the data run, as will be shown in Figure~\ref{fig:mirror_gradient}. This divergence makes it harder to characterize a cavity temperature and consequently makes the extracted noise temperature have more uncertainty.

%% file: experimental_design_operation.tex
\chapter{Electronics, Data Acquisiton System, and Operations}\label{ch:operations}
This chapter focuses on how measurements in the Orpheus experiment are taken. The chapter talks about the radio frequency (RF) and intermediate frequency (IF) electronics, the steps taken for each data-taking cycle, and the software that controls and monitors the experiment. The chapter ends with a narrative of the real-time operation of the first science run.

\section{Cold Electronics}
\begin{figure}
  \centering
  \includegraphics[width=\textwidth]{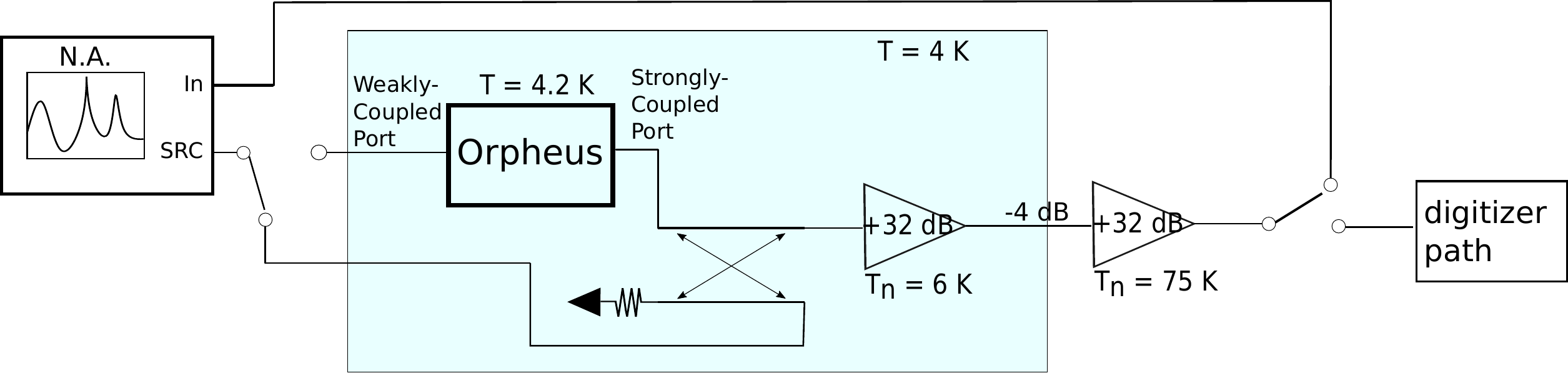}
  \caption{A diagram of how transmission, reflection, and power measurements are taken. Room temperature Teledyne switches are used to switch between transmission and reflection measurements, and from VNA measurements to power measurements.}
  \label{fig:cold_electronics}
\end{figure}

The diagram for the cold electronics is shown in Figure~\ref{fig:cold_electronics}. The cavity has a weakly-coupled port and a strongly-coupled port. The strongly-coupled port is connected to a WR-62 \SI{20}{dB} crossguide coupler (PEWCP1047). The crossguide coupler is attached to a waveguide to coax adapter (PE9803). The coax adaptor connects directly to the cryogenic low noise amplifier (LNF-LNC6\_20C). The cryogenic amplifier output is connected to an RG405 coaxial cable that connects directly to the room-temperature SMA bulkhead\footnote{This direct connection causes a large heat leak but was easy to implement. Future runs will take steps to mitigate this heat leak.}. 

During a science run, the strongly-coupled port transmission and reflection coefficients need to be measured for each data-taking cadence. For a transmission measurement, the VNA port 1 is connected to the weakly-coupled port. For a reflection measurement, the VNA port 1 is connected to the crossguide coupler coupled port, as shown in Figure~\ref{fig:cold_electronics}.  The signal then travels to the strongly-coupled port, gets reflected, and reaches the input of the cryogenic amplifier. The coupler\footnote{A broadwall coupler would have been better than a crossguide coupler because it has a coupling value frequency response is much flatter, and its directivity is far higher. Unfortunately, a broadwall coupler doesn't fit inside the \SI{4}{in} insert tube. A circulator may have also been more attractive. But they are far more expensive and require magnetic shielding. Also, I didn't find an appropriate cryogenic \SI{18}{GHz} circulator from Pasternack or QuinStar.} is needed because the VNA source signal needs to bypass the cryogenic amplifier to reach the strongly-coupled port. The remaining port in the crossguide coupler is terminated with a waveguide terminator (PE6804\footnote{PEWTR1004 has a much better VSWR. Unfortunately, there was no room for it inside the vacuum vessel.}). A Teledyne SMA switch is used to switch between a reflection measurement and transmission measurement.

\section{Warm Electronics}
\begin{figure}
  \centering
  \includegraphics[width=\textwidth]{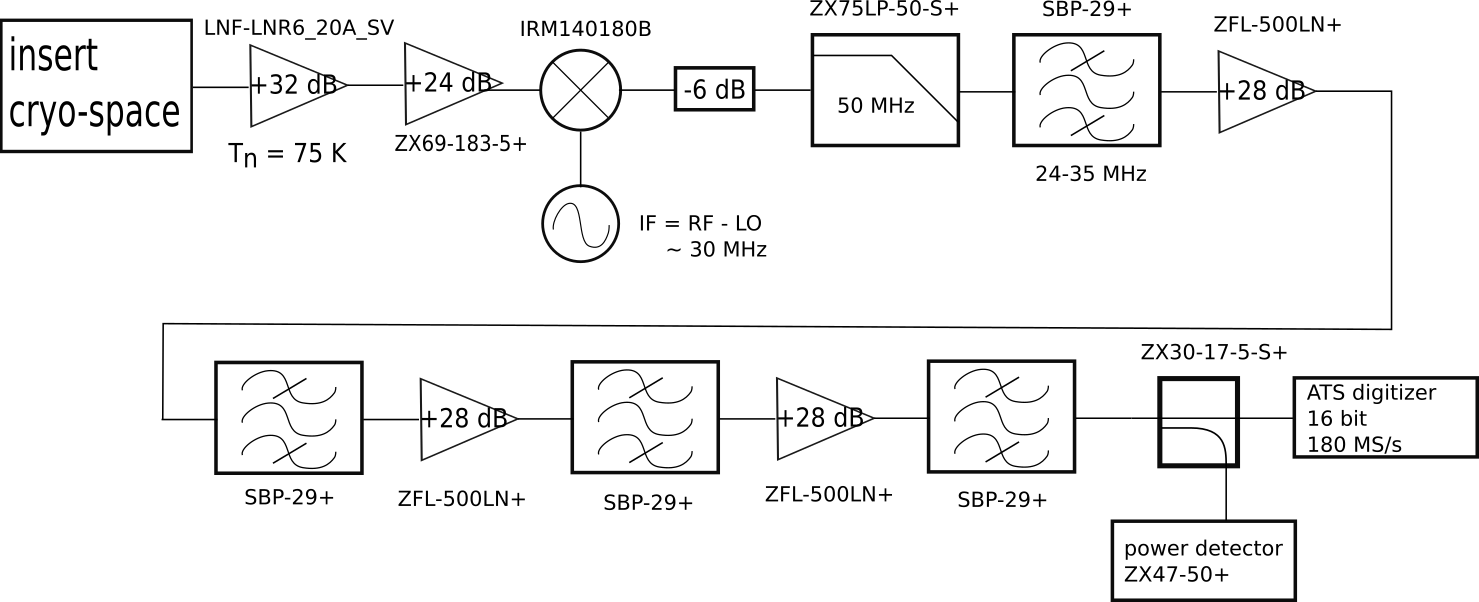}
  \caption{Orpheus measurement scheme. There are 3 paths: transmission, reflection, and digitization.}
  \label{fig:warm_box_schematic}
\end{figure}

\begin{figure}
  \centering
  \includegraphics[width=0.8\textwidth]{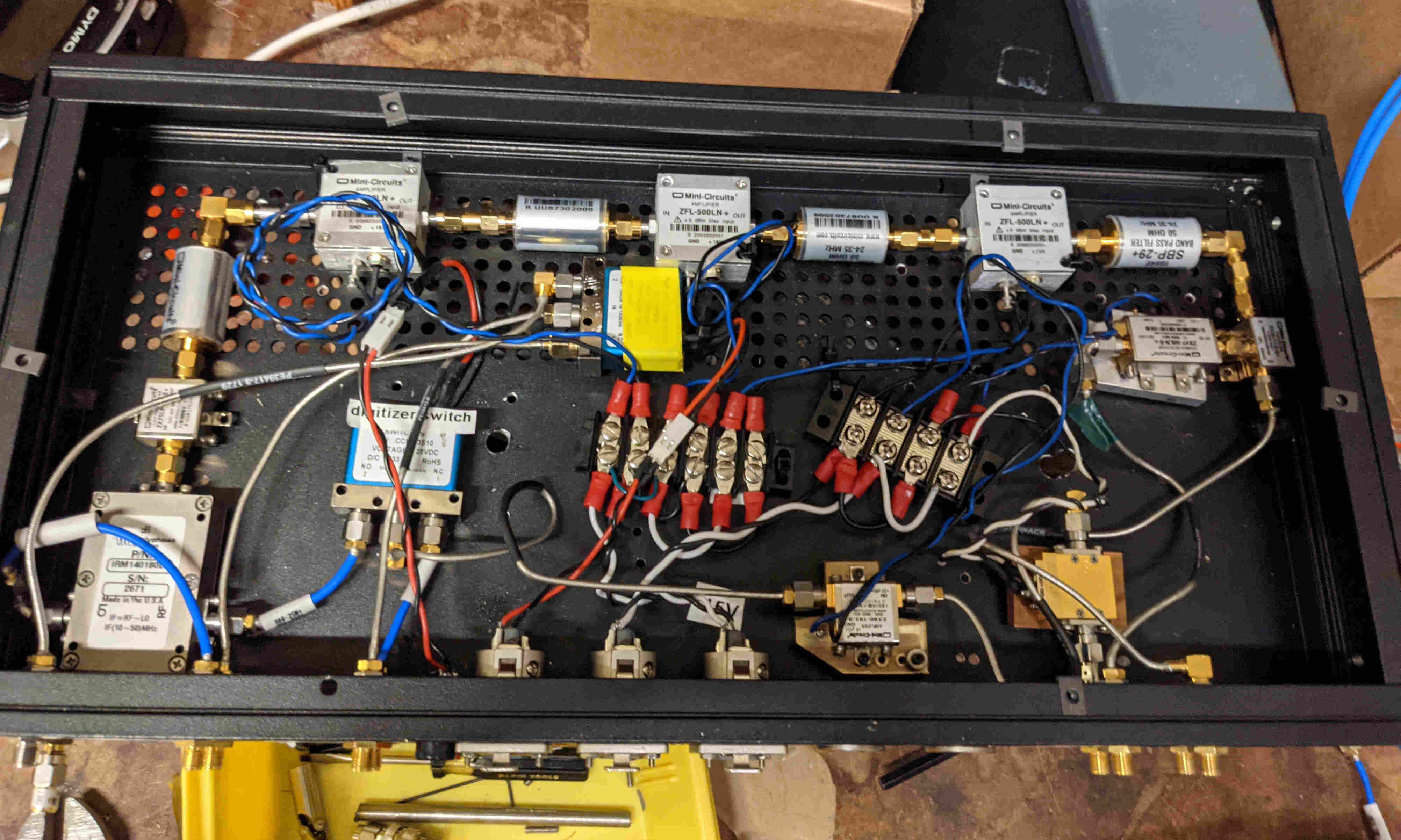}
  \caption{RF/IF electronics box. Contains warm RF stage, IF stage, and coaxial switches to switch between a VNA transmission measurement, VNA reflection measurement, and a power spectrum measurement.}
  \label{fig:warm_box_real}
\end{figure}

The schematic for warm electronics is shown in Figure~\ref{fig:warm_box_schematic}, a photograph is shown in Figure~\ref{fig:warm_box_real}, and the list of components with the associated cascade analysis is shown in Table~\ref{tab:cascade}.

After the signal makes it out of the vacuum space, it is directed to a room-temperature low noise amplifier (LNF-LNR6\_20A SV) to ensure that the gain is high enough so the noise from the rest of the noisy high-power electronics doesn't reduce the SNR. The signal is amplified again by the ZX60-183-S. The \SI{18}{GHz} signal needs to be downsampled so that it can be digitized by a \SI{180}{MSPS} digitizer card. Downsampling is achieved by sending the RF signal to an Image-Reject Mixer (IRM140180B). The mixer uses the nonlinear semiconductor properties to multiply the RF signal with angular frequency $\omega_{RF}$ by a local oscillator signal with angular frequency $\omega_{LO}$. The result is a signal with spectral components with both the sum and difference of the two frequencies (and higher-order harmonics), i.e. $\sin(\omega_{RF}t)\sin(\omega_{LO}t) \propto \sin((\omega_{RF}-\omega_{LO})t) + \sin((\omega_{RF}+\omega_{LO})t)$. The signal component with ${\omega = \omega_{RF}+\omega_{LO}}$ is removed using a low pass filter, and the signal component with ${\omega = \omega_{RF}-\omega_{LO}}$ is digitized. The downsampled frequency $2\pi(\omega_{RF}-\omega_{LO})$ is known as the Intermediate Frequency (IF). The IF is chosen to be about \SI{30}{MHz} to be compatible with the IRM140180B. 

\begin{figure}
  \centering
  \includegraphics[width=0.6\textwidth]{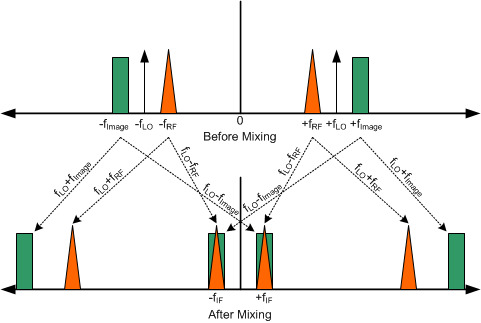}
  \caption{A demonstration of image noise after mixing. The negative image frequency folds into the positive IF frequency, and the positive image frequency folds into the negative IF frequency. Figure from~\cite{mixer_image}.}
  \label{fig:image_noise}
\end{figure}

Image rejection is important because otherwise, the noise would unnecessarily double. To see why this is true, one can examine Figure~\ref{fig:image_noise}. One can keep track of all the combinations of beat frequencies to see that the negative image frequency folds into the positive IF frequency and the positive image frequency folds into the negative IF frequency. The image rejection works using phasing techniques, which can be read about in Edward Daw's thesis\cite{daw2018search}. 

After the RF signal is multiplied by the local oscillator signal, the resulting signal passes through a low pass filter, and only the IF frequency survives. There is an attenuator between the mixer and low pass filter to eliminate standing waves between the two\footnote{From Christian Boutan:  Mixers are sometimes temperamental with impedance matching and reflections off the mixer can actually cause filters to detune. The fix is having a few dB of attenuation between mixers and filters to kill standing waves.}. After this, the IF signal passes through a series of IF amplifiers and bandpass filters. The IF amplifiers amplify the IF signal to the dynamic range of the digitizer card, and the bandpass filters attenuate noise outside the frequency of interest. The amplified IF signal then goes to a directional coupler. The coupled port is connected to a power detector (ZX47-50+), allowing for the real-time measurement of the IF signal power. The directional coupler's through port is connected to the Alazartech digitizer card (ATS9462).

The Alazartech digitizer is set to a sampling rate of \SI{125}{MSPS}. This rate is chosen so that the Nyquist frequency is well above the IF band. Each subspectrum contained \num[group-separator={,}]{50000} time samples, resulting in a \SI{2.5}{kHz} bin width. Depending on the run settings, each spectrum is either the average of \num[group-separator={,}]{75000} (30 seconds) subspectra or \num[group-separator={,}]{250000} (100 seconds) subspectra. No windowing is applied to each subspectrum. It is a 16-bit digitizer with a dynamic range of $\pm\SI{200}{mV}$ to $\pm\SI{16}{V}$ 

The cascade analysis in Table~\ref{tab:cascade} shows that the gain from the RF and IF electronics is about \SI{140}{dB}. For $T_n \approx \SI{11}{K}$ and bin width $b = \SI{2.5}{kHz}$, the noise power without the system gain is $P_n = k_B T_nb  = \SI{-118}{dBm}$. Thus the noise power seen by the digitizer is $P_n = \SI{21.8}{dB} = \SI{8}{V_{p-p}}$. That's well within the dynamic range of the digitizer.

\begin{table}[]
\centering
\resizebox{\textwidth}{!}{%
  \begin{tabular}{lllp{3cm}lp{3cm}l}
\textbf{Part Description}                                                          & \textbf{vendor}     & \textbf{part number} & \textbf{Relative power gain} & \textbf{absolute gain} & \textbf{device noise temperature} & \textbf{cascaded noise temperature} \\
LNF-LNC6\_20C s/n 1556Z                                                            & Low Noise Factory   &                      & 33                           & 33                     & 6                                 & 10.5                                \\
SMA Male to SMA Male Cable Using RG405 Coax, RoHS                                  & Pasternack          & PE3818LF-72          & -7.6                         & 25.4                   & 0                                 & 10.5                                \\
C3146 SMA Hermetic Bulkhead Adapter 18Ghz                    & Centric RF          & C3146                & -0.2                         & 25.2                   & 0                                 & 10.5                                \\
Low Loss Test Cable 12 Inch Length, PE-P142LL Coax & Pasternack          & PE341-12             & -0.82                        & 24.38                  & 0                                 & 10.5                                \\
Hand-Flex Interconnect, 0.086" center diameter, 18.0 GHz                           & Minicircuits        & 086-SBSMR+           & -0.3                         & 24.08                  & 0                                 & 10.5                                \\
LNF-LNR6\_20A\_SV s/n 1257Z                                                        & Low Noise Factory   &                      & 32                           & 56.08                  & 100                               & 10.8908408957924                    \\
Right Angle Semi-Flexible Cable                               & Pasternack          & PE39417-6            & -2                           & 54.08                  &                                   &                                     \\
Wideband Microwave Amplifier 6 to 18 GHz                                           & Minicircuits        & ZX60-183-S+          & 24                           & 78.08                  & 1378.8                            & 10.8942410708306                    \\
Semi-Flexible Cable                               & Pasternack          & PE39417-9            & -3                           & 75.08                  &                                   &                                     \\
Hand-Flex Interconnect, 0.086" center diameter, 18.0 GHz                           & Minicircuits        & 086-4SM+             & -0.38                        & 74.7                   & 0                                 & 10.8942410708306                    \\
Hand-Flex Interconnect, 0.086" center diameter, 18.0 GHz                           & Minicircuits        & 086-4SM+             & -0.38                        & 74.32                  & 0                                 & 10.8942410708306                    \\
DC–18 GHz/DC-22 GHz SPDT Coaxial Switch                          & Teledyne            & CCR-33S/CR-33S       & -0.4                         & 73.92                  & 0                                 & 10.8942410708306                    \\
Hand-Flex Interconnect, 0.086" center diameter, 18.0 GHz                           & Minicircuits        & 086-2SM+             & -0.32                        & 73.6                   & 0                                 & 10.8942410708306                    \\
IMAGE-REJECT MIXER 14.0 – 18.0 GHz                                                 & Polyphase Microwave & IRM140180B           & -8.5                         & 65.1                   & 3360.9                            & 10.8943877794366                    \\
3 dB Fixed Attenuator                                                              & Pasternack          & PE7005-3             & -3                           & 62.1                   & 0                                 & 10.8943877794366                    \\
Low Pass Filter                                                                    & Minicircuits        & ZX75LP-50-S+         & -1.38                        & 60.72                  & 0                                 & 10.8943877794366                    \\
Hand-Flex Interconnect, 0.086" center diameter, 18.0 GHz                           & Minicircuits        & 086-2SM+             & -0.32                        & 60.4                   & 0                                 & 10.8943877794366                    \\
Lumped LC Band Pass Filter, 24 - 35 MHz, 50$\Omega$                                       & Minicircuits        & SBP-29+              & -0.88                        & 59.52                  & 0                                 & 10.8943877794366                    \\
Low Noise Amplifier, 0.1 - 500 MHz, 50$\Omega$                                            & Minicircuits        & ZFL-500LN+           & 28.21                        & 87.73                  & 275                               & 10.8946949168298                    \\
Lumped LC Band Pass Filter, 24 - 35 MHz, 50$\Omega$                                       & Minicircuits        & SBP-29+              & -0.88                        & 86.85                  & 0                                 & 10.8946949168298                    \\
Low Noise Amplifier, 0.1 - 500 MHz, 50$\Omega$                                            & Minicircuits        & ZFL-500LN+           & 28.21                        & 115.06                 & 275                               & 10.8946954848093                    \\
Lumped LC Band Pass Filter, 24 - 35 MHz, 50$\Omega$                                       & Minicircuits        & SBP-29+              & -0.88                        & 114.18                 & 0                                 & 10.8946954848093                    \\
Low Noise Amplifier, 0.1 - 500 MHz, 50$\Omega$                                            & Minicircuits        & ZFL-500LN+           & 28.21                        & 142.39                 & 275                               & 10.8946954858597                    \\
Lumped LC Band Pass Filter, 24 - 35 MHz, 50$\Omega$                                       & Minicircuits        & SBP-29+              & -0.88                        & 141.51                 & 0                                 & 10.8946954858597                    \\
17.5 dB Directional Coupler, 5 - 2000 MHz, 50$\Omega$                                     & Minicircuits        & ZX30-17-5-S+         & -0.7                         & 140.81                 & 0                                 & 10.8946954858597                   
\end{tabular}%
}
\caption{List of electronics and cascade analysis.}
\label{tab:cascade}
\end{table}

\FloatBarrier
\section{System Noise Temperature}\label{sec:thermal_model}
The system noise power can be described as the sum of the blackbody radiation of the cavity and the added Johnson noise of each electrical component, and is written as $P_n=G k_b b T_{sys}$, where $G$ is the system gain that depends on both frequency and physical temperature, $b$ is the frequency bandwidth, and $T_{sys}$ is the system noise temperature referenced to the cavity. 

From examining Figure~\ref{fig:reflection_measurement}, one can determine that 
\begin{align}
  T_{sys} = T_{cav}(1-|\Gamma|^2) + T_{amp, input}|\Gamma|^2 + T_{rec}
  \label{eqn:orpheus_tsys}
\end{align}
where $T_{cav}$ is the physical temperature of the cav, $T_{amp, input}$ is the noise temperature coming from the input of the cryogenic amplifier, $T_{rec}$ is the noise temperature of the receiver chain from the output of the cryogenic amplifier outward (discussed in Section~\ref{sec:receiver_noise_temp}), and $\Gamma$ is the reflection coefficient and $|\Gamma|^2$ is the fraction of power reflected. As a reminder of Equation~\ref{eqn:lorentzian_fits}, $|\Gamma|^2$ depends on both the cavity coupling coefficient $\beta$ and the detuning factor $\Delta$. 

To understand why Equation~\ref{eqn:orpheus_tsys} is plausible, consider several limiting cases. When the receiver is critically coupled to the cavity, the cavity on resonance looks like a blackbody. Thermal photons are in thermal equilibrium with the cavity, and $T_{sys} = T_{cav} + T_{rec}$. If the receiver is poorly coupled or if the RF frequency is far off resonance, the cavity looks more like a mirror. In this scenario, thermal photons are emitted from the input of the amplifier. These thermal photons reach the flat mirror and are reflected back into the amplifier. Thus, $T_{sys} = T_{amp, input} + T_{rec}$. Generally, the cavity is only partially reflecting, and the noise temperature is described by Equation~\ref{eqn:orpheus_tsys}.

\section{Receiver Noise Temperature}\label{sec:receiver_noise_temp}
\begin{figure}
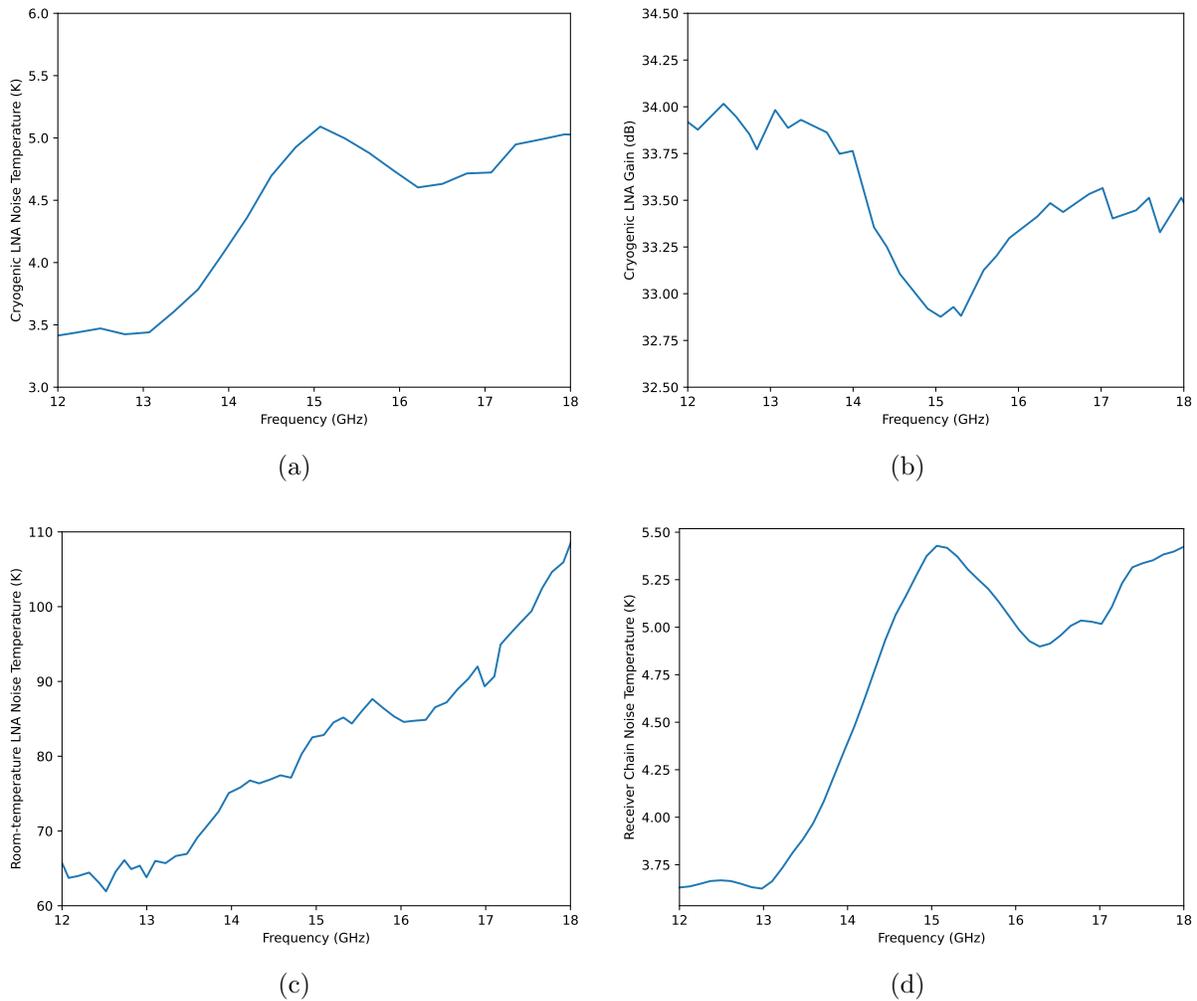

  \centering
  \subfloat[]{\includegraphics[width=0.48\textwidth]{cryo_lna_noise_temperature.pdf}}\hfil
  \subfloat[]{\includegraphics[width=0.48\textwidth]{cryo_lna_gain.pdf}}\\
  \subfloat[]{\includegraphics[width=0.48\textwidth]{roomtemp_lna_noise_temperature.pdf}}\hfil
  \subfloat[]{\includegraphics[width=0.48\textwidth]{receiver_noise_temperature.pdf}}
  \caption{The derived receiver noise temperature. The Friis equation is used to derive the receiver noise temperature (d) from the noise temperature (a) and gain (b) of the 1st stage cryogenic LNA and the noise temperature of the 2nd stage room-temperature amplifier (c). These values are obtained from data sheets from Low Noise Factory. The cavity temperature can be added to the receiver noise temperature to obtain the system noise temperature, as referenced to the cavity.}
  \label{fig:receiver_noise_temperature}
\end{figure}

The noise temperature of the receiver is derived from datasheets from Low Noise Factory and the Friis cascade equation~\cite{friis} 
\begin{align}
  T_{rec} = T_1 + \frac{T_2}{G_1} + \frac{T_3}{G_1G_2} + ...
\end{align}
$T_1$ and $G_1$ are the noise temperature and gain of the 1st stage cryogenic amplifier. $T_2$ and $G_2$ are the noise temperature and gain of the 2nd stage room-temperature LNA. $T_3$ is the noise temperature of the 3rd stage amplifier. The 3rd term is negligible because of the gain from the first two amplifiers. The derived receiver noise temperature is shown in Figure~\ref{fig:receiver_noise_temperature}.

\FloatBarrier
\section{Software Stack}
The control software stack consists of Python, Postgresql, RabbitMQ, Dripline, Docker, Kubernetes, Helm, and Grafana. Except for the digitizer driver, all software is open source. 

\begin{figure}
  \centering
  \includegraphics[width=0.7\textwidth]{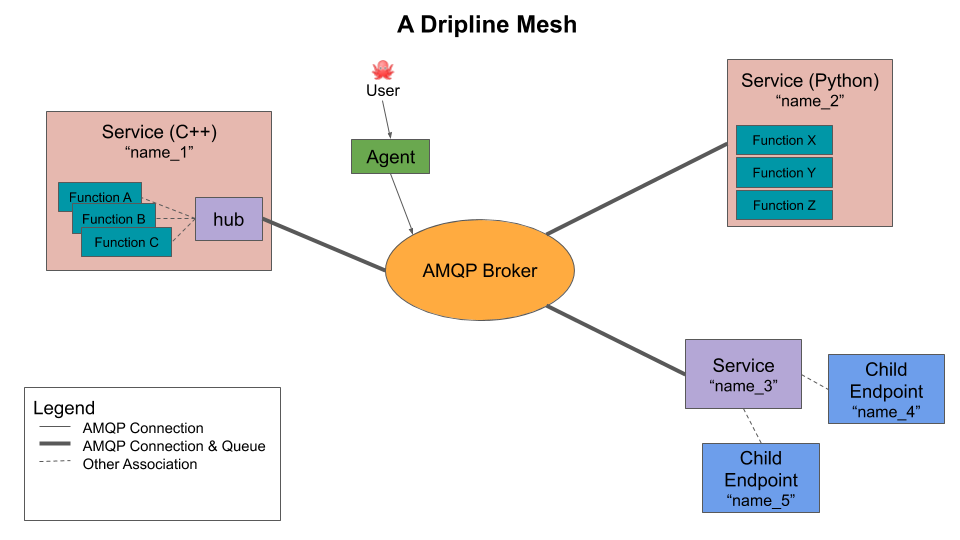}
  \caption{Dripline software architecture. Figure from~\cite{dripline}.}
  \label{fig:dripline}
\end{figure}

The software consists of modular, self-healing, loosely-coupled services with a standardized messaging protocol for all serial communication between devices. The software stack consists of:

\begin{itemize}
  \item Dripline: The standardized messaging protocol developed by Project 8 and maintained by Pacific Northwest National Laboratory. Dripline allows for a standardized way to talk to different devices running on different servers, as long as they accept socket commands. One can use the same messaging protocol to talk to stepper motors, digitizer cards, and databases. A fundamental design philosophy of Dripline is RESTfulness and that software services should run without making assumptions of what other services are doing. This philosophy allows for a continuously evolving DAQ system. More information can be found on their website~\cite{dripline}. A schematic of the Dripline architecture is shown in Figure~\ref{fig:dripline}.
  \item Python: The scripting language of choice. Python scripts control the data-taking cadences and online data analysis.
  \item RabbitMQ: A messaging broker. It's like the post office for Dripline messages. 
  \item Postgresql: An object-relational database.
  \item Docker containers: Provides OS-level virtualization. Packages an application and its dependencies in a virtual container.
  \item Kubernetes: Manages the lifecycle of Docker containers. For example, Kubernetes can be configured to always have a certain Docker container at all times. If that container crashes, Kubernetes will detect that crash and create another container to replace the crashed container.
  \item Helm: A package manager for Kubernetes configuration files.
  \item Grafana: Real-time data visualization. Integrates really well with Postgresql.
\end{itemize}

Since Orpheus took data only for a few days, the overhead of implementing this software stack was likely not worth it. But Dripline is used for the ADMX flagship experiment, where data is collected continuously for six months, and a fault-tolerant DAQ system is required. The flagship experiment uses Dripline v2. Orpheus is the first implementation of Dripline v3 for science data-taking. Orpheus uses much of the same equipment as the flagship experiment, so implementing the newly developed Dripline v3 allowed for stress testing, debugging, and development of new services in a smaller test stand.

\section{Data-taking cadence}
The basic strategy is to tune the cavity to scan for dark photons with different masses. For each cavity length, a series of ancillary measurements are taken to extract a noise power calibration and expected dark photon signal. The power is then measured out of the cavity in search of non-statistical peaks that may correspond to a dark matter signal. The data-taking cadence is as follows:

\begin{enumerate}
  \item The state of the system is recorded. The temperature sensors and motor encoders are logged into the database. These measurements allow the extraction of the noise power, cavity length, and dielectric plate positions.
  \item The transmission and reflection coefficient is measured \SI{20}{MHz} around the \tem mode. Parameters $f_0$, $Q_L$ and $\beta$ are extracted from these measurements. The extracted parameters are logged to the database.
  \item The VNA output is disabled, and the power spectrum is measured. The power spectrum is integrated for either \SI{30}{s} or \SI{100}{s}.
  \item The motors are tuned in a coordinated way. This means the curved mirror is moved by an amount specified in a configuration file. Simultaneously, the top dielectric plate is moved by $1/5$ the number of steps of the curved mirror, and the bottom dielectric plate is moved by $4/5$ the number of steps as the curved mirror.
  \item Repeat data-taking cadence for new cavity length.
  \item Every 20 data-taking cycles, the transmission coefficient is measured with a frequency range of \SI{15}{GHz} to \SI{18}{GHz}. This is known as a wide scan.
\end{enumerate}
\FloatBarrier
\section{September 1st Commissioning Narrative}\label{sec:sep1_narrative}
I will end this chapter with a real-time narrative of the September science run, as it may explain the quirks in the data.

\begin{figure}
  \centering
  \includegraphics[width=0.99\textwidth]{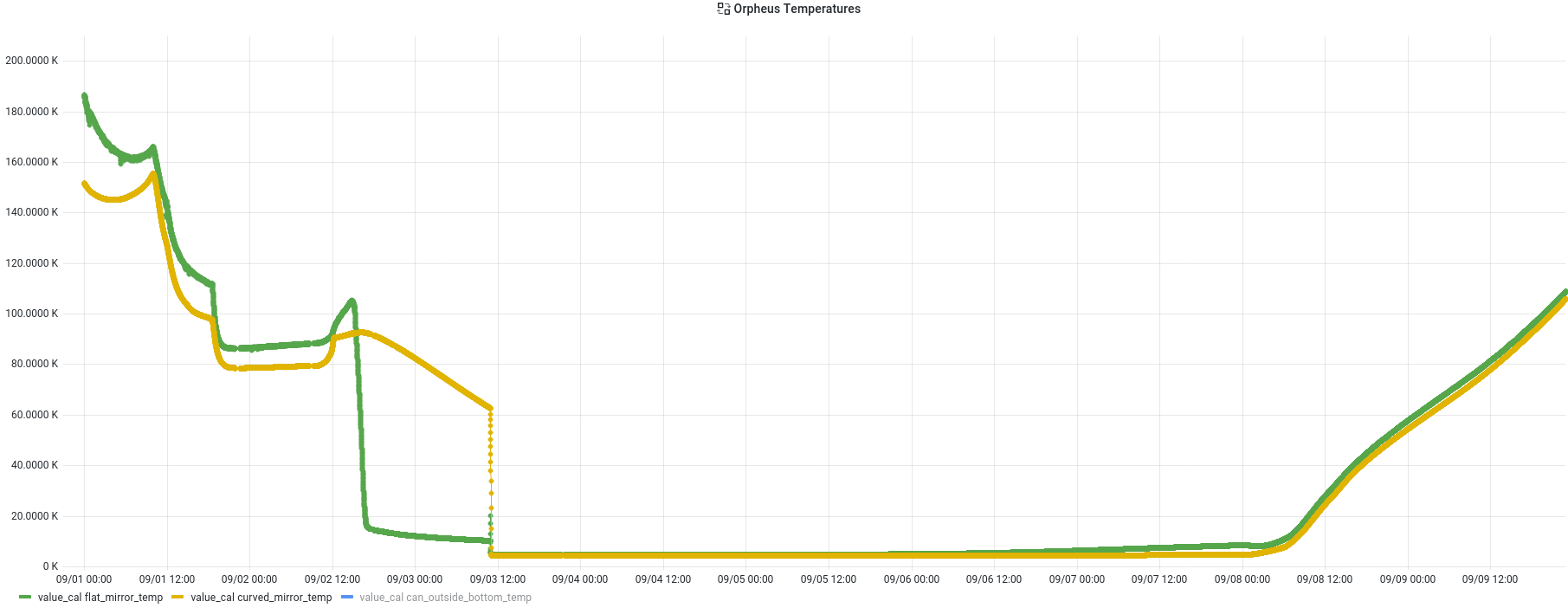}
  \caption{Flat mirror and curved mirror temperatures throughout the entire commissioning of the experiment, including warmup and cooldown.}
  \label{fig:mirror_gradient}
\end{figure}

On 9/1/2021, I tuned the cavity so that the \tem mode was at \SI{16.15}{GHz}. I knew this to be a frequency region with good Q and coupling. We had \SI{350}{L} of liquid helium at our disposal, and I wanted to make the best use of it. Charles, Grant, and I began pre-cooling Orpheus with liquid nitrogen in the reservoir space and introduced a nitrogen exchange gas in the vacuum space. The cooling and temperature history is shown in Figure~\ref{fig:mirror_gradient}.

On 9/2/2021, we boiled off the LN2 and pumped out the nitrogen exchange gas from the vacuum space. This was so that we didn't waste LHe freezing LN2. Then we filled the reservoir with liquid helium. 

By the morning of 9/3/2021, the top mirror cooled down to \SI{10}{K}, but the curved mirror was cooling very slowly and had only reached about \SI{60}{K}. Furthermore,  most of the liquid helium in the reservoir had boiled off, and we didn't have any liquid helium left to refill the reservoir. It seemed like the bottom of the cavity was thermally detached from the flat mirror, and we couldn't wait for the curved mirror to cool to \SI{4}{K}. Around noon, we introduced helium exchange gas into the vacuum space, and things cooled to below \SI{5}{K} within minutes. Once the cavity was cooled, we pumped out some of the exchange gas. No one knows how much exchange gas we introduced or how much we pumped out, but the procedure worked well.

\begin{figure}
  \centering
  \includegraphics[width=0.7\textwidth]{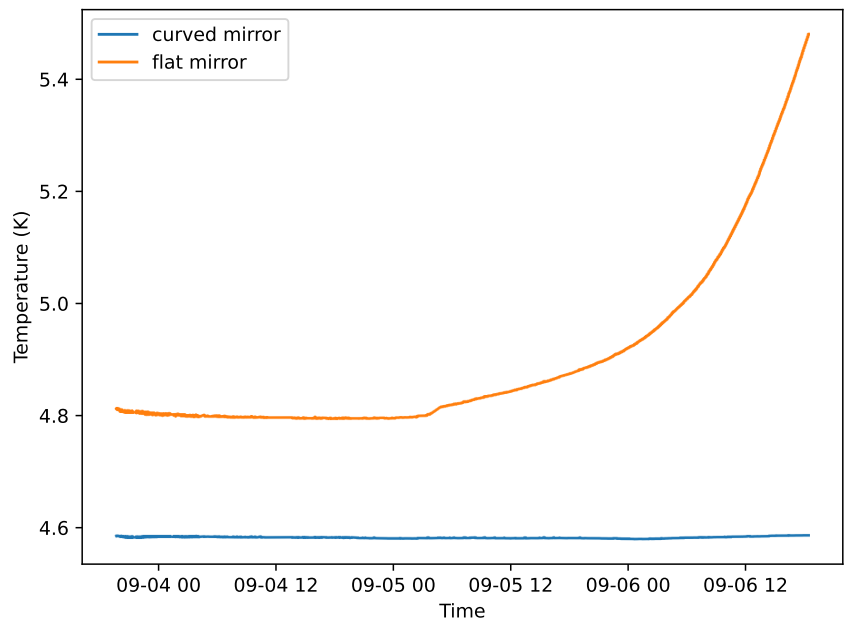}
  \caption{Flat mirror and curved mirror temperatures throughout the data-taking run. Physics data-taking stopped because the flat mirror temperature was deviating significantly from the curved mirror temperature, causing a significant increase in the uncertainty in the cavity temperature.}
\end{figure}

\begin{figure}
  \centering
  \includegraphics[width=0.7\textwidth]{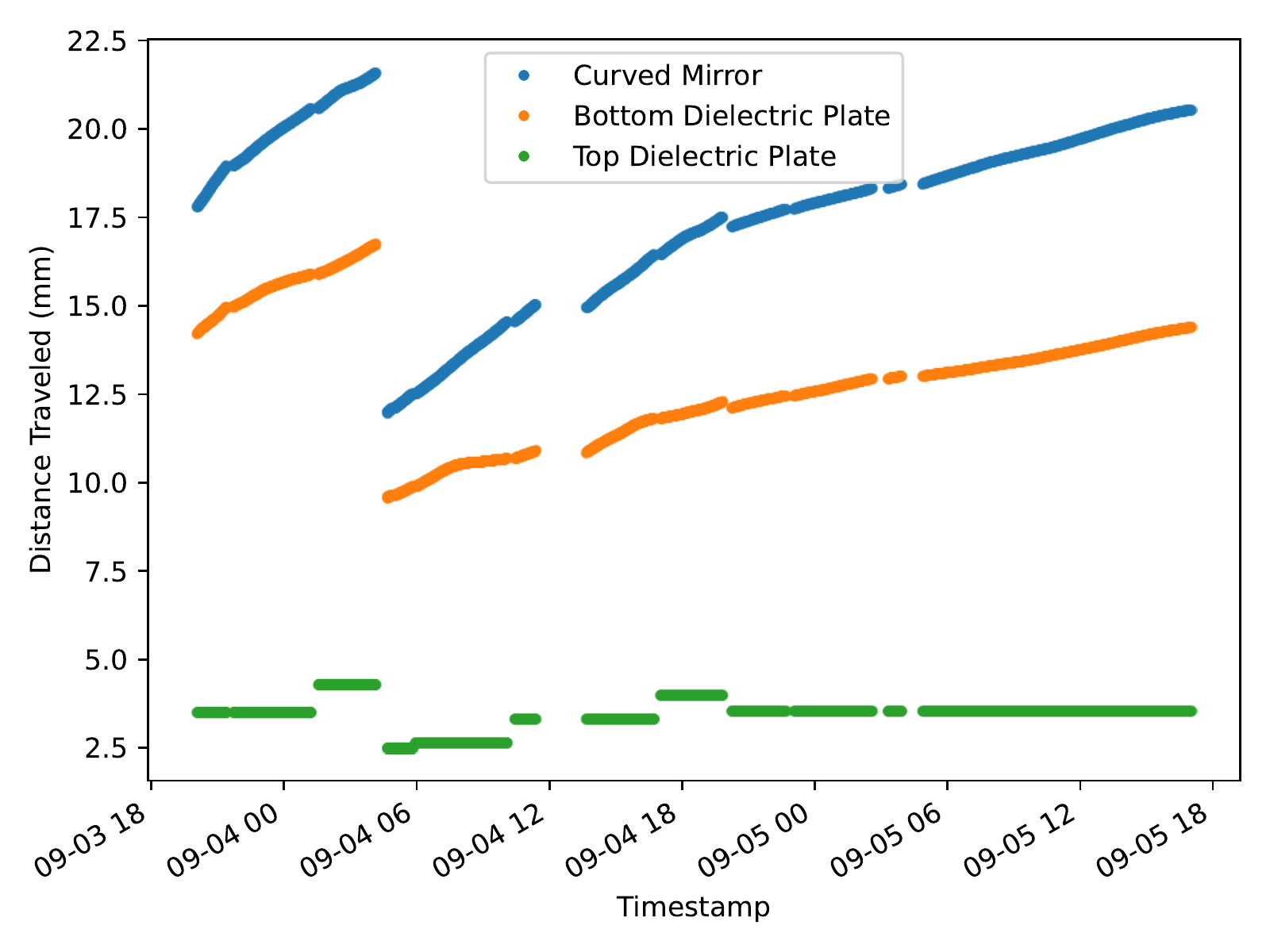}
  \caption{The motor steps as recorded by the encoder. From the motor steps, one derives the distance traveled by the curved mirror, bottom dielectric plate, and top dielectric plate. The top dielectric plate motor stalled after cooldown, preventing the top dielectric plate from being tuned remotely and in a continuous manner. I needed to turn the motor feedthroughs by hand periodically (explaining the discrete tuning steps for the top dielectric plate.}
  \label{fig:dpsearch_motor_steps_v_time}
\end{figure}

\begin{figure}
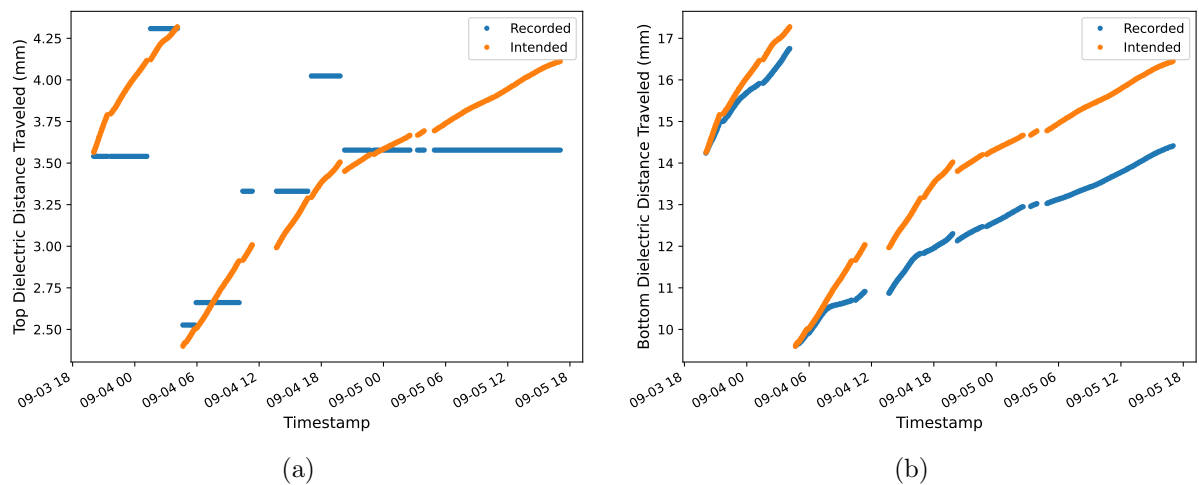

  \centering
  \subfloat[]{\includegraphics[width=0.48\textwidth]{top_dielectric_plate_motor_steps_dark_photon_search.pdf}}\hfil
  \subfloat[]{\includegraphics[width=0.48\textwidth]{bottom_dielectric_plate_motor_steps_dark_photon_search.pdf}}
  \caption{Due to a cascade of mechanical failures, software design flaws, and operation errors, the top dielectric plate and bottom dielectric plate deviated from their intended positions throughout the data-taking run.}
  \label{fig:dpsearch_motor_actual_vs_intended}
\end{figure}

Once the cavity was cold, I tried tuning, but the top dielectric plate motor was stalled. I found that with great force, I could tune the top dielectric plate by hand. So I decided to skip the top dielectric plate tuning during a data-taking cadence and to tune by hand every few hours to correct for the top dielectric plate lagging behind. The top dielectric plate needed to be tuned with 1/5 of the steps of the curved mirror. I was able to tune the \tem mode without tuning the top plate, so I decided that this top plate position error would just amount to a perturbation in $V_{eff}$. Luckily, the motor encoders kept count of the steps, even if the motor was turned manually. The recorded motor positions are shown in Figure~\ref{fig:dpsearch_motor_steps_v_time}. The actual motor positions vs intended motor positions are shown in Figure~\ref{fig:dpsearch_motor_actual_vs_intended}.

\begin{figure}
  \centering
  \includegraphics[width=0.7\textwidth]{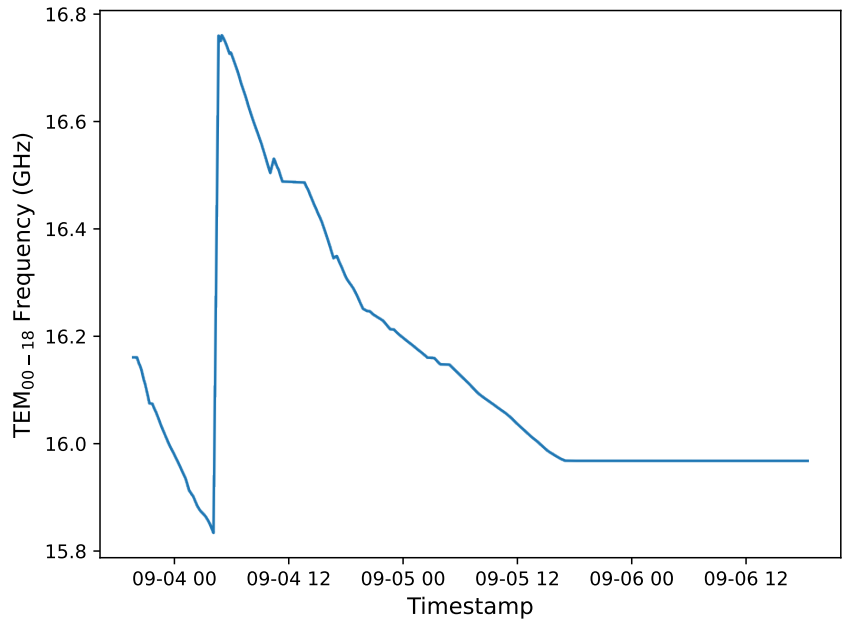}
  \caption{The \tem resonant frequency vs time.}
  \label{fig:dpsearch_freq_v_time}
\end{figure}

I set the digitization time to 30 seconds and tuned downwards until I reached about \SI{15.8}{GHz} (Figure~\ref{fig:dpsearch_freq_v_time}). I stopped here because the $Q$ and cavity coupling were decreasing. I decided to tune the cavity in the opposite direction so that $f_{00-18}\approx \SI{16.8}{GHz}$. Unfortunately, all three motors stalled while tuning in the reverse direction. So I had to turn each motor by hand. Ironically, the Orpheus experiment couldn't look back without great effort. Once the cavity was set to \SI{16.8}{GHz}, I programmed the cavity to tune downwards continuously.

At 9/4/21 4:20 AM, I decided to take a two-hour nap. Coincidentally, at 4:26 AM, the power supply that controlled the RF switches stopped working, and so the digitizer was not connected to the cavity while measuring power spectra. So the resulting data was useless. I went back to the lab at around 7:30 AM to manually reverse the motors so that I could rescan the affected frequency range.

At 12PM, I noticed that Orpheus was closer to becoming critically coupled. So I decided the cavity had returned to a more optimal frequency range, and I increased the digitization time to \SI{100}{s}.

\begin{figure}
  \centering
  \includegraphics[width=0.7\textwidth]{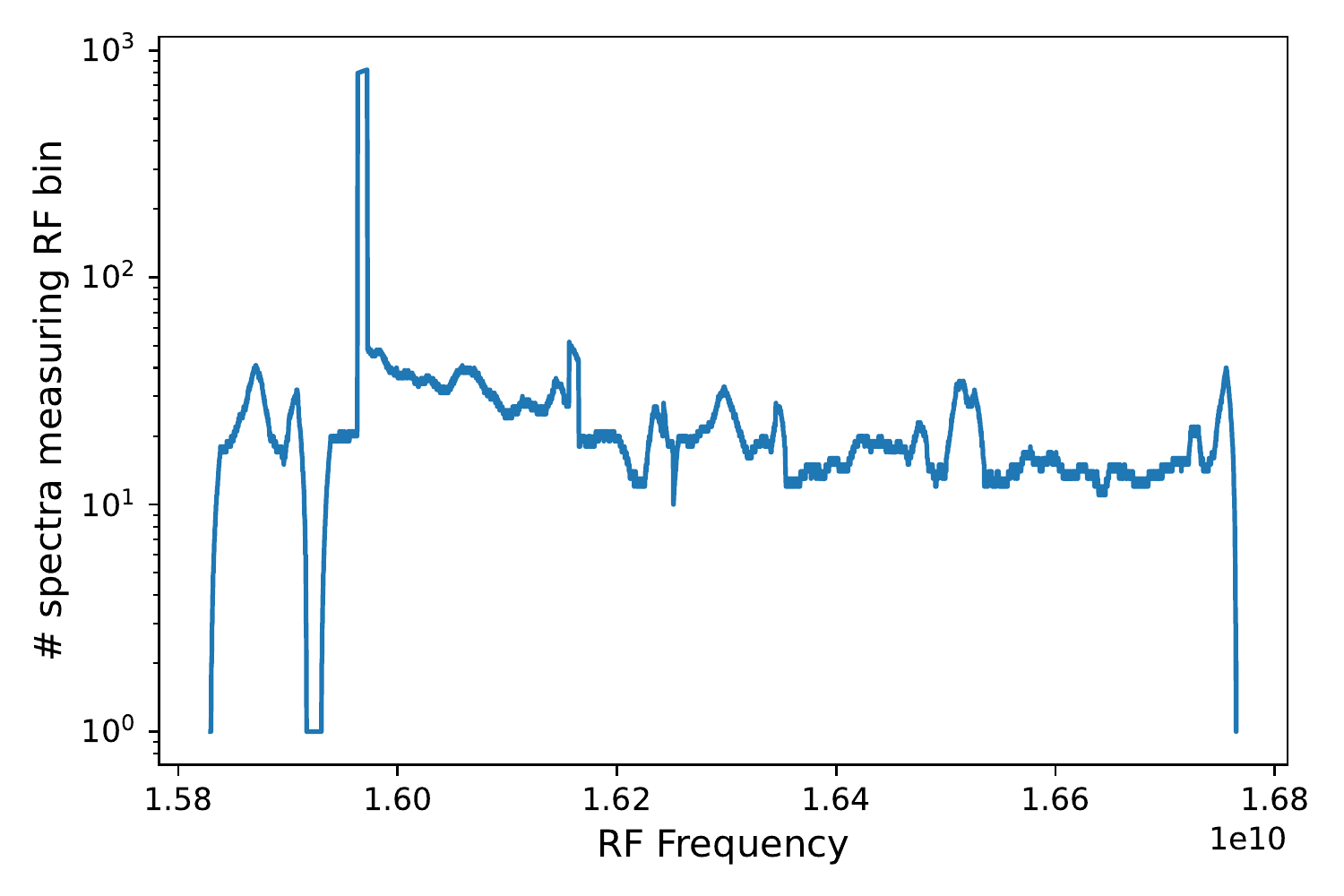}
  \caption{Number of spectrum containing a particular RF bin.}
  \label{fig:if_rf_bin}
\end{figure}

At 10PM, I noticed that the flat mirror's temperature was increasing while the curved mirror temperature stayed constant. I feared I was running out of liquid helium. Because I wanted to sleep and I was unsure how long the experiment could keep running, I decided to stop tuning. I would remain at \SI{15.96}{GHz} until the end of the run. At the time, I didn't know the dark photon coupling sensitivity would be, so I decided that I could have world-leading sensitivity for at least one frequency. This was a bad decision since dark photon sensitivity scales poorly with integration time ($\chi_{exc} \propto t^{-1/4}$). This extended integration time at \SI{15.96}{GHz} is shown in the histogram of the number of IF bins corresponding to an RF bin, shown in Figure~\ref{fig:if_rf_bin}.

At 9/5/2021 6PM, I noticed that the flat mirror was \SI{0.5}{K} hotter than the curved mirror. I suspected that it would be difficult to interpret a cavity temperature from this data. So that's why I ended the run.

%% file: analysis.tex
\chapter{Dark Photon Search Analysis}\label{ch:analysis}
The data collected between 9/3/2021 and 9/7/2021 was used to search for dark photons between \SI{65.5}{\mu eV} (\SI{15.8}{GHz}) and \SI{69.5}{\mu eV} (\SI{16.8}{GHz}). No candidate for a dark photon was found, so a 90\% confidence level exclusion limit was placed on the kinetic mixing angle $\chi$ in this mass range. The procedure for deriving the exclusion limits in this chapter follows the procedure developed by ADMX and HAYSTAC~\cite{PhysRevD.64.092003, PhysRevD.96.123008, PhysRevD.103.032002}, and is adapted for dark photon searches~\cite{caputo2021dark, Ghosh:2021ard}. Perhaps one of the more innovative aspects of this analysis is producing dark photon limits directly from haloscope data rather than rescaling dark photon limits from axion limits.

The dark photon signal would show up as spectrally narrow power excess over a noise background (Section~\ref{sec:haloscope_search}). The dark photon signal power is 

\begin{align}
  & P_{S} = 
  \begin{cases}
    \eta \chi^2 m_{\ap} \rho_{\ap} V_{eff} Q_L \betaterm L(f, f_0, Q_L) & \quad \text{natural units} \\
     \frac{1}{\hbar}\eta \chi^2 (m_{\ap}c^2) (\rho_{\ap}c^2) V_{eff} Q_L \betaterm L(f, f_0, Q_L) & \quad \text{with physical constants} 
  \end{cases}\\
  & L(f, f_0, Q_L) = \frac{1}{1+4\Delta^2} \hspace{2cm} \Delta \equiv Q_L \frac{f-f_0}{f_0}
  \label{eqn:dark_photon_power}
\end{align}
where $\chi$ is the kinetic mixing angle between the dark photon and Standard Model (SM) photon, $\eta$ is a signal attenuation factor, $m_{\ap}$ is the dark photon mass, $\rho_{\ap}$ is the local density of dark matter, $V_{eff}$ is the detector's effective volume, $Q_L$ is the loaded quality factor of the relevant mode, and $\beta$ is the cavity coupling coefficient. $L(f, f_0, Q_L)$ is the Lorentzian term and depends on the SM photon frequency $f$, cavity resonant frequency $f_0$, and $Q_L$. The effective volume is a focal point of this thesis and is discussed in various other sections, including Section~\ref{sec:orpheus_simulations}. It is calculated by 
\begin{align}
  V_{eff} = \frac{\left (\int dV \vec{E}(\vec{x}) \vdot \vec{X}(\vec{x})\right )^2}{\int dV |\vec{E}(\vec{x})|^2|\vec{X}(\vec{x})|^2}
\end{align}
where $\vec{E}$ is the relevant mode's electric field and $\vec{X}$ is the dark photon field. Let $\theta$ be the angle between the cavity field and the dark photon field, such that $\vec{X}\vdot\vec{E} = |\vec{X}||\vec{E}|\cos\theta$. Orpheus is only sensitive to dark photons polarized in the $\hat{y}$ direction (along the short side of the rectangular waveguides). If the dark photon is polarized, then $\langle \cos^2\theta \rangle_T$ can be as low as 0.0025.
If the dark photon is unpolarized, Orpheus can detect a third of the dark photons and $\langle \cos^2\theta \rangle_T = 1/3$~\cite{caputo2021dark}.   

The dark photon search strategy is to look for a spectrally narrow power excess over the noise floor. In broad strokes, the strategy is to first remove the low-frequency structure from each power spectrum, such that the population mean of each bin is zero and deviation from zero is either from statistical fluctuation due to the noise temperature of the detector or from a coherent RF signal. This results in a unitless power excess normalized to system noise power. In searching for potential dark photon candidates, power excesses with high SNR are sought. Thus the power excess is rescaled such that the power excess is in units of single-bin dark photon power. That leads to the population mean of a bin of 1 in the presence of a single-bin dark photon signal. To account for the dark photon power being spread across many bins, a lineshape filter, i.e., matched filter with dark photon kinetic energy distribution, is applied to the rescaled power excess to increase the SNR of the potential signal. The different, partially overlapping spectra are then combined using a maximum likelihood weighting procedure. In the absence of any dark photon signal candidates, the sample mean and sample standard deviation of each bin in the grand spectrum can be used to place a 90\% confidence exclusion limit on the scanned dark photon mass ranges.

\section{Confirmation that the DAQ Followed the Correct Mode}
\begin{figure}
  \centering
  \subfloat[]{\includegraphics[width=0.49\textwidth]{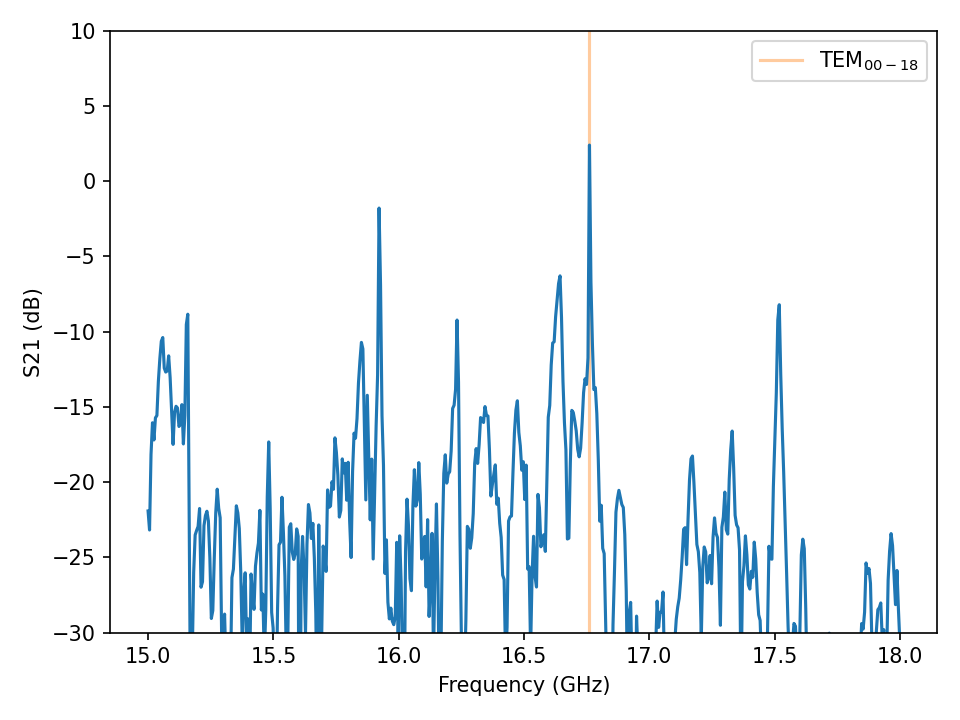}}\hfil
  \subfloat[]{\includegraphics[width=0.49\textwidth]{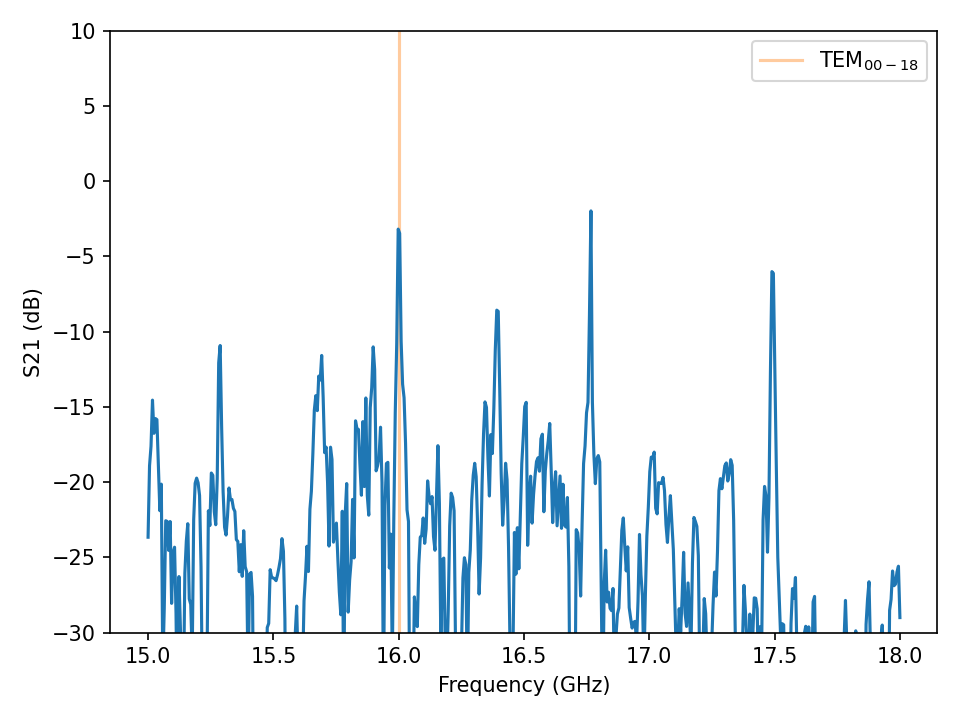}}
  \caption{Wide scan transmission measurements with the simulated mode overlaid for different cavity lengths. Because the resonator is encased in a stainless steel can, there are low-Q non-tuning modes that make the transmission measurement messier than in Figure~\ref{fig:tabletop_orpheus_widescan}.}
  \label{fig:dpsearch_widescan}
\end{figure}

\begin{figure}
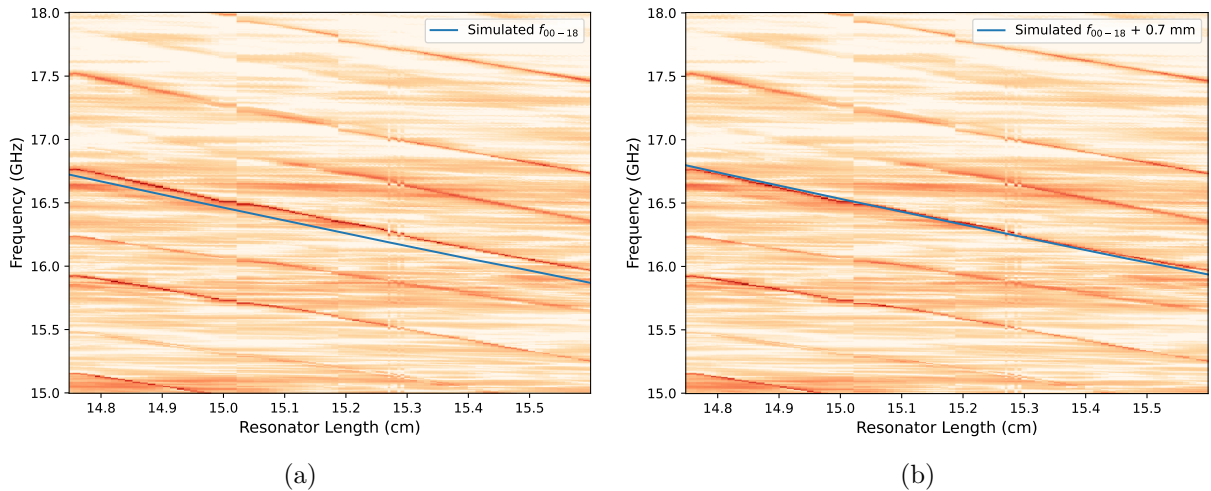

  \centering
  \subfloat[]{\includegraphics[width=0.49\textwidth]{dark_photon_search_modemap.pdf}}\hfil
  \subfloat[]{\includegraphics[width=0.49\textwidth]{dark_photon_search_modemap_adjusted_mode.pdf}}
  \caption{Piecing together wide scan measurements to form a mode map. The discontinuities near \SI{15}{cm} and \SI{15.3}{cm} are from adjusting the top dielectric plate every once in a while. (b) is the same as (a), but the simulated frequency is shifted over by \SI{0.7}{mm} to show the systematic error in measuring cavity length.}
  \label{fig:dpsearch_modemap}
\end{figure}

\begin{figure}
  \centering
  \includegraphics[height=0.3\textheight]{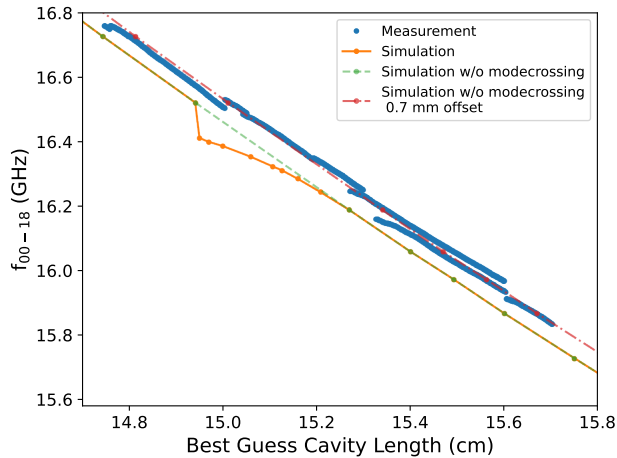}
  \caption{The \tem mode frequency vs the cavity length. The cavity length is determined from the stepper motor encoder values.}
  \label{fig:dpsearch_modefreq}
\end{figure}

An ideal way to confirm whether digitization was centered on the dark photon-coupling mode is to measure the mode map, similar to the tabletop measurement Figure~\ref{fig:tabletop_orpheus_modemap}. From the mode map, one determines if the measured cavity mode structure matches expectation, leading to greater confidence that the DAQ followed the correct mode. Unfortunately, the motor mechanism could not tune smoothly throughout the entire range of motion. Manual intervention was needed to handle portions of the tuning range where the motor would stall, so an automated mode map was infeasible. Instead, VNA wide scans were taken once every 20 data-taking cycles. The wide scans measured the transmission coefficient S$_{21}$ from \SI{15}{GHz} to \SI{18}{GHz}, as shown in Figure~\ref{fig:dpsearch_widescan}. These wide scan measurements for various cavity lengths can be stitched together to form the mode map shown in Figure~\ref{fig:dpsearch_modemap}. The blue line in Figure~\ref{fig:dpsearch_modefreq} is the frequency the DAQ determined to be the \tem mode in real-time, and digitization was always centered around this frequency. The simulated \tem mode is overlaid in both Figure~\ref{fig:dpsearch_modemap} and Figure~\ref{fig:dpsearch_modefreq} and is shown to be offset from the measured mode directly above by about \SI{0.7}{mm}. This can be attributed to a systematic error in measuring the absolute length of the cavity, likely caused by the motor system's backlash. Overall, Figures~\ref{fig:dpsearch_modemap} and~\ref{fig:dpsearch_modefreq} show that the \tem mode is easy to follow and that the DAQ digitized around the correct frequency.

The motor encoder values are used to calculate the cavity length. The motor encoders are first set to 0 when the cavity length is \SI{13.6}{cm} long. The steps counted by the motor encoders are then used to calculate relative changes in distance. One full revolution corresponds to 20,000 motor steps, and twenty revolutions correspond to a plate moving one inch along the 1/4''-20 threaded rod.

There are some features and oddities in Figures~\ref{fig:dpsearch_widescan},~\ref{fig:dpsearch_modemap}, and~\ref{fig:dpsearch_modefreq} that must be explained. The wide scans in Figure~\ref{fig:dpsearch_widescan} are more feature-rich than the wide scans of the tabletop setup shown in Figure~\ref{fig:tabletop_orpheus_widescan}. That is because the cavity sits inside a stainless steel rectangular vessel, and the non-tuning modes of the vessel pollute the mode structure. These non-tuning modes are also evident as horizontal lines in the mode map on Figure~\ref{fig:dpsearch_modemap}. The rectangular cavity was not included in the simulations because it would have been computationally expensive. Although it's possible these rectangular cavity modes can mix with the \tem mode and affect $V_{eff}$, I believe this to be a negligible effect. The quality factors of these rectangular cavity modes are around 200, which is much lower than \tem Q of 5000-10,000. Because of the difference in Q, these modes will hardly mix, and I will use the $V_{eff}$ from the simulations. 

Another oddity is that the $f_{00-18}$ in Figure~\ref{fig:dpsearch_modefreq} seems double-valued and discontinuous. The discontinuities were created when the top dielectric plate was tuned by hand every few hours to compensate for the fact that the motor would stall under normal operation. Each time the top dielectric plate was tuned, $f_{00-18}$ shifted by more than \SI{10}{MHz}. The discrete steps are shown in Figure~\ref{fig:dpsearch_motor_steps_v_time}. The discontinuities are evident in the mode map in~Figure~\ref{fig:dpsearch_modemap} and the measured mode frequency in Figure~\ref{fig:dpsearch_modefreq}.  

The resonant frequencies in Figure~\ref{fig:dpsearch_modefreq} look double-valued because the cavity lengths are swept through twice. The same VNA measurements taken with the same motor encoder values at different times don't result in the same resonant frequency because the motor backlash would lead to different cavity lengths.  The mode map avoids the double-valued resonant frequencies because timestamps below 2021-09-03 21:30:00 UTC are filtered out, so the mode map only shows the second sweep.

The last oddity is mentioned in Section~\ref{sec:orpheus_simulations}. The simulated frequencies near the simulated mode crossing deviate from the straight line. But tabletop mode map measurements (Figure~\ref{fig:tabletop_orpheus_modemap}) shows that the \tem mode tunes smoothly and continuously. This deviation in simulation is likely because the simulations don't have enough precision to deal with the mode crossing. To deal with this, I interpolate the resonant frequency and $V_{eff}$ through the simulated mode crossing. Section~\ref{sec:parameter_extraction} will describe how $\veff$ is determined for data analysis.

\FloatBarrier
\section{Confirmation of a Healthy Receiver}
Two in-situ measurements were performed to make sure that the receiver was healthy, i.e., no compression and every component is operating within the desired dynamic range.

\begin{figure}
  \centering
  \includegraphics[height=0.3\textheight]{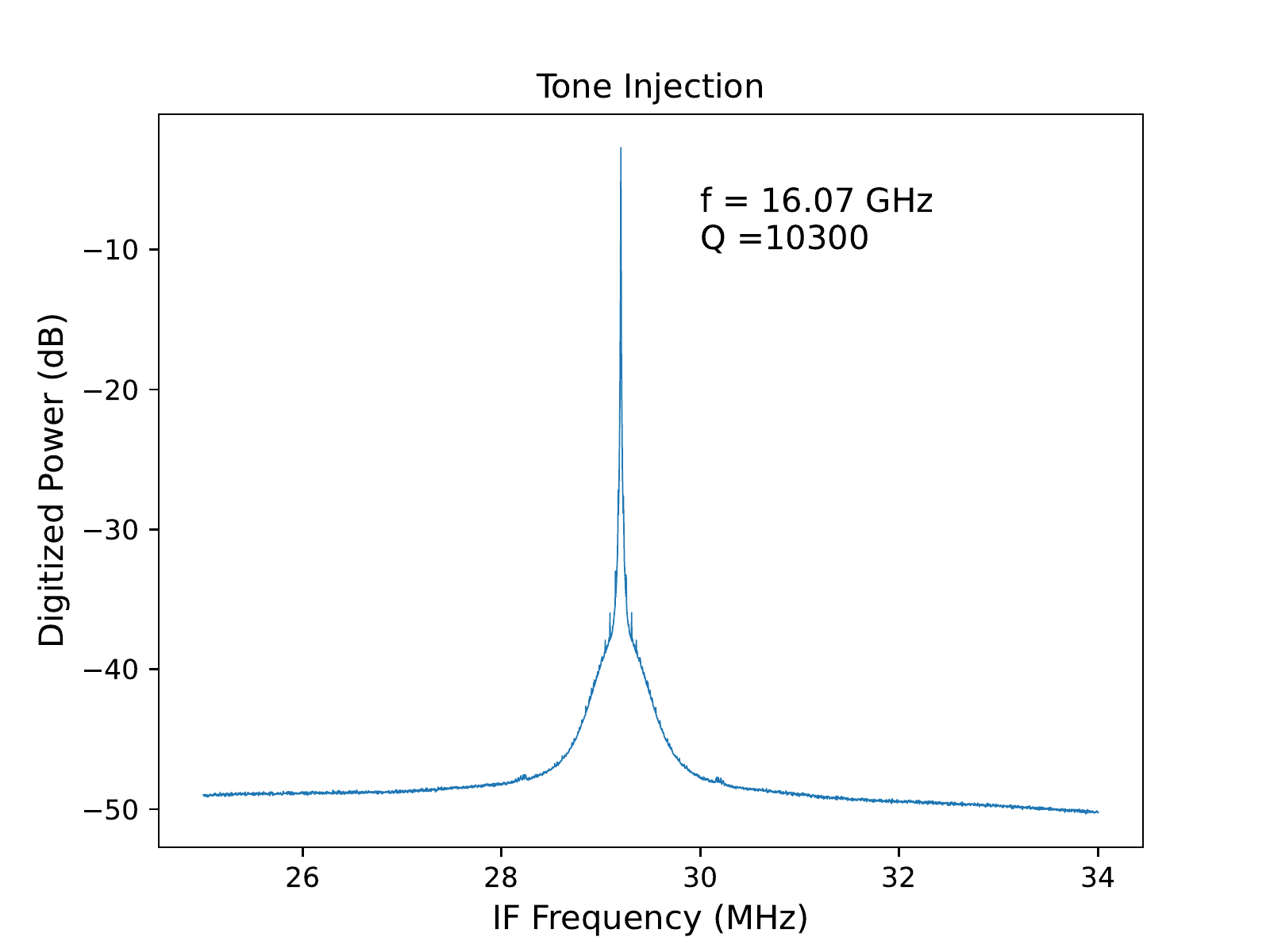}
  \caption{Use the VNA to inject a tone into the cavity. Both the tone frequency and the resonant frequency was at \SI{16.07}{GHz}. The loaded Q was about 10,000.}
  \label{fig:dpsearch_tone_injection}
\end{figure}

One measurement is to record the power spectrum as one injects a sine wave, i.e., tone injection, into the cavity's weak port. If the receiver is healthy, the injected RF tone should show up as a tone in the digitized spectrum. A signal generator wasn't available for this injection, so the output of the VNA was used as a tone generator. The VNA behaves like a tone generator if the frequency span and IF bandwidth are set to zero. The center of the VNA sweep was set to the resonant frequency of the cavity. For this measurement (Figure~\ref{fig:dpsearch_tone_injection}), the resonant frequency was $f_{00-18}=\SI{16.0742}{GHz}$, the loaded Q is $Q_L=10300$, and the cavity was critically coupled. 

Another measurement is to record the system noise power as a function of cavity temperature. This measurement can be performed during a cool down or warm up. The system noise power can be described as the sum of the blackbody radiation of the cavity and the added Johnson noise of each electrical component, and is written as $P_n=G k_b b T_{sys}$, where $G$ is the system gain that depends on both frequency and physical temperature, $b$ is the frequency bandwidth, and $T_{sys}$ is the system noise temperature referenced to the cavity. The thermal model is described in Section~\ref{sec:thermal_model}.
\begin{align}
  T_{sys} = T_{cav}(1-|\Gamma|^2) + T_{amp, input}|\Gamma|^2 + T_{rec}
\end{align}

\begin{figure}
  \centering
  \includegraphics[height=0.3\textheight]{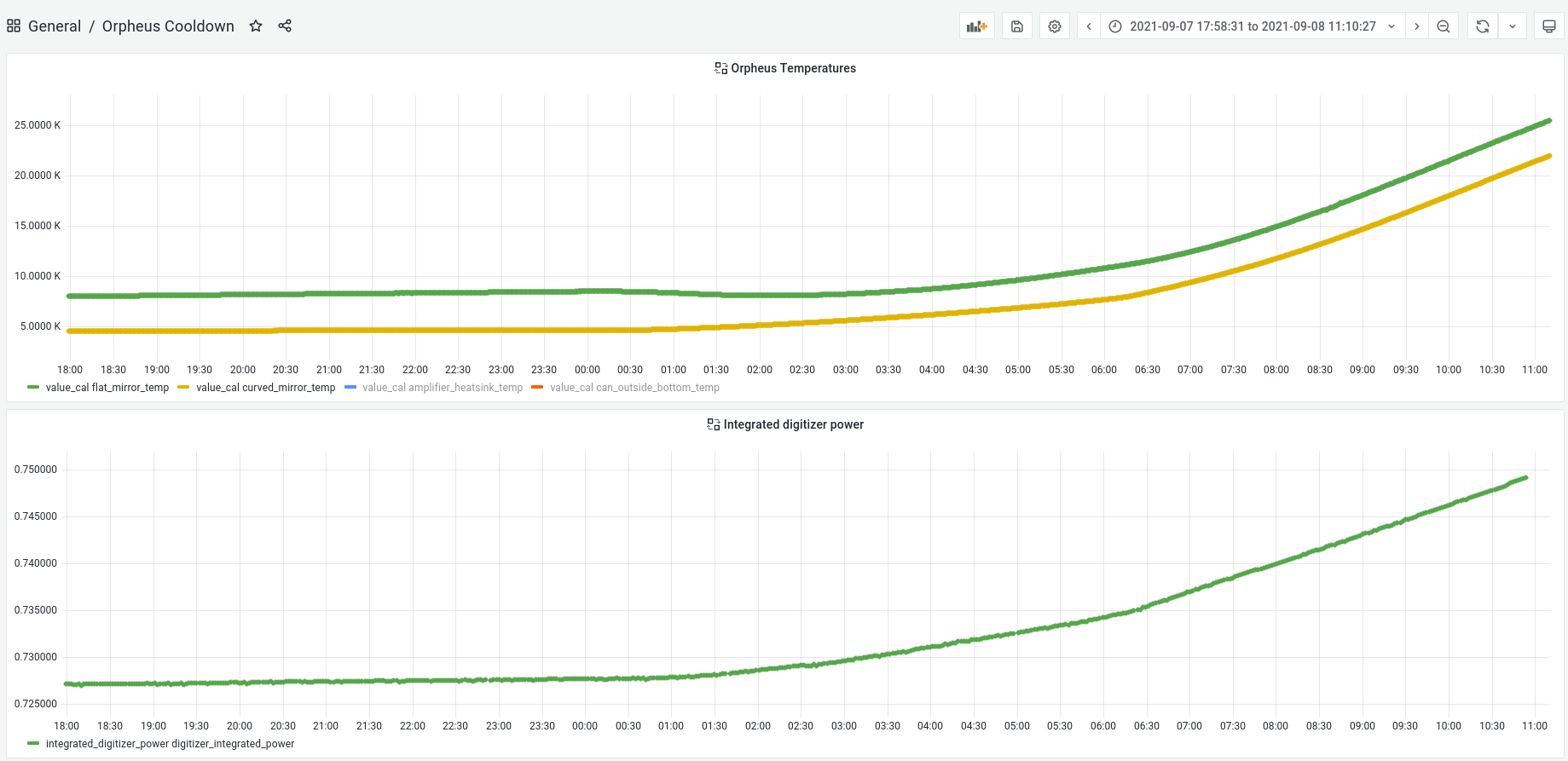}
  \caption{Plotted integrated digitizer power as the cavity is warming up. For reasons outlined in the text, this measurement is difficult to interpret as a Y-factor measurement to obtain the receiver noise temperature. But the digitized power rises more-or-less linearly with the cavity temperature. This suggests a healthy, low-noise receiver chain.}
  \label{fig:dpsearch_warmup}
\end{figure}

Figure~\ref{fig:dpsearch_warmup} shows that the integrated digitized power (the sum of the bins in a power spectrum) tracked linearly with the cavity temperature. For this measurement, the resonant frequency was about \SI{16}{GHz} and the cavity was critically coupled. The digitized spectrum had a frequency span of six cavity widths and was centered on a frequency two cavity widths away from resonance\footnote{I was trying to do a Y-factor measurement off-resonance because I was trying to avoid the uncertainty in the cavity temperature. But there is also a lot of uncertainty in the noise temperature coming from the input of the cryogenic amplifier.}. This linear dependence demonstrates a healthy, non-compressed receiver.

$T_{rec}$ may be determined from the Y-factor measurement at \SI{16}{GHz} from the relationship between the on-resonance power and cavity temperature. The thermal model for this Y-factor measurement is $T_{sys} = T_{cav} + T_{rec}$. Unfortunately, $\sigma_{T_{cav}}\sim \SI{1}{K}$ during the warm up because of the temperature gradient. So $T_{rec}$ is determined from the measurements performed by Low Noise Factor (Section~\ref{sec:receiver_noise_temp}). 

I also confirmed the amplifiers were not being compressed during operation by adjusting the VNA output power and seeing that the transmission measurement remained the same. I did not check if the IF amplifiers were compressed, but the cascade analysis suggests this to be highly unlikely\footnote{In future work, I recommend checking compression of the IF amplifiers by varying the tone power during the tone injection studies.}. All this suggests a healthy, happy, low-noise receiver chain.

\FloatBarrier
\section{Parameter extraction}\label{sec:parameter_extraction}
For each data-taking cadence, ancillary measurements were taken to extract the parameters needed to determine the noise temperature and the dark photon power. The necessary parameters are the cavity temperature, resonant frequency, loaded Q, cavity coupling, and effective volume.

\begin{figure}
  \centering
  \includegraphics[height=0.4\textheight]{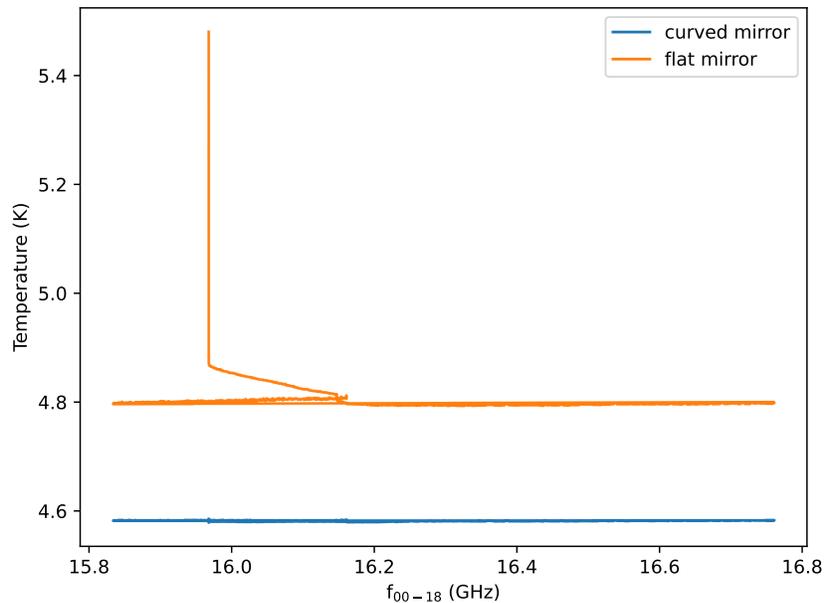}
  \caption{Cavity temperature as a function of mode frequency.}
  \label{fig:temp_v_freq}
\end{figure}

The cavity temperature $T_{cav}$ is taken to be the mean of the flat mirror and curved mirror temperature shown in Figure~\ref{fig:temp_v_freq}, and the uncertainty is taken from a continuous uniform distribution, $\sigma_{T_{cav}} = \frac{T_{flat} - T_{curved}}{\sqrt{12}}$. This temperature $T_{cav}$ is thought to be an overestimate of the mean temperature of the thermal photons coming from the cavity and is therefore a conservative choice. The flat mirror was hotter than the rest of the system because it is thermally sunk to the amplifier, which deposits heat. The flat mirror is also subject to a greater heat leak than the curved mirror because the flat mirror is connected to two copper coaxial cables that connect directly to a room temperature SMA vacuum feedthrough, whereas the curved mirror is only connected to one such copper coaxial cable. Due to issues with cooling down described in Section~\ref{sec:sep1_narrative}, I believe the flat mirror is in poor thermal contact with the rest of the cavity. I also believe that the temperature of the dielectrics is closer to the temperature of the curved mirror than to the flat mirror, and most of the cavity's thermal photons come from the curved mirror and dielectric plates. 

\begin{figure}
  \centering
  \includegraphics[height=0.4\textheight]{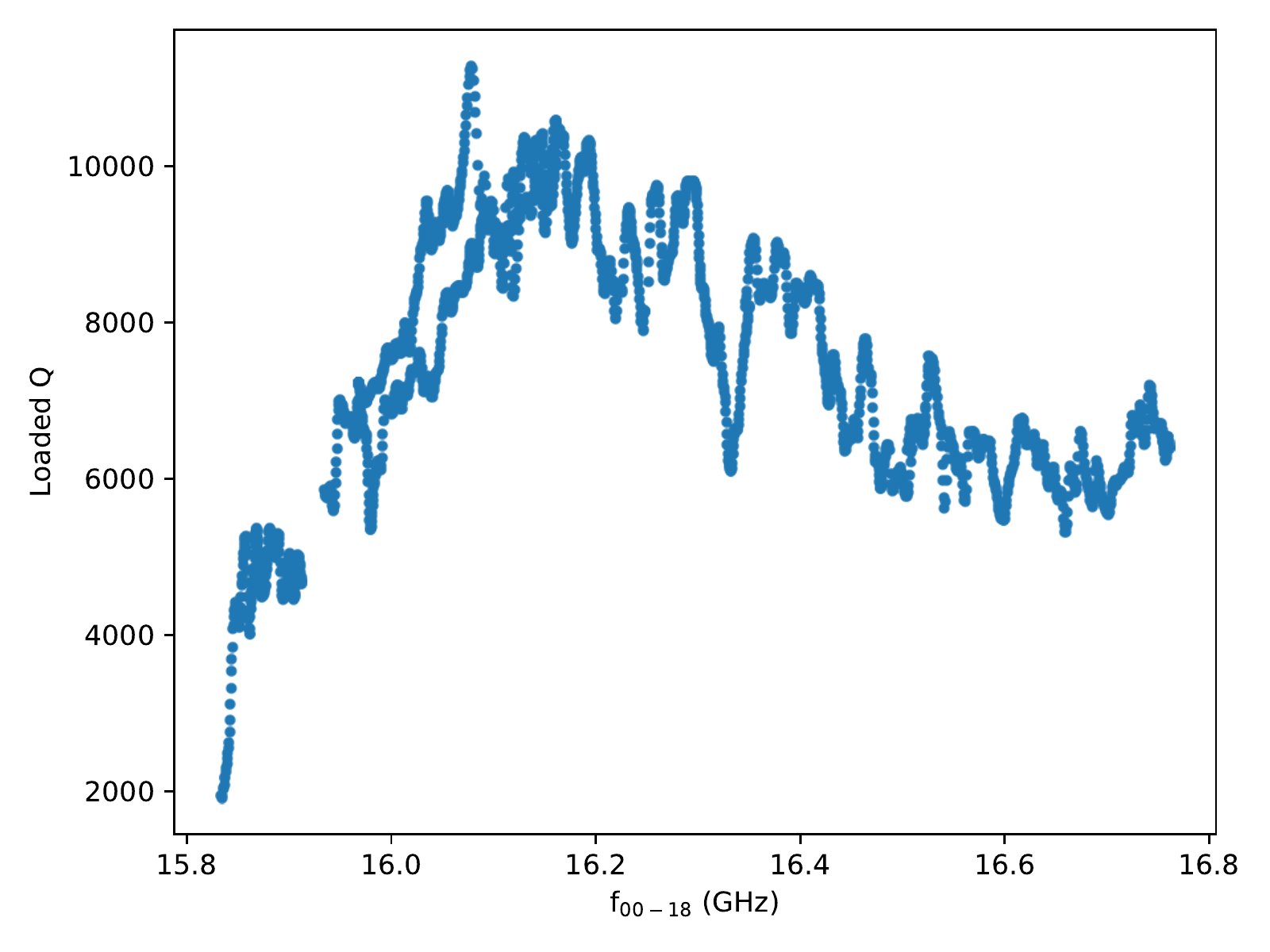}
  \caption{Loaded quality factor as a function of mode frequency.}
  \label{fig:dpsearch_q_v_freq}
\end{figure}

\begin{figure}
  \centering
  \includegraphics[height=0.4\textheight]{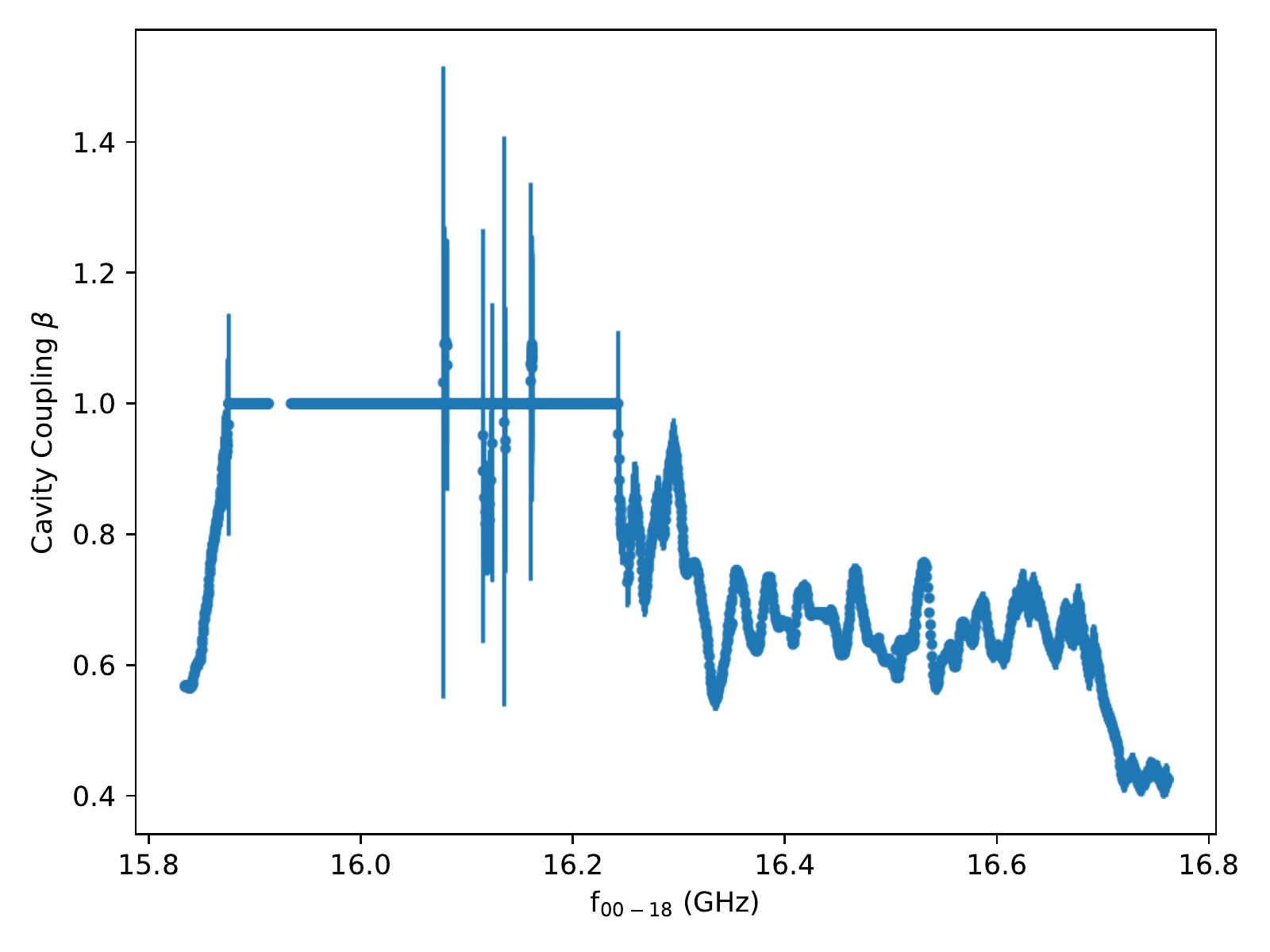}
  \caption{Cavity coupling as a function of mode frequency.}
  \label{fig:dpsearch_beta_v_freq}
\end{figure}

The resonant frequency, loaded Q, and cavity coupling coefficient were derived using the same type of transmission and reflection measurements used in Chapter~\ref{ch:rf_design}. The Lorentzian fits were performed online and offline\footnote{I will refer to online fits as fits that were performed during the data-taking run and offline as fitting performed after the data-taking run.} while building the combined spectrum. For this analysis, both online and offline fitted parameters were used to overcome different data quirks. The online fitted resonant frequencies were used because the starting and ending frequency for each VNA measurement were improperly logged into the database\footnote{But I know that the frequency span is \SI{20}{MHz}. This helped with the offline Q and cavity coupling fits.}. The DAQ also logged the online fitted parameters but didn't log the fitting uncertainties. So the fit was reapplied offline to transmitted and reflected power to obtain these fitting uncertainties ($\frac{\sigma_{f_0}}{f_0}$ is small compared to $\frac{\sigma_{Q_L}}{Q_L}$, so I deemed it ok to just use the online fitted $f_0$). $Q_L$ and $\sigma_{Q_L}$ is determined from the Lorentzian fit of the transmission measurement, and $\beta$ and $\sigma_{\beta}$ is determined from the Lorentzian dip of the reflection measurement (Section:~\ref{sec:empty_measurement}). It's possible to determine $Q_L$ from the reflection measurement, but crosstalk-like effects in the waveguide coupler distort the Lorentzian shape of the reflection measurement\footnote{An example of a distorted Lorentzian is found in Figure~\ref{fig:complicated_fit}.}, making the reflection $Q_L$ fit unreliable. It's possible to add extra parameters to the reflection fitting measurement to better fit the distorted Lorentzian shape and extract a more accurate $Q_L$ (as is done in Appendix~\ref{sec:reflection_coupler}), but the extra fitting parameters have a degeneracy, and extracting a coupling coefficient seems infeasible. The offline fitted $Q_L$ and $\beta$ are shown in Figures~\ref{fig:dpsearch_q_v_freq} and~\ref{fig:dpsearch_beta_v_freq}.

Because the Lorentzian fit doesn't capture the full physics of a reflection measurement, there are often fits, particularly when the cavity is critically-coupled, when the fitted reflected power is negative on resonance. When this occurs, the Lorentzian dip is ofter greater than \SI{15}{dB}, which is very close to critical coupling. In this case, the coupling coefficient is taken to be 1 with no uncertainty. This choice seems safe because near critical coupling, the dark photon power is insensitive to uncertainty in $\beta$, as shown in Figure~\ref{fig:dgamma_dbeta}.
\begin{figure}
  \centering
  \includegraphics[height=0.4\textheight]{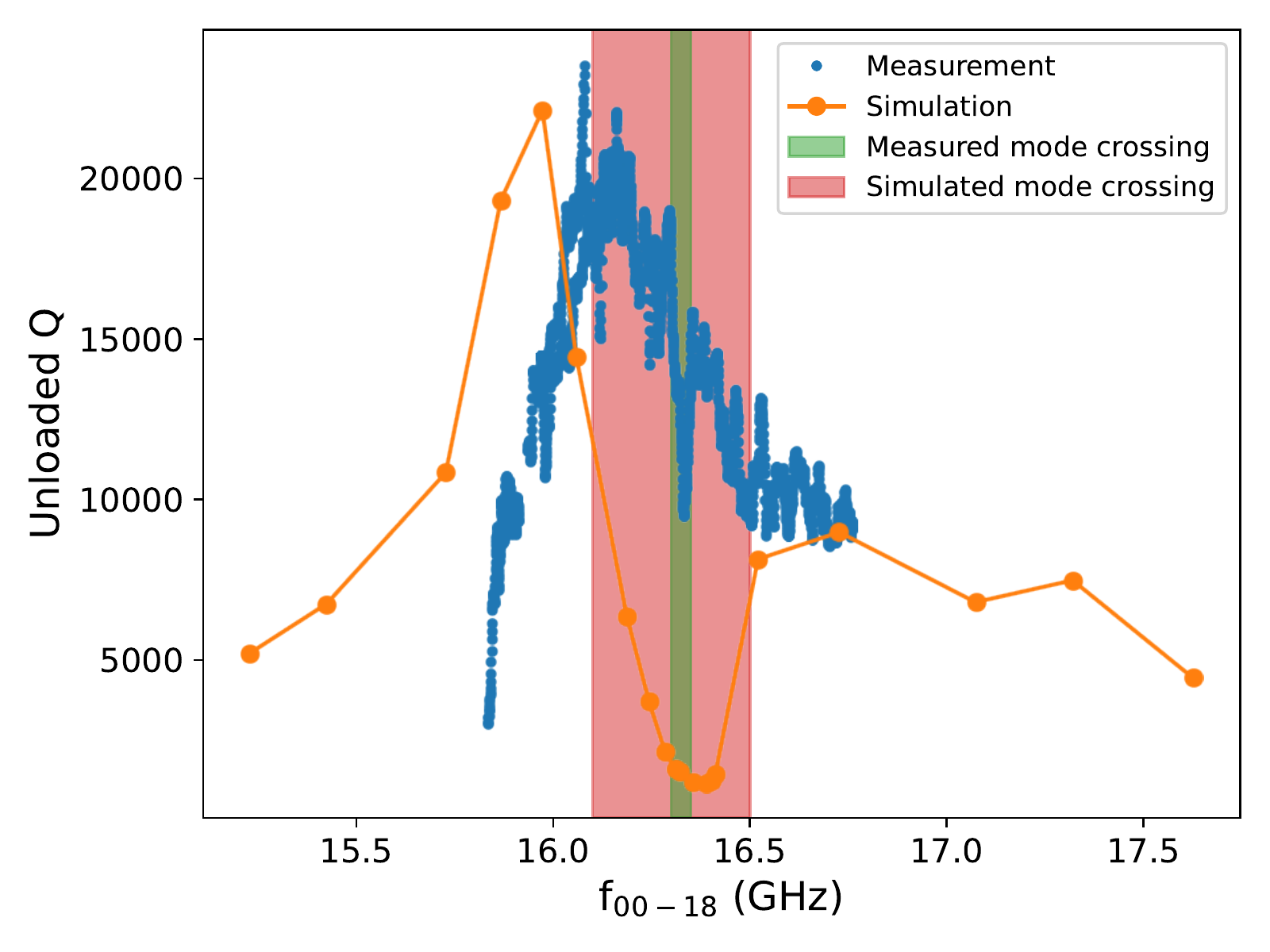}
  \caption{Unloaded quality factor as a function of frequency, as derived from the measured loaded quality factor and cavity coupling. The results from the simulation are overlaid for comparison and to identify the mode crossing in the data. The notch in the simulated Q is much wider and steeper than the measured Q. This discrepancy suggests the mode crossing in the simulation is much larger than the mode crossing of the physical cavity.}
  \label{fig:Qu_v_freq}
\end{figure}

\begin{figure}
  \centering
  \includegraphics[height=0.4\textheight]{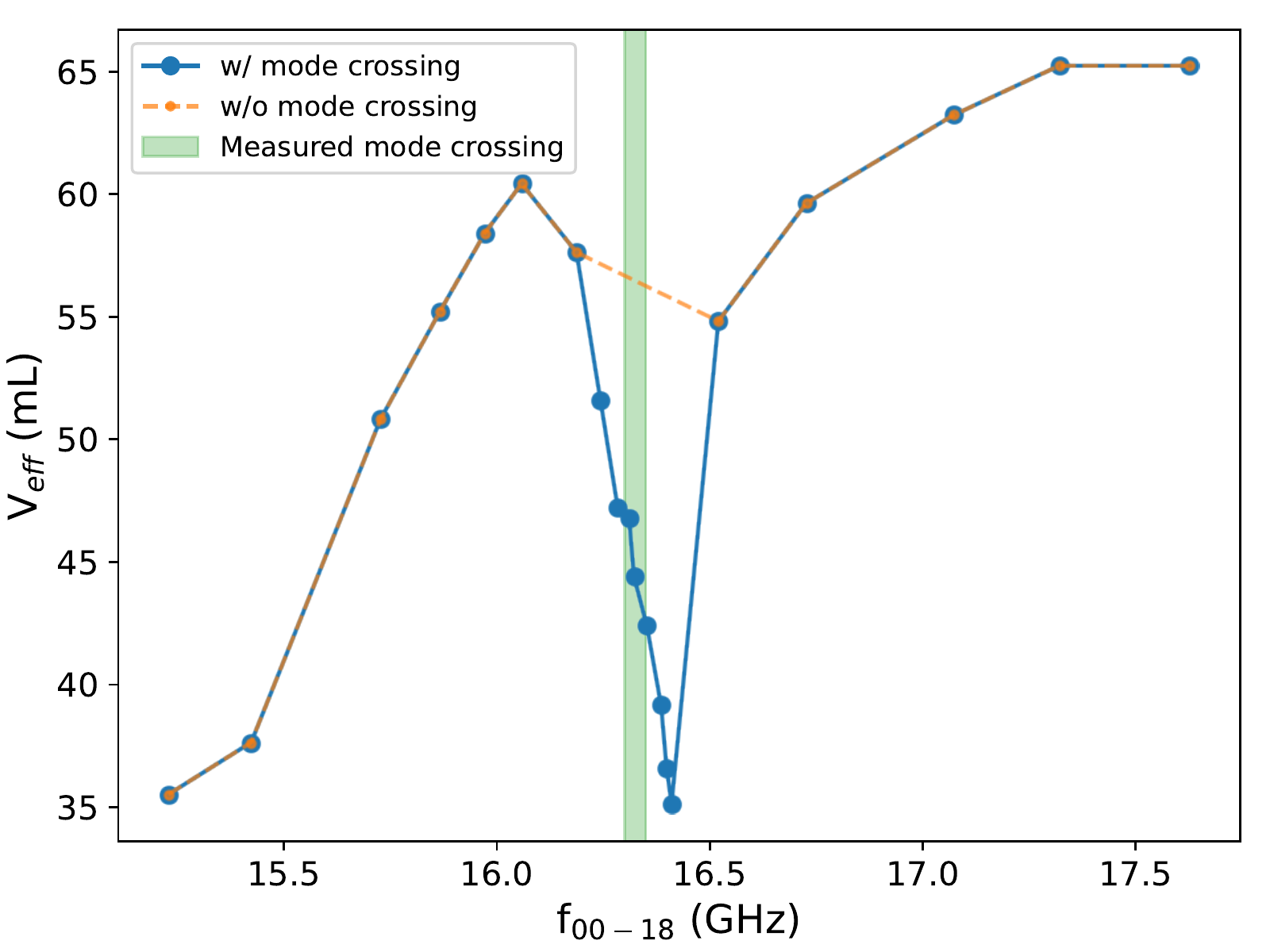}
  \caption{The simulated effective volume a function of mode frequency. $V_{eff}$ is plotted with and without the mode crossing because I will have to take care of how I handle the mode crossing in the data analysis.}
  \label{fig:veff_v_frequency}
\end{figure}

$V_{eff}$ is determined from simulation because directly measuring the electric field is infeasible. The simulations are described in detail in Section~\ref{sec:orpheus_simulations}. This section will focus on dealing with the mode crossing near \SI{16.35}{GHz}. As has been discussed in Sections~\ref{sec:orpheus_simulations} and~\ref{sec:tem_characterization}, a measured dip in quality factor can be used as a proxy for a dip in $\veff$. Figure~\ref{fig:Qu_v_freq} compares the measured unloaded quality factor $Q_u = Q_L \left ( \beta + 1 \right ) $ and the simulated $Q_u$. The dip in $Q_u$ corresponds to the mode crossing, but the comparison shows that the measured mode crossing is much narrower than the simulated mode crossing. This suggests that the actual mode crossing does not affect $\veff$ as dramatically as simulations suggest. To deal with this discrepancy, $V_{eff}$ is interpolated through the mode crossing, as shown by the orange curve in Figure~\ref{fig:veff_v_frequency}. Within the measured mode crossing region between \SI{16.3}{GHz} and \SI{16.35}{GHz}, the effective volume is set to $V_{eff} = \SI{35}{mL}$, the smallest $\veff$ predicted by simulation. Mode crossings are often omitted from haloscope science limits, but I don't think that's necessary for Orpheus. The intruder mode has a much smaller Q than the \tem mode, and the mode crossing in the measured mode map lacks the avoided crossing structure that's ubiquitous to mode crossings in other haloscopes. $V_{eff}=\SI{35}{mL}$ is a conservative estimate considering that the measured mode crossing is milder than the simulated mode crossing. Outside the measured mode crossing, the orange curve in Figure~\ref{fig:veff_v_frequency} is used.

\begin{figure}
  \centering
  \includegraphics[height=0.4\textheight]{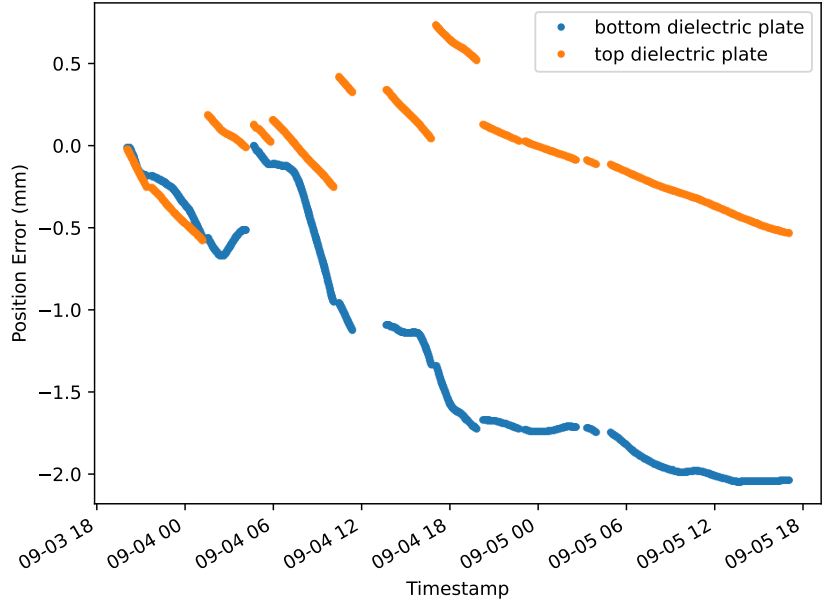}
  \caption{The deviation of the plate positions from their targeted positions. This position error was used to estimate the uncertainty in $V_{eff}$.}
  \label{fig:motor_step_error}
\end{figure}

$V_{eff}$ is simulated for when the dielectrics are evenly spaced throughout the cavity. Unfortunately, neither the top nor bottom dielectric plate was in this configuration the entire run. A mechanical failure caused the top dielectric plate motor to stall, so automated tuning was not possible. A cascade of DAQ design flaws and operating errors caused the bottom dielectric plate motor to undershoot the intended position. The position error is shown in Figure~\ref{fig:motor_step_error}. Fortunately, the simulations in Section~\ref{sec:position_error} show that at \SI{15.8}{GHz}, a negative position error actually increases the form factor. The position error is always negative for the bottom dielectric plate and almost always negative for the top dielectric plate.  It would take more time than I have to simulate $\veff$ for the position errors of the entire run. But there is a compelling physical intuition for why a negative position error increases the form factor, and this ought to hold true for all frequency ranges scanned. 

The position error is used to place an uncertainty on $V_{eff}$. The position error for the bottom plate is always negative. If the position error of the top dielectric plate is negative, then the analysis uses $V_{eff}$ for the evenly configured case with no uncertainties. Since negative position errors increase $V_{eff}$, this is a conservative estimate. However, positive position errors decrease $\veff$. The top dielectric plate is rarely above \SI{5}{mm}, and Figure~\ref{fig:veff_position_err} suggests that a \SI{5}{mm} error corresponds to a 7.4\% decrease in $V_{eff}$. In the absence of a full simulation campaign, I use the heuristic that any positive position error corresponds to a relative uncertainty of $\frac{\sigma_{V_{eff}}}{V_{eff}}=0.074$.

For this analysis, the attenuation factor $\eta = 1$. The attenuation factor is comprised of both the insertion loss of the waveguide and electronic components and analysis attenuation from removing the spectral baseline. The insertion loss is largely ohmic, and the resistivity of the waveguide components diminishes at \SI{4}{K}. As a concrete example, the waveguide-to-coaxial adaptor has an insertion loss of \SI{0.3}{dB} at room temperature and is made out of aluminum. The residual resistance ratio of aluminum is around 32~\cite{Fickett2019ElectricalPO}, so the insertion loss would be about \SI{0.01}{dB}, subdominant to other sources of uncertainty. The signal attenuation induced by the baseline removal method (discussed in Section~\ref{sec:baseline_removal}) has not yet been studied but should before publication.

\FloatBarrier
\section{Estimating expected sensitivity}\label{sec:expected_sensitivity}
Before assembling the combined spectrum to derive an exclusion limit for $\chi$, I first use the operating parameters to estimate the detector sensitivity. I will use natural units for this calculation. The operating parameters are shown can be read in the plots from the previous section and are shown in Table~\ref{tab:operating_parameters}.

\begin{table}[ht]
\begin{tabular}{l|l|l}
parameter   & value               & natural units conversion if needed \\ \hline 
$\beta$     & 1                   &                                    \\\hline
$V_{eff,max}$   & \SI{60}{cm^3}       &                                    \\\hline
$\rho_{\ap}$ & \SI{0.45}{GeV/cm^3} &                                    \\\hline
$m_{\ap}$    & \SI{65e-15}{GeV}    &                                    \\\hline
t           & \SI{100}{s}         & \SI{1.5e26}{GeV^{-1}}              \\\hline
Q           & 10000               &                                    \\\hline
$T_n$       & \SI{10}{K}          & \SI{8.617e-13}{GeV}                \\\hline
SNR         & 3                   &
\end{tabular}
\caption{Operating parameters for Orpheus used to estimate Orpheus science reach.}
\label{tab:operating_parameters}
\end{table}

\begin{align*}
  P_{\ap} &=  \chi^2 m_{\ap} \rho_{\ap} V_{eff} Q_L \betaterm \\
  P_n &= b T_n;\hspace{5mm} \sigma_{P_n} = \frac{P_n}{\sqrt{N}} = \frac{P_n}{\sqrt{t b }}\\
  \snr &= \frac{Ps}{\sigma_{P_n}} = \frac{Ps}{P_n}\sqrt{t b}\\
  \chi &= \sqrt{\frac{\beta+1}{\beta}\frac{\snr \times b T_n}{m_{\rho}V_{eff}Q_L}}\left ( \frac{1}{b t} \right )^{1/4}
\end{align*}
So 
\begin{align*}
  \chi_{max} &= \SI{4.94e-14}{}; \hspace{5mm}\chi = \frac{\chi_{max}}{\sqrt{\langle \cos^2\theta \rangle_T}}\\
  \chi_{unpolarized} &= \frac{\chi_{max}}{\sqrt{1/3}} = \SI{8.56e-14}{}\\
  \chi_{polarized} &= \frac{\chi_{max}}{\sqrt{0.0025}} = \SI{9.88e-13}{}
\end{align*}

\FloatBarrier
\section{Processing Individual Raw Spectrum}
\subsection{Removing Spectral Baseline}\label{sec:baseline_removal}
\begin{figure}
  \centering
  \subfloat[]{\includegraphics[width=0.33\textwidth]{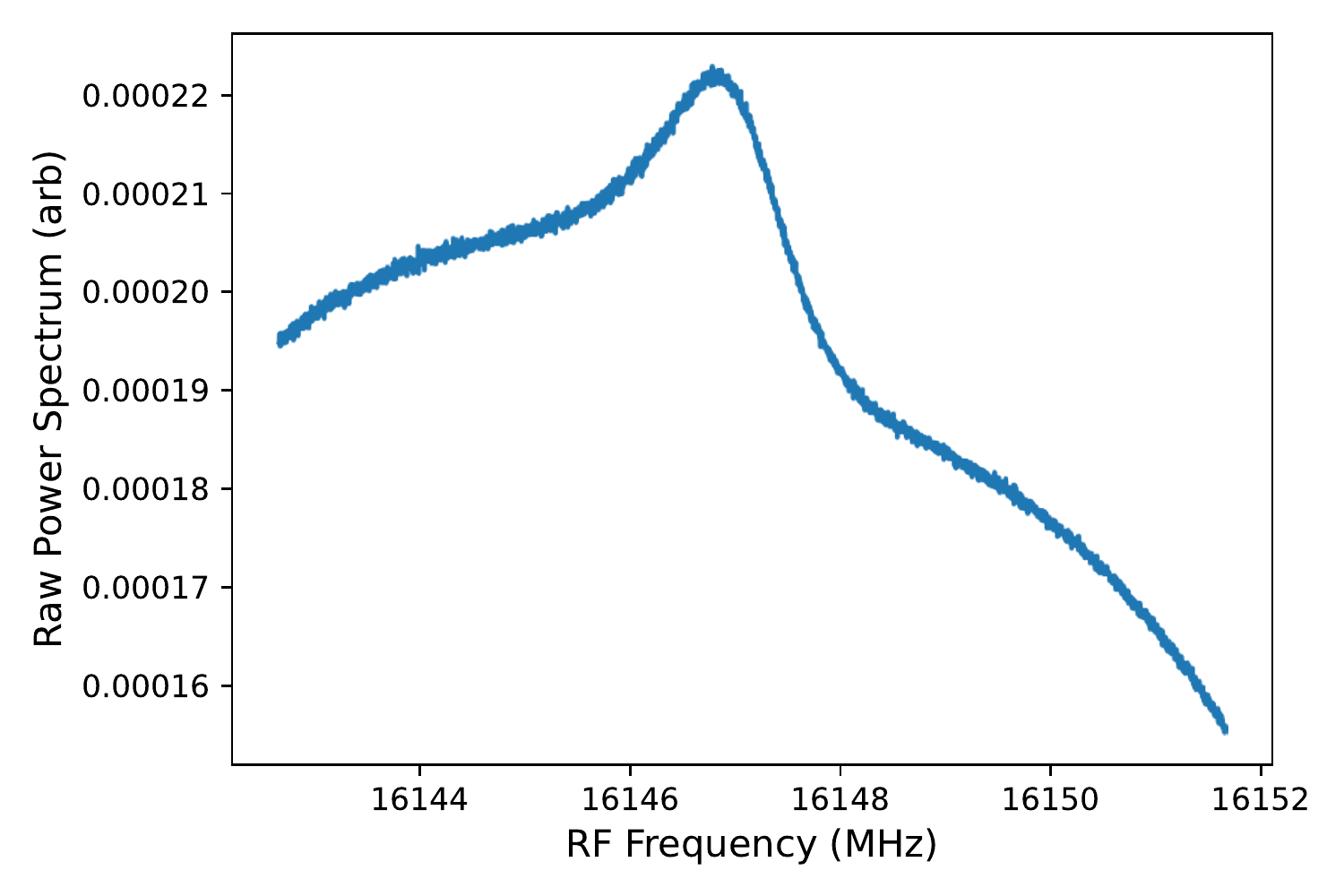}}\hfil
  \subfloat[]{\includegraphics[width=0.33\textwidth]{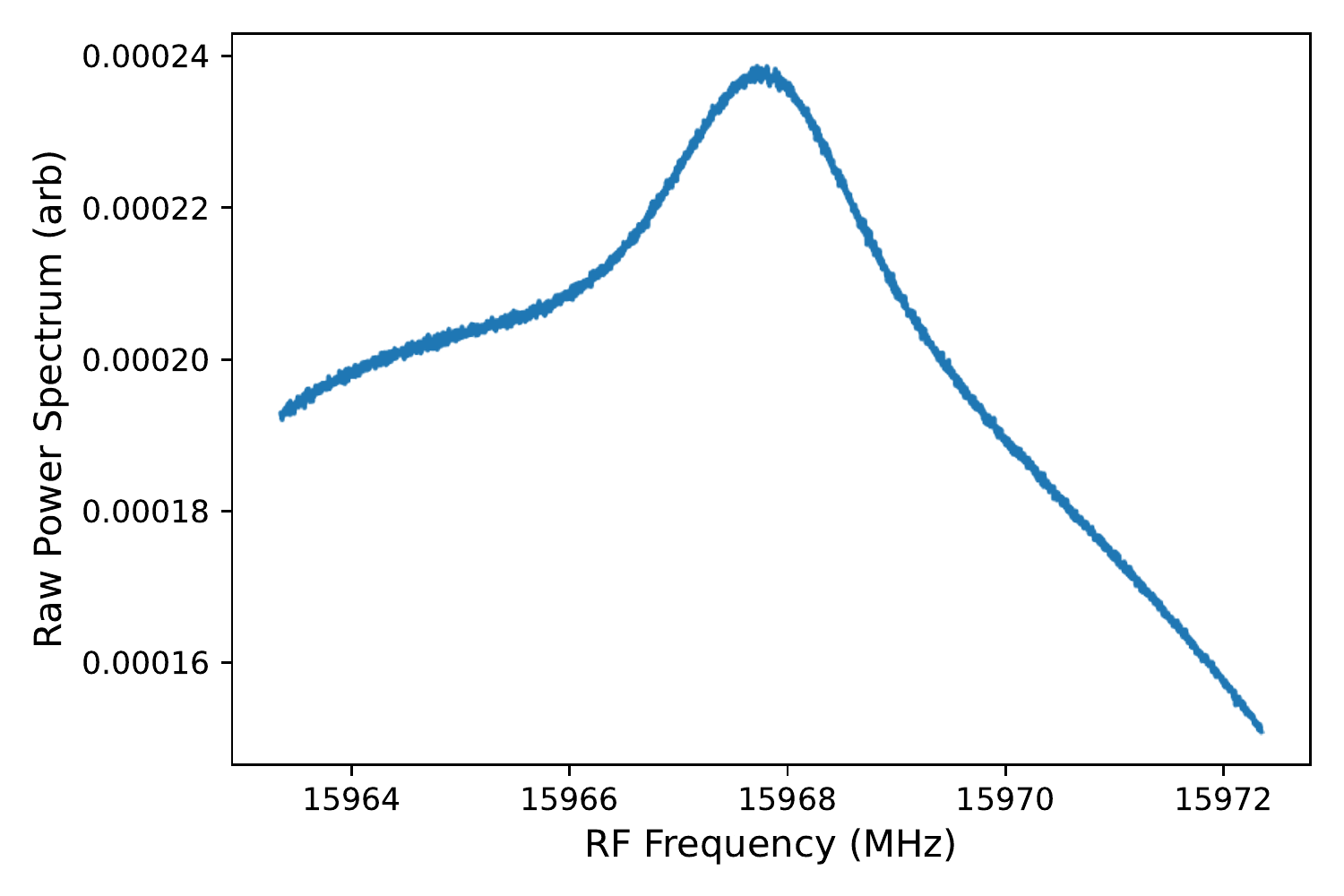}}\hfil
  \subfloat[]{\includegraphics[width=0.33\textwidth]{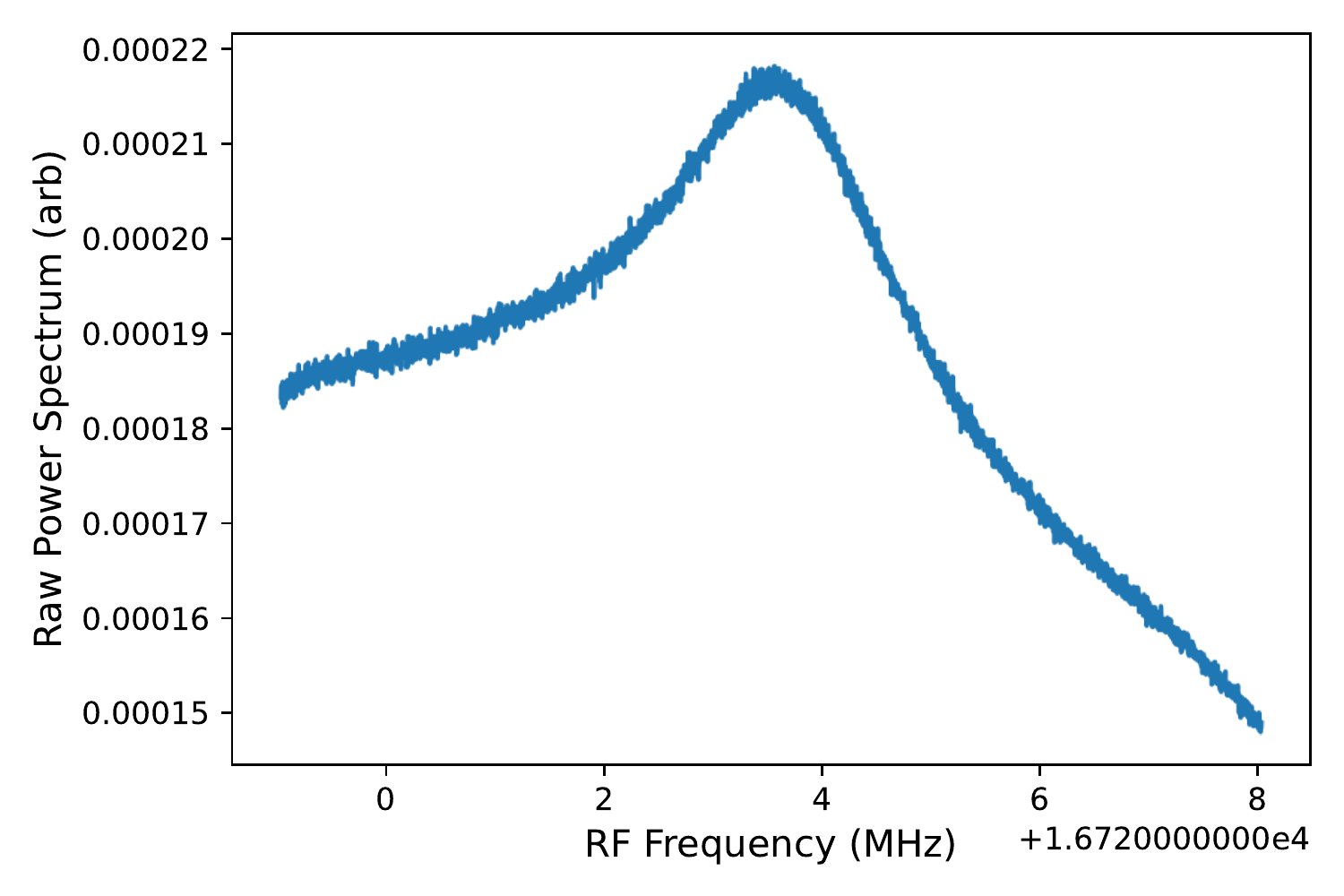}}
  \caption{Raw spectrum as measured by the digitizer. Assuming the spectrum is just noise, the raw spectrum is the noise power of the system.}
  \label{fig:raw_spectrum}
\end{figure}

The raw spectrum consists (Figure~\ref{fig:raw_spectrum}) of the system noise power $P_n = G b k_b T_n$, plus fluctuations about this noise power. $T_n$ affects the SNR of the dark photon data. The system gain, i.e., transfer function, does not affect the SNR of the dark photon data, but causes the large-scale gain variation seen in Figure~\ref{fig:raw_spectrum}. The first step of processing the raw spectra is to remove this gain variation. The gain variation is caused by both the IF electronics and RF electronics. The IF gain and RF gain are thought of as separate because the IF gain applies equally to all spectra, whereas the RF gain depends on the frequency of the cavity. So to remove the gain variation, we first divide the spectrum by $G_{IF}$, and then by $G_{RF}$. 

\begin{figure}
  \centering
  \subfloat[]{\includegraphics[width=0.33\textwidth]{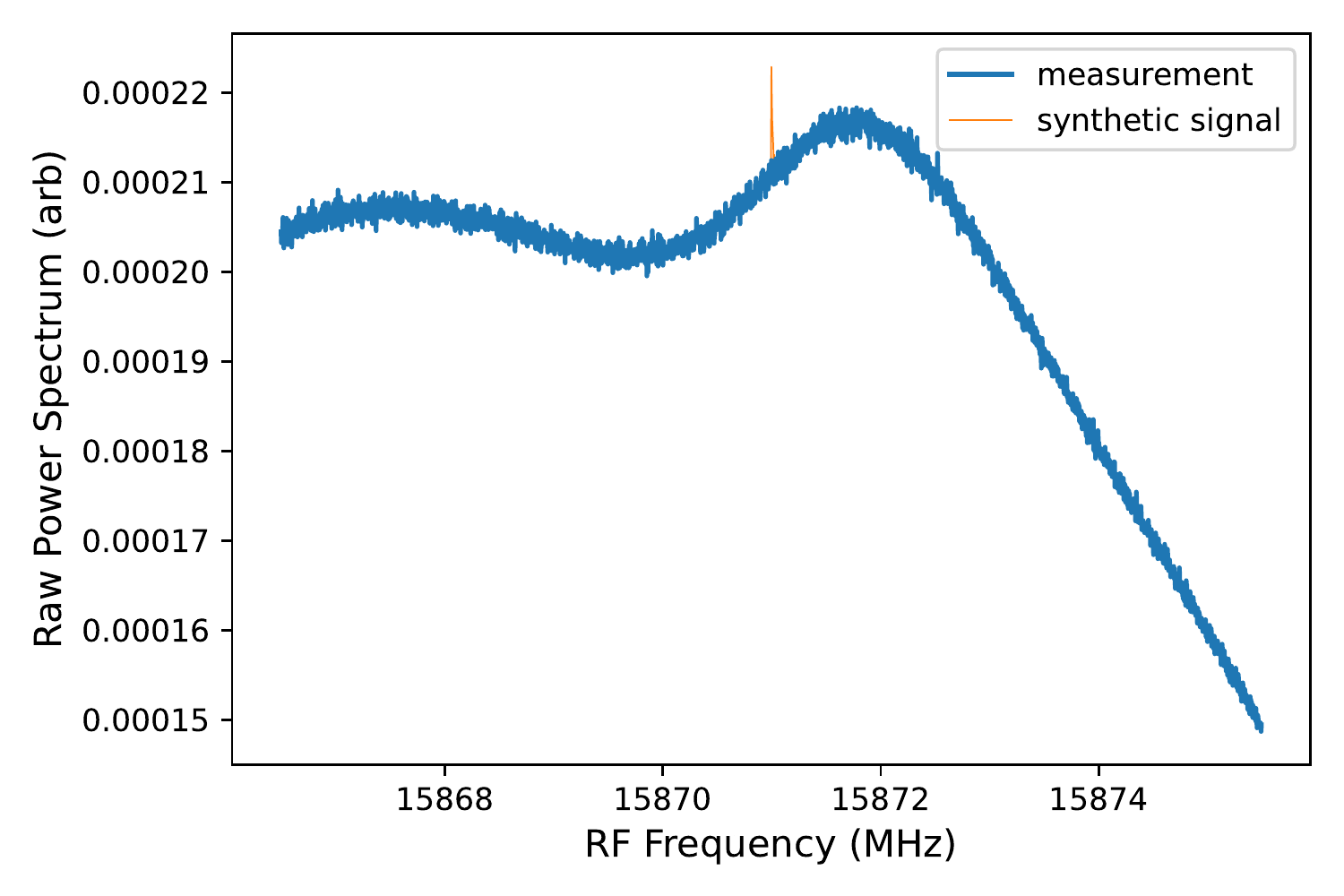}}\hfil
  \subfloat[]{\includegraphics[width=0.33\textwidth]{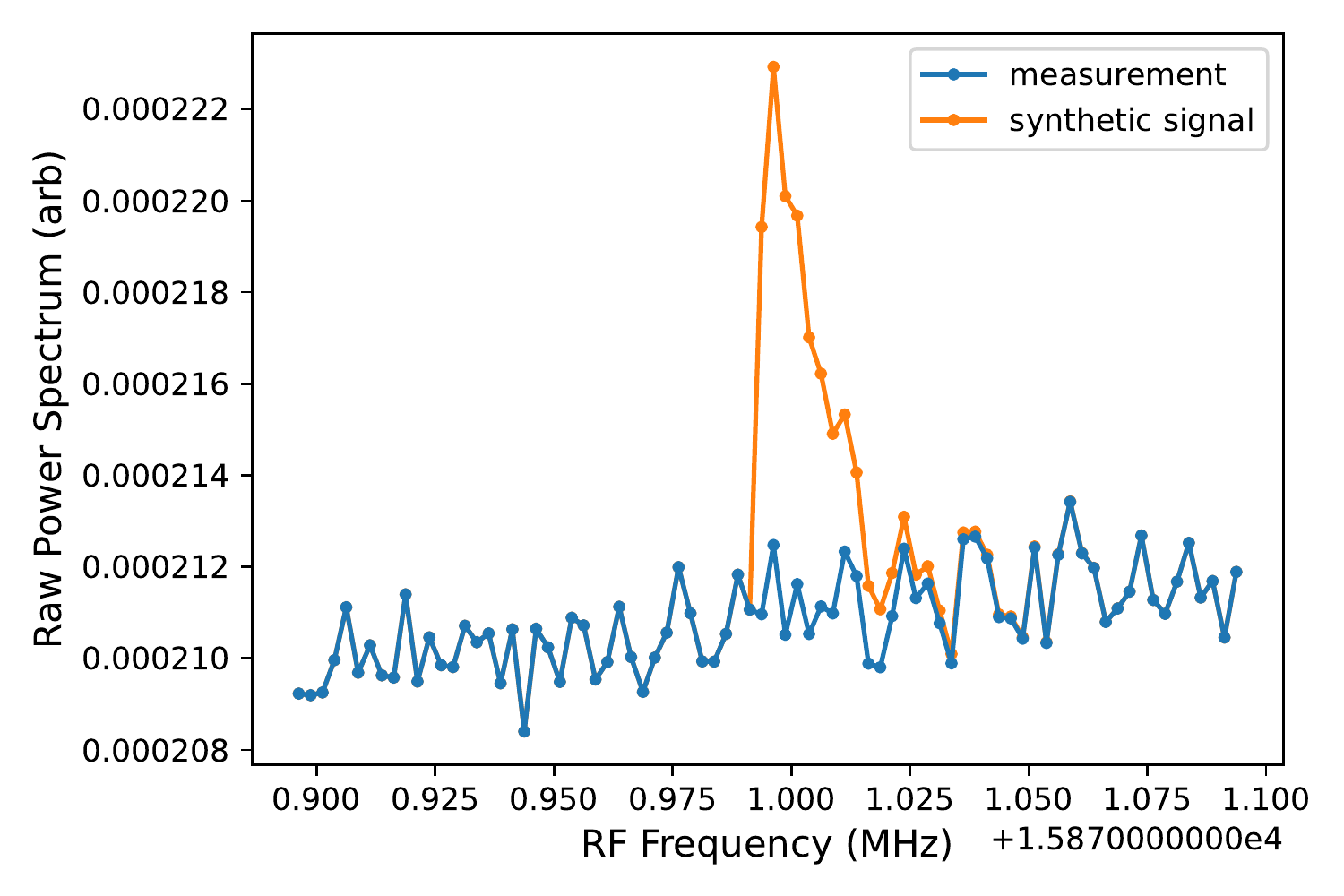}}
  \caption{Raw spectrum with a hypothetical dark photon signal of arbitrary power superimposed. (a) The dark photon signal is much narrower than the large-scale structure of the noise background. Thus a low-pass filter can be safely applied to each power spectrum. (b) Same as (a), but zoomed into the dark photon signal. The hypothetical dark photon signal covers about 8 RF bin widths.}
  \label{fig:raw_spectrum_with_cartoon_signal}
\end{figure}

Figure~\ref{fig:raw_spectrum_with_cartoon_signal} shows the same raw spectrum but with a hypothetical dark photon signal of arbitrary power superimposed. The dark photon signal is much narrower than the large-scale gain variation caused by the electronics. Thus it is safe to apply filters to remove this gain variation. 

\begin{figure}
  \centering
  \includegraphics[height=0.3\textheight]{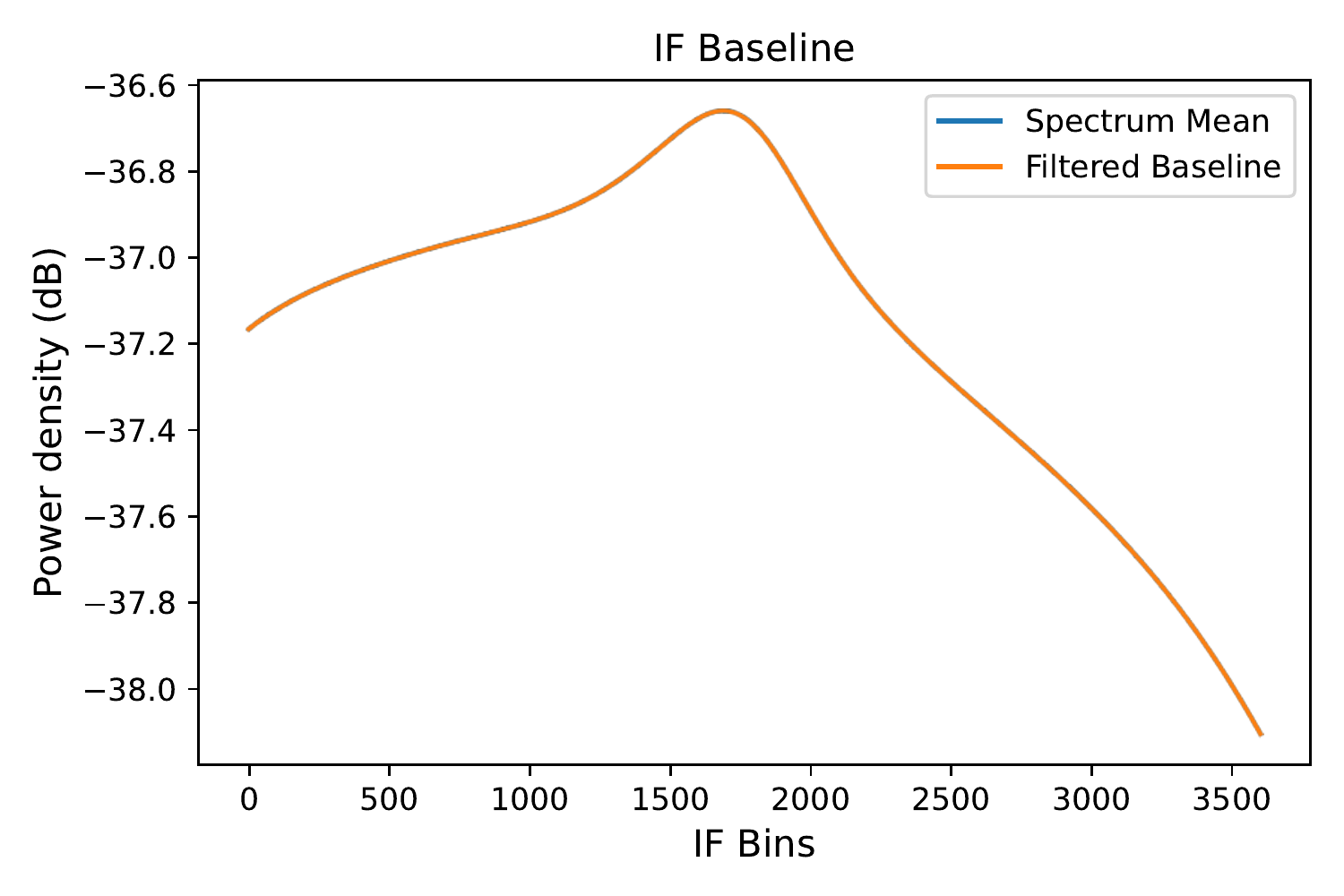}
  \caption{The average of all the recorded spectrum. This is thought of as an good estimation of the gain variation caused by the IF electronics. The estimation is less good if you don't have even and continuous tuning through your ffrequency range.}
  \label{fig:if_baseline}
\end{figure}

\begin{figure}
  \centering
  \subfloat[]{\includegraphics[width=0.33\textwidth]{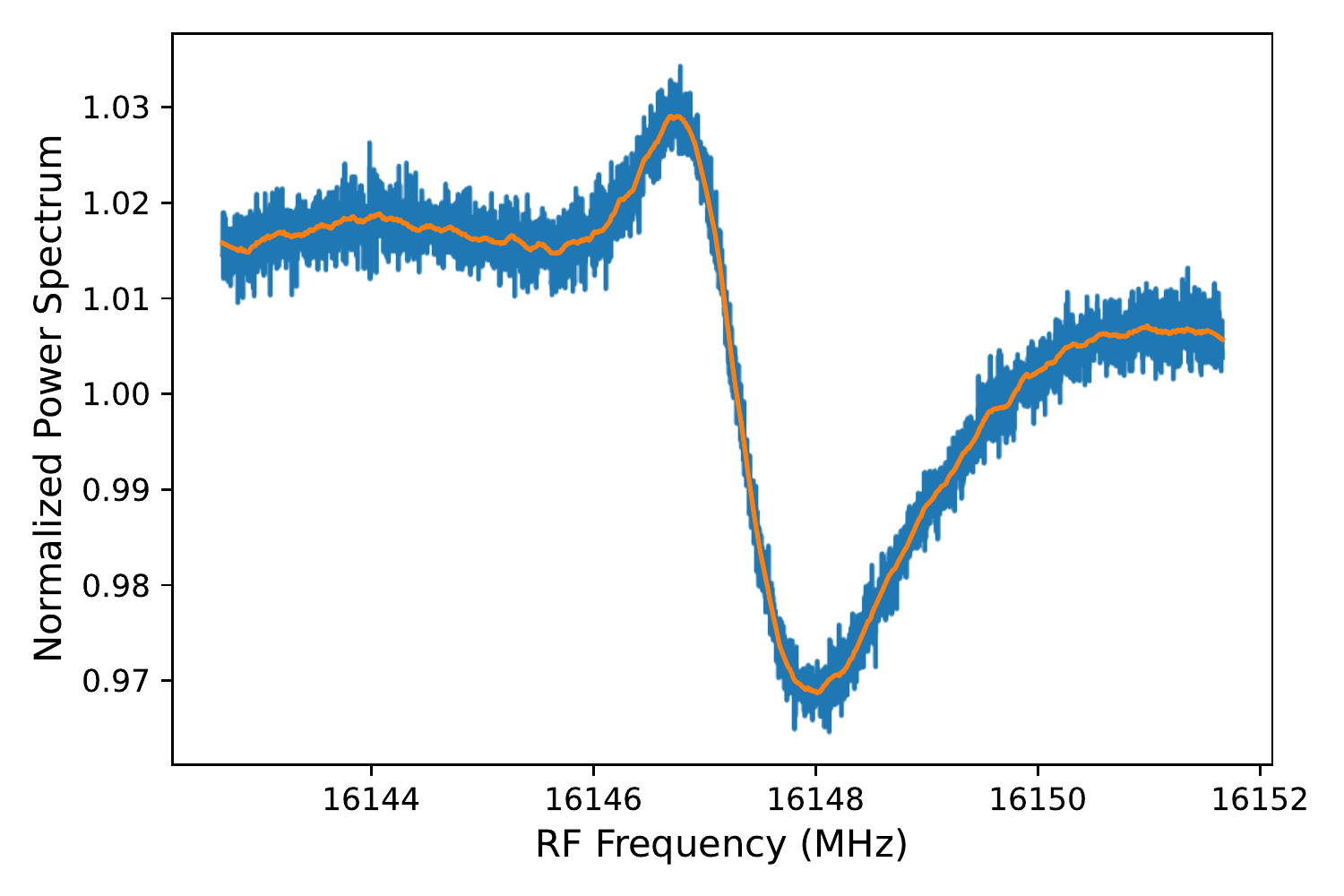}}\hfil
  \subfloat[]{\includegraphics[width=0.33\textwidth]{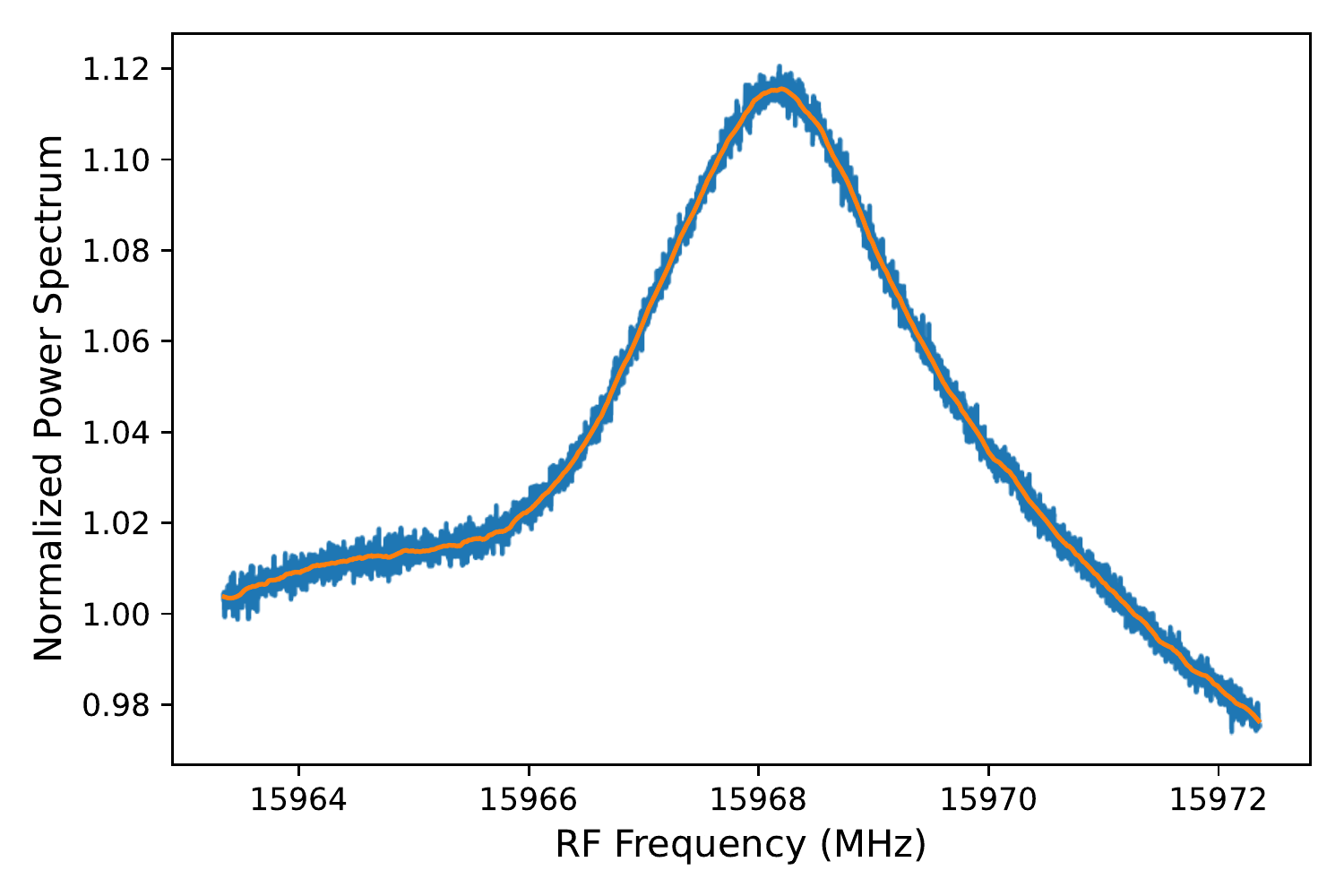}}\hfil
  \subfloat[]{\includegraphics[width=0.33\textwidth]{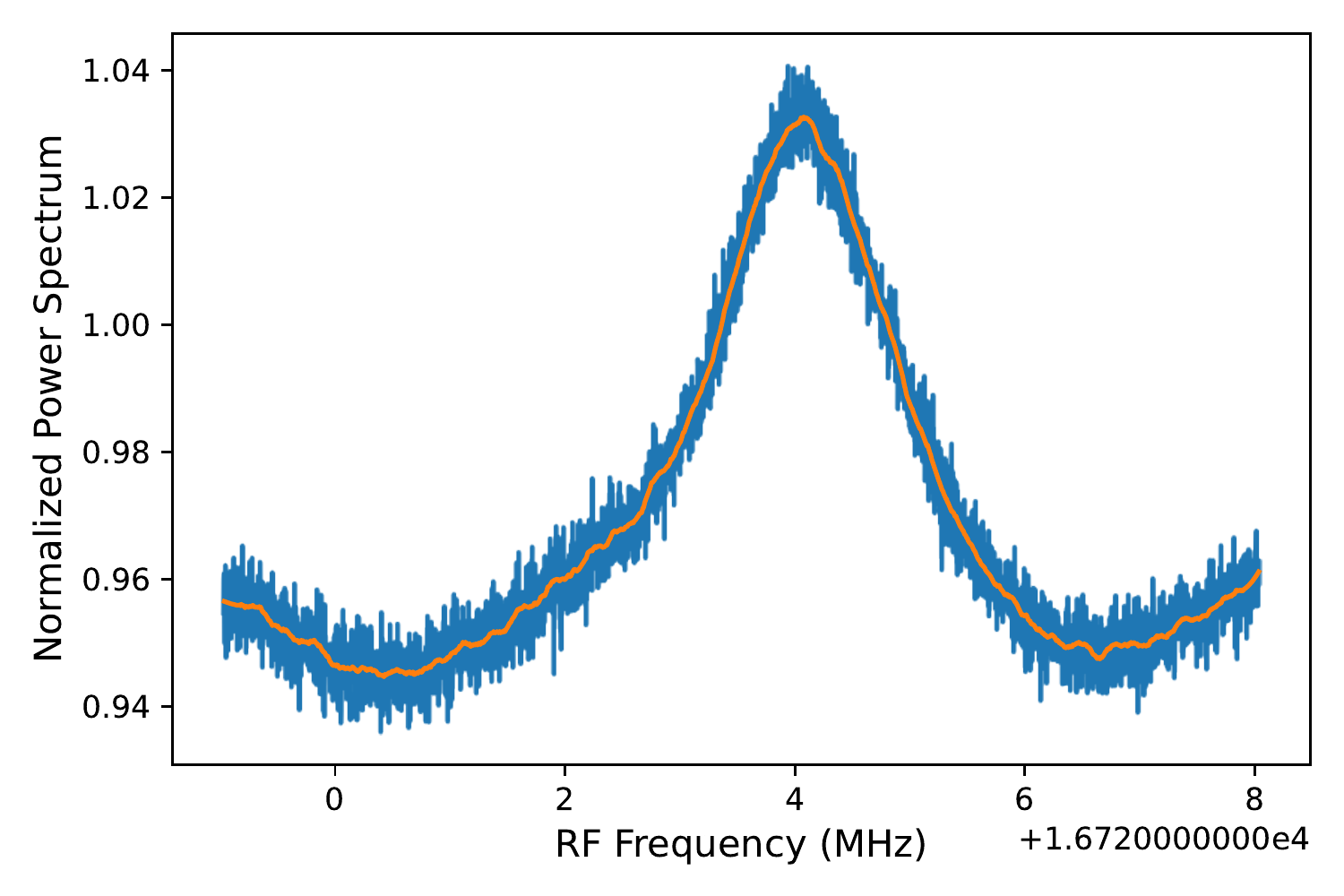}}\hfil
  \caption{The raw spectra are all divided by the IF baseline to form some normalized spectrum. A Savitzky-Golay filter is then applied (orange) to obtain gain variation from the RF electronics.}
  \label{fig:spectra_if_baseline}
\end{figure}

The IF baseline is estimated by averaging all the power spectrum while the cavity was actively tuning. Power spectra with timestamps after 2021-09-05 17:00:00 UTC were excluded because the cavity rested at \SI{15.8}{GHz}, and including those power spectrum in this baseline would have heavily weighted that RF frequency\footnote{In practice, neglecting this cut didn't seem to make much difference.}. The IF baseline is then smoothed using a fourth-order Savitzky-Golay software filter with a window length of 501 bins. Note that the Savitzky-Golay window is much larger than the hypothetical dark photon signal in Figure~\ref{fig:raw_spectrum_with_cartoon_signal}, which is about eight bins wide. The resulting IF baseline is shown in Figure~\ref{fig:if_baseline}. Every spectrum is divided by this IF baseline, and the result is a unitless power spectrum that is still affected by the RF gain variation, as shown in Figure~\ref{fig:spectra_if_baseline}.

\begin{figure}
  \centering
  \subfloat[]{\includegraphics[width=0.33\textwidth]{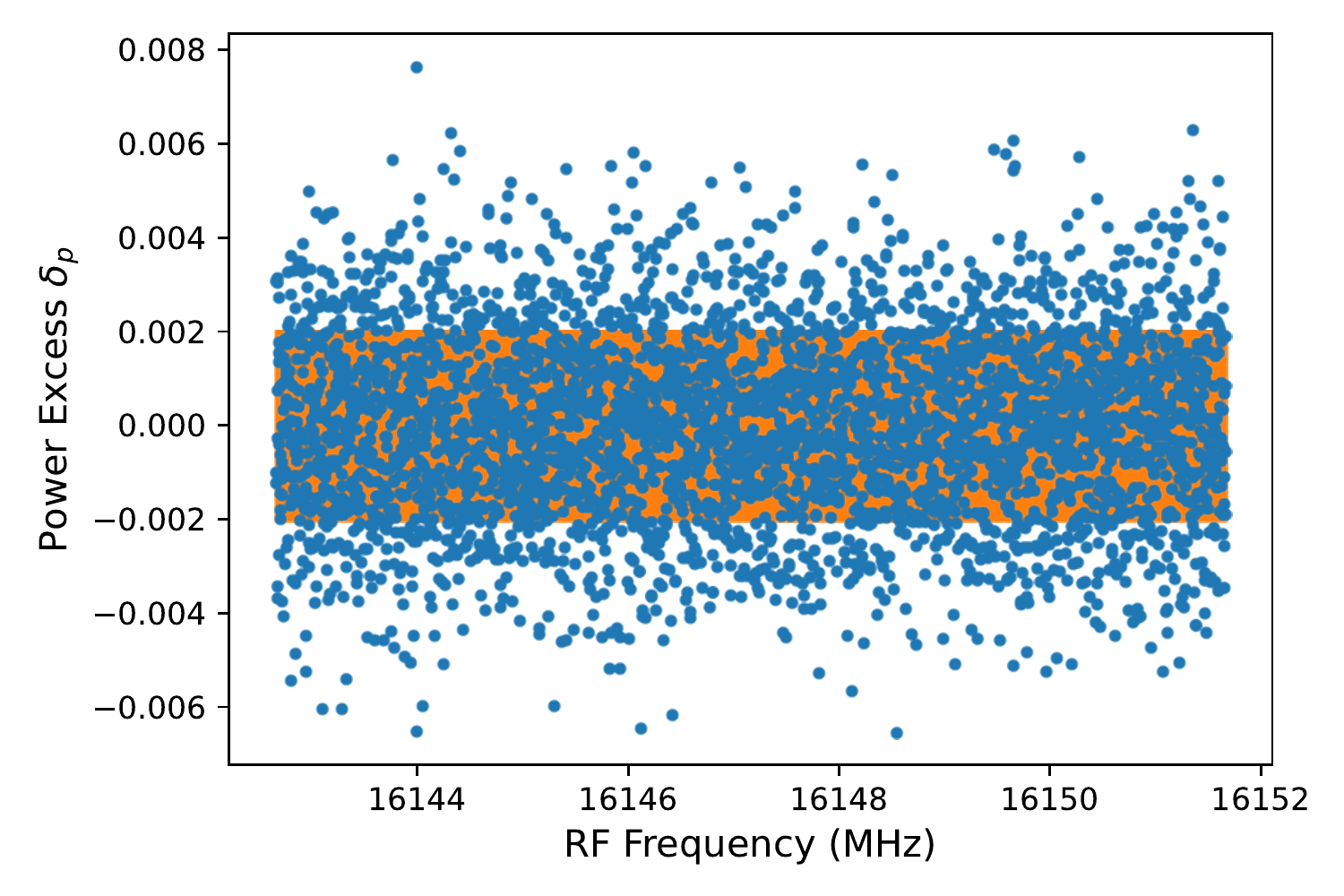}}\hfil
  \subfloat[]{\includegraphics[width=0.33\textwidth]{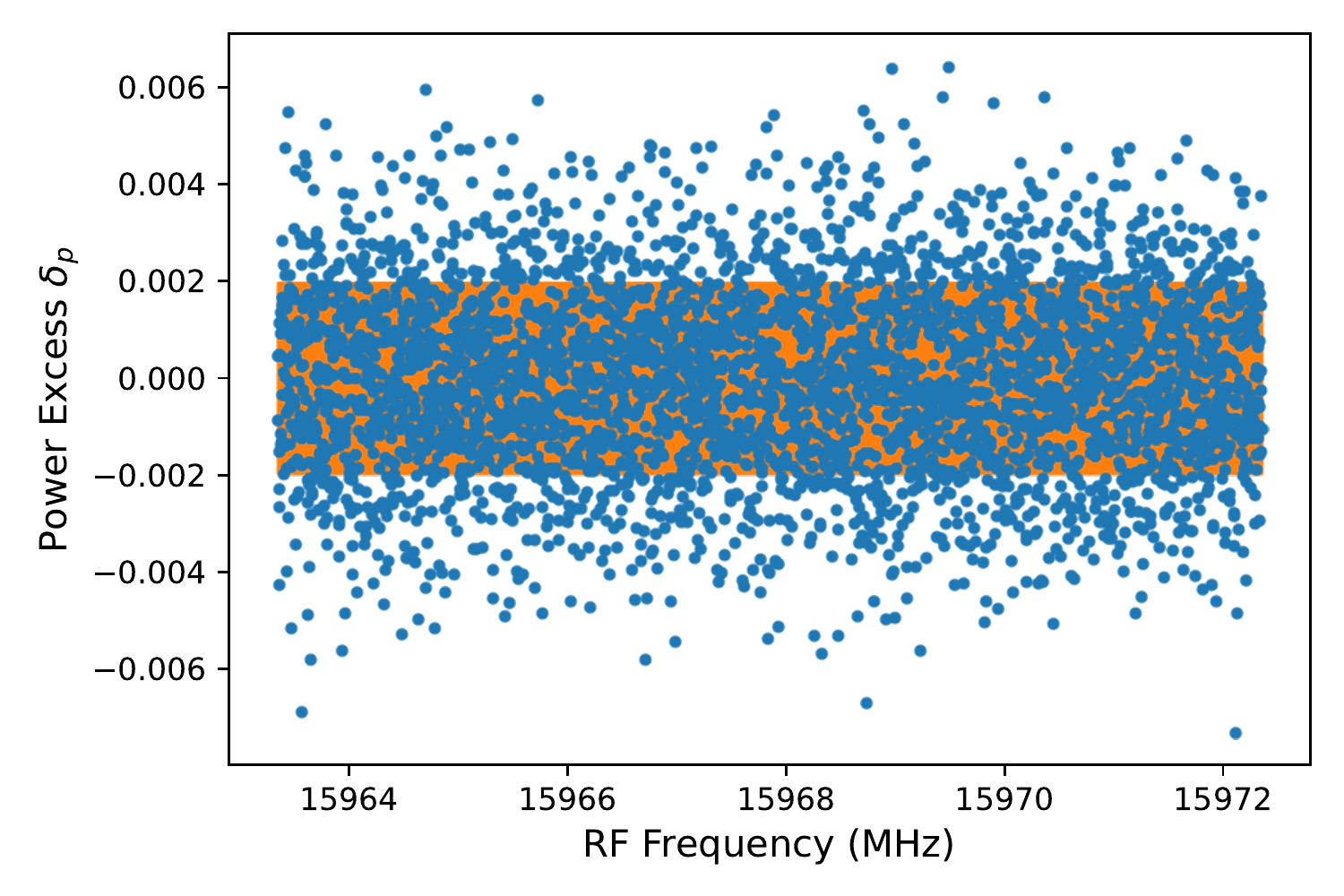}}\hfil
  \subfloat[]{\includegraphics[width=0.33\textwidth]{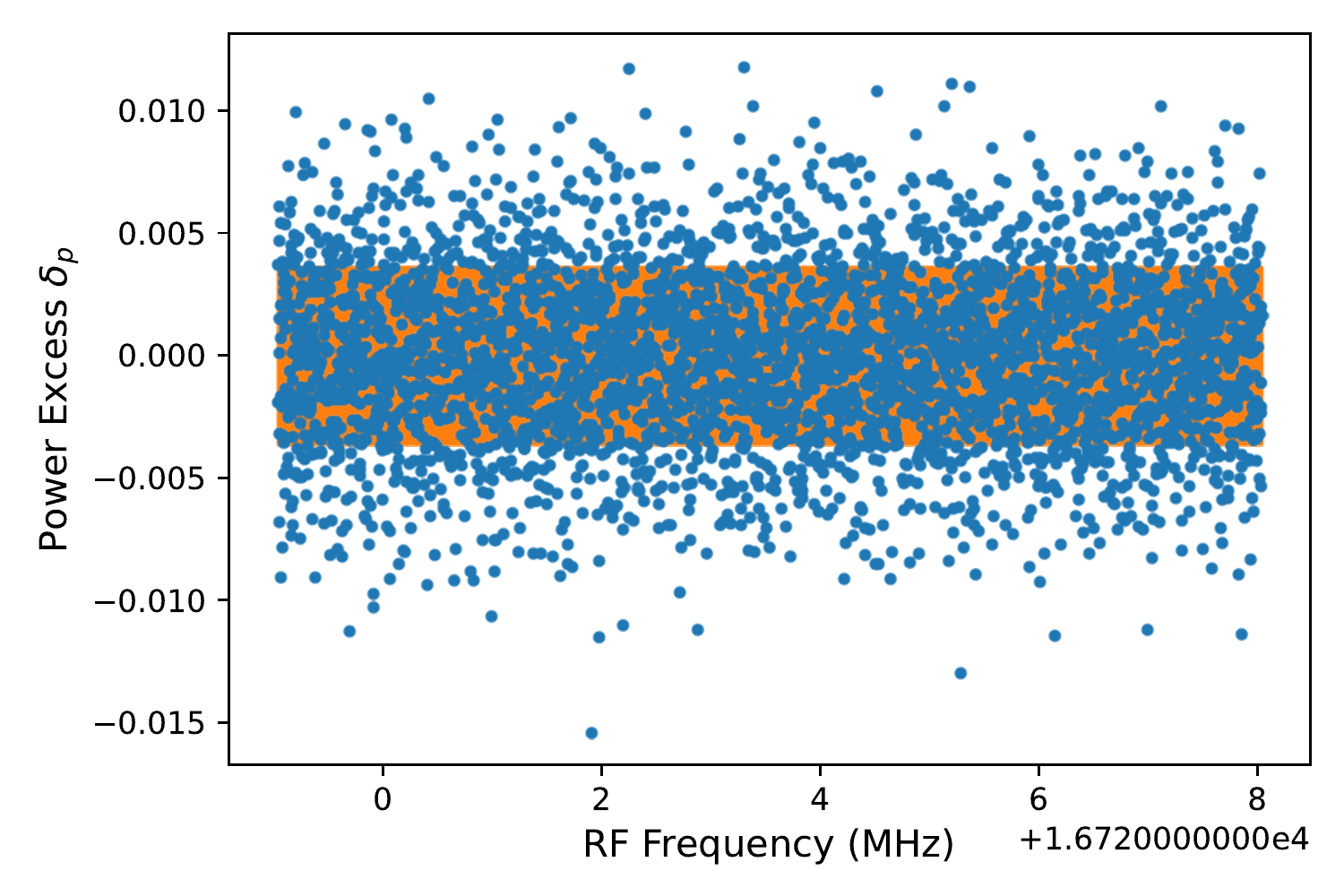}}\hfil
  \caption{The power excess in units of system noise power.}
  \label{fig:delta_p}
\end{figure}

To remove the RF gain variation, a second-order Savitzky-Golay software filter with a window length of 101 bins is applied to the individual spectrum\cite{PhysRevD.96.123008}, as shown in the orange curves in Figure~\ref{fig:spectra_if_baseline}. The spectra are then divided by the filtered curve to remove the RF gain variation. The processed spectra are now normalized such that the mean of the bins is one. The dark photon signal would show up as a power excess, so the process spectra are subtracted by one to yield the flat spectra fluctuating about zero, shown in Figure~\ref{fig:delta_p}. Each bin is either an average of \num[group-separator={,}]{75000} spectra or \num[group-separator={,}]{250000} spectra, and so the fluctuations about zero are Gaussian by the Central Limit Theorem. The power excess for each bin in the processed spectrum is denoted as $\delta_p$. All bins in a spectrum are pulled from the same Gaussian distribution, and so the uncertainty of each bin $\sigma_{p}$ is taken as the standard deviation of all the bins in a spectrum. $\delta_p$ is the power excess normalized to units of system noise power $P_n$.

\FloatBarrier
\subsection{Rescaling Spectra to be in units of dark photon power}
Next, the data is rescaled to be in units of dark photon power. Different operating conditions, such as different temperatures or $Q_L$, will change the noise temperature or expected signal power. The data is rescaled so that a single-bin dark photon signal would have one unit of excess power, regardless of the operating conditions. This rescaling allows for different spectra to be more directly comparable. This rescaling also makes the SNR (the signal being that from the dark photon) the true figure of merit for which potential dark photon candidates are sought. Currently, the spectra are in units of noise power. To rescale, multiply each bin by the noise power and divide by the hypothetical single-bin dark photon power,
\begin{align}
  \delta_s = \delta_p \frac{k_b b T_n}{P_{s}}
\end{align}
There is an ambiguity in Equation~\ref{eqn:dark_photon_power} because $\chi$ and $\theta$ parameters that are not determined by experimental conditions. In particular, there is no natural benchmark model for dark photon (like there are with axion models) to use as a baseline. So for spectrum processing, $\chi=1$ and $\theta=0$. The analysis procedure will self-correct for the arbitrarily chosen $\chi=1$, and the limits can be rescaled at the very end to match any cosmological model that claims a value for $\theta$. 

The uncertainty is determined using standard uncertainty propogation techniques.
\begin{align}
  \frac{\sigma_s^2}{\delta_s^2} = \frac{\sigma_p^2}{\delta_p^2} + \frac{\sigma_{P_N}^2}{\delta_{P_N}^2} + \frac{\sigma_{P_S}^2}{\delta_{P_S}^2}
\end{align}

\begin{figure}
  \centering
  \subfloat[]{\includegraphics[width=0.33\textwidth]{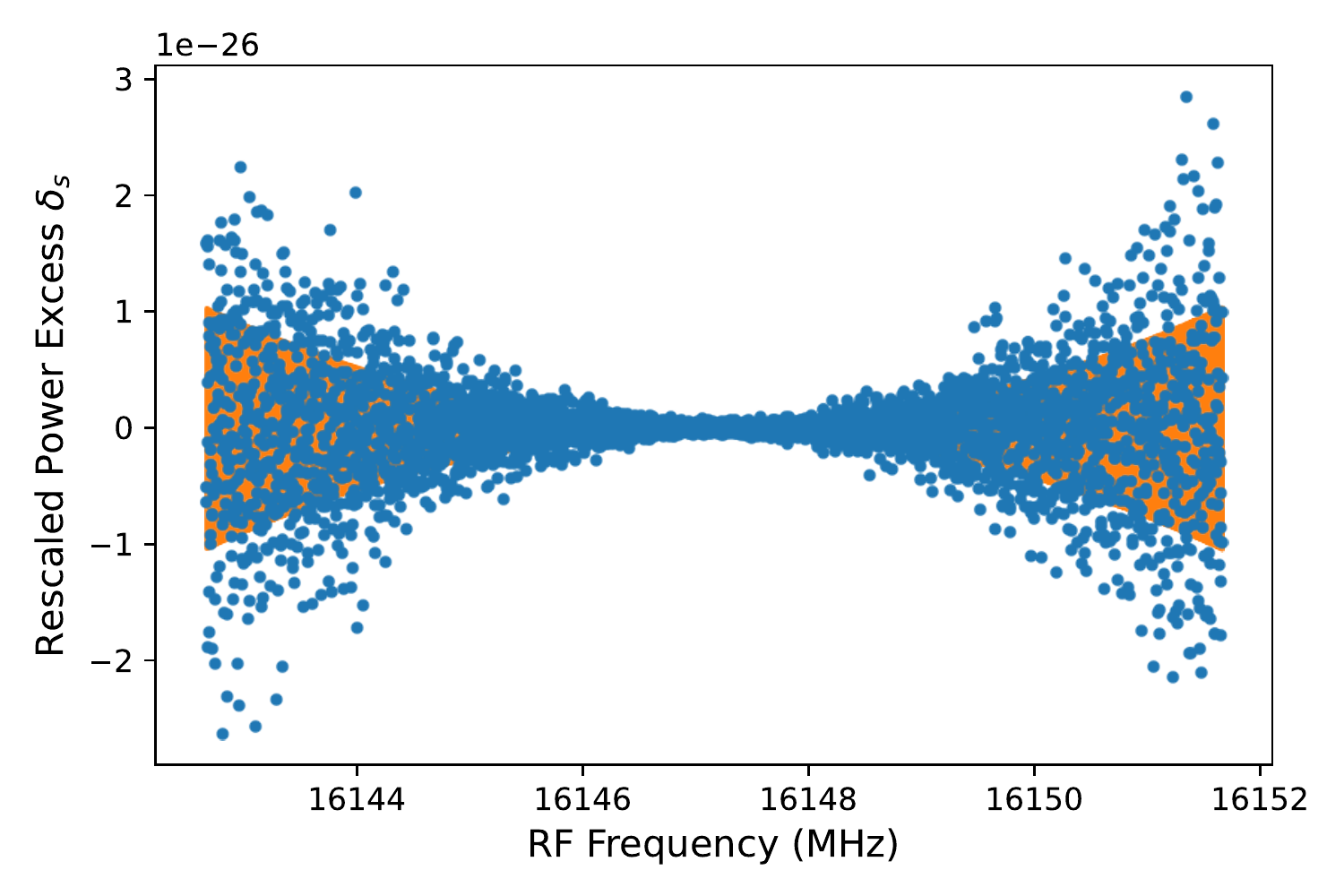}}\hfil
  \subfloat[]{\includegraphics[width=0.33\textwidth]{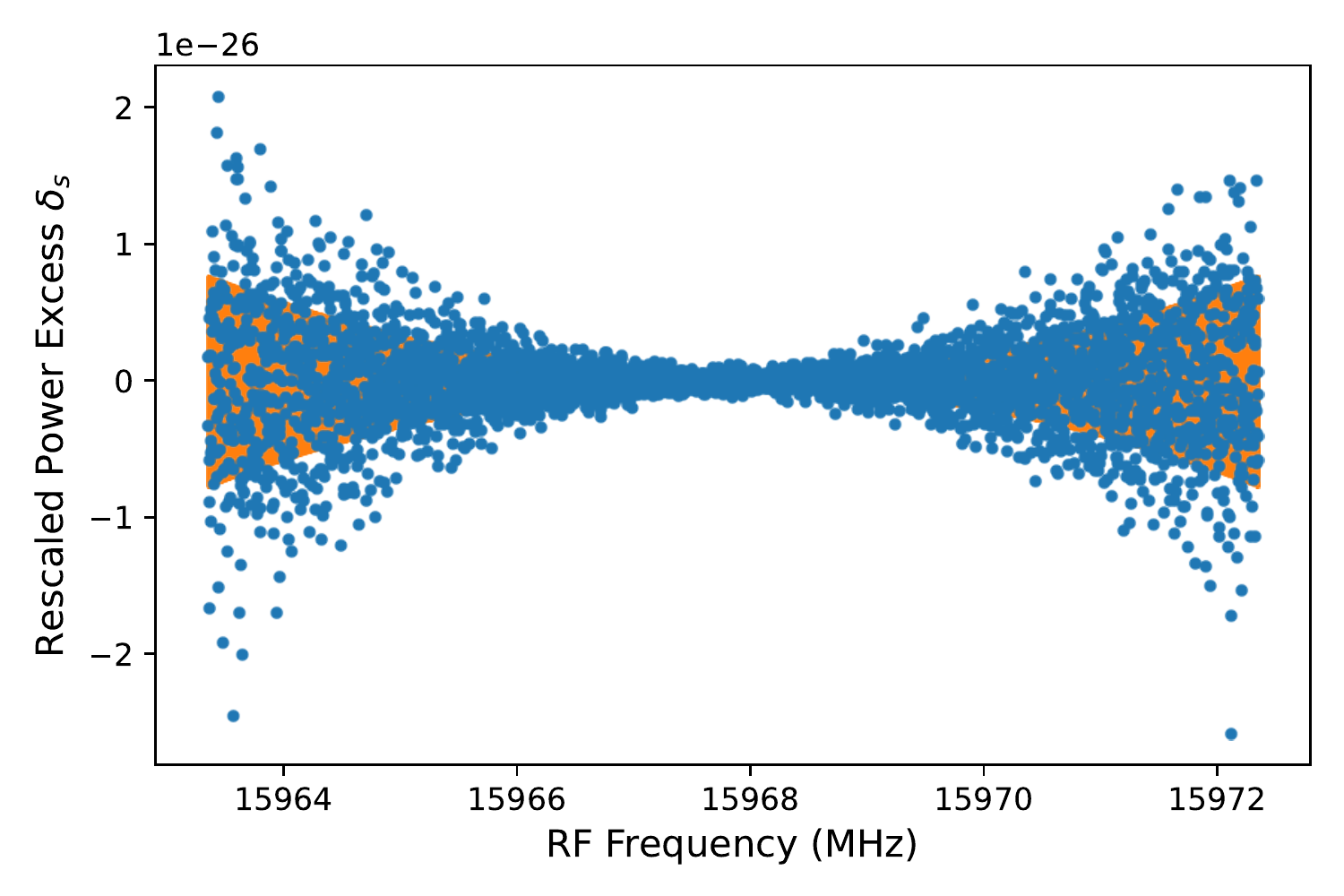}}\hfil
  \subfloat[]{\includegraphics[width=0.33\textwidth]{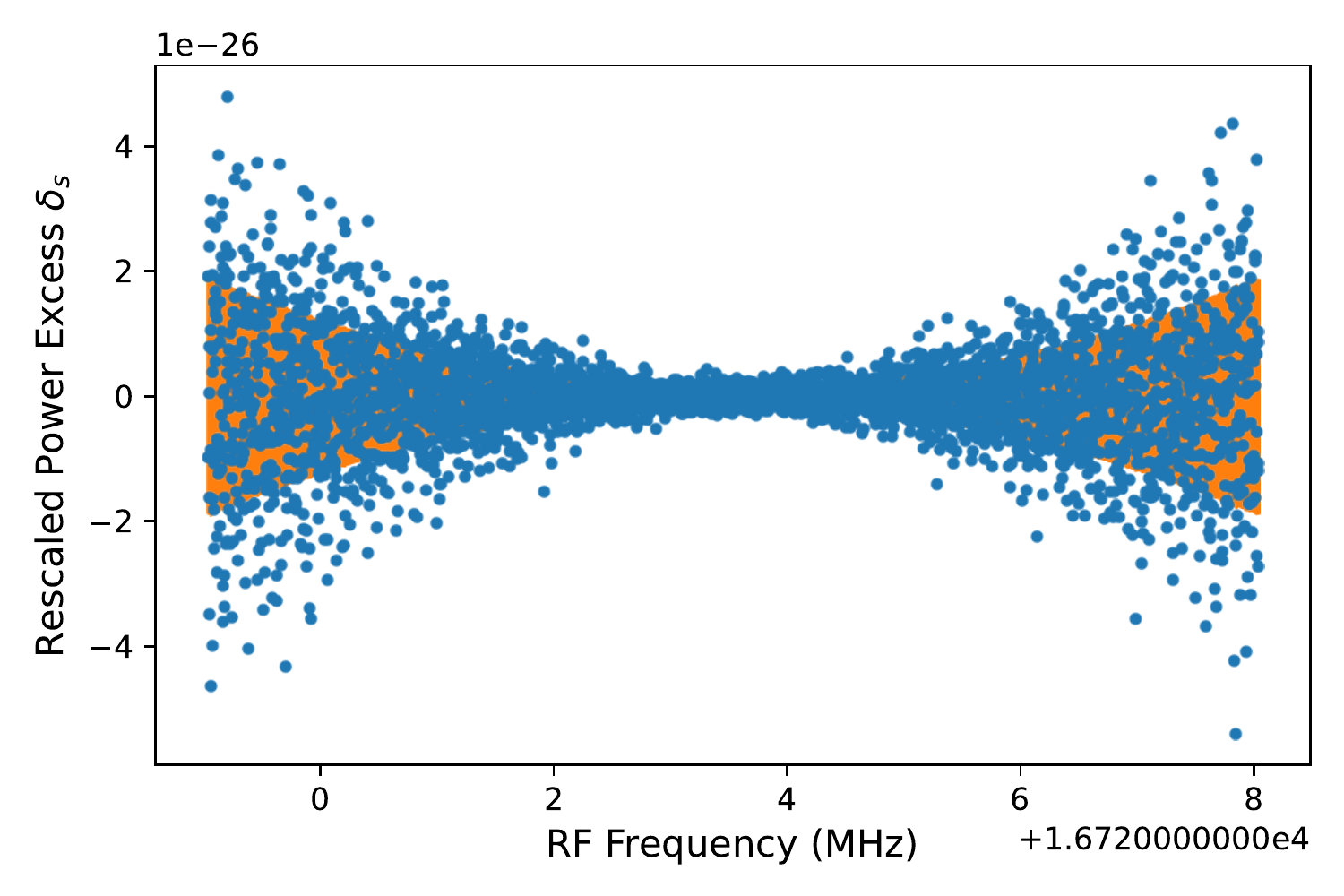}}\hfil
  \caption{The power excess rescaled to units of single-bin dark photon power.}
\end{figure}

\FloatBarrier
\subsection{Taking into account the dark photon lineshape}
The spectra were rescaled so that a dark photon signal would have unit height if the dark photon signal was confined to a single bin. However, Figure~\ref{fig:raw_spectrum_with_cartoon_signal} shows that the expected hidden photon lineshape is spread over about eight bin widths, and this reduces the SNR of each bin. 

The spectral shape of the dark photon signal is proportional the the dark photon kinetic energy distribution. The most conservative energy distribution assumes a virialized, isothermal halo that obeys a Maxwell-Botzmann distribution,

\begin{align}
  f(f) = \frac{2}{\sqrt{\pi}}\sqrt{f - f_a} \left ( \frac{3}{f_a/ \langle \beta^2 \rangle }\right )^{3/2} \exp \frac{-3(f-f_a)}{f_a\langle \beta^2 \rangle}
  \label{eqn:lineshape}
\end{align}
where $f$ is the measured frequency, $f_a$ is the frequency of the associated SM photon, $\langle \beta^2 \rangle = \frac{\langle v^2 \rangle}{c^2}$, and $\langle v^2 \rangle$ is the rms velocity of dark matter halo. The ``quality factor'' of the lineshape is $\mathcal{O}(10^6)$.
In the lab frame, the lineshape takes a more complicated form, but is well approximated by the Equation~\ref{eqn:lineshape} if $\langle \beta^2 \rangle \rightarrow 1.7\langle \beta^2 \rangle$~\cite{PhysRevD.96.123008}.

\begin{figure}
  \centering
  \includegraphics[width=0.75\textwidth]{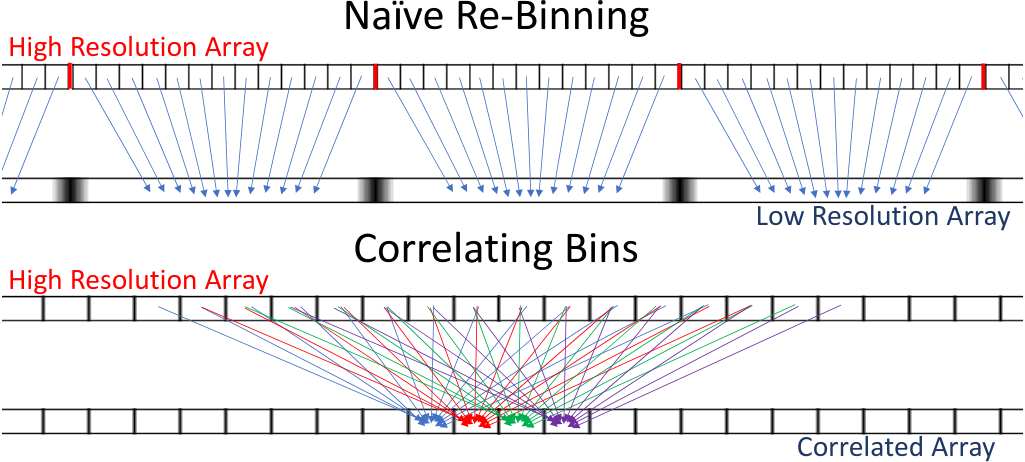}
  \caption{A diagram of how convolution can enhance the signal of a multi-bin dark photon signal. Figure courtesy of Christian Boutan~\cite{2017PhDT24B}.}
  \label{fig:convolution_cartoon}
\end{figure}

One naive approach is to rebin the spectrum so that it has lower frequency resolution, as shown in Figure~\ref{fig:convolution_cartoon}. However, the dark photon power can still be divided into two neighboring bins. At worst, the power will be divided equally into two bins.

A more elegant approach that increases the SNR is to convolve the spectrum, such that each bin is a weighted sum of the neighboring bins. The weight is determined by the lineshape. This has the effect of enhancing bumps that span multiple neighboring bins, increasing the SNR of a potential dark photon signal. The convolution is shown at the bottom of Figure~\ref{fig:convolution_cartoon}. In digital signal processing, this is known as applying a matched filter.

There is a subtlety. How does one perform a convolution when neighboring bins have different uncertainties associated with them, and what is the resulting uncertainty after the convolution? The issue is resolved by treating it like a $\chi^2$ minimization problem. Let 

\begin{align}
  \chi^2 = \sum \frac{(Ak_i - \delta_{si})^2}{\sigma_{si}}
\end{align}
where $\delta_{si}$ ($\sigma_{si})$ is $\delta_{s}$ ($\sigma_{s}$) corresponding to bin $i$, $k_i$ is the element of the convolution kernel corresponding to bin i, and A is a amplitude of the signal shape to which we are fitting our data. The more the bins $\delta_{si}$ match the shape of $k_i$, the larger the amplitude $A$. $\chi^2$ is minimized by setting the derivative of $\chi^2$ to zero and solving for A.

\begin{align}
  A = \frac{\sum \frac{\delta_{si} k_i}{\sigma_{si}^2}}{\sum \frac{k_i^2}{\sigma_{si}^2}}
\end{align}

A is the value for each bin in the optimally convolved spectrum. The uncertainty for parameters derived from $\chi^2$ minimization fits is known to be $\sigma_A^2 = \left ( \eval{\dv[2]{\chi^2}{A} }_{\chi^2_{min}} \right )^{-1}$, resulting in

\begin{align}
  \sigma_A^2 = \frac{1}{\sum \frac{2 k_i^2}{\sigma_i^2}}
\end{align}
In other words, a moving chi-square minimization fit is performed on the data. This is reminiscent of pulse shape fitting done in other analyses.

\begin{figure}
  \centering
  \subfloat[]{\includegraphics[width=0.33\textwidth]{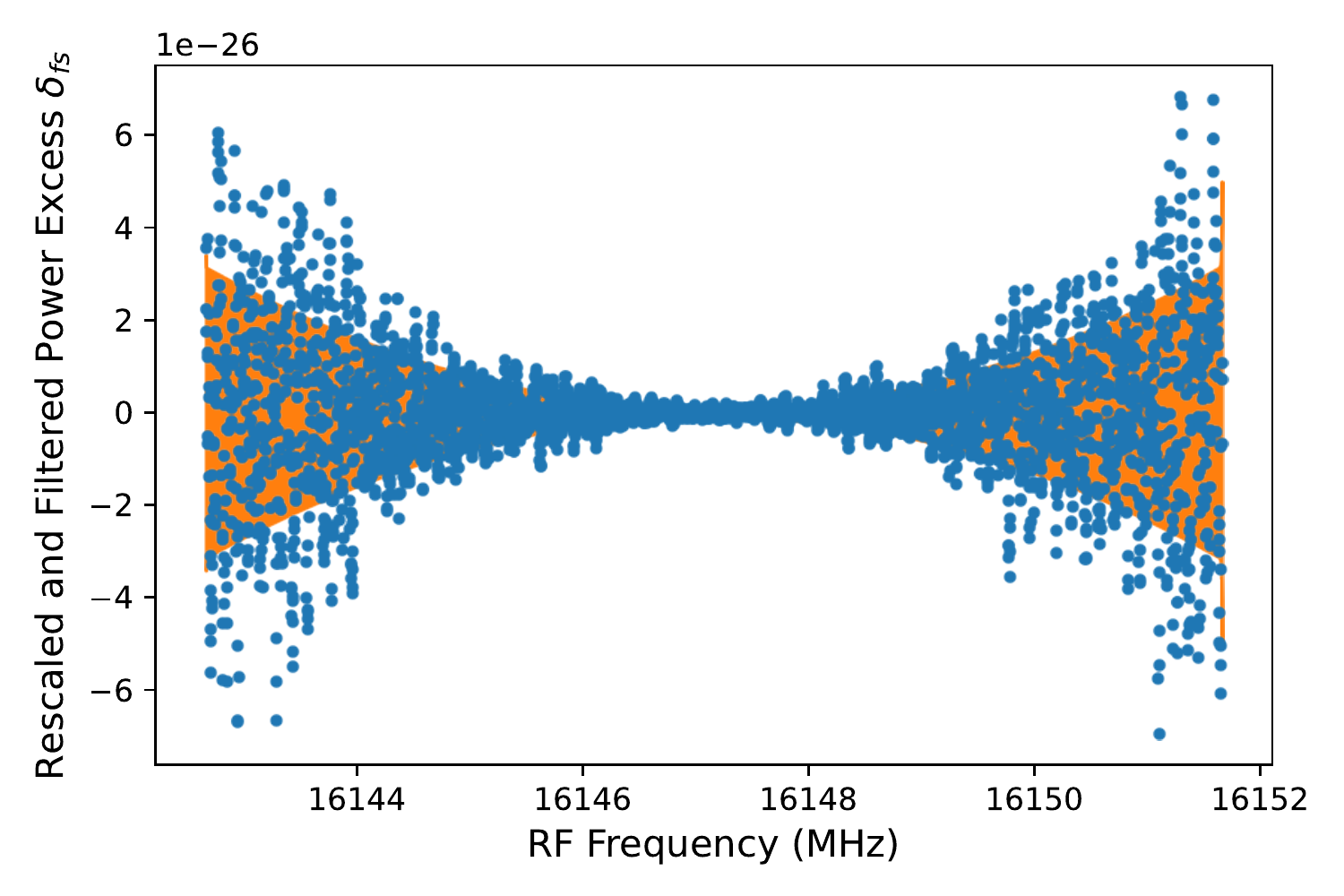}}\hfil
  \subfloat[]{\includegraphics[width=0.33\textwidth]{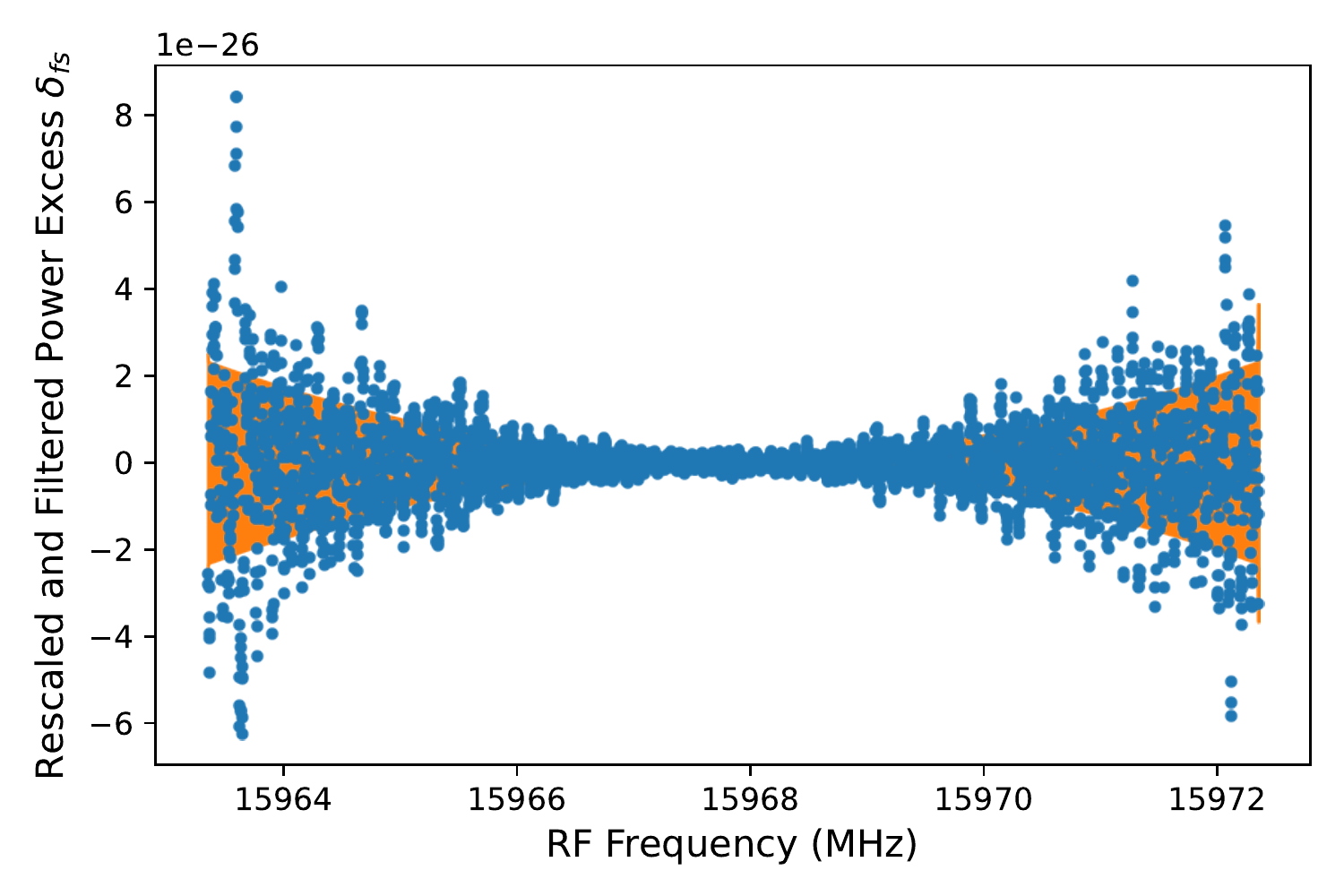}}\hfil
  \subfloat[]{\includegraphics[width=0.33\textwidth]{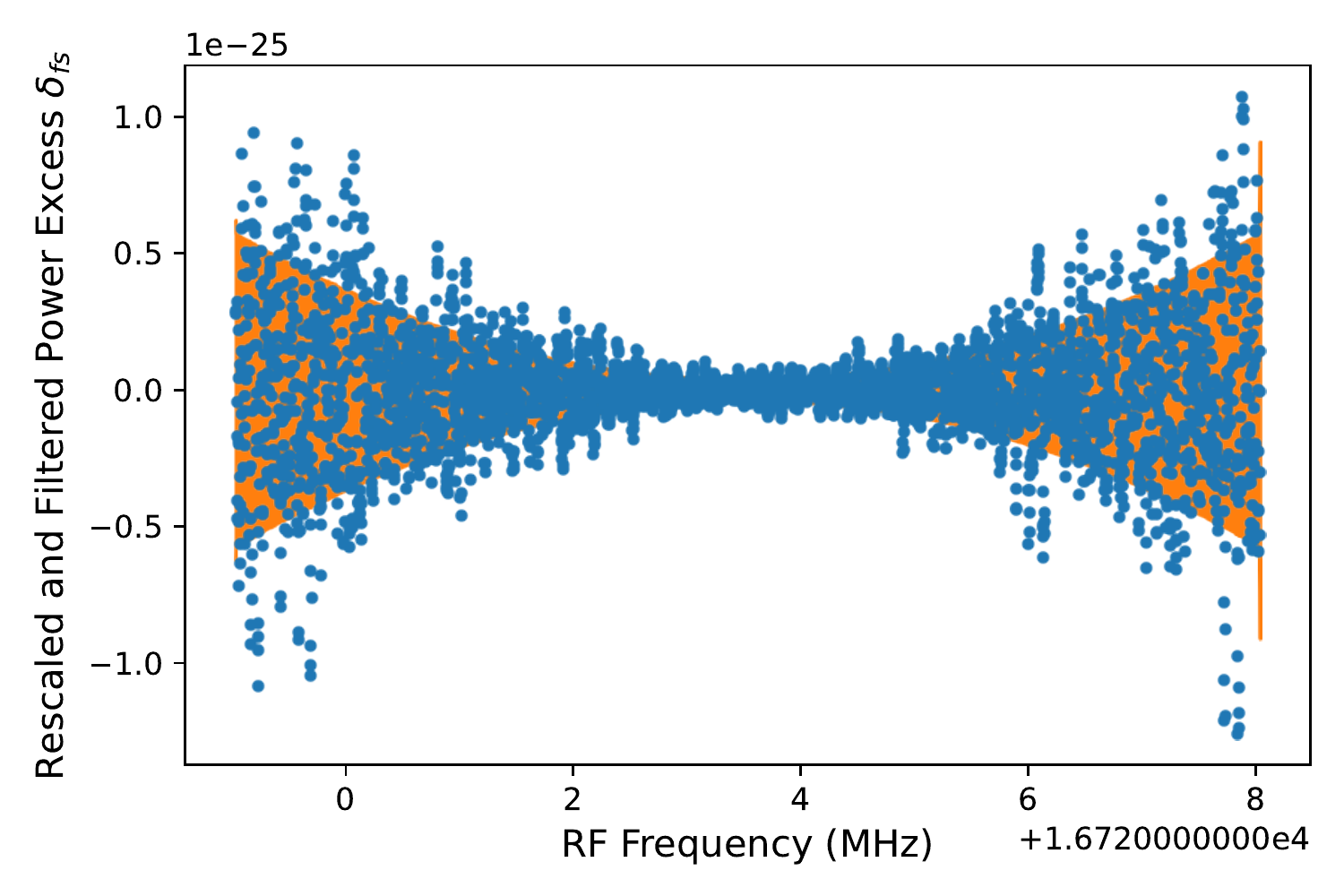}}\hfil
  \caption{The rescaled power excess after the dark matter halo lineshape filter has been applied.}
  \label{fig:filtered_spectra}
\end{figure}

Applying the matched filter with the lineshape kernel leads to the spectra in Figure~\ref{fig:filtered_spectra}. 

\begin{figure}
  \centering
  \subfloat[]{\includegraphics[width=0.33\textwidth]{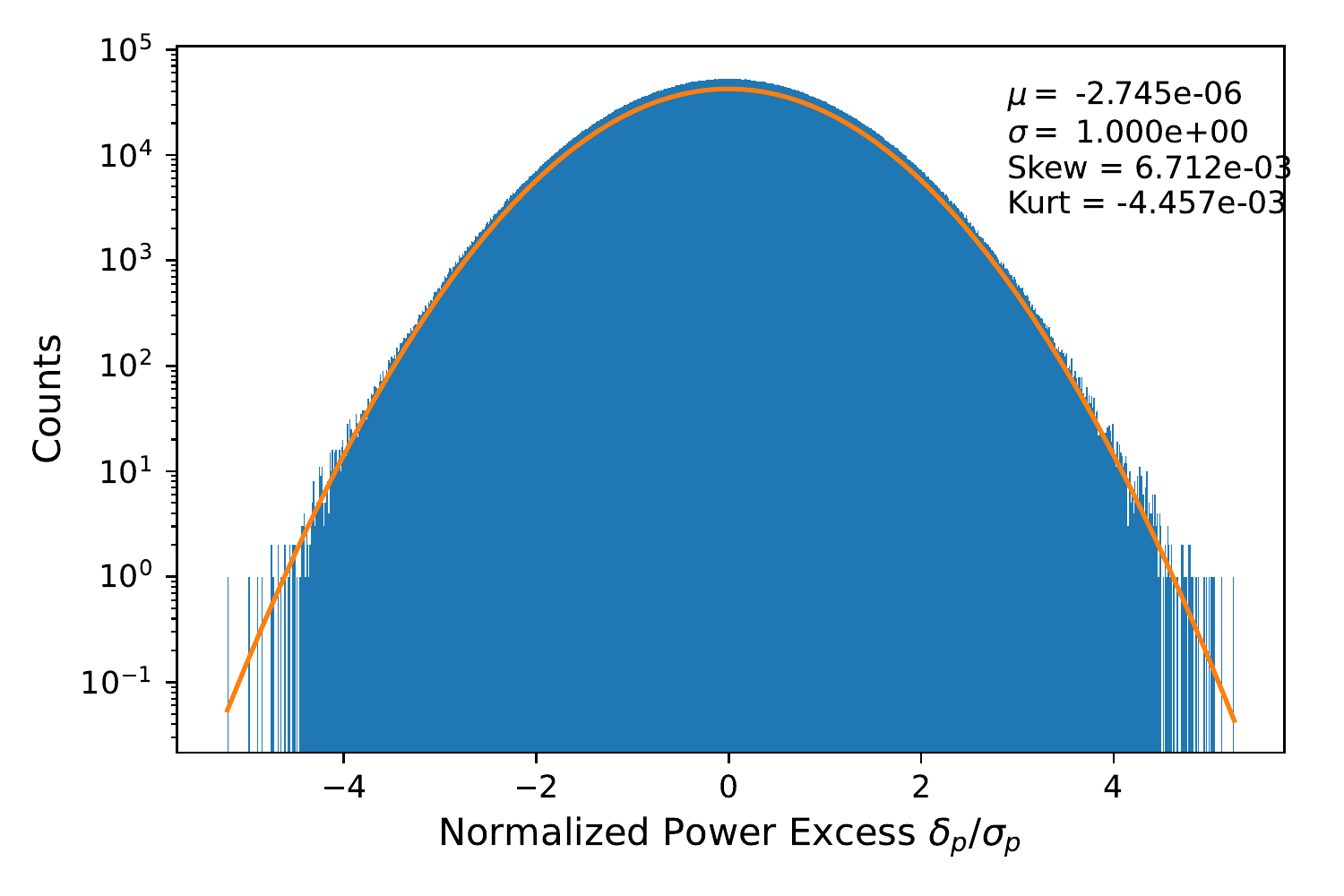}}\hfil
  \subfloat[]{\includegraphics[width=0.33\textwidth]{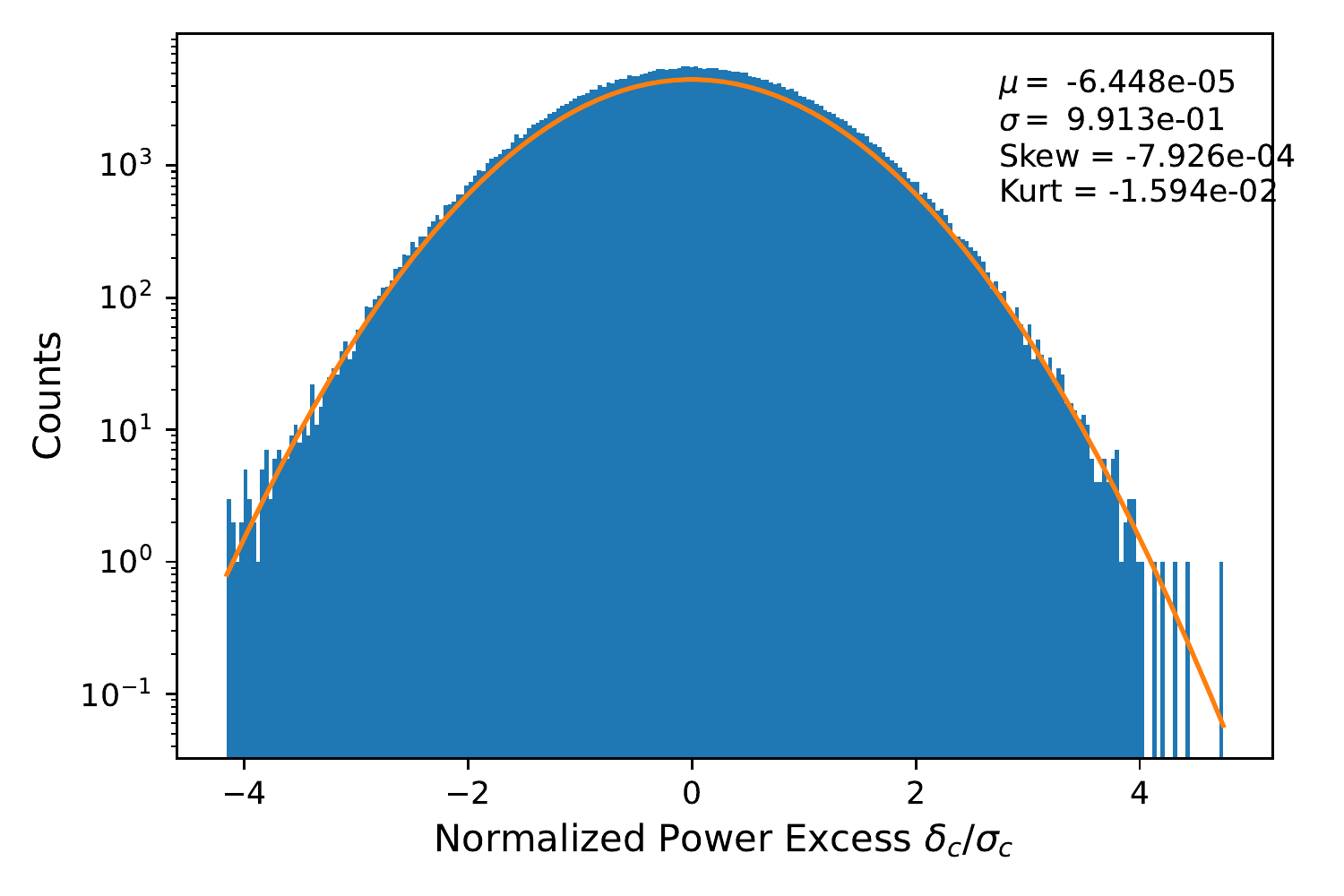}}\hfil
  \subfloat[]{\includegraphics[width=0.33\textwidth]{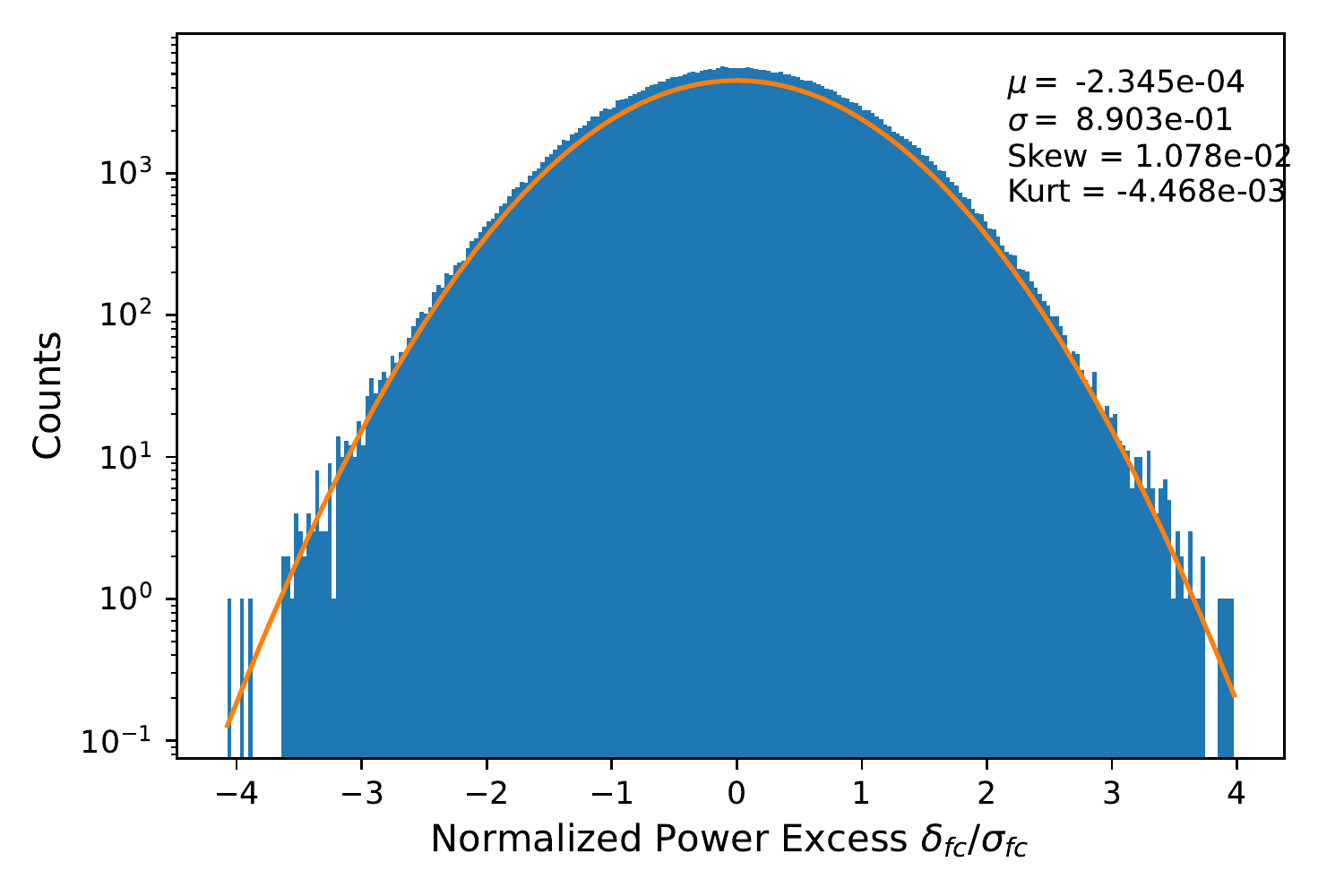}}
  \caption{The (a) power excess $\delta_p$ and (b) rescaled power excess $\delta_s$ has a Gaussian distribution. The filtered power excess has a smaller standard deviation than what is expected of Gaussian noise. This can be attributed to correlations induced by the convolution.}
  \label{fig:spectra_histogram}
\end{figure}

Once all the spectra are processed, one can obtain a histogram of the power fluctuations normalized to the standard deviation, as shown in Figure~\ref{fig:spectra_histogram}. The normalized processed power fluctuations $\frac{\delta_p}{\sigma_p}$ and rescaled process spectrum $\frac{\delta_s}{\sigma_s}$ follow the normal distribution beautifully, with zero mean and unit standard deviation. The normalized filtered spectrum $\frac{\delta_{fs}}{\sigma_{fs}}$, follows the shape of a standard normal distribution, but has a standard deviation of 0.89. This is because the convolution correlates neighboring bins.

\FloatBarrier
\section{Combined Spectrum}
The individual spectra corresponding to different RF frequencies have been processed and are now ready to be combined into a single combined spectrum. The Maximum-Likelihood estimate of the mean and uncertainty of the combined spectrum is obtained by a weighted average of all IF bins corresponding to a particular RF bin. The weights of each contributing bin are the inverse variance.

\begin{align}
  \delta_c = \frac{\sum_i \frac{\delta_{si}}{\sigma^{2}_{si}}}{\sum_i \frac{1}{\sigma^{2}_{si}}}
\end{align}
The standard deviation for each bin in the combined spectrum is 
\begin{align}
  \sigma_c = \sqrt{\frac{1}{\sum_i \frac{1}{\sigma_{si}^2}}}
\end{align}

This method of combining spectrum is well-established in haloscope data analysis~\cite{PhysRevD.96.123008, PhysRevD.103.032002, PhysRevD.64.092003}. Finding the Maximum-Likelihood estimations of means and standard deviations is a general and ubiquitous problem in experimental data analysis.

\begin{figure}
  \centering
  \includegraphics[width=0.85\textwidth]{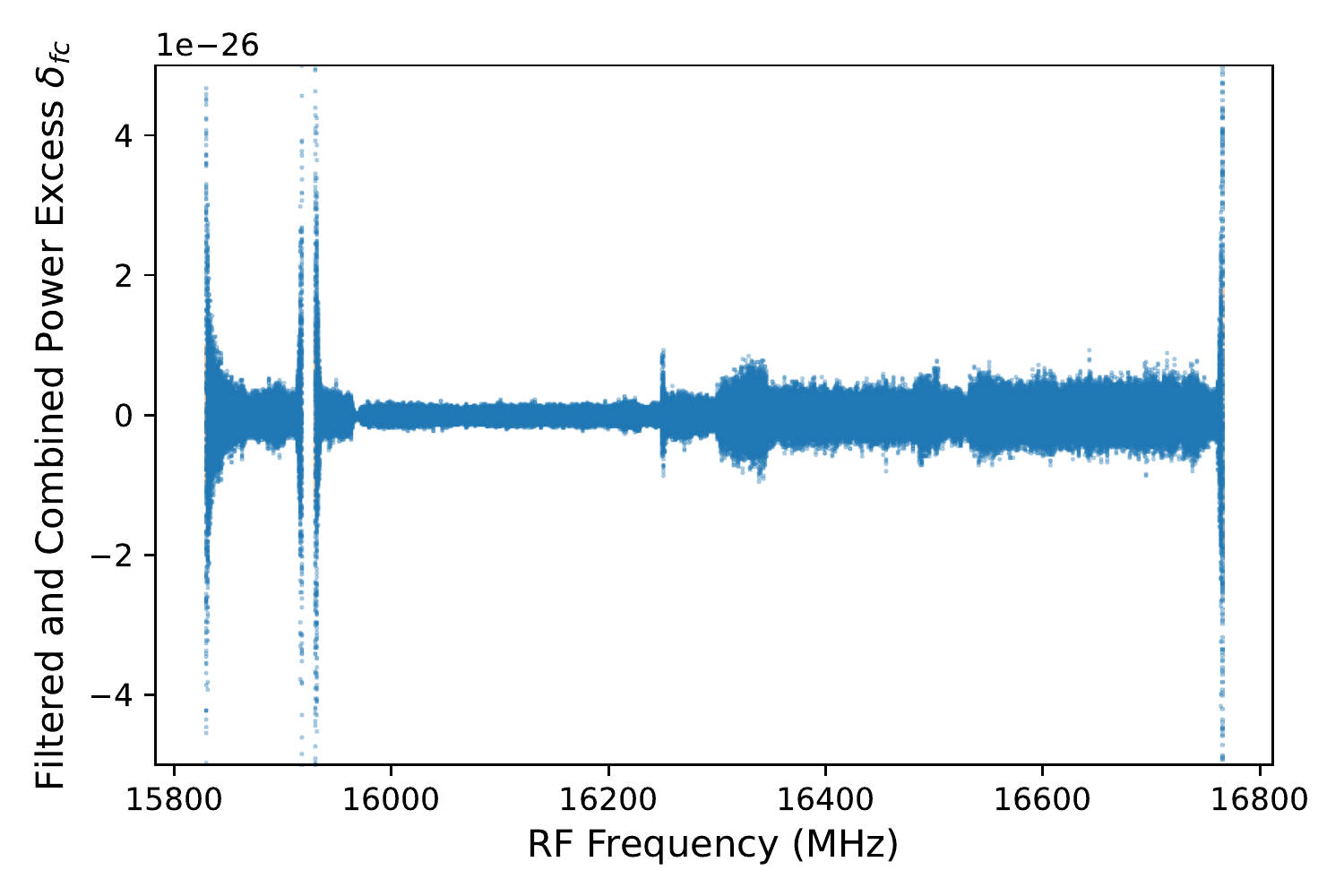}
  \caption{The combined spectrum.}
  \label{fig:combined_spectrum}
\end{figure}

The combined power excess is shown in Figure~\ref{fig:combined_spectrum}. The gap near \SI{15.95}{GHz} is a result of manually tuning the top dielectric plate.  

\FloatBarrier
\section{Placing 90\% Exclusion Limit}
Figure~\ref{fig:spectra_histogram} suggests that there are no signals that are astatistical. In the absence of any obvious dark photon candidates, I choose to set an upper limit on what dark photon signals could exist based on the statistics of the combined spectrum.

The probability distribution of measuring a power excess $\delta_c$ given a haloscope signal power is a normal distribution centered around the dark photon signal power (the signal power can be 0).
\begin{align}
  P(\delta_c|S) = \frac{1}{\sqrt{2\pi\sigma_{c}}}\exp\left ( -\frac{(\delta_{c} - S)^2}{2\sigma_{c}^2} \right )
\end{align}
where $\delta_{c}$, $\sigma_{c}$ are the mean and standard deviation of the combined power excess, and S is the dark photon signal power. I've measured power excess and want to know the probability distribution of the dark photon signal. In other words, I want to know probability distribution of S given that I know the power excess $\delta_c$, $P(S|\delta_c)$. Using Bayes Theorem
\begin{align}
  P(S|\delta_c) = P(\delta_c|S) \frac{P(S)}{P(\delta_c)}
  \label{eqn:bayes}
\end{align}
where $P(S)$ is the probability of measuring dark photon signal power S, and $P(\delta_c)$ is the probability of measuring power excess $\delta_c$. $\delta_c$ in Equation~\ref{eqn:bayes} is a parameter and not a continuous variable, so $P(\delta_c)$ is just a constant. Our only prior is that the signal power can not be negative\footnote{S also has an upper limit based on previous exclusions, but this limit has a negligible effect on the subsequent analysis.}. In the absence of any other information, any value for S above zero is as good as any other. Consequently, 
\begin{align}
  \frac{P(S)}{P(\delta_c)} \propto H(x)
\end{align}
where $H(x)$ is the unit step function. That means $P(\delta_s|S)$ is just a normal distribution truncated at zero\footnote{\cite{PhysRevD.101.123011, PhysRevD.57.3873} talk about similar subjects.}.
\begin{align}
  P(S|\delta_c) \propto \frac{1}{\sqrt{2\pi\sigma}}\exp\left ( -\frac{(\delta_c - S)^2}{2\sigma_c^2} \right ) H(x)
\end{align}

\begin{figure}
  \centering
  \subfloat[]{\includegraphics[width=0.33\textwidth]{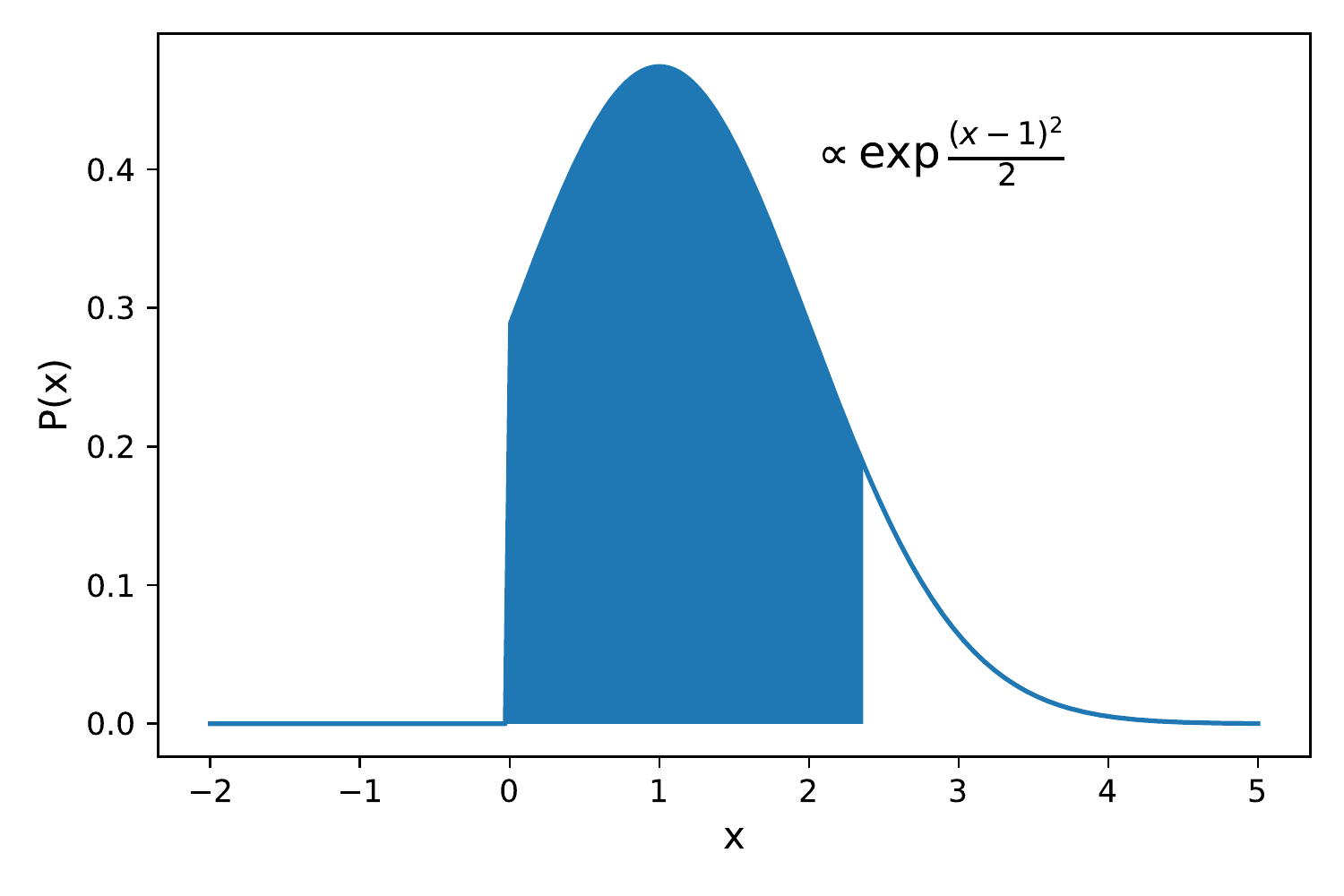}}\hfil
  \subfloat[]{\includegraphics[width=0.33\textwidth]{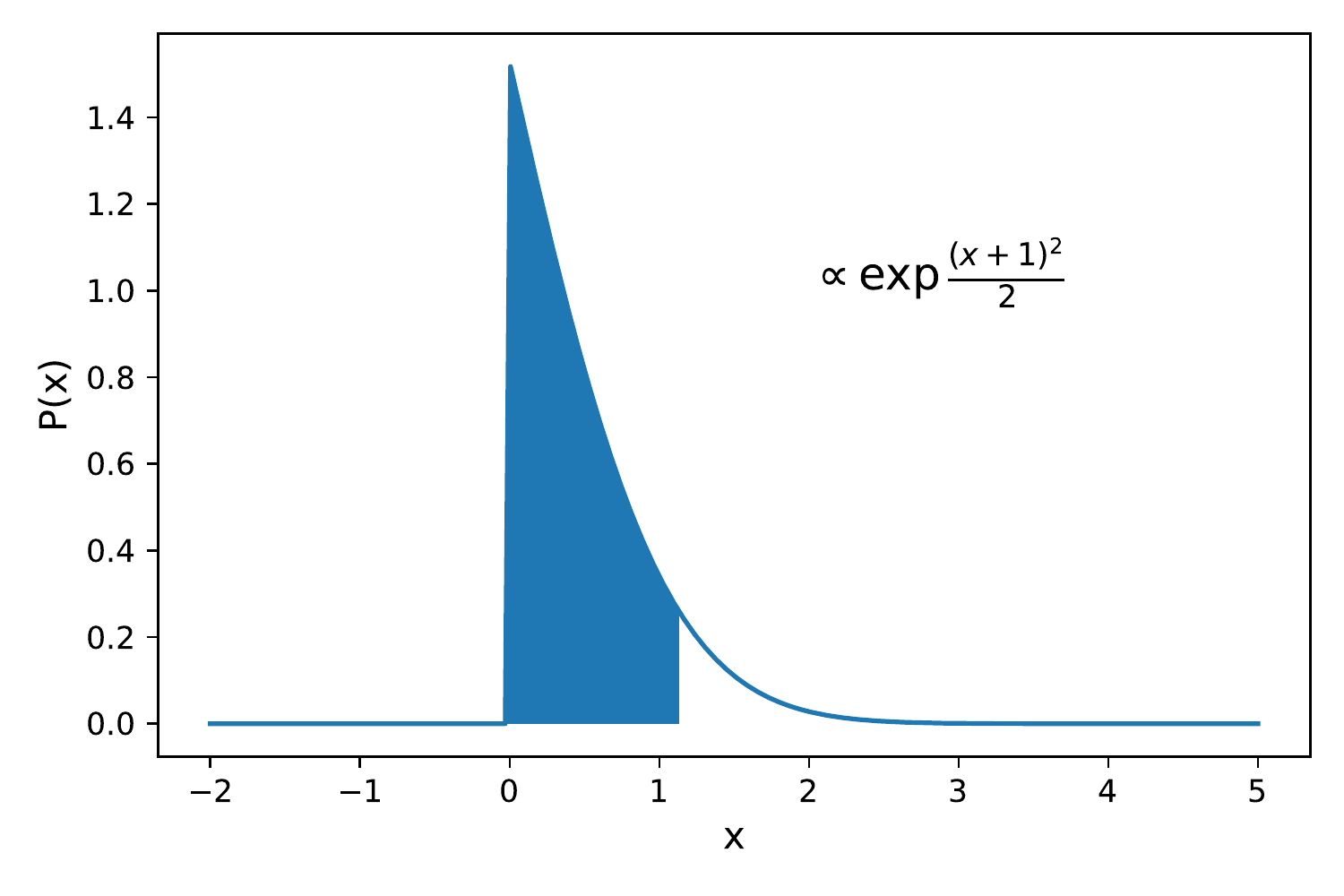}}\hfil
  \subfloat[]{\includegraphics[width=0.33\textwidth]{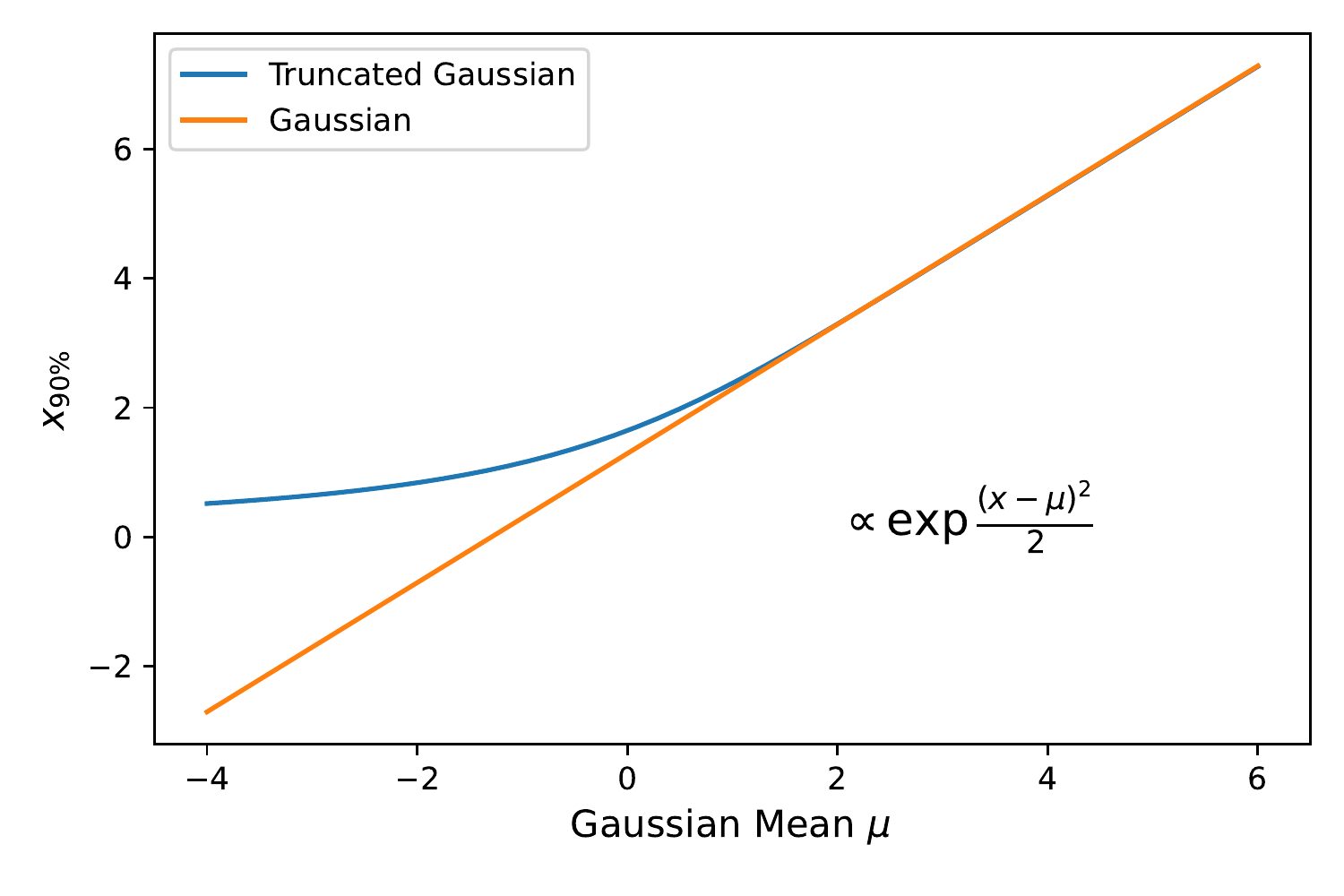}}\hfil
  \caption{The probability distribution of finding a dark photon signal, given a measured power excess, is represented as a truncated Gaussians. The truncated Gaussians are plotted for (a) $\delta_{fc}/\sigma_{\delta_{fc}}=1$ and (b) $\delta_{fc}/\sigma_{\delta_{fc}}=-1$. The 90\% confidence interval is shaded in the plots. The value representing the upper limit of the confidence interval is plotted for different values of $\delta_{fc}/\sigma_{\delta_{fc}}$. The confidence interval for a full normal distribution is plotted in orange for comparison.}
  \label{fig:truncated_gaussian}
\end{figure}

\begin{figure}
  \centering
  \includegraphics[width=0.85\textwidth]{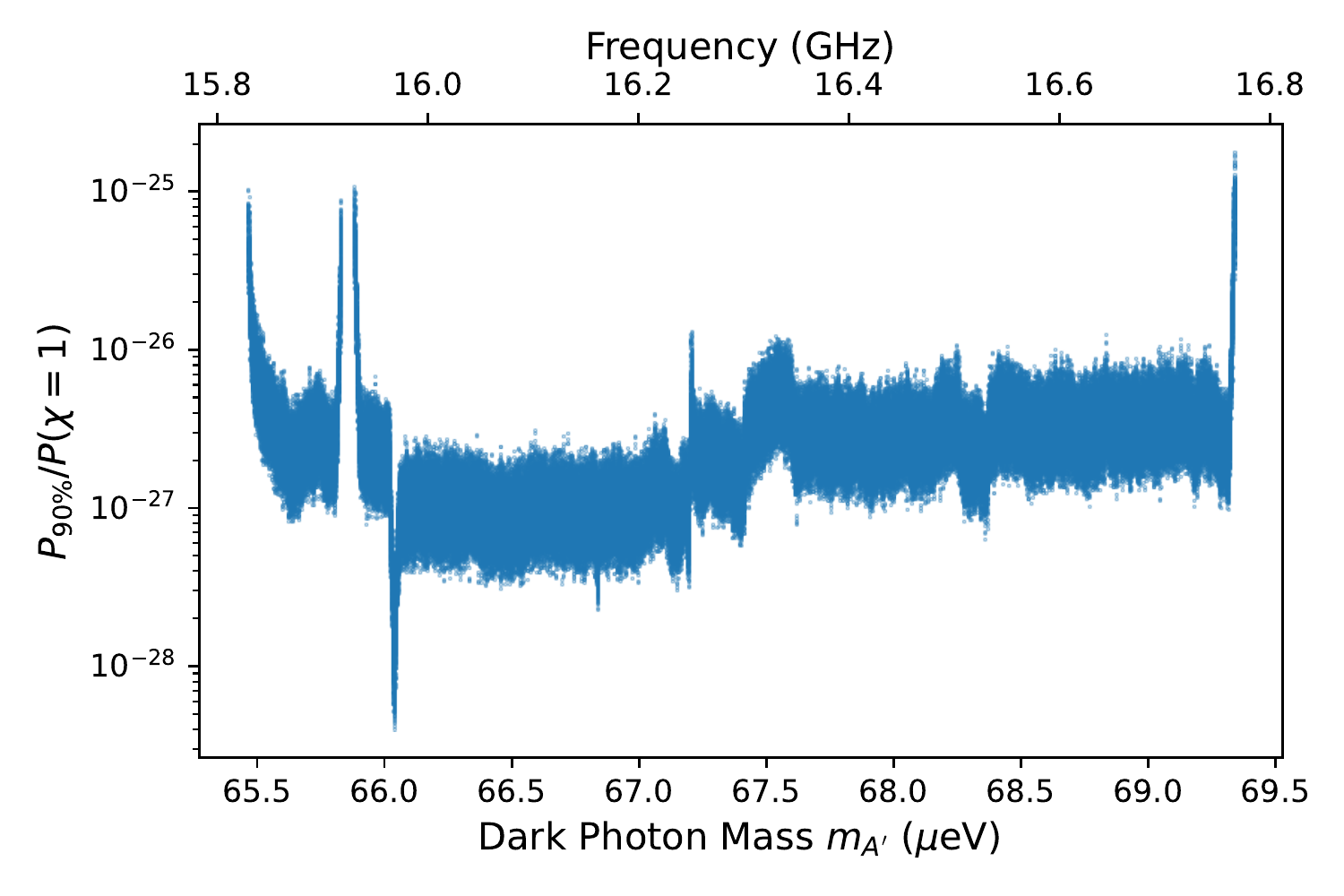}
  \caption{The 90\% confidence limit on the dark photon power, normalized to dark photons signal power of mixing angle $\chi=1$.}
  \label{fig:dp_power_exclusion}
\end{figure}

The 90\% confidence limit for the dark photon signal power is the value of $\delta_c$ for which 90\% of the probability distribution lies below. The inverted, but equivalent, statement is that signal powers above this $\delta_c$ are excluded with 90\% confidence. The confidence interval means that if the same experiment were repeated, there is a 90\% chance the dark photon signal (even if that signal is 0) would lie in this interval. The confidence interval for truncated normal distributions with different means is shown in Figure~\ref{fig:truncated_gaussian}. The 90\% confidence limits for dark photon signal power is determined by applying the percent point function to the truncated normal distribution\footnote{Python's SciPy module has a native function for this.}. The result is shown in Figure~\ref{fig:dp_power_exclusion}.

\begin{figure}
  \centering
  \includegraphics[width=0.85\textwidth]{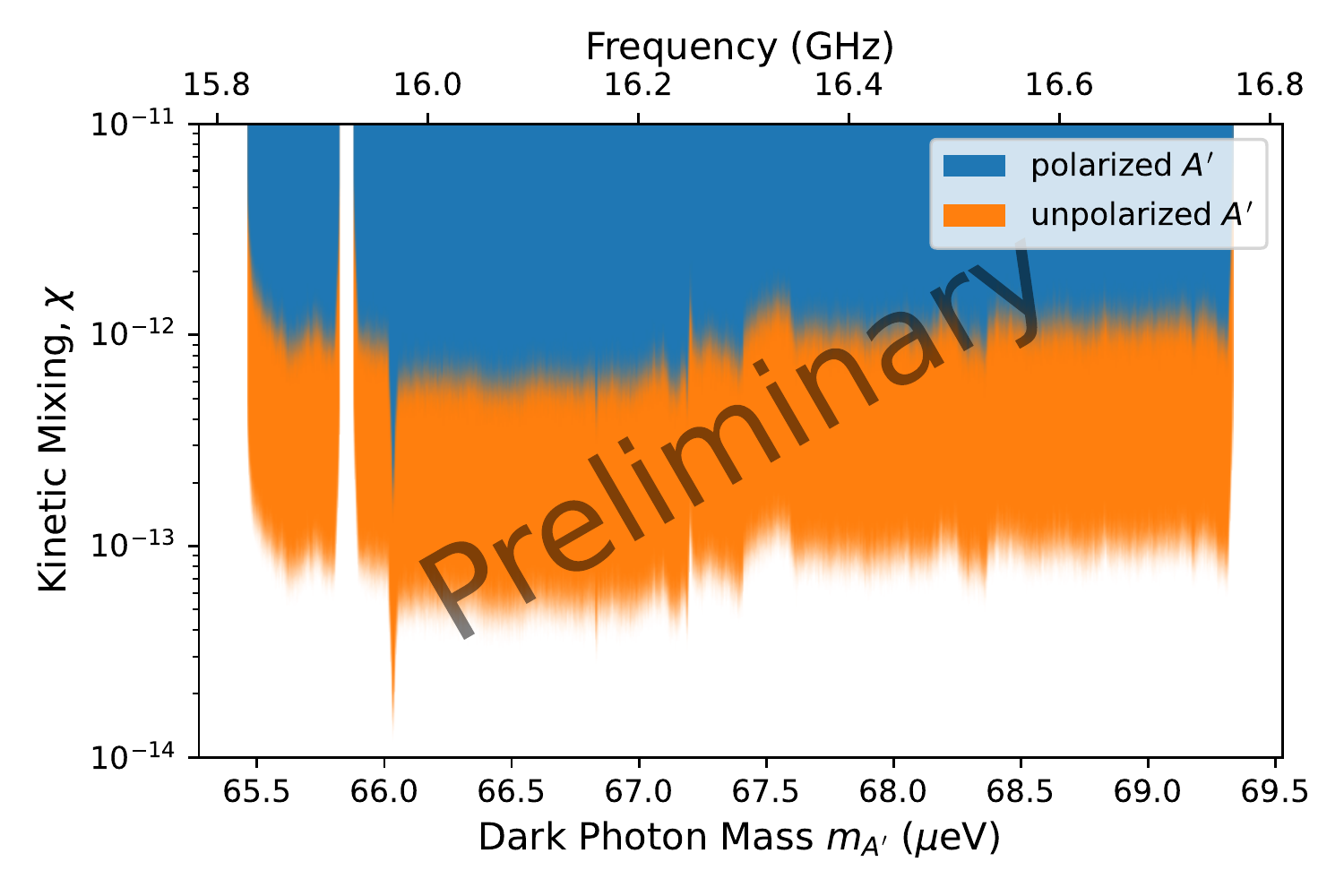}
  \caption{A 90\% exclusion on the mixing angle parameter space.}
  \label{fig:chi_exc}
\end{figure}

\begin{figure}
  \centering
  \includegraphics[width=\textwidth]{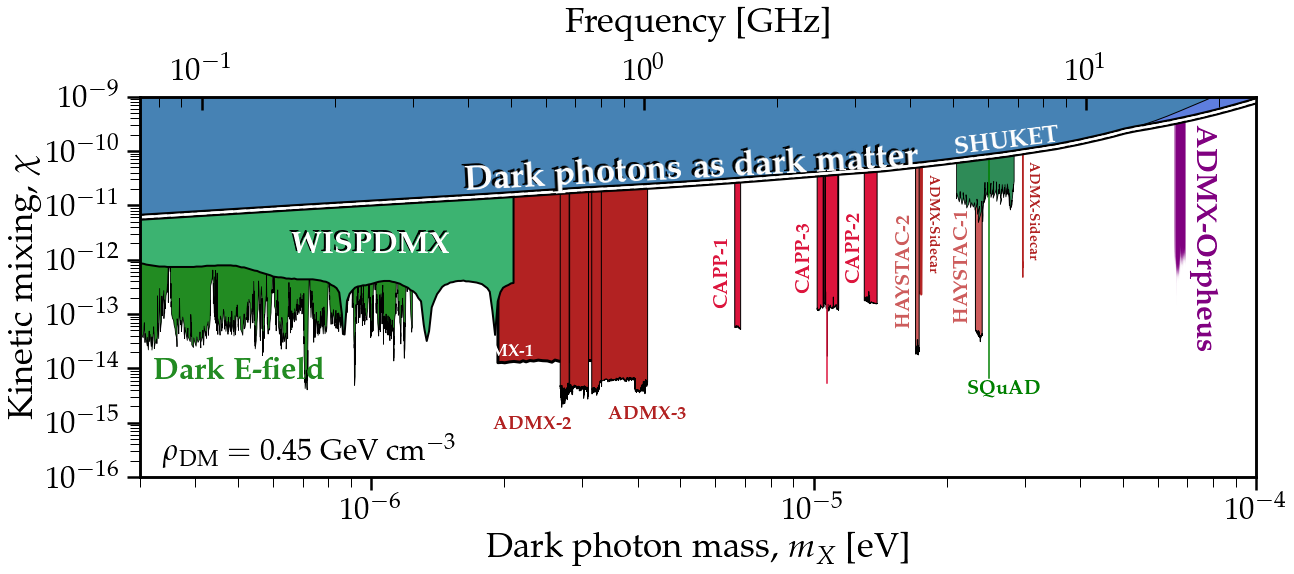}
  \caption{Orpheus science reach in the context of other microwave cavity results. The limits are for the uniformly polarized case. Figure adapted from~\cite{ciaran_o_hare_2020_3932430}.}
  \label{fig:chi_exc_haloscopes}
\end{figure}

By examining Equation~\ref{eqn:dark_photon_power}, one sees that 
\begin{align}
  \frac{\chi_{90\%}}{\chi=1} = \sqrt{\frac{P_{90\%}}{P(\chi=1)}}
\end{align}
Finding $\chi_{90\%}$ amounts to taking the taking the square root of Figure~\ref{fig:dp_power_exclusion}. However, this $\chi_{90\%}$ corresponds to the maximum $V_{eff}$. This is the maximally benevolent model where the photon polarization always aligns with the polarization of the \tem mode, even as the Earth spins and orbits through the cosmos. The realistic mixing angle is $\chi_{90\%} = \frac{\chi_{90\%,max}}{\langle \cos^2 \theta \rangle_T}$. $\langle \cos^2 \theta \rangle_T = 1/3$ if the photons are unpolarized, and $\langle \cos^2 \theta \rangle_T \geq 0.0025$ if they are polarized. Between $\SI{65.5}{\mu eV}$ and $\SI{69.3}{\mu eV}$, the excluded dark photon mixing angle is $\chi_{90\%} \sim \SI{1e-12}{}$ for the polarized dark photon case and $\chi_{90\%} \sim \SI{1e-13}{}$ for the unpolarized dark photon case. The limits are shown in Figure~\ref{fig:chi_exc} and are consistent with expectations laid out in Section~\ref{sec:expected_sensitivity}. The limits are compared to other haloscope experiments in Figure~\ref{fig:chi_exc_haloscopes}. Orpheus achieves similar sensitivities at substantially higher frequencies than other larger, more well-funded haloscope experiments while maintaining a respectable tuning range.

The limits in the polarized case could be improved by perhaps an order of magnitude for selected frequencies by using the fact that the same frequencies were rescanned at different times of the day\cite{caputo2021dark}. This may be accounted for in a future analysis.

%% file: future_directions.tex
\chapter{Future Direction}
I spent three years simulating, prototyping, designing, characterizing, debugging, and commissioning Orpheus. The result was two very fruitful days of collecting dark photon data. The Orpheus test stand remains and is ready to gather more science data as needed. This chapter describes how to get more science out of Orpheus.

\section{Upcoming Axion Run}
\begin{figure}
  \centering
  \includegraphics[width=0.9\textwidth]{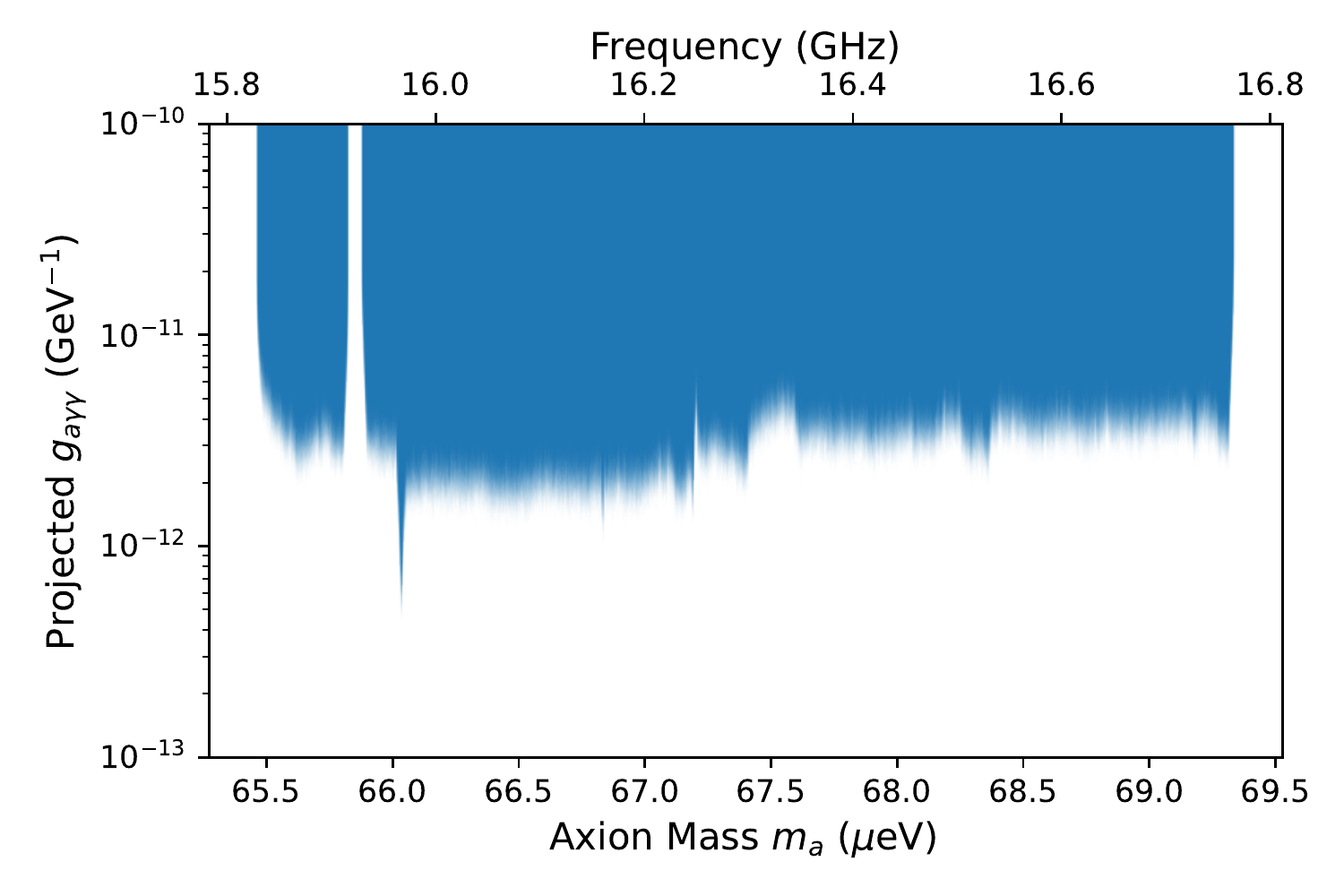}
  \caption{A 95\% exclusion on the axion-photon coupling $g_{a\gamma\gamma}$ had there been a \SI{1.5}{T} dipole magnet.}
  \label{fig:axion_projection}
\end{figure}

\begin{figure}
  \centering
  \includegraphics[width=0.9\textwidth]{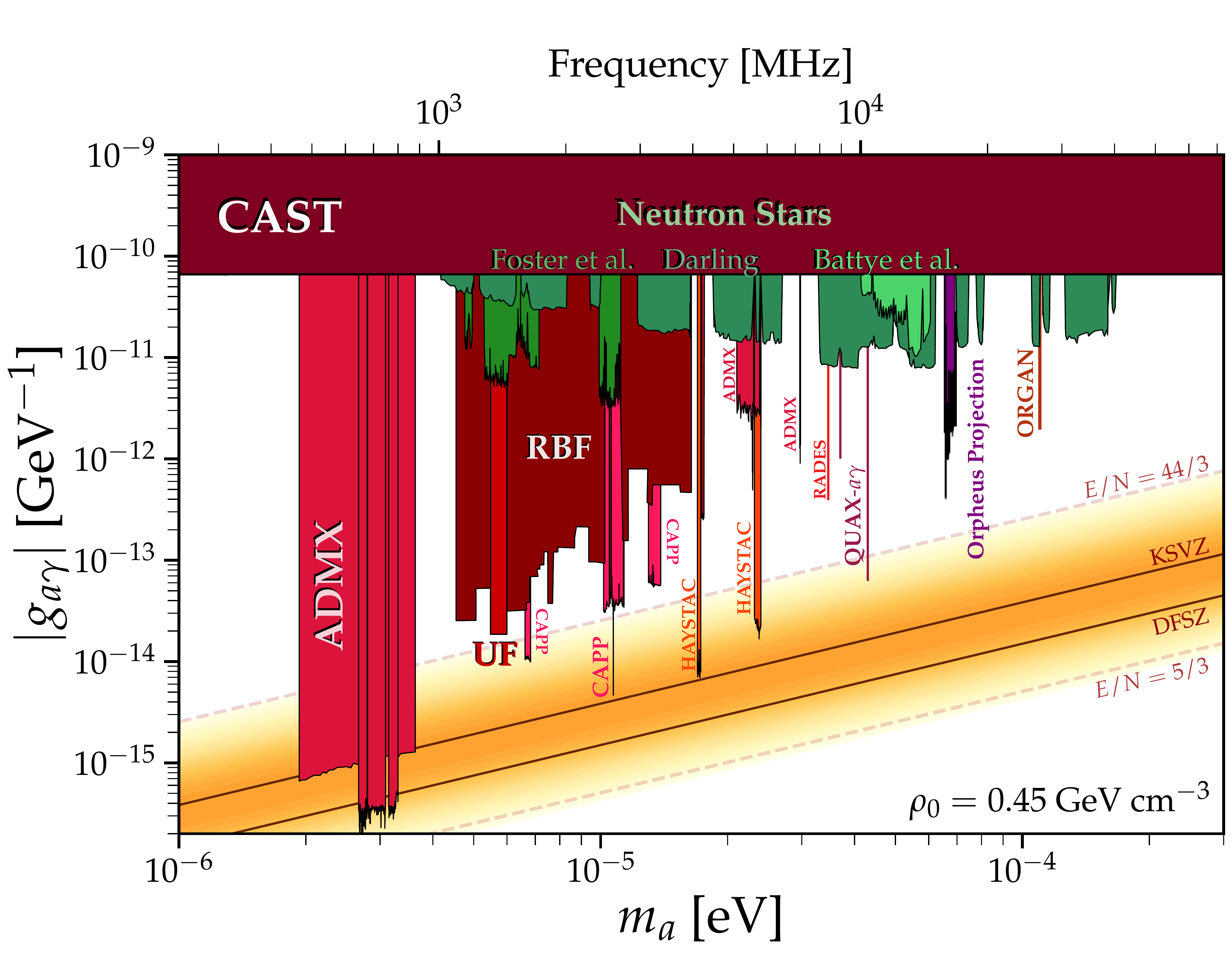}
  \caption{The projected axion limits in the context of other microwave cavity searches. Figure adapted from~\cite{ciaran_o_hare_2020_3932430}.}
  \label{fig:big_axion_projection}
\end{figure}

The last run was without a magnetic field, and so Orpheus was only sensitive to dark photons. But adding an imaginary \SI{1.5}{T} magnet to the very real haloscope data demonstrates that Orpheus would have easily been sensitive to $\gagg\sim\SI{3e-12}{}$ from \SI{15.8}{GHz} to \SI{16.8}{GHz} (Figure~\ref{fig:axion_projection}), over an order of magnitude more sensitive than CAST. That would be the first demonstration of a tunable haloscope beyond \SI{7}{GHz}, as shown in Figure~\ref{fig:big_axion_projection}.

Orpheus is currently preparing for an axion data-taking run, with James Sinnis at the helm. James Sinnis is winding a \SI{1.5}{T} superconducting dipole magnet and has spent the Summer of 2021 developing and practicing the winding technique. See Appendix~\ref{appendix:magnet} to learn more about the magnet design and winding technique.

There are near-term hardware upgrades to improve the automated tuning mechanism so the motors don't stall. Mainly, the coupled shafts need to be better aligned, the motor vacuum ports need to be straightened, and more appropriate shaft couplers are needed to handle any remaining misalignments. Section~\ref{sec:bad_shafts} explains the myriad of motor problems in need of fixing. Thermal issues also limited the science reach of this past run. The thermal issues described in Section~\ref{sec:thermal_issues} should be addressed by clamping the mirror to its holder so that it won't lose thermal contact during cool down. The cold finger will also be thermally connected to other portions of the cavity for more robust cooling in case the flat mirror loses thermal contact during cooldown. Temperature sensors will be added to the dielectric plates to better understand the thermal gradient along the cavity. The cavity must also be better thermally anchored to itself for better temperature uniformity.

A series of DAQ bugs mentioned in Chapters~\ref{ch:operations} and~\ref{ch:analysis} made data analysis awkward and will be fixed for the next run.

\FloatBarrier
\section{More Science with Same Apparatus}
Beyond this coming axion run, there is more science one can do with the same apparatus. 
\subsection{Electrodynamic Optimizations}
Orpheus has lots of degrees of freedom in its design, and one can spend time optimizing the cavity properties. Many design choices were just reasonable first guesses from what I knew as a third-year grad student. I chose the dielectric thickness to be about $\lambda/2$ at \SI{16.5}{GHz} and mirror radius of curvature to be about twice the cavity's optical length\footnote{This is the confocal configuration for a Fabry-Perot cavity.} near \SI{18}{GHz}. Simulations were difficult, and I had little computing power to simulate different geometries. The simulations disagreed with what I expected from physical optics, and I didn't know if the simulations were accurate. I built the Orpheus cavity as soon as I could to validate the simulations and to just have something working as quickly as I could\footnote{Software developers have a maxim, "Make it work. Make it right. Make it fast." In my context, I wanted to get something working before I thought about optimizations.}. Before building the cavity, no one knew if the cavity design would work, and many around me were skeptical. It was not until the Orpheus cavity was assembled and the modes matched the simulation that I felt comfortable trusting my simulation.

Now that the cavity concept has been verified and simulations have been validated, there are a lot of degrees of freedom to optimize to improve the quality factor and effective volume. As currently configured, the Orpheus $V_{eff}$ is about $2\%$ of the physical volume, and that's not good. Surprisingly, moving the dielectrics away from the curved mirror by \SI{1}{mm} improved $\veff$ by more 7\% (Section~\ref{sec:position_error}). Now that the simulations can be trusted, one can perform a parametric sweep of the dielectric positions, mirror radius of curvature, and dielectric thickness to try and optimize the $Q_u$ and $\veff$. One could also increase $\veff$ by placing dielectrics at every other half-wavelength instead of every fourth half-wavelength. But I suspect that reduces the bandwidth of the cavity. 

For scaling the Orpheus concept to larger sizes, it's important to manage the diffraction losses. Diffraction is a dominant source of loss\footnote{I haven't studied whether diffraction losses are greater than losses from dielectric dissipation. But it's more dominant than ohmic losses.} and will only become worse as the cavity length increases relative to the mirror size. Diffraction losses can be mitigated by optimizing the mirror radius of curvature (likely by decreasing it). One can also reduce diffraction by increasing the mirror size, but that increases the physical volume of your cavity, which would necessitate a larger magnet, and large dipole magnets are difficult to fabricate\footnote{Orpheus is an open resonator, so it \emph{may} be possible to design mirrors that extend outside the volume of the magnet. But then the dipole magnet may distort the electric field.}. It may be possible to mitigate diffraction losses by curving the surface of the dielectric plate so that they act as lenses that collimate the field. However, given that the dielectrics are only a few mm thick, its radius of curvature would have to be about \SI{100}{in}, and may be difficult to machine. One workaround is to have dielectrics with a smaller diameter than the mirrors, but that's another parameter to optimize. It's also likely that curving the dielectrics reduces the cavity's bandwidth. 

The dielectric losses can be reduced by using sapphire instead of alumina. Sapphire has a dielectric loss tangent $\tan_d\sim0.00002$ compared to alumina's $\tan_d \sim 0.0001$. But one would have to deal with sapphire's birefringence.

Developing a reliable and adjustable cavity coupling mechanism is important for scaling the concept. For a fixed coupling hole size, the cavity coupling coefficient reduces for longer cavities. Impedance matching is a theoretically solved problem (see Pozar 4th ed. Ch. 5~\cite{pozar}). However, my prototyping with a waveguide tuner proved unfruitful (Section~\ref{sec:orpheus_coupling}). 

Throughout this thesis, I've focused on the \tem mode. It's possible that other modes have substantial coupling to the axion or dark photon. The $V_{eff}$ of these modes should be simulated. The electronics in the Orpheus DAQ have a pretty broad frequency range and can digitize around various modes. If multiple modes couple to dark matter, then collecting data around these modes would be an easy way to increase the scanned mass range of a data-taking run. 

\subsection{Mechanical Optimizations}
One can also think about how to redesign the cavity's mechanical design if one were to start from scratch. I would redesign it so that the tuning is less dependent on the flexures. The flexures reduce friction while the plates move vertically, but they also allow for misalignments. The combination of springs with the correct stiffness constant and correct length of bearings mitigates tilting while tuning. But that solution isn't viable when the dielectrics are placed at every other half-wavelength instead of every quarter of a wavelength. A trained mechanical engineer with cryogenic experience would be beneficial in redesigning the cavity. But I believe each plate should have three points of contact. Perhaps one could modify the current design so that it uses three guiding rods and three scissor jacks. Each of the thrust bearings should also have 2\degree misalignment capability.

The cavity has to be better thermally sunk to itself, such that each mirror and dielectric plate are close to the same temperature.

The gear ratio can also be increased to reduce the backlash.
\subsection{Scanning More Dark Matter Parameter Space}
Even without the mentioned optimizations, the same detector can be used to scan through more parameter space. The limiting factors are that the VNA's upper limit is \SI{18}{GHz}, the WR62 waveguides have a frequency range between \SI{12.4}{GHz} and \SI{18}{GHz}, and the Image Reject Mixer has a range between \SI{14}{GHz} and \SI{18}{GHz}. Without changing anything, the electronics are capable of searching for axions between \SI{14}{GHz} and \SI{18}{GHz} at possibly similar sensitivities. We can then scan between \SI{12.4}{GHz} and \SI{14}{GHz} if we swap out the Image Reject Mixer. 

Scanning from \SI{12.4}{GHz} to \SI{18}{GHz} with the same detector would require several runs with different dielectric thicknesses and mirror curvatures.  It's possible that going to lower frequencies would increase sensitivity to the axion since lower frequencies need longer cavities, leading to an increased $V_{eff}$. However, it's not clear to me how decreasing the frequency would change the quality factor or cavity coupling coefficient. 

\begin{figure}
  \centering
  \subfloat[]{\includegraphics[width=0.48\textwidth]{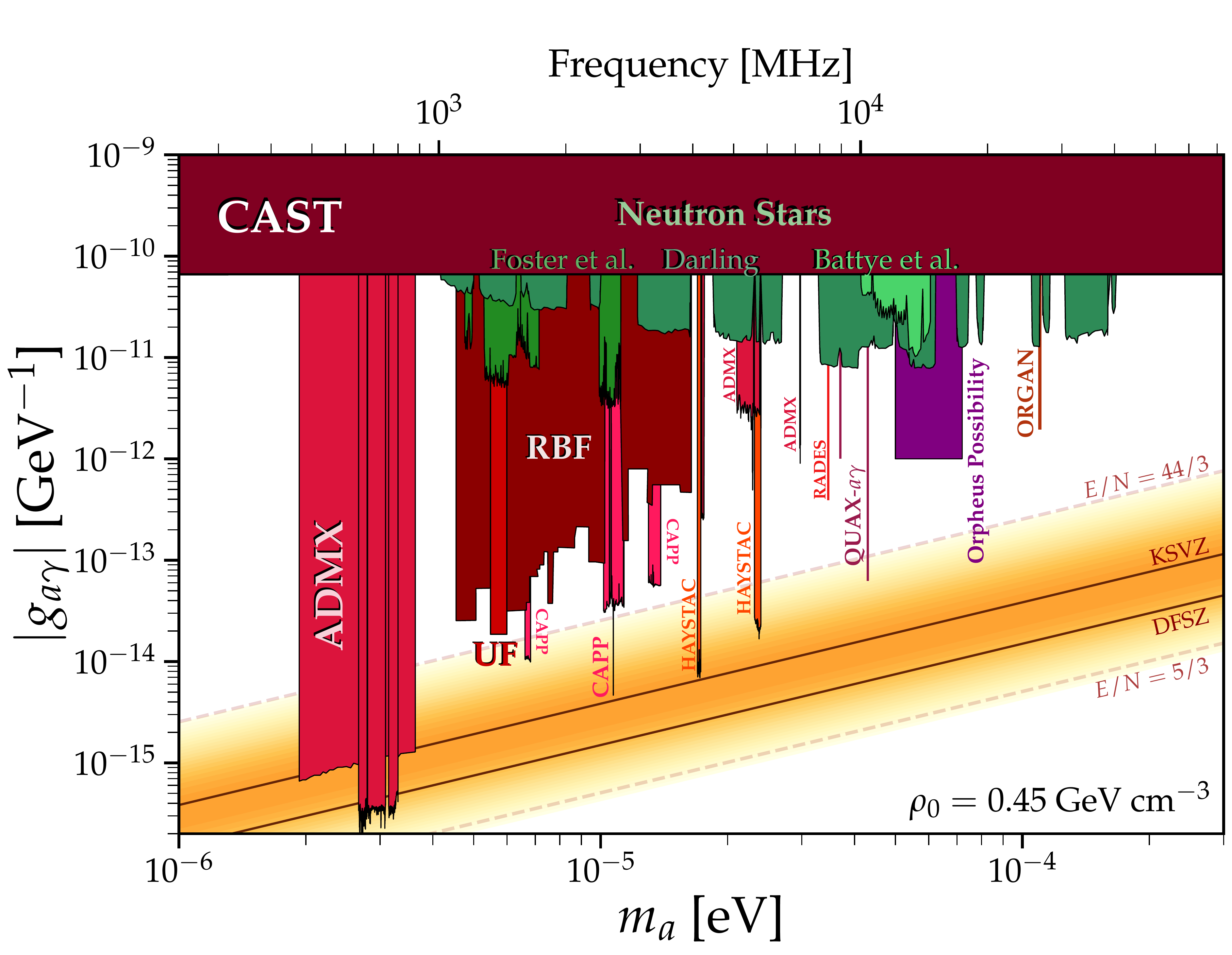}}\hfil
  \subfloat[]{\includegraphics[width=0.48\textwidth]{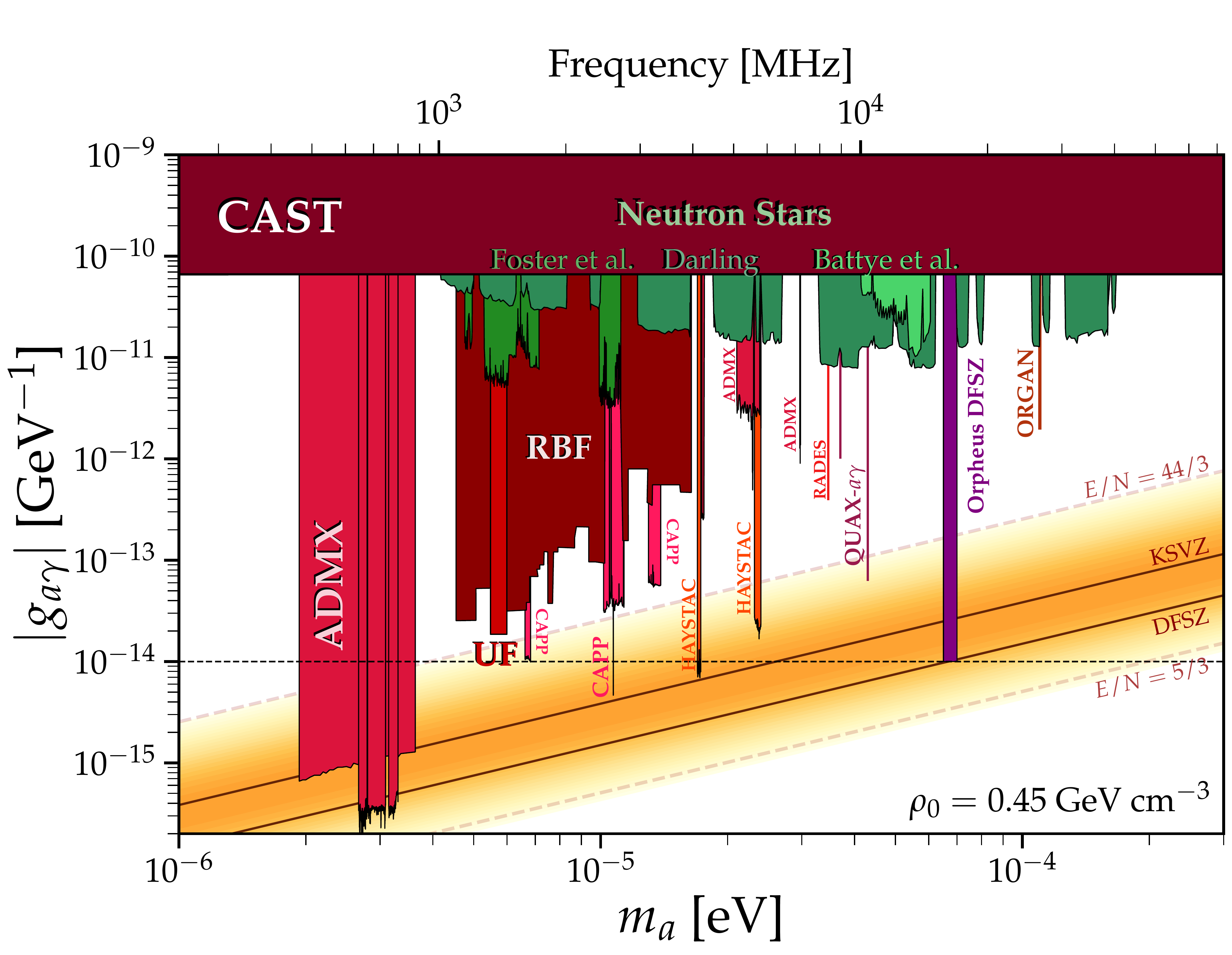}}
  \caption{(a) It is likely that we can achieve sensitivity with the same insert and electronics by swapping out dielectrics of different thicknesses and mirrors with different radius of curvature. (b) DFSZ can be reached with a similar cavity design at similar frequency ranges if the magnetic field was increased to \SI{10}{T}, the effective volume increased by a factor of 10, the quality factor doubled, and the noise temperature reduced by a factor of 10.  Figures adapted from~\cite{ciaran_o_hare_2020_3932430}.}
  \label{fig:extended_orpheus}
\end{figure}

If we swap out dielectric thicknesses and mirror curvatures at the appropriate frequency ranges, the Orpheus detector may be used to scan axion masses at photon couplings of $\gagg \sim 10^{-12}$ for axion masses between \SI{45}{\mu eV} and \SI{75}{\mu eV}, as shown in Figure~\ref{fig:extended_orpheus}. All that is required is additional funding and a graduate student lifecycle.
\section{How to reach KSVZ and DFSZ}
In natural units
\begin{align*}
  P_{a} &=  \frac{\gagg^2}{ m_{a}} \rho_{a} B_o^2 V_{eff} Q_L \betaterm \\
  \gagg &= \frac{1}{B_0}\sqrt{\frac{\beta+1}{\beta}\frac{\snr \times \Delta f T_n m_{a}}{\rho_a V_{eff}Q_L}}\left ( \frac{1}{\Delta f t} \right )^{1/4}
\end{align*}

I will use the projected results in Figure~\ref{fig:axion_projection} as a baseline. To reach KSVZ sensitivity, $V_{eff}$ and $Q_L$ need to double and the noise temperature needs to reduce ten-fold ($T_n\sim\SI{1}{K}$). That would require the optimizations I mentioned in this chapter, cooling the cavity with a dilution refrigerator, quantum noise limited amplifiers, and technological advances in winding superconducting dipole magnets. Except for the dipole magnet, all this seems tractable with current technology.

From the KSVZ estimates, one could reach DFSZ by increasing $V_{eff}$ more. So increase $V_{eff}$ ten-fold instead of two-fold. Sub quantum-noise-limited amplification with photon counters would also be viable in this regime.

%% file: conclusion.tex
\chapter{Conclusion}
Orpheus is a pathfinder experiment that demonstrates that a dielectrically loaded structure can be implemented and tuned in a cryogenic environment to search for axions and dark photons.

In the past three years, I spearheaded this experiment from the design\footnote{Rich Ottens designed the vast majority of the mechanical structure and cryogenic infrastructure before I joined ADMX.} phase through construction, commissioning, and data analysis. I excluded parameter space that is comparable to that of larger, more well-funded experiments. I figured out how to simulate the dielectrically-loaded cavity reliably on a desktop, which was difficult because the cavity is electrically large and has a high mode density. I demonstrated that the cavity had promising sensitivity to wavelike dark matter, and that the mode of interest had good tuning capabilities and forgiving tolerances to mechanical perturbations. I commissioned this experiment at \SI{4.2}{K} with a low noise HFET receiver.

Between $\SI{65.5}{\mu eV}$ and $\SI{69.3}{\mu eV}$, the excluded dark photon kinetic mixing angle is ${\chi_{90\%} \sim \SI{1e-12}{}}$ for the polarized dark photon case and ${\chi_{90\%} \sim \SI{1e-13}{}}$ for the unpolarized dark photon case. With modest alterations and several iterations, the same apparatus may be used to exclude larger parameter space from \SI{45}{\mu eV} to \SI{80}{\mu eV}  with similar sensitivities. 

This is a pathfinder experiment with a limited scope. It's my hope that the hard-earned knowledge gained from Orpheus proves useful to other experiments that implement dielectric haloscopes, such as MADMAX and DBAS.

This work was supported by the U.S. Department of Energy through Grants No. DE-SC0011665 and by the Heising-Simons Foundation.

%% file: appendix.tex
\chapter{Reflection Fitting}
\section{First pass at a reflection fit}\label{sec:reflection_fit_procedure}
\begin{figure}[!htbp]
    \centering
    \includegraphics[width=0.8\textwidth]{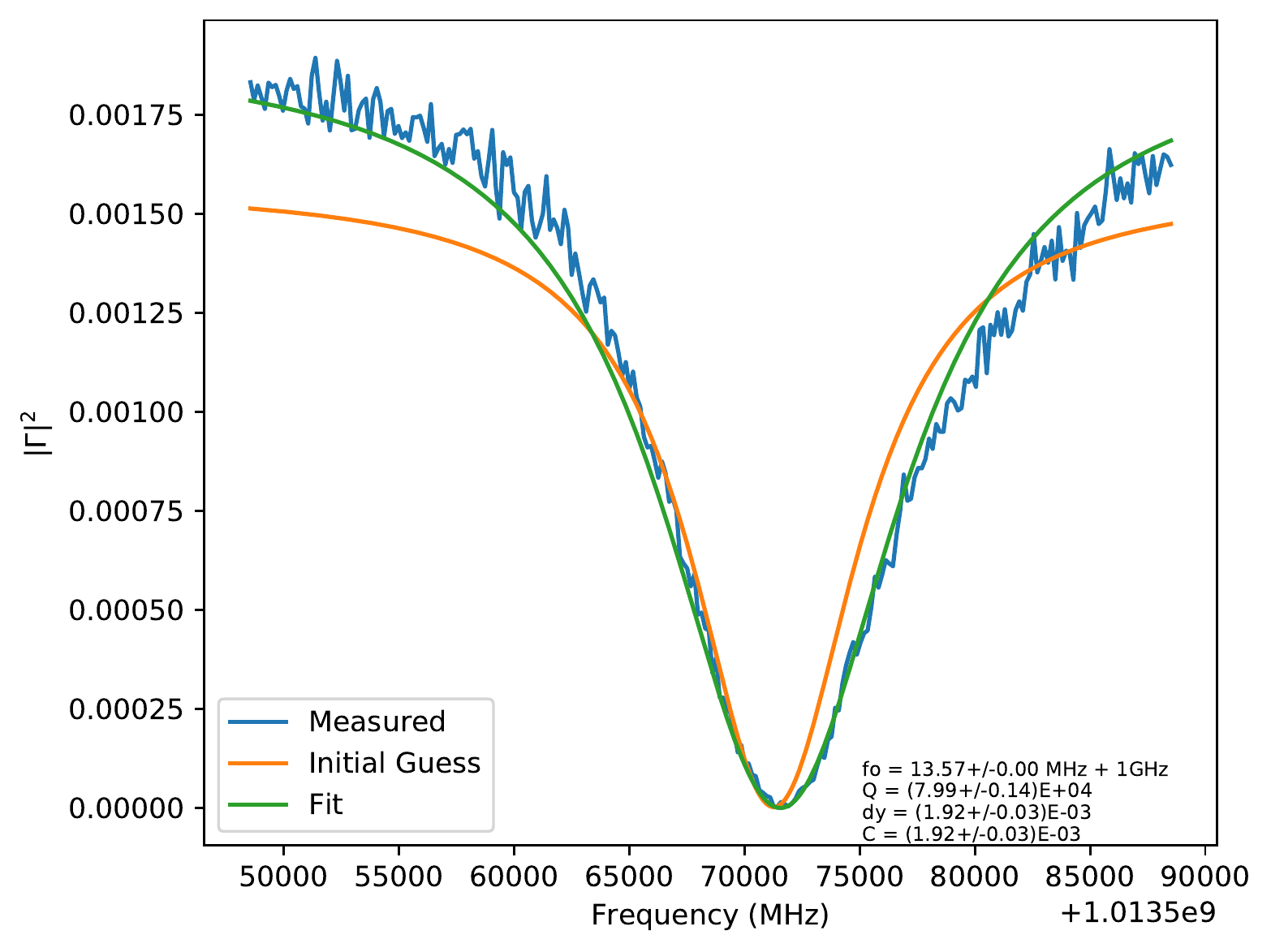}
		\caption{The process of fitting a Lorentzian to the reflection coefficient to extract the resonant frequency and quality factor. The blue curve is the square of the measured reflected coefficient. The orange line is the initial guess of the Lorentzian. The green line is the curve fit, which tries to minimize the $\chi^2$.}
    \label{fig:reflection_fitting}
\end{figure}

The measured reflected power  is modeled as
\begin{equation}
  \abs{\Gamma}^2 = C-\frac{\delta y}{1+4\Delta^2}; \hspace{1cm} \Delta \equiv Q\frac{f-f_0}{f_0}
		\label{eqn:reflection_model}
\end{equation}
where $\Gamma$ reflection coefficient, $f_0$ ($Q$) is the resonant frequency (quality factor) of the cavity, C is some constant offset, and $\delta y$ is the depth of the Lorentzian. One can check that $\Gamma(f_0) = C-\delta y$ and $\Gamma(f_0 - \delta f/2)=C-\delta y/2$, where the bandwidth $\delta f = f_0/Q$. 
To obtain the resonant frequency and Q of the cavity, we fit Equation~\ref{eqn:reflection_model} to the measured reflected power, as shown in Figure~\ref{fig:reflection_fitting}.
Non-linear fits are often sensitive to the initial guess of fit parameters, so the initial guesses should be close to the correct answer (aim for an order of magnitude). One can come up with reasonable guesses from abstracting features of the reflected power. Below is a list of methods to obtain the initial guesses.

\newcommand{\gu}{\text{guess}}

\begin{enumerate}
	\item $f_{0,\gu}$: Find the frequency that corresponds to $\Gamma^2_{min}$

	\item $C_{\gu}$: Take the median of the upper 2/3 of the $\abs{\Gamma}^2$ values. The lower 1/3 probably corresponds to the reflection on resonance. 

	\item $\delta y_{\gu} = C_{\gu} - \abs{\Gamma}^2_{\text{min}}$ 

	\item $Q_{\gu}$: Find bandwidth $\delta f$ by determining frequencies where $\Gamma = C_{\gu} - \delta y_{\gu}/2$, i.e., find the full-width-half-max. $Q_{\gu} = f_{0,\gu}/\delta f$.\\
    Of course, if you already roughly know the Q of the system, then you can manually put in that guess.
\end{enumerate}

\begin{figure}[!htbp]
    \centering
    \includegraphics[width=0.8\textwidth]{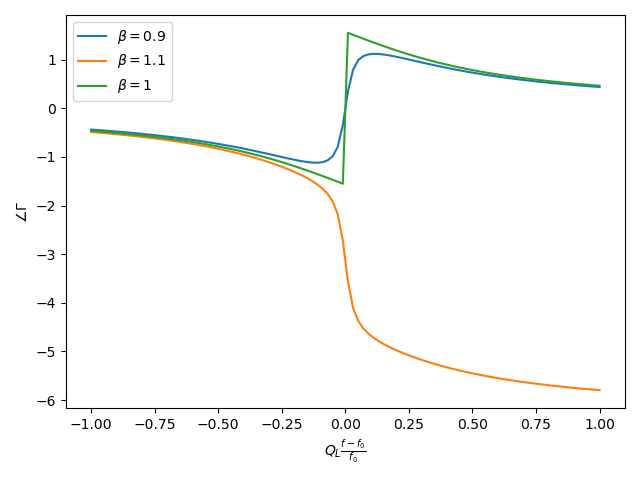}
  \caption{The phase change of the reflection coefficient for the undercoupled case (blue), critically-coupled case (green), and overcoupled case (orange). The derivative on resonance is positive for the undercoupled case and negative for the overcoupled case.}
    \label{fig:gamma_angle}
\end{figure}

\FloatBarrier
Now let us extract the antenna coupling coefficient $\beta$. In later sections, I derive that the reflection coefficient of the cavity is $\Gamma_{c} = \frac{\frac{1-\beta}{1+\beta} +2i\Delta}{1+2i\Delta} $. Figure~\ref{fig:gamma_angle} shows the phase response for an undercoupled cavity ($\beta<1$), overcoupled cavity ($\beta>1$), and critically-coupled cavity ($\beta=1$). We can check the phase of the measured IQ data around resonance to determine if we are undercoupled or overcoupled. As one option, one can check if $\angle \Gamma(f_0^+) > \angle \Gamma(f_0^-)$ to check if the cavity is undercoupled.

We can calculate the coupling coefficient from the reflected power. It's best to use the value from the fitted function since the measured values at low power tend to be noisy.

\begin{align}
 \beta = 
  \begin{cases} 
   \frac{1-{\abs{\Gamma_c(f_0)}}}{1+\abs{\Gamma_c(f_0)}} & \text{if } \beta \leq 1 \\
   \frac{1+\abs{\Gamma_c(f_0)}}{1-\abs{\Gamma_c(f_0)}} & \text{if } \beta \geq 1
  \end{cases}
\end{align}

But we do not measure the cavity reflection coefficient directly. The signal is amplified and then attenuated by cables. To estimate $\abs{\Gamma_c(f_0)}$, we note that the cavity acts like a reflector off resonance, so $\abs{\Gamma_c(\Delta)} \rightarrow 1$ for large $\Delta$. After normalizing the reflected power to estimate $\abs{\Gamma_c(\Delta)}^2$, we can calculate $\abs{\Gamma_c(f_0)}$ using the fitted parameters $\abs{\Gamma_c(f_0)} = \sqrt{1-\frac{\delta y}{C}}$.I

\FloatBarrier
A proper chi-square minimization requires inputting accurate relative uncertainties of each datapoint in the reflected power. The uncertainty of each data point can be derived as $\sigma_P^2 = 4 \sigma_V^2 P$, where $\sigma_V $ is the uncertainty of the IQ voltages. $\sigma_V$ depends on the VNA settings such as the number of averages and the IF bandwidth. $\sigma_V $ can be estimated by measuring the standard deviation of the recorded IQ voltages from the VNA noise floor. 

The derivation for the uncertainty in reflected power follows:
\begin{proof}
	\begin{align*}
		P &\propto V_{I}^2 + V_{R}^2
	\end{align*}
	Let $Re \equiv V_{R}^2$ and $Im \equiv V_{I}^2$.

	\begin{align*}
		P &\propto R + I \\
		\sigma_P^2 &= \sigma_{R}^2 + \sigma_{I}^2\\
		\left(\frac{\sigma_{R}^2}{R}\right)^2 &= \left(\frac{2\sigma_{V_{R}}}{V_{R}}\right)^2\\
		\sigma_{R}^2 &= (2V_{R}\sigma_{V_{R}})^2
	\end{align*}
	Similarly
	\begin{align*}
		\sigma_{I}^2 &= (2V_{I}\sigma_{V_{I}})^2\\
		\sigma_{P}^2 &= (2V_{R}\sigma_{V_{R}})^2 + (2V_{I}\sigma_{V_{I}})^2\\
		\sigma_{V_R} &= \sigma_{V_I} \equiv \sigma_V\\
		\sigma_{P}^2 &= (2\sigma_V)^2 (V_R^2 + V_I^2)\\
		\sigma_{P}^2 &= 4\sigma_V^2 P 
	\end{align*}
	
\end{proof}

\section{Deriving the reflection off of the cavity}\label{sec:reflection_derivation}
In this section, we derive why the reflected power off a cavity follows a Lorentzian profile.

A cavity can be modeled as a series RLC circuit (Pozar 4th ed. Ch. 6), and the cavity impedance $Z_c$ is
\begin{align*}
  Z_c &= R + \frac{1}{i\omega C} + i\omega L \\
      &= R + \frac{1-\omega^2 L C }{i \omega C}
\end{align*}

For an RLC circuit, $\omega_o = \frac{1}{LC}$. Define $\omega \equiv \omega_o - \delta \omega$. Then $\omega_o^2 \approx \omega_o^2-2\omega_o\delta\omega$.

\begin{align*}
  Z_c-R &\approx  \frac{1-\frac{1}{\omega_o^2}(\omega_o^2 -2\omega_o\delta\omega)}{i \omega_o C} \\
          &= \frac{-2\frac{\delta\omega}{\omega_o}}{i\omega_o^2 C}
\end{align*}

Define $Q_o \equiv \frac{1}{\omega_o R C} = \frac{1}{R}{\sqrt{\frac{L}{C}}} = \frac{\omega_o L}{R}$.

\begin{align*}
  Z_c = R\left(1+2i\frac{\delta\omega Q_o}{\omega_o}\right)
\end{align*}

Let's connect the cavity to a load with impedance $Z_o=R_o$. Assume that the transmission line connecting the cavity to the load is impedance-matched to the load. Then the reflection coefficient, from Pozar 4th ed. Equation~2.35, is
\begin{align*}
  \Gamma_{c}=\frac{Z_c-Z_o}{Z_c+Z_o}
\end{align*}

Define an antenna coupling coefficient $\beta=\frac{Q_o}{Q_e} = \frac{Z_o}{R}$. 
\begin{align*}
  \frac{1}{Q_L} &= \frac{1}{Q_o}+\frac{1}{Q_e}\\
  \frac{1}{Q_L} &= \frac{1}{Q_o}\left(1+\frac{Q_o}{Q_e}\right)\\
  Q_L &= \frac{Q_o}{1+\beta}
\end{align*}

On resonance, $Z_c = R$ and the magnitude of the reflection coefficient depends on the coupling coefficient.
\begin{align}
  \Gamma_{c}(f_o) &= \frac{1-\frac{Z_o}{R}}{1+\frac{Z_o}{R}} \nonumber\\
  \Gamma_{c}(f_o) &= \frac{1-\beta}{1+\beta} 
\end{align}

Let's now solve for the reflection coeffient around the resonance.
\begin{align*}
  \Gamma_{c} &= \frac{Z_c-Z_o}{Z_c+Z_o}\\
   &= \frac{1+2i\frac{\delta\omega Q_o}{\omega_o}-\frac{Z_o}{R}}{1+2i\frac{\delta\omega Q_o}{\omega_o}+\frac{Z_o}{R}}\\
   &= \frac{1+2i\frac{\delta\omega Q_o}{\omega_o}-\beta}{1+2i\frac{\delta\omega Q_o}{\omega_o}+\beta}\\
   &= \frac{1+2i\frac{\delta\omega Q_o}{\omega_o}-\beta}{1+2i\frac{\delta\omega Q_o}{\omega_o}+\beta}\times\left(\frac{1+\beta}{1+\beta}\right)\\
  \Gamma_{c} &= \frac{\frac{1-\beta}{1+\beta} +2i\frac{\delta\omega Q_L}{\omega_o}}{1+2i\frac{\delta\omega Q_L}{\omega_o}}\\
\end{align*}
Define $\Delta \equiv Q_L \frac{\delta \omega}{\omega_o}$
\begin{align}
  \Gamma_{c} &= \frac{\frac{1-\beta}{1+\beta} +2i\Delta}{1+2i\Delta}
  \label{eqn:reflection_coefficient}
\end{align}

The reflected power is the magnitude-squared of the reflection coefficient.
\begin{align*}
  |\Gamma_{c}|^2 = \frac{\frac{1-\beta}{1+\beta}^2 + 4\Delta^2}{1+4\Delta^2}\\
  \frac{4\Delta^2}{1+4\Delta^2} = 1-\frac{1}{1+4\Delta^2}\\
  |\Gamma_{c}|^2 = 1-\frac{\frac{4\beta}{(1+\beta)^2}}{1+4\Delta^2}\\
\end{align*}
Let $L(f) \equiv \frac{1}{1+4\Delta^2}$. This is the functional form of the Lorentzian. Then 
\begin{align}
  |\Gamma_{c}|^2 = 1 - \frac{4\beta}{(\beta+1)^2}L(f)
\end{align}

\FloatBarrier
\section{Model of a resonator connected to directional coupler and amplifier}\label{sec:reflection_coupler}
In the previous section, we modeled a cavity connected to a load, and the transmission line was impedance-matched to the load. More realistically, the transmission line is composed of many components with various sources of impedance mismatches.

\begin{figure}[!htbp]
    \centering
  \includegraphics[height=0.3\textheight]{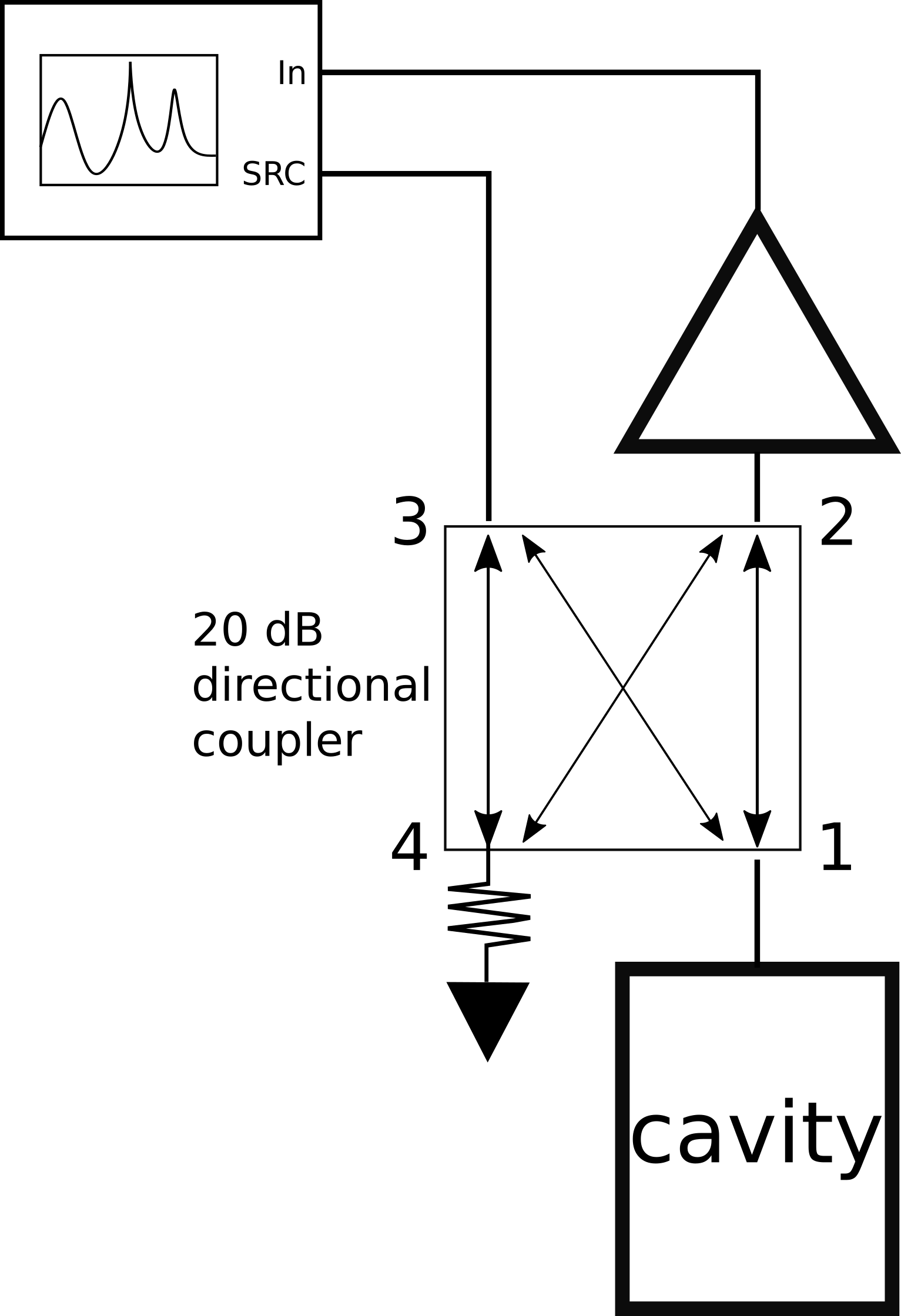}
  \caption{The reflection measurement off of the Orpheus cavity. The VNA sends a signal to port 3 of the directional coupler. The signal travels to port 1 and reflects off the cavity. Then the signal travels from port 1 to port 2, and then through the amplifier and back to the VNA.}
  \label{fig:orpheus_reflection_measurement}
\end{figure}

Figure~\ref{fig:orpheus_reflection_measurement} demonstrates the reflection measurement off of the Orpheus cavity. The VNA sends a signal to port 3 of the directional coupler. The signal travels to port 1 and reflects off the cavity. Then the signal travels from port 1 to port 2, and then through the amplifier and back to the VNA. However, the amplifier could be mismatched, allowing for standing waves to form in the transmission path. Furthermore, there might be additional crosstalk-like effects when one considers the directivity and isolation of the coupler. 

\begin{figure}[!htbp]
    \centering
  \includegraphics[height=0.3\textheight]{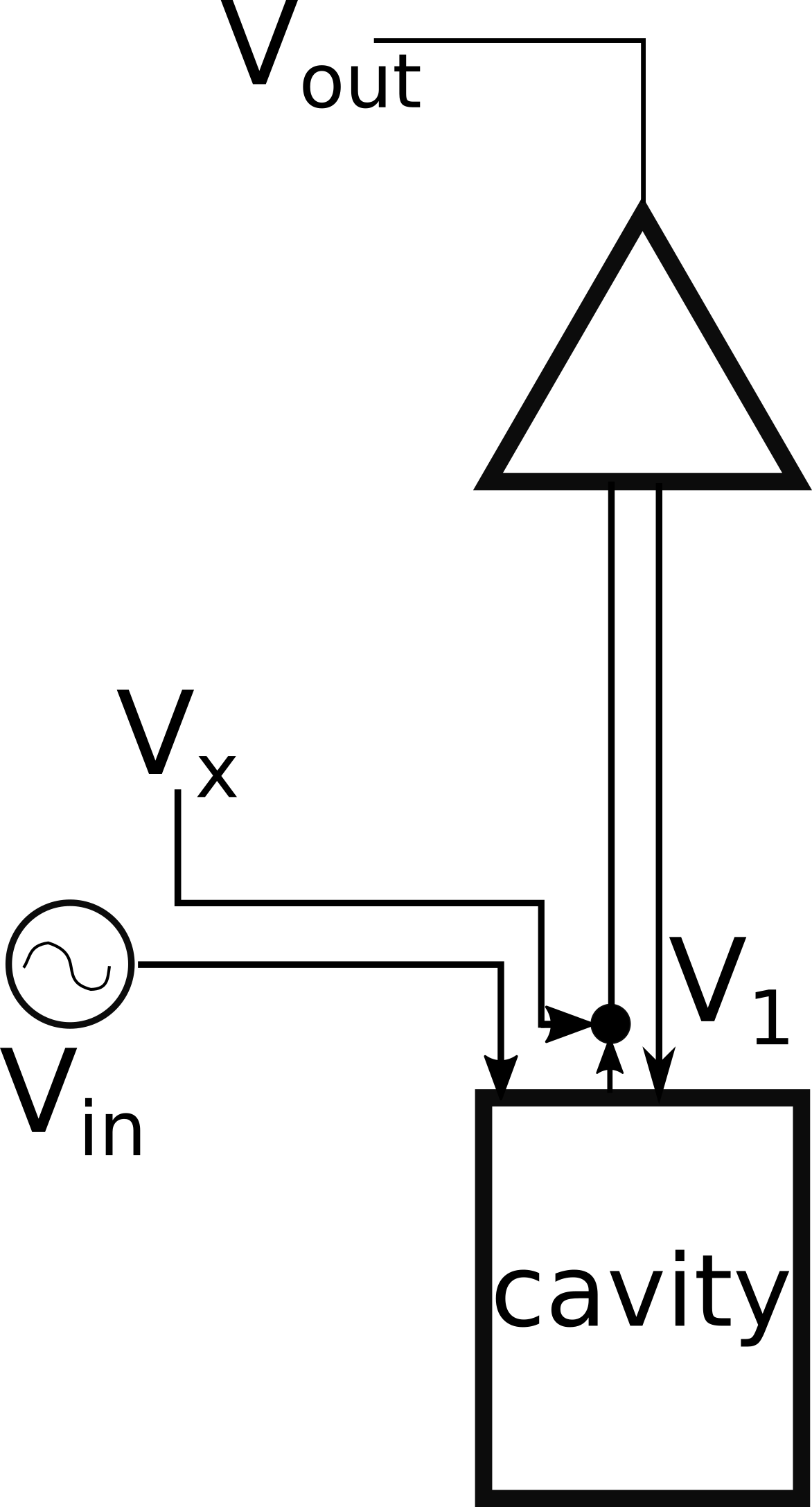}
  \caption{A simplified mathematical model for a reflection measurement. A signal is injected into the cavity. $V_1$ is the signal reflected off the cavity heading towards the direction of the amplifier. $V_1$ reaches the amplifier and is partially reflected back to the cavity. That signal is then reflected again off the cavity and is added to $V_1$. There is an additional $V_x$ term that represents the complications from the injected signal interacting with the four different ports in the directional coupler.}
  \label{fig:coupling_math_model}
\end{figure}

To simplify the analysis, I've simplified the reflection model in Figure~\ref{fig:orpheus_reflection_measurement} to Figure~\ref{fig:coupling_math_model}. Our VNA measures $\frac{V_{out}}{V_{in}}$. Let $V_1$ be the voltage reflected off of the cavity heading towards the amplifier, and let $\Gamma_a$ be the reflection coefficient of the amplifier. Furthermore, let $V_x$ represent the crosstalk-like effects of the directional coupler that are abstracted away. $V_{out} \propto V_{in}$, so let's solve for $V_{in}$

\begin{align*}
  V_1 = V_{in}\Gamma_c + V_1 e^{i\theta}\Gamma_c \Gamma_a + V_x\\
  \frac{V_1}{V_{in}} = \frac{\Gamma_c + \frac{V_x}{V_{in}}}{1-e^{i\theta}\Gamma_c \Gamma_a}\\
  \end{align*}

  The denominator is reminiscent of the mismatched generator problem in Pozar 4th ed. 2.6. Let's say $\frac{V_x}{V_{in}}=e^{i\phi}\chi$ and assume that $\chi$ is constant across the resonance of the cavity.

  Now, let's assume that the amplifier is matched well enough that reflections are minimal and set $\Gamma_c=0$

\begin{align*}
  \frac{V_1}{V_{in}} \approx \Gamma_c + e^{i\phi}\chi\\
\end{align*}

We want to see how this additional term perturbs the reflected power away from the Lorentzian.

\begin{align*}
  \abs{\frac{V_1}{V_{in}}}^2 = \abs{\Gamma_c}^2 + \abs{\chi}^2 + \chi (\Gamma_c e^{-i\phi} + \Gamma_c^* e^{i\phi})\\
\end{align*}

By using Equation~\ref{eqn:reflection_coefficient} and expanding, one can show
\begin{align}
  \Gamma_c e^{-i\phi} + \Gamma_c^* e^{i\phi} = 2\cos\phi - \frac{4\beta}{1+\beta}\cos\phi L(f) + \frac{8\beta}{1+\beta}\sin\phi\Delta L(f)
  \label{eqn:complex_expansion}
\end{align}

Using Equation~\ref{eqn:complex_expansion}, we get

\begin{align}
  \abs{\frac{V_1}{V_{in}}}^2 &= 1 - \frac{4\beta}{(\beta + 1)^2}L(f) + \abs{\chi}^2 + 2\chi\cos\phi - \frac{4\chi\beta}{1+\beta}\cos\phi L(f) + 8\chi\sin\phi\frac{\beta}{1+\beta}\Delta L(f)\nonumber\\
                             &= 1 + \abs{\chi}^2 + 2\chi\cos\phi - 4\left(\frac{\beta}{(\beta + 1)^2} + \frac{\chi\beta}{1+\beta}\cos\phi \right)L(f) + 8\chi\sin\phi\frac{\beta}{1+\beta}\Delta L(f)
\end{align}

Let $C\equiv 1-2\chi^2\cos\phi+\chi^2$, $D\equiv \chi \cos\phi$, $F\equiv 8\chi\sin\phi$.
\begin{align*}
  \abs{\frac{V_1}{V_{in}}}^2 = C - 4\left(\frac{\beta}{(\beta + 1)^2} + D\frac{\beta}{1+\beta}\right)L(f) + F\frac{\beta}{1+\beta}\Delta L(f)
\end{align*}
Finally,
\begin{align}
  \abs{\frac{V_{out}}{V_{in}}}^2 = A\left(C - 4\left(\frac{\beta}{(\beta + 1)^2} + D\frac{\beta}{1+\beta}\right)L(f) + F\frac{\beta}{1+\beta}\Delta L(f)\right)
  \label{eqn:complicated_fit}
\end{align}
where $A$ represents the amplification of the amplifier and the attenuation from the amplifier to the VNA.

If we had not set $\Gamma_a=0$, then Equation~\ref{eqn:complicated_fit} becomes
\begin{align*}
  \abs{\frac{V_{out}}{V_{in}}}^2 = A\left(\frac{C - 4\left(\frac{\beta}{(\beta + 1)^2} + D\frac{\beta}{1+\beta}\right)L(f) + F\frac{\beta}{1+\beta}\Delta L(f)}{1+\Gamma_a(2\cos\theta-4\frac{\beta}{\beta+1}(\cos\theta+\Delta\sin\theta)) + \Gamma_a^2 \left(1-\frac{4\beta}{(\beta+1)^2}L(f)\right)}\right)
\end{align*}

\begin{figure}[ht]
    \centering
  \includegraphics[height=0.4\textheight]{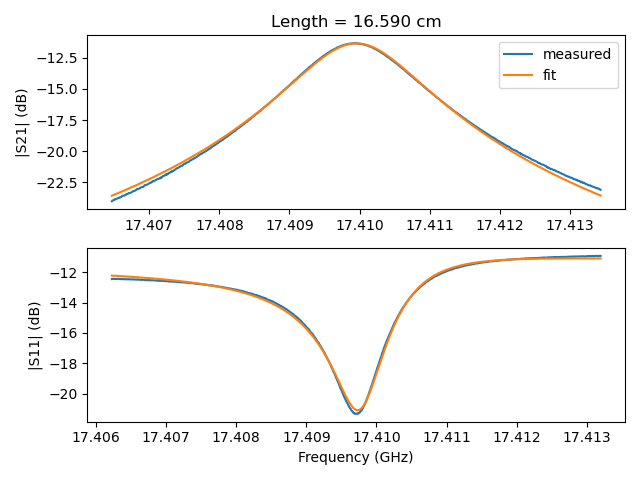}
  \caption{The fitted transmitted (standard Lorentzian fit) and reflected power (perturbed Lorentzian fit, Equation~\ref{eqn:complicated_fit}).}
  \label{fig:complicated_fit}
\end{figure}

\begin{figure}[ht]
    \centering
  \includegraphics[height=0.4\textheight]{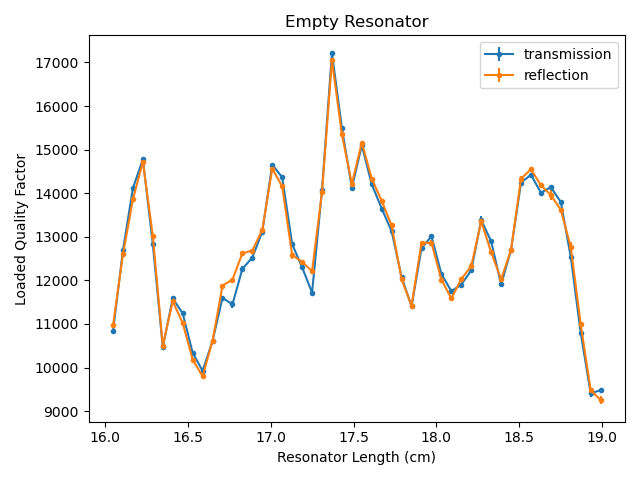}
  \caption{The fitted loaded quality factors for transmission (standard Lorentzian fit) and reflection (perturbed Lorentzian fit, Equation~\ref{eqn:complicated_fit}).}
  \label{fig:fitted_qs}
\end{figure}

Equation~\ref{eqn:complicated_fit} is a promising candidate to fit reflected powers that deviate from the standard Lorentzian shape. This model fits Orpheus data better than a Lorentzian (Figure~\ref{fig:complicated_fit}) and more accurately extracts the Q parameter, as evident by the fact that the reflection $Q_L$ matches the transmission $Q_L$ (Figure~\ref{fig:fitted_qs}). Unfortunately, there is a large degeneracy between $\beta$, $\chi$, and $\phi$. The resulting elements in the covariance matrix are huge. So Equation~\ref{eqn:complicated_fit} hasn't proved useful for extracting $\beta$. One may be able to make progress by constraining $\chi$ to be appropriately small. 

\FloatBarrier
\section{Axion coupling sensitivity to uncertainty in antenna coupling coefficient}
A resonators sensitivity to the axion is related to the antenna coupling coefficient

\begin{align*}
  g_{a\gamma \gamma} \propto \left(\frac{\beta+1}{\beta}\right)^{1/2}
\end{align*}

The uncertainty in $g_{a\gamma \gamma}$ is related to the uncertainty in $\beta$ by
\begin{align*}
  \Delta g_{a\gamma\gamma} &= \abs{\pdv{g_{a\gamma\gamma}}{\beta}} \Delta\beta\\
  \pdv{g_{a\gamma\gamma}}{\beta} &= -\frac{1}{2\beta^2\sqrt{\frac{\beta+1}{\beta}}}
\end{align*}

\begin{figure}[!htbp]
    \centering
  \includegraphics[height=0.3\textheight]{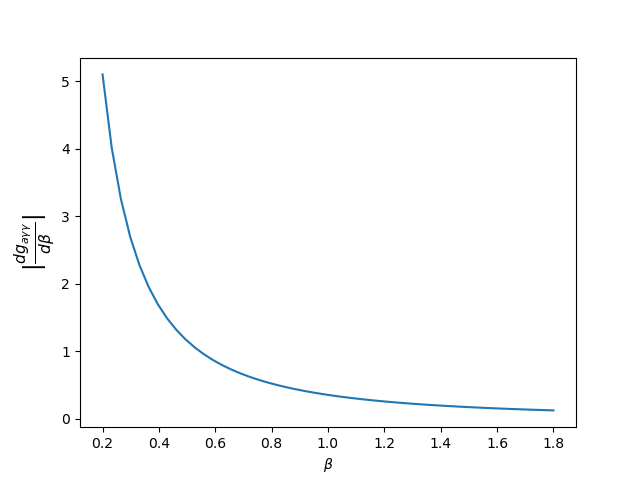}
  \caption{The derivative of the axion-photon coupling constant vs antenna coupling coefficient. This relationship determines how the uncertainty in $\beta$ affects the uncertainty in $g_{a\gamma\gamma}$.}
  \label{fig:dgamma_dbeta}
\end{figure}

$\pdv{g_{a\gamma\gamma}}{\beta}$ is plotted in Figure~\ref{fig:dgamma_dbeta}. This suggests that near critical-coupling, the uncertainty in axion sensitivity is insensitive to the uncertainty in antenna coupling. Then it becomes very sensitive when we are weakly coupled.

\section{Magnet Design}\label{appendix:magnet}
\begin{figure}
  \centering
  \subfloat[Along the axis of the cavity]{\includegraphics[height=0.25\textheight]{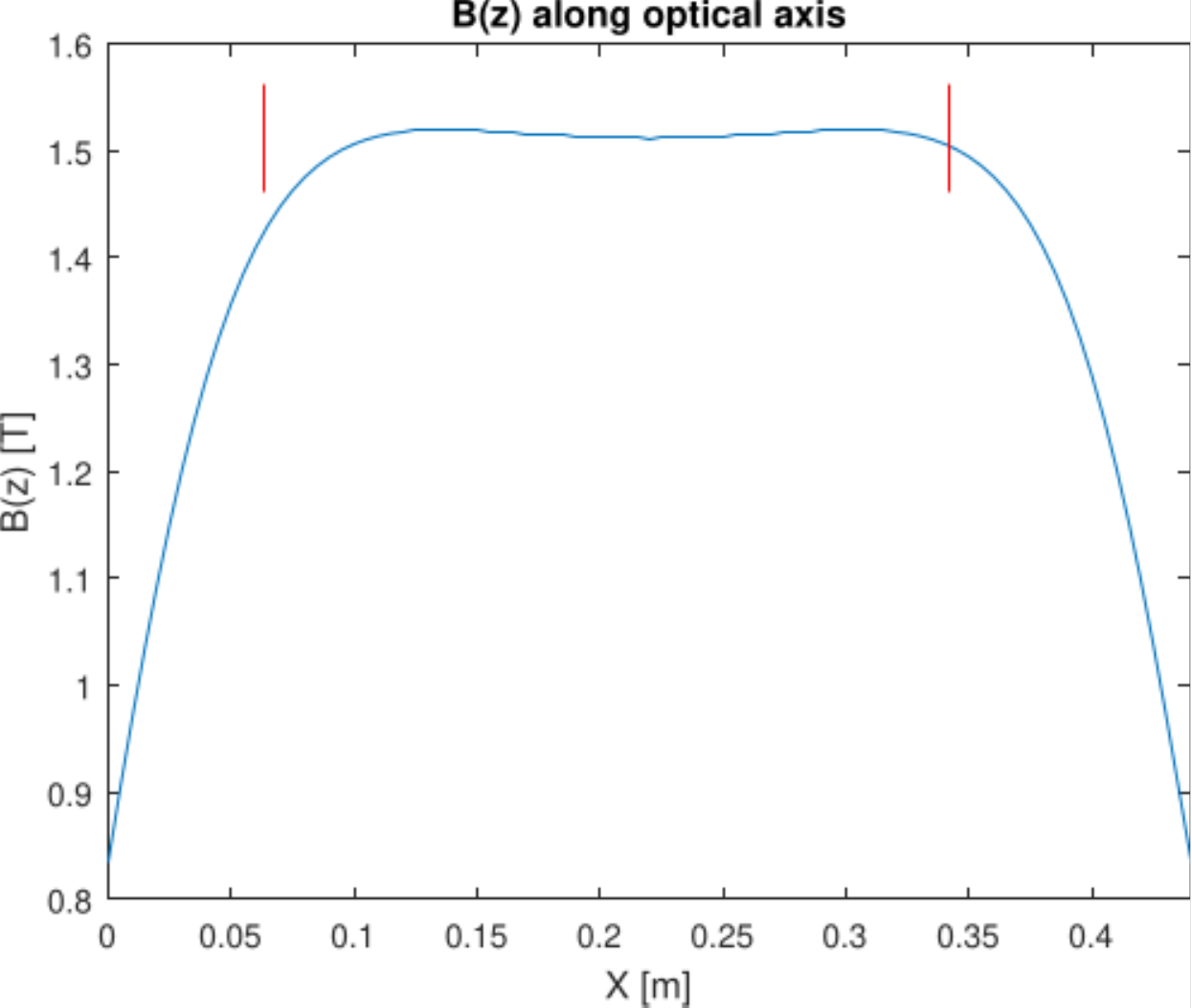}}\hfil
  \subfloat[XY Plane]{\includegraphics[height=0.25\textheight]{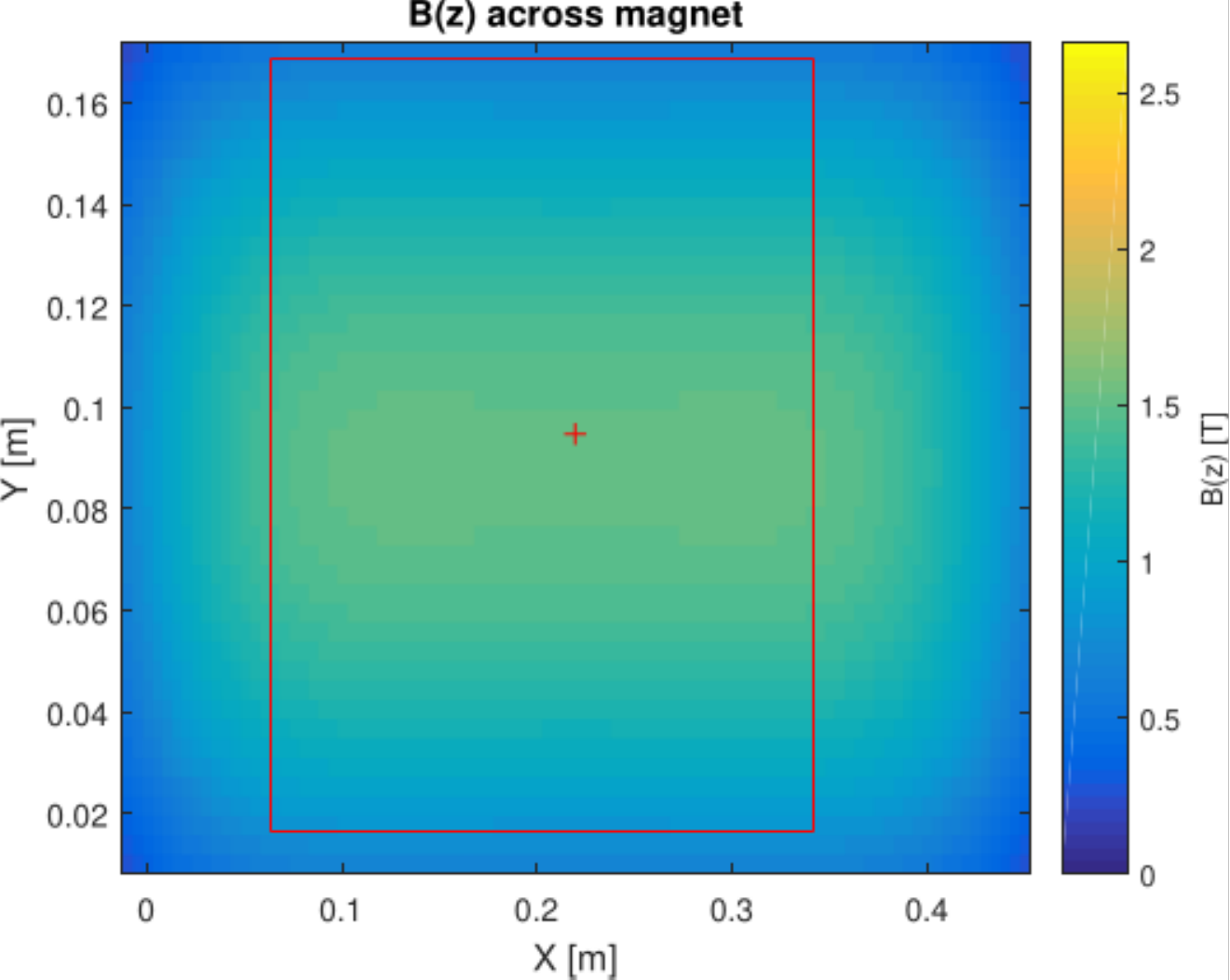}}\\
  \subfloat[XZ Plane]{\includegraphics[height=0.25\textheight]{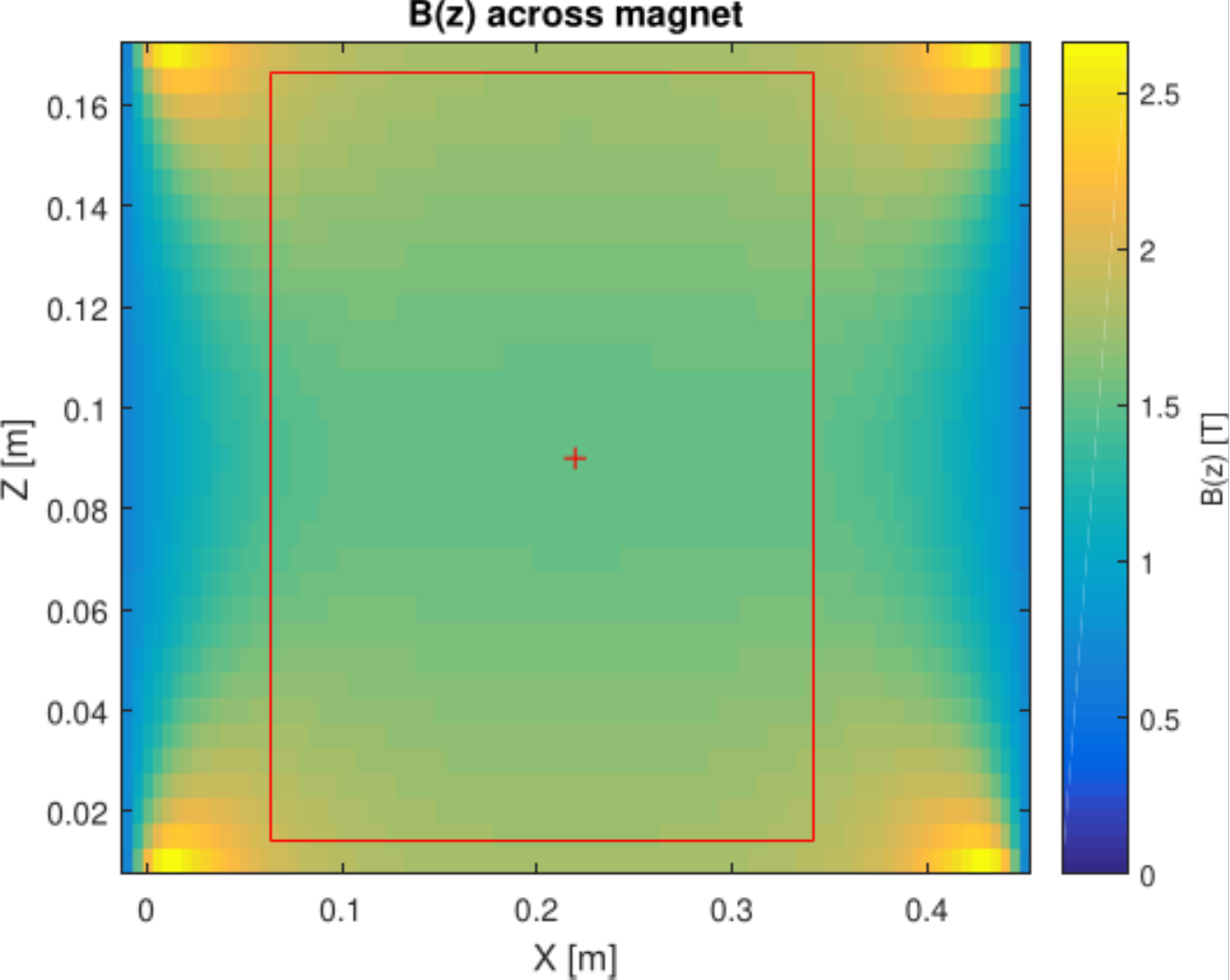}}\hfil
  \subfloat[YZ Plane]{\includegraphics[height=0.25\textheight]{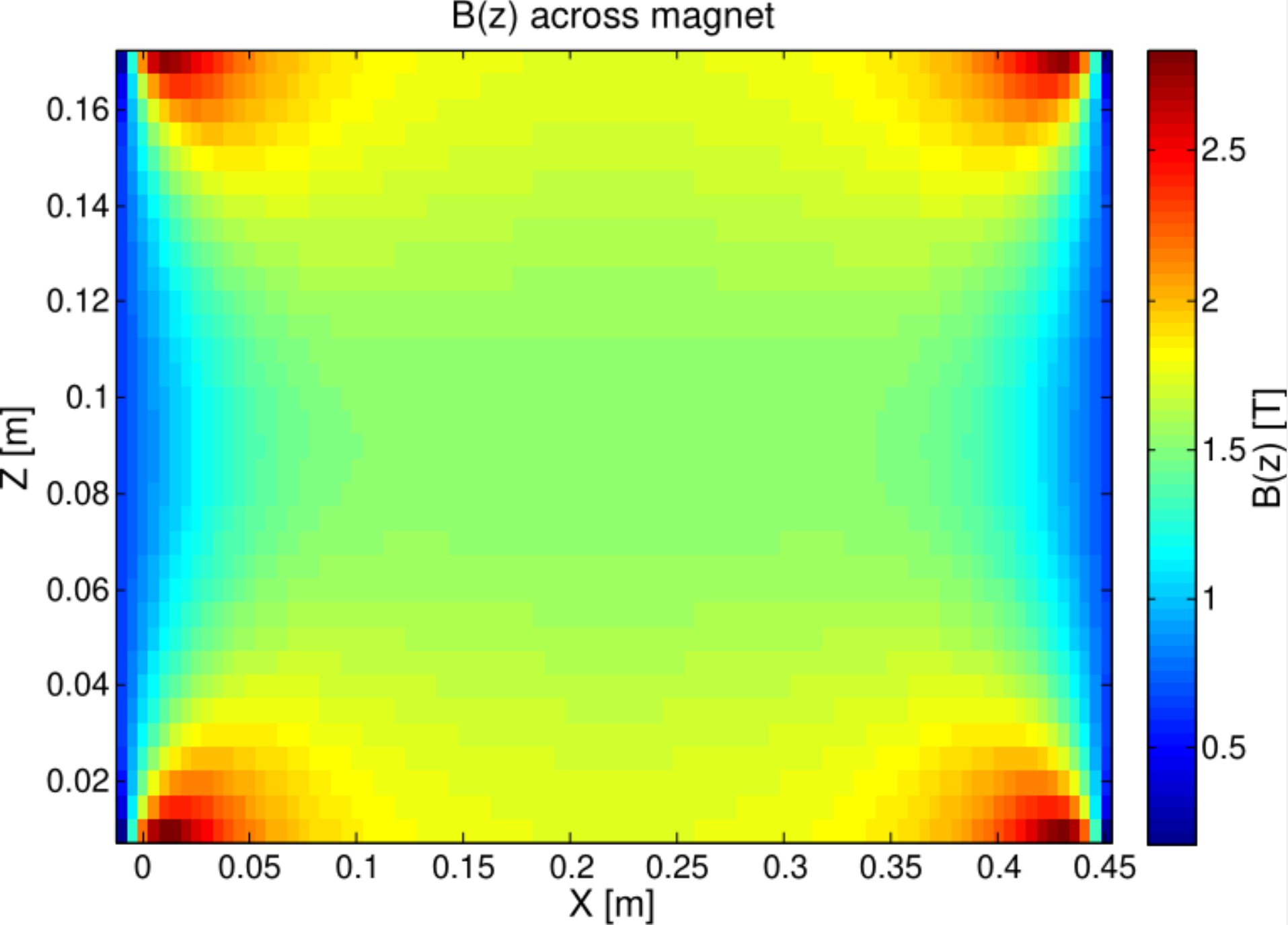}}
  \caption{Calculated magnetic field of the planned Orpheus magnet. For this calculation, Z is the axis of the magnet, X is along the long side of the dipole magnet, and Y is along the short side of the dipole magnet.}
  \label{fig:magnet_field}
\end{figure}

Seth Kimes and Richard Ottens designed the magnet. The planned dipole magnet has dimensions $44\times18.5\times18$ \SI{}{cm^3}. The NbTi superconducting is from Supercon, Inc. The model is 56S53, the diameter is \SI{0.30}{mm} bare and \SI{0.33}{mm} with the insulation. The critical current is \SI{125}{A} at \SI{3}{T}. Each dipole side will have 57 layers of winding, and each layer will have 57 turns. The dipole magnet will run \SI{103}{A} to produce a \SI{1.5}{T} field inside the detector volume. The magnetic field profile is shown in Figure~\ref{fig:magnet_field}. The force between the wires of the dipole magnet, depending on the direction, is $\mathcal{O}(\SI{2000}{lbs.})$ of force. Seth Kimes designed the vacuum vessel to withstand that force with a certain safety factor.

\begin{figure}
  \centering
  \subfloat[]{\includegraphics[height=0.2\textheight]{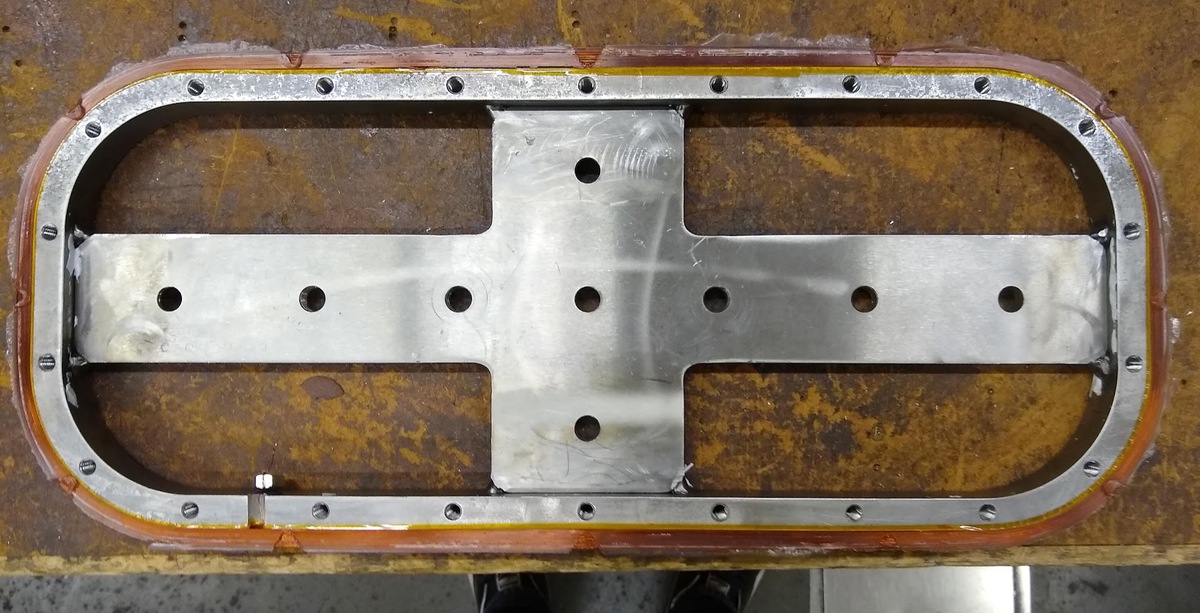}}\hfil
  \subfloat[]{\includegraphics[height=0.2\textheight]{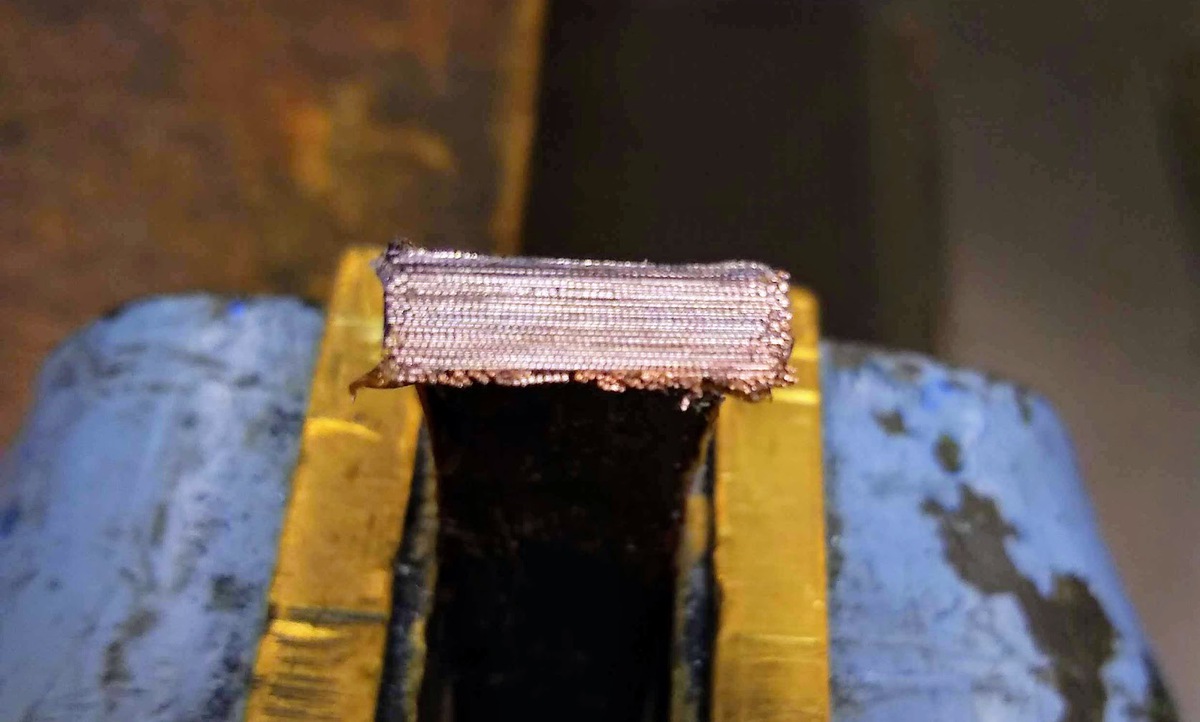}}
  \caption{Practicing Orpheus magnet widning with copper wire.}
  \label{fig:magnet_winding}
\end{figure}

We are currently in the process of winding the superconducting dipole magnet. James Sinnis spent the Summer of 2021 practicing the winding and potting technique, and the results in Figure~\ref{fig:magnet_winding} look satisfactory. We are now waiting for cryogenic epoxy to arrive, but the supply chains are currently in chaos.
